\documentclass[]{aa} 
 
\usepackage{graphicx, times, float, rotating, color, lscape, fancyvrb} 
\usepackage[figuresright]{rotating}
\usepackage[varg]{txfonts}
\usepackage{url}
\usepackage{bm}
\usepackage{placeins}

\newcommand{\less}{\raisebox{-1.1mm}{$\stackrel{<}{\sim}$}} 
\newcommand{\more}{\raisebox{-1.1mm}{$\stackrel{>}{\sim}$}} 
\newcommand{\msol}{\mbox{M$_{\odot}$}} 
 
\newcommand{\msolyr}{{M$_{\odot}$}\,yr$^{-1}$} 
\newcommand{\mdot}{$\dot{M}$}
\newcommand{\lsol}{\mbox{L$_{\odot}$}}

\newcommand{\ks}{km s$^{-1}$} 
 
\newcommand{\mum}{$\mu$m}

\begin{document}

\title{
A WISE view on extreme AGB stars 
\thanks{
Tables~\ref{Tab-Cres},  \ref{WISEREALSam}--\ref{WISEPernotREAL}, and \ref{Tab-Ores} are available at the CDS via anonymous ftp to  \protect\url{cdsarc.u-strasbg.fr} (130.79.128.5)
or via \protect\url{http://cdsarc.u-strasbg.fr/viz-bin/cat/J/A+A/vol/page}.
Figures~\ref{Fig:WISE}--\ref{Fig:ZTF}, and \ref{Fig-SED-Cstars} are available at \protect\url{https://doi.org/10.5281/zenodo.5825878}.
}
%\fnmsep
%\thanks{
%Based on observations 
%} 
}  
 
\author{ 
M.~A.~T.~Groenewegen 
}

\institute{ 
Koninklijke Sterrenwacht van Belgi\"e, Ringlaan 3, B--1180 Brussels, Belgium \\ \email{martin.groenewegen@oma.be}
} 
 
\date{received: 2021, accepted: 2021} 
 
\offprints{Martin Groenewegen}

\abstract {
Variability is a key property of stars on the asymptotic giant branch (AGB). Their pulsation period is related to the luminosity
and mass-loss rate (MLR) of the star. Long-period variables (LPVs) and Mira variables are the most prominent of all types of variability of
evolved stars. However, the reddest, most obscured AGB stars are too faint in the optical and have eluded large variability surveys.
}
{Our goal is to obtain a sample of LPVs with large MLRs by analysing WISE W1 and W2 light curves (LCs) for  about 2000 sources, photometrically selected
 to include
 known C-stars with the 11.3~$\mu$m silicon carbide dust feature in absorption, and Galactic O-stars with periods longer than 1000~days.}
{Epoch photometry was retrieved from the AllWISE and NEOWISE database  and fitted with a sinus curve.
 Photometry from other variability surveys was also downloaded and fitted. 
 For a subset of 316 of the reddest stars, spectral energy distributions (SEDs) were constructed, and, together with mid-infrared (MIR) spectra
 when available, fitted with a dust radiative transfer programme in order to derive MLRs.
}
{WISE based LCs and fits to the data are presented for all stars. Periods from the literature and periods from refitting other literature data are presented.
 The results of the spatial correlation with several (IR) databases is presented.
About one-third of the sources are found to be not real, but it appears that these cannot be easily filtered out by using WISE flags.
Some are clones of extremely bright sources, and in some cases the LCs show the known pulsation period.
Inspired by a recent paper, a number of non-variable OH/IRs are identified.
Based on a selection on amplitude, a sample of about 750 (candidate) LPVs is selected of which 145 have periods $>$1000~days, many of them being new.
For the subset of the stars with the colours of C-rich extremely red objects (EROs) the fitting of the SEDs (and available MIR spectra) separates
them into C- and O-rich objects.
Interestingly, the fitting of MIR spectra of mass-losing C-stars is shown to be a powerful tracer of interstellar reddening when $A_{\rm V}$ \more 2~mag.
The number of Galactic EROs appears to be complete up to about 5~kpc and a total dust return rate in the solar neighbourhood for this class is determined.
In the LMC 12 additional EROs are identified. Although this represents only about 0.15\% of the total known LMC C-star population adding their
MLRs increases the previously estimated dust return by 8\%.
Based on the EROs in the  Magellanic Clouds, a bolometric period luminosity is derived.
It is pointed out that due to their faintness, EROs and similar O-rich objects are ideal targets for a NIR version of {\it Gaia} to obtain distances,
observing in the $K$-band or, even more efficiently, in the $L$-band.
}
{}

\keywords{stars: variables: general -- infrared: stars  -- 
          stars: AGB and post-AGB   -- Stars: mass-loss -- Magellanic Clouds} 

\maketitle

\section{Introduction} 

At the end of their lives almost all low- and intermediate-mass stars (with initial masses from
$\sim 0.9$ to $\sim 10$~\msol) will go through the (super)-asymptotic giant branch ((S-)AGB) phase.
They end up as $\sim 0.55-1.4$~\msol\ white dwarfs which implies that a large fraction of the initial mass of a star
is returned to the interstellar medium (ISM).
Pulsation is an important characteristic of AGB stars, and they are typically divided into stars with small
amplitudes (the semi-regular variables, SRVs) and the large-amplitude Mira variables. 
The term long-period variable (LPV) is now commonly used  for a pulsating AGB star, regardless of pulsation amplitude. 
The most promising mechanisms to explain wind driving are pulsation-induced shock waves and radiation pressure on dust,
especially regarding the more evolved AGB stars with low effective temperatures, large pulsation amplitudes,
and high mass-loss rates (MLRs; see the review by \citealt{HO18}).

Analysis of the MLRs of essentially complete samples of AGB stars in the Magellanic Clouds (MCs) has shown that the
gas and dust return to the ISM is dominated by a small percentage of stars with the highest MLRs
(e.g. \citealt{Matsuura09,Boyer12,Nanni19}, and references therein).
These stars are often characterised by the longest pulsation periods.

Current surveys in the optical domain (including OGLE and {\it Gaia}) will, however, miss the reddest, most
obscured AGB stars. At the very end of the AGB the (dust) MLR may become so high that the object becomes very faint,
beyond the OGLE $I$-band detection limit of about 21~mag or the {\it Gaia} $G$-band limit of about 21.5~mag.
The dust grains in the circumstellar envelope (CSE) scatter and
absorb the emission in the optical to re-emit it in the NIR and mid-infrared (MIR), where they become bright sources.     
These stars are known to exist in the MCs. They were initially selected and identified as having
{\it Infrared Astronomical Satellite} (IRAS) colours similar to obscured AGB stars in our Galaxy, and later on
photometric and spectroscopic observations with the {\it Spitzer Space Telescope} (SST, \citealt{Werner04}) confirmed this and added
additional examples of this class of extreme mass-losing objects, mostly being carbon-rich AGB stars \citep{Gruendl08,Sloan16,GS18}.

In an earlier related work, \cite{Gr20} presented a sample of 217 likely LPVs in the MCs.
This paper investigated the variability of 1299 objects in the $K$-band, based on VISTA Magellanic Cloud (VMC) survey
data \citep{Cioni11}, supplemented with literature data. The aim of that paper was also to find red AGB stars with long periods,
although potentially not as red as the sources studied here as the very reddest sources will also be faint or invisible even
in the $K$-band. 
Although the VMC data are of high quality the sampling is not optimal for detecting LPVs (typically 15 data points
spread over 6 months ordinarily). Although $K$-band data from the literature was added
(e.g. 2MASS \citep{Cutri_2MASS}, 2MASS 6X \citep{Cutri_2MASS6X}, IRSF \citep{Kato_IRSF}, DENIS \citep{DENIS05},
as well as the pioneering monitoring works of \citealt{Wood1992}, \citealt{Wood1998}, and \citealt{Whitelock2003}), in some
cases, no unique period could be derived and several periods could fit the $K$-band data.

The
Wide-field Infrared Survey Explorer (WISE) \citep{Wright10}
and the
Near-Earth Object WISE (NEOWISE) and NEOWISE Reactivation mission \citep{Mainzer11,Mainzer14}
are ideal surveys to study LPVs. The total time span covered is about nine years which covers two or more
pulsation cycles even for extremely long periods.
Other advantages are that they survey at wavelengths where the reddest objects are the brightest, and 
they survey the entire sky.

Previous studies already explored the time variability offered by the WISE mission.
\citet{Chen18} presented a catalogue of $\sim$ 50~000 periodic variables with periods shorter than 10 days,
\citet{Petrosky20} presented a similar catalogue of $\sim$ 63~500 periodic variables with periods shorter than 10 days (using different criteria), 
while
\citet{Uchiyama19} studied the MIR variability in massive young stellar objects (YSOs).

The outline of the paper is as follows.
Section~2 introduces the sample of known very red C- and O-rich AGB stars that will serve as templates to select candidates based on
photometric selection criteria using AllWISE data.
Section~3 describes the selection of the time series data both from the WISE mission and other literature data,
and the analysis and fitting of the time series data.
Section~4 outlines the results of an extensive literature study into the classification of the objects.
Section~5 briefly describes  the various tables that contain the results of the literature search and the period analysis.
Section~6 discusses these results by addressing various topics in more detail, including the discovery of new LPVs with periods
over 1000~days and new AGB stars with extremely large MLRs.

\begin{table*}
\setlength{\tabcolsep}{1.2mm}

  \caption{Template sample of extreme AGB stars}
%\footnotesize
  \begin{tabular}{lrrrrrrrrrrrrrr}
  \hline
  Name         & W1 & error & S/N &   W2 & error & S/N &    W3 & error & S/N &   W4 & error & S/N  \\
                & (mag) & (mag) &  & (mag) & (mag) & & (mag) & (mag) &  & (mag) & (mag)  \\
  \hline
\multicolumn{13}{c}{known EROs in the Galaxy} \\

AFGL 190        &  7.445 & 0.025 & 42.7 &  3.264 & 0.255 &  4.3 & -1.449 & 0.346 & 3.1 & -3.137 & 0.002 & 575. & \\ % 19.4639421 +67.2314488
AFGL 3068       &  4.689 & 0.288 &  3.8 & -0.085 & -     &  0.9 & -3.063 & -     & 0.6 & -3.975 & 0.002 & 625. & \\
AFGL 3116       & -0.480 & -     &  0.4 &  1.721 & -     & -9.3 & -2.966 & -     & 1.3 & -3.493 & 0.002 & 703. & \\ 
IRAS 08171$-$2134 &  7.340 & 0.053 & 20.5 &  3.795 & 0.313 &  3.5 & -0.755 & 0.395 & 2.8 & -2.766 & 0.001 & 816. & \\
IRAS 19075$+$0921 &  6.802 & 0.064 & 16.9 &  3.044 & 0.416 &  2.6 & -1.165 & 0.354 & 3.1 & -3.105 & 0.002 & 462. & \\
IRAS 15471$-$5644 &  6.063 & 0.097 & 11.2 &  1.907 & -     &  1.9 & -1.333 & 0.386 & 2.8 & -3.191 & 0.001 & 865. & \\
IRAS 21318$+$5631 &  6.276 & 0.037 & 29.3 &  1.707 & -     &  1.8 & -1.859 & 0.349 & 3.1 & -3.263 & 0.002 & 580. & \\

\multicolumn{13}{c}{Known EROs in the LMC} \\

ERO 0502315      &   17.233 & 0.067 & 16.2 & 12.693 & 0.021 & 51.7 & 5.553 & 0.012 & 73.7 & 3.131 & 0.015 & 72.7 & \\
ERO 0504056      &   18.692 & -     &  1.9 & 13.058 & 0.023 & 47.0 & 5.919 & 0.014 & 77.5 & 3.653 & 0.014 & 77.9 & \\
ERO 0518117      &   14.752 & 0.026 & 42.6 & 11.481 & 0.020 & 53.9 & 5.366 & 0.014 & 76.2 & 2.887 & 0.011 & 98.8 & \\
ERO 0518484      &   16.293 & 0.476 &  2.3 & 12.770 & 0.036 & 30.2 & 5.597 & 0.014 & 79.2 & 3.406 & 0.020 & 55.0 & \\
ERO 0525406      &   16.831 & -     & -0.5 & 13.427 & 0.040 & 27.1 & 6.136 & 0.010 & 105. & 3.855 & 0.020 & 54.5 & \\      
ERO 0529379      &   13.672 & 0.023 & 47.0 & 10.259 & 0.020 & 54.4 & 5.491 & 0.014 & 76.0 & 3.649 & 0.017 & 64.2 & \\ % 82.407959 -72.831322
ERO 0550261      &   14.939 & 0.031 & 35.3 & 10.787 & 0.020 & 54.9 & 4.848 & 0.015 & 74.6 & 2.793 & 0.012 & 89.6 & \\
IRAS 05133$-$6937 &   19.082 & -     & -48. & 14.289 & 0.042 & 26.0 & 5.955 & 0.018 & 59.7 & 3.444 & 0.023 & 47.7 & \\        
IRAS 05315$-$7145 &   13.910 & 0.026 & 41.2 & 11.992 & 0.022 & 49.9 & 5.779 & 0.014 & 76.2 & 2.939 & 0.013 & 82.5 & \\ % 82.684006 -71.716766
IRAS 05495$-$7034 &   15.733 & 0.030 & 36.1 & 13.558 & 0.024 & 45.7 & 5.864 & 0.012 & 87.7 & 2.608 & 0.009 & 116. & \\             
IRAS 05568$-$6753 &   11.113 & 0.023 & 47.6 &  8.238 & 0.020 & 53.1 & 4.316 & 0.014 & 77.0 & 2.787 & 0.011 & 95.5 & \\ % 89.161500 -67.892889 

\multicolumn{13}{c}{Known Galactic O-stars with $P>$1000~days \citep{Menzies19} } \\

V1360 Aql, OH 30.7$+$0.4 &  5.360 &  0.075 &    14.5 &  1.309 &   -     &     1.4 &  0.237 &  0.423 &   2.6 & -2.138 &  0.009 &  119.6 & \\ % 281.4685864 -01.7787028 
V1362 Aql, OH 30.1$-$0.7 &  6.935 &  0.143 &     7.6 &  1.625 &   -     &     1.6 & -0.659 &  0.477 &   2.3 & -3.453 &  0.006 &  188.8 & \\ % 282.1747832 -02.8413567 
V1363 Aql, OH 32.0$-$0.5 &  7.609 &  0.027 &    40.1 &  3.626 &  0.173 &     6.3 &  0.282 &  0.083 &  13.1 & -1.654 &  0.003 &  325.9 & \\ % 282.8593295 -01.0645807 
V1365 Aql, OH 32.8$-$0.3 &  6.908 &  0.062 &    17.4 &  3.010 &  0.356 &     3.0 &  0.085 &  0.406 &   2.7 & -2.751 &  0.010 &  103.7 & \\ % 283.0927753 -00.2367139 
V1366 Aql, OH 39.7$+$0.5 &  1.522 &    -    &     1.2 & -1.304 &   -     &     0.8 & -1.603 &  0.368 &   3.0 & -3.138 &  0.003 &  345.8 & \\ % 284.6252968 +06.7159668 
V1368 Aql, OH 42.3$-$0.1 &  7.686 &  0.024 &    44.7 &  3.814 &  0.040 &    26.9 &  0.407 &    0.030   &  35.9 & -1.480 &  0.006 &   186.1 & \\ % 287.2846276 +08.2761164
V669 Cas, OH 127.8$-$0.0 &  3.829 &  0.334 &     3.3 & -0.393 &   -     &     0.8 & -1.031 &  0.311 &   3.5 & -2.948 &  0.003 &  421.2 & \\ % 023.4633422 +62.4481603 
OH104.9$+$2.4, AFGL 2885 &  2.695 &  0.016 &    68.2 &  1.647 &  0.028 &    39.3 & -1.963 &  0.235 &   4.6 & -4.075 &  0.001 & 1601.2 & \\ % 334.864610 +59.856052  WISE, not in ALWISE
IRAS 03293$+$6010, OH 141.7$+$3.5 &  4.602 &  0.215 & 5.0 &  1.675 & -     &     2.0 & -0.209 &  0.380 &   2.9 & -2.347 &  0.001 &  768.4 & \\ % 053.3774898 +60.3359569 
IRAS 05131$+$4530, AFGL 712     &  3.291 &  0.527 & 2.1 &  0.190 &  -    &     0.8 & -0.214 &  0.359 &   3.0 & -2.307 &  0.002 &  674.7 & \\ % 079.1977472 +45.5678613 
IRAS 07222$-$2005       &  4.518 &  0.257 &     4.2 &  3.240 &  0.236 &     4.6 &  1.656 &  0.018 &  59.0 &  0.435 &  0.007 &  150.3 & \\ % 111.1017382 -20.1987916 
V1185 Sco, OH 357$-$1.3 AFGL 5379 &  4.524 &  0.275 &   4.0 & -1.250 & -  &     0.5 & -3.008 &     -   &   1.1 & -3.513 &  0.011 &  103.1 & \\ % 266.0996994 -31.9276239 
V437 Sct, OH 26.5$+$0.6   & -0.614 &   -     &     0.3 & -2.449 &  -      &     0.5 & -2.837 &   -     &   1.4 & -3.879 &  0.016 &   69.2 & \\ % 279.3854407 -05.3997526  
V438 Sct, OH 26.2$-$0.6   &  4.213 &  0.273 &     4.0 &  0.749 &   -     &     1.2 & -0.448 &  0.345 &   3.1 & -2.539 &  0.003 &  320.2 & \\ % 280.3097807  -06.2501488 
V441 Sct, OH 21.5$+$0.5   &  9.035 &  0.025 &    44.3 &  4.047 & 0.051  &    21.2 &  0.130 &  0.022 &  49.5 & -1.761 &  0.010 &  109.1 & \\ % 277.128905 -09.970644 WISE, not in ALWISE
IRAS 03206$+$6521, OH 138.0$+$7.2 &  4.186 &  0.267 & 4.1 &  0.443 &  -    &     0.9 & -0.341 &  0.334 &   3.3 & -2.540 &  0.001 &  815.0 & \\ % 051.2851985 +65.5353592 
%HV 888  x            &  6.330 &  0.084 &    12.9 &  6.143 &  0.033 &    33.0 &  4.395 &  0.015 &  72.9 &  3.364 &  0.018 &   60.8 & \\ % 076.0588769 -67.2706758 
%MSX LMC 807          &  8.217 &  0.023 &    48.1 &  7.085 &  0.020 &    54.2 &  4.803 &  0.015 &  72.4 &  3.069 &  0.020 &   53.8 & \\ % 083.1548719 -67.1156680 

\hline
\end{tabular}
\label{Tab-Known}
\end{table*}

\section{The template sample and source selection} 

This section describes the template sources of very long-period variables and very red sources that were used to
create a WISE colour-selected sample of candidate very evolved AGB stars.

\subsection{Carbon-rich AGB stars} 

The term extreme AGB star is not well defined. It was probably first used by \citet{Volk92} in connection
with carbon-rich AGB stars (hereafter C-stars).
Their investigation was spurred by the fact that previous surveys in the infrared, such as the 
Two-micron sky survey \citep{IRC69} and the Air Force Geophysics Lab (AFGL, \citealt{AFGL76}) survey discovered C-stars with unusually 
thick dust shells, such as IRC +10~216 (CW Leo) or AFGL 3068.
They selected a group of 31 stars based on certain spectral characteristics observed in
8-23~$\mu$m Low Resolution Spectrograph (LRS) data taken during the
%{\it Infrared Astronomical Satellite} (IRAS)
IRAS mission.
Independently, \citet{Groenewegen92} listed eight sources (out of 109, their  ‘group V’ class) in their flux-limited (IRAS $S_{12} > 100$~Jy) 
sample of C-stars with very similar properties to those in  \citet{Volk92} based on the IRAS colour-colour diagram and LRS types.
Later, \citet{Speck09} studied ten of these sources (one new) using superior {\it Infrared Space Observatory} (ISO)
Short Wavelength Spectrometer (SWS) data. Many of these sources displayed the silicon carbide (SiC) 11.3 $\mu$m dust feature in
absorption, indicating a very large optical depth as the feature is normally seen in emission in C-stars.
The sample of seven known C-stars with SiC in absorption is listed in Table~\ref{Tab-Known}, together with
the WISE magnitude, the error in the magnitude, and the signal-to-noise 
(some being negative) in the four bands of WISE (W1 at 3.4 $\mu$m, W2 at 4.6 $\mu$m, W3 at 12 $\mu$m, and W4 at 22 $\mu$m).

Then, \citet{Gruendl08} discussed a dozen sources in the direction of the Large Magellanic Cloud (LMC)
characterised by extremely red MIR colours ([4.5]-[8.0]~$>$~4.0) based on SST colours
and spectral energy distributions (SEDs), peaking between 8 and 24~$\mu$m.
% [8.0]−[24]= 3–5
Seven of those show a flat red continuum or SiC in absorption based on Infrared Spectrograph (IRS; Houck et al.\ 2004) data.
They introduced the term extremely red objects (EROs). They did not discuss any link with the known similar objects in the Milky Way.

Table~\ref{Tab-Known} lists the properties of those seven sources together with four other LMC sources with SiC in absorption based
on other IRS programmes, see \citet{Sloan16} and \citet{GS18}.
Interestingly, no known EROs exist in the SMC.
\citet{Ventura16} explained the fact that the reddest C-stars in the LMC are redder than the reddest C-stars in the Small Magellanic
Cloud (SMC), which is related to a difference in initial mass (2.5--3~\msol, respectively, $\sim$1.5~\msol), consistent with the difference
in star formation histories between the two galaxies.

To complete the description of the terminology, the term extreme AGB stars (often designated x-AGB stars or X-stars)
is also used in the literature based on photometric criteria (and it can refer to C-stars or oxygen-rich AGB stars
(hereafter O-stars)), for example \citet{Blum06} who used a limit of $J$--[3.6] $>3.1$. As x-AGB stars can be invisible in the $J$-band,
other criteria have been adopted, for example [3.6]--[8.0] $>$ 0.8 \citep{Boyer11} or [3.6]--[4.5] $>$ 0.1 \citep{Boyer15}.
As discussed in \citet{Sloan16}, this terminology is something of a misnomer as sources with such colours produce an appreciable
amount of dust, but this is a common phenomenon as stars evolve on the AGB, and they are not `extreme' in that sense.
The C-stars with a red flat continua or SiC in absorption are a subset of x-AGB stars, and they represent the
reddest colours, for example [3.6]--[4.5] $\more$ 1.5 \citep{Sloan16}.

\subsection{Oxygen-rich AGB stars}
\label{S-Orich}

As a class, the OH/IR stars (see \citealt{Hyland74} for an early review) come closest to being called extreme O-stars, as they can be
very red ($K-L > 7$, \citealt{Jones82}) and are recognised as the O-stars with the largest MLRs \citep{Herman85}.
In C-stars SiC is a minor dust species compared to amorphous carbon and so for the 11.3~\mum\ to go into absorption very high dust densities
are required.
On the other hand, silicates are the dominant species in the CSEs around O-stars and therefore stars
with the  9.8~\mum\ silicate feature in absorption are not uncommon (although in most O-rich stars it is seen in emission).
At even larger densities the silicate 18~\mum\ feature also goes into absorption, and these sources have been called
extreme OH/IR stars \citep{Justtanont15}, and OH 26.5+0.6 is a prime example \citep{Etoka07}.

Early on it was also recognised that (extreme) OH/IR stars are associated with Mira-like large-amplitude variability with
(very) long periods (\citealt{Engels83}; e.g. OH~26.5+0.6 has a period of 1559~days, \citealt{Suh02}).
The template sample for extreme O-stars that is used to define selection criteria in WISE colours is the compilation
of known Galactic O-stars with periods over 1000~days from \citet{Menzies19}, and the WISE properties of these stars
are listed in Table~\ref{Tab-Known}.
\citet{Menzies19} also lists SMC and LMC variables with periods over 1000~days.
These have not been used to define WISE colour-based criteria, but are all included in the final sample (except for two supergiants).

\subsection{WISE colour selection} 

\VerbatimFootnotes

The AllWISE source catalogue \citep{Cutri_Allwise} contains over 747 million sources.
Based on the colours and  signal-to-noise ratios (SNs) in Table~\ref{Tab-Known}, a
query\footnote{The SQL query was: 
\begin{verbatim}
WHERE 
((( (w2mpro-w3mpro) >2.2 and (w3mpro-w4mpro) >1.4 
     and w2snr >20. and w3snr >50. and w4snr >40.)
  or
  ( (w1mpro-w2mpro) >2.8 and (w2mpro-w3mpro) >2.2 
     and (w3mpro-w4mpro) >1.4 and w1snr >2.0 
     and w2snr >0.75 and w3snr >0.5 and w4snr >100.)
  or
  ( (w1mpro-w4mpro) >3.0 and w4snr >60.)))
\end{verbatim}
}
was run on the
AllWISE source catalogue as available through the
IPAC Infrared Science Archive (IRSA)\footnote{\url{https://irsa.ipac.caltech.edu/Missions/wise.html}}
to select a sub-sample of about 60~000 sources, containing all of the 34 sources in Table~\ref{Tab-Known}).

\begin{figure}

\begin{minipage}{0.485\textwidth}
\resizebox{\hsize}{!}{\includegraphics{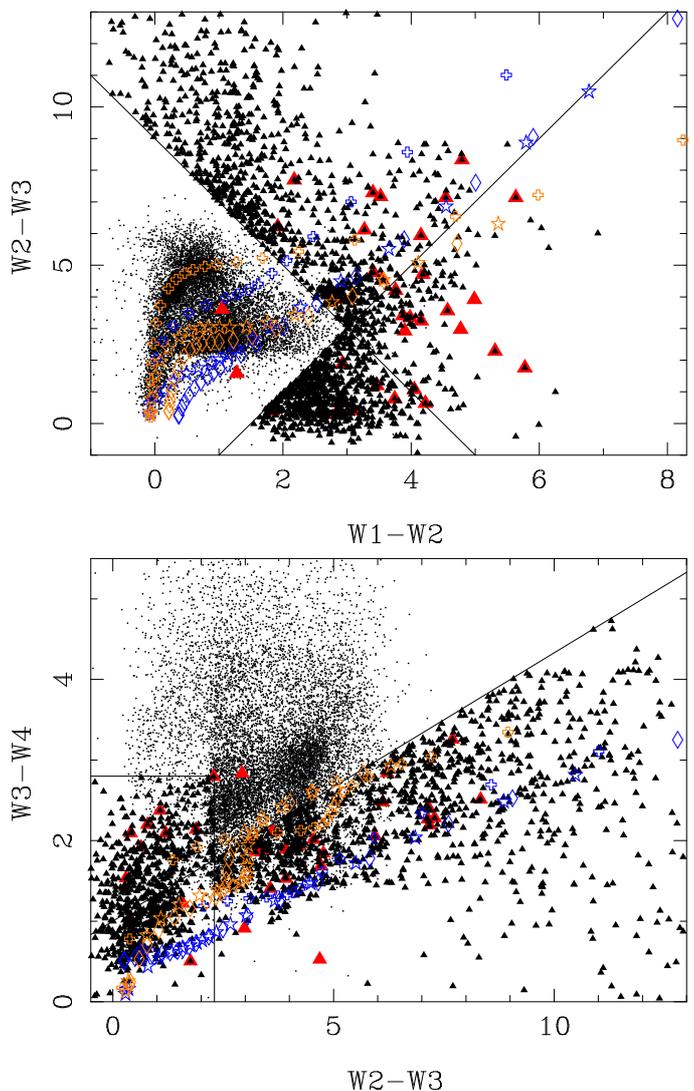}}
\end{minipage}

\caption[]{ 
[W2--W3]$-$[W1--W2] and [W3--W4]$-$[W2--W3] CCDs.
Red triangles are the sources from the template sample in Tab.~\ref{Tab-Known}.
Small black triangles are selected sources (see main text).
Small dots are non-selected sources; to avoid cluttering stricter SN criteria were applied for them to be plotted (see main text).
Other symbols indicate sequences of increasing MLRs (see main text).
Blue colours represent C-star models while orange colours represent O-star models for
($T_{\rm eff}, T_{\rm inn}$) combinations of (2600, 1000), (3300, 800), and (4000, 400~K)
in diamonds, five-pointed stars, and open plus signs, respectively.
} 
\label{Fig-CCD} 
\end{figure}

Figure~\ref{Fig-CCD} shows the colour-colour diagrams (CCDs) of that sample in WISE colours.
The sources from Table~\ref{Tab-Known} are plotted as red triangles.
To help identify the location of $\mbox{(post-)}$ AGB (P-AGB) stars in these CCDs dust radiative transfer calculations were performed 
with the code More of DUSTY (MoD, \citealt{Gr_MOD}), which is an extension of the radiative transfer code DUSTY \citep{Ivezic_D}.
This was done by using combinations of the effective temperature and temperature at the inner dust radius of
($T_{\rm eff}, T_{\rm inn}$) (2600, 1000), (3300, 800), and (4000, 400~K), representative of late-AGB and early P-AGB evolution.
This was done for C-stars, with model atmospheres from \citet{Aringer09} and a dust mixture of SiC and amorphous carbon,
and O-stars, with MARCS model atmospheres \citep{Gustafsson_MARCS} and a dust mixture of silicate and metallic iron,
for 20 optical depths at 0.55~$\mu$m ranging from 0.001 to 1000. The WISE magnitudes were calculated
from the SEDs and the resulting colours were plotted using different colours and symbols (see the figure caption).
As expected the sequences start at blue colours and then become increasingly red as the optical depth increases.

Based on the location of the known sources and the sequence of theoretical colours, the following further selection was applied:

\begin{itemize}

  \item
    (W2 - W3) $> -2.0$  $\cdot$ (W1 - W2) $+9.0$
  \item[]
   or
  \item
    (W2 - W3) $< +2.0$  $\cdot$ (W1 - W2) $-3.0$ 
\end{itemize}
and
\begin{itemize}

  \item
    (W3 - W4) $< 0.333$ $\cdot$ (W2 - W3) $+1.0$
  \item[]
   or
  \item
    (W2 - W3) $< 2.3$ and (W3 - W4) $< 2.8$.
\end{itemize}

\begin{figure}

\begin{minipage}{0.485\textwidth}
\resizebox{\hsize}{!}{\includegraphics{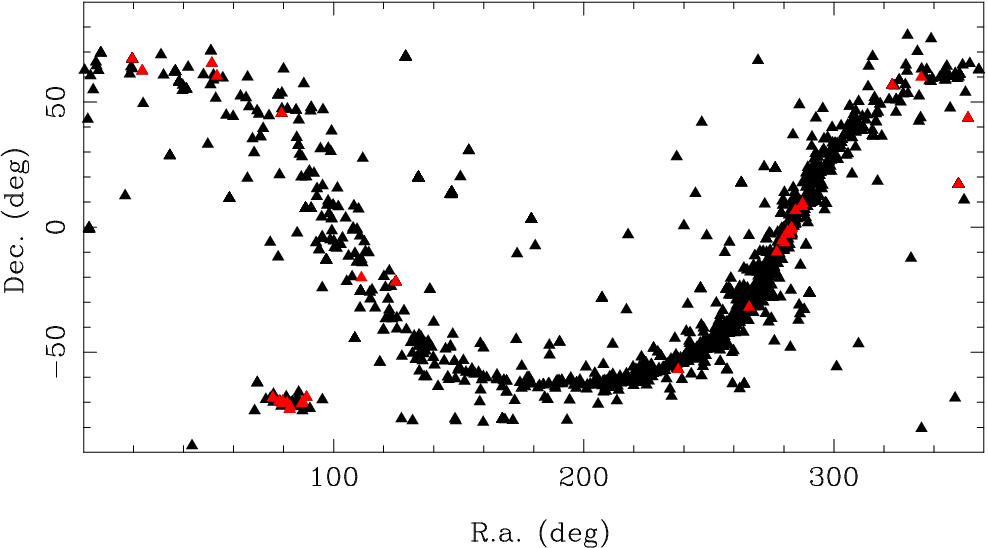}}
\end{minipage}

\caption[]{ 
Distribution on the sky of the sample. Symbols are as in Fig.~\ref{Fig-CCD}
} 
\label{Fig-Sky} 
\end{figure}

The sources fulfilling these conditions were plotted as small triangles in Fig.~\ref{Fig-CCD}.
To avoid cluttering in the plot, the non-selected sources (small dots) were only plotted %in Fig.~\ref{Fig-CCD}
when additional criteria were fulfilled (SN $> 45$ in all four WISE filters).
As one can notice, some known sources (big red triangles) were not selected by these conditions (they are not over plotted by a
small black triangle).
In most cases, this is due to the extreme brightness of these sources (e.g. OH~21.5 and AFGL~3068), corrupting their colours.
These sources were added to the sample manually.
As discussed below, the (NEO)WISE epoch databases often do not contain useful data for these types of very bright sources, but they often have
parasitic sources for which a period can be derived.
Figure~\ref{Fig-Sky} shows the distribution on the sky with the sources in the direction of the MCs and in the Galactic plane
showing up prominently.

To this pure WISE colour-selected all-sky sample the sample of 217 likely LPVs in the MCs from \cite{Gr20} was added.
As mentioned in the introduction, this sample is based on the analysis of the $K$-band from the VMC survey \citep{Cioni11},
supplemented with literature data. In some cases no unique period could be derived and several periods could fit the $K$-band data.
The WISE time series data will allow one to independently determine these periods.

The total sample for which the WISE time series will be studied is 1992 objects.
It is stressed that the sample (in particular the sample of about 1750 Galactic objects) should not be considered
as a complete sample. The selection on the colour and SN will introduce biases.

\section{Time series data and analysis}
\label{timeseriesdata}

\subsection{WISE}

From the AllWISE multi-epoch photometry table and the NEOWISE-R single exposure source table all entries within 1\arcsec\ of the
AllWISE coordinates were downloaded in the W1 and W2 filters with the additional constraint that the error bars on the magnitudes
were $<0.25$~mag and applying the flags {\it saa\_sep} $>5$ and {\it moon\_masked} $=0$ (e.g. \citealt{Uchiyama19}).
No other flags were applied (see Sect.~\ref{S-NR}).

At the bright end the WISE and NEOWISE data suffer from saturation that influences the photometry, and a correction was applied.
Table~2 in Sect.~II.1.c.iv.a of the NEOWISE Explanatory Supplement\footnote{\url{http://wise2.ipac.caltech.edu/docs/release/neowise/expsup/sec2_1civa.html}}
contains correction tables in the W1 and W2 filters for all phases of the mission.
The corrections are negligible to small at W1 and W2 $\sim 7-8$~mag and reach almost 1.5~mag in W2 in NEOWISE-R.
For stars brighter than the brightest entry in the tables the corresponding correction was kept without attempting any extrapolation.
The Explanatory Supplement furthermore states that even with a correction, no useful information is available for sources brighter
than W1 $\less$ 2 and W2 $\less$ 0~mag.

\subsection{SAGE-VAR}

For stars in the direction of the MCs the WISE W1 and W2 data were combined with IRAC 3.6~$\mu$m and
IRAC 4.5~$\mu$m data, respectively, from 
the SAGE-VAR survey \citep{Riebel15} that adds four epochs from the warm SST mission
at 3.6 and 4.5~$\mu$m for portions of the LMC and SMC.
The filters of the IRAC 3.6 and 4.5 bands are similar but not identical to the WISE W1 and W2 filters and the
transformation from \citet{Sloan16} was used to bring the IRAC photometry to the WISE system.

\subsection{Other time series data}

As will be discussed below in detail, the literature was searched for known periods of the stars in the sample.
However, next to quoting these periods, it turned out useful or even necessary to refit the original data in many cases for
several reasons.
In some cases the period in the literature seemed inconsistent with that expected for an LPV or inconsistent with that derived
from the WISE data.
In cases with multiple periods available from the literature these were sometimes inconsistent with each other.
Also in the case of no or insufficient WISE data it seemed valuable to provide a pulsation period based on other data to the community.
The major sources of additional time series photometry are described below.

The All-Sky Automated Survey for SuperNovae (ASAS-SN) \citep{Shappee14,Kochanek17,Jayasinghe18} identified 666~502 variables.
From the survey website\footnote{\url{https://asas-sn.osu.edu/variables}} the basic data of these variables was downloaded which included
the coordinates and the pulsation period. A search radius of 2\arcsec\ was used to correlate it with the target list.
In case the LC was refitted the $V$-band data were retrieved from this website as well.

The Asteroid Terrestrial-impact Last Alert System (ATLAS) \citep{Tonry18} published data on over 4.3 million candidate variable
objects \citep{Heinze18}.
This dataset is available through the VizieR database\footnote{J/AJ/156/241/table4} and it was correlated with the target list using a
search radius of 2\arcsec. The ATLAS team derived periods in several different ways and two are quoted, called
{\it fp-LSper}     ('original period from fourierperiod's Lomb-Scargle periodogram') and
{\it fp-lngfitper} ('final master period from the long-period Fourier fit').
The original data were retrieved via a website\footnote{\url{http://mastweb.stsci.edu/ps1casjobs/}} following the instructions
in Appendix~B in \citet{Heinze18}. ATLAS observed in two bands and the redder one (the $o$-band peaking at 0.68~$\mu$m) is used to
refit the LC.

Data from the Zwicky Transient Facility (ZTF) \citep{Masci19,Bellm19} was downloaded following the suggestions on their
website\footnote{\url{https://irsa.ipac.caltech.edu/docs/program_interface/ztf_lightcurve_api.html}}.
This involved user-customised scripts using {\tt wget} and a query to select the
data\footnote{For example 
wget \url{
    https://irsa.ipac.caltech.edu/cgi-bin/ZTF/nph_light_curves?POS=CIRCLE+352.573853+53.883614+0.00028&NOBS_MIN=3&BAD_CATFLAGS_MASK=32768&FORMAT=ipac_table
    } -O 352.573853.tbl
to select data within 0.00028 degree (1\arcsec) around (Ra, Dec)= (352.573853, 53.883614) filtering out bad data and with at least three observations.}.
For the sources in the target list data is available in the Sloan $g$- and $r$-filters, and the redder one was used to refit the LC.

The VISTA Variables in the V\'{i}a L\'{a}ctea (VVV) ESO Public Survey \citep{Minniti2010}
has been mapping the NIR variability in the $K_{\rm s}$-band of the Milky Way Bulge and the adjacent southern disk.
Recently, \citet{FerreiraLopes20} published the VVV Infrared Variability Catalogue (VIVA-I) containing data on almost 45 million
variable star candidates. The catalogue contains periods based on five different methods, and also a 'best period', {\tt bestPeriod}.
From the VISTA Science Archive (VSA)\footnote{\url{http://surveys.roe.ac.uk/vsa/index.html}} the basic data of the 6.7 million
sources in VIVA-I with  a {\tt bestPeriod} $>$ 0.5~days were downloaded, which was then cross-matched with the target sample using
a search radius of 3\arcsec.
In case the LC was refitted, the timeseries data were downloaded from a website\footnote{\url{http://horus.roe.ac.uk:8080/vdfs/VcrossID_form.jsp?disp=adv}} using
a dedicated query\footnote{After preparing a file with Ra and Dec, choosing a pairing radius of 3\arcsec, and selecting `all nearby sources' as option
  the query is:
\begin{verbatim}
SELECT #upload.*, #proxtab.distance, d.RA, d.Dec,
d.filterID, d.mjd,  d.aperMag3, d.aperMag3err, 
d.ppErrBits FROM #upload left  
outer join  #proxtab on #upload.upload_id=upid 
left outer join   vvvDetection  as d on 
d.objID=archiveID left outer join  multiframe on  
multiframe.multiframeId=d.multiframeID where frametype 
like  'tilestack' order by upload_id  
\end{verbatim}
} to obtain the publically available data.

The analysis of 3.6 years of data from the Diffuse Infrared Background Experiment (DIRBE) provided a list 597 (candidate) variables \citep{Price10}.
The data are available through VizieR\footnote{\url{J/ApJS/190/203/var}} and this includes coordinates, mean magnitudes and errors, amplitudes,
and periods in four photometric bands. A pairing radius of 8\arcsec\ was used.
The VizieR table also includes a link to the time series data, which is used when the DIRBE data are refitted. In those cases the data
at 4.9~$\mu$m was used as they are typically the brightest for the sources in the target list.

Data from other surveys has been analysed for a handful of sources, namely 
the Optical Monitoring Camera (OMC) data on board {\sc INTEGRAL} (\citealt{AlfonsoGarzon12}; one source), from the
Catalina Sky Survey (CSS; \citealt{Drake09}; three sources), the 
Bochum Galactic Disk Survey (GDS; \citealt{Hackstein15}; two sources), and
$K$-band photometry from \citet{Kerschbaum06} with photometry from the literature being added (two sources).
In addition, VMC $K$-band data from \citet{Gr20} were refitted with an improved initial period from the present work.

\subsection{Period analysis and LC modelling}
\label{modelling}

The automatic analysis of the LCs was carried out with the Fortran codes available in {\it numerical recipes} \citep{Press1992} as
described in Appendix~A of \citet{Groenewegen04} and modified to analyse the VMC $K$-band data as described in \citet{Gr20}.
The Fourier analysis was done using the subroutine {\sc fasper}. However, as a cross-check, most of the stars in the sample were analysed manually
with the code {\sc Period04} \citep{Period04} as well.
After an initial guess for the period was determined (either through the automatic routine, a period found in the literature, or
from the manual fitting of the LC), a function of the form

\begin{equation}
\label{Eq-fit}
m(t) = m_0 +
%\left(
A \sin (2 \pi \; t \; e^{f}) +
B \cos (2 \pi \; t \; e^{f}) % \right)
\end{equation}
was fitted to the LC using the weighted linear least-squares fitting routine {\sc mrqmin}.
This results in the parameters listed in Tables~\ref{WISEREALPer}, \ref{noWISEPerbutREAL}, and \ref{WISEPernotREAL}, namely 
mean magnitudes ($m_0$), periods ($\exp (-f)$), and amplitudes ($\sqrt{A^2 + B^2}$) with their associated uncertainties.
Equation~\ref{Eq-fit} implies that the LC can be described by a single period. It is well known that the LCs of LPVs are not
strictly single-periodic (as many of the fitted LCs show). However with the limited number of data points available one is in general
not able to comment on the presence of more than one period.

A comparison of the LC with the fit sometimes suggested that alternative periods may be possible as well.
These cases were inspected by the manual fitting of the LC using {\sc Period04}, and alternative periods (denoted Palt) are sometimes
indicated in the comments for Tables~\ref{WISEREALPer}, \ref{noWISEPerbutREAL}, and \ref{WISEPernotREAL}. 
The tables also include the reduced $\chi^2$, 
defined as $\chi_{\rm r}^2 =  \sum_{\rm i} \left(((m_{\rm i} - o_{\rm i})/\sigma_{\rm o_{\rm i}})^2 \right)/ (N-N_{\rm p}),$ with $m$, $o$ and
$\sigma_{\rm o}$ indicating the model magnitude, the observed magnitude, and the error, respectively, $N$ is the number of data points,
and $N_{\rm p}$ = 1 or 4, depending on whether Eq.~\ref{Eq-fit} is fitted without or with the period.

The Fourier analysis and the LC modelling were done on the W1 and W2 data separately.
If the total time span of the data is less than 400~days, or the total number of data points is less than eight, the LC fitting process was terminated.
This resulted in cases where only one LC was generated and these cases were inspected more closely.
In most cases ($\sim100$ stars) the data refer to a fake source in the AllWISE catalogue (see Sect.~\ref{S-NR}) and that source was removed
from the fitting all together; in $\sim15$ cases, the fit seemed reliable and in those cases the fitting process was carried out
in the other filter using the available data, even if there were fewer than eight data points covering a shorter time span.

\section{Literature data}
\label{S-Lit}

To characterise the sources better, the target list was correlated with other databases.
From the SIMBAD database, some common names and the object type were retrieved using a search radius
of 3\arcsec\footnote{With an exception for one well-known source which was located at 4.1\arcsec\ from the AllWISE coordinates.
The SIMBAD query was done in June 2020.}.

Real sources that were detected in the W1 and W2 filters are expected to have been detected in other IR bands as well.
To verify that, the target list was correlated with the following photometric catalogues which are all available through VizieR:
2MASS \citep{Cutri_2MASS} using a search radius of 1.0\arcsec;
the {\it Akari}/IRC MIR all-sky survey  \citep{Ishihara10}  using a search radius of 3.0\arcsec;
the {\it Midcourse Space Experiment} (MSX) \citep{Egan03} using a search radius of 5.0\arcsec;
the {\it Herschel} infrared Galactic Plane Survey (Hi-GAL) 70~$\mu$m catalogue \citep{Molinari16}, using a search radius of 4.5\arcsec;
the Galactic Legacy Infrared Midplane Survey Extraordinaire (GLIMPSE) \citep{Benjamin03,GLIMPSE} using a search radius of 3.0\arcsec, and 
the MIPSGAL survey at 24~$\mu$m \citep{Gutermuth15} using a search radius of 2.0\arcsec.
The first three surveys are all-sky, while the latter three are surveys mainly of the galactic plane.

Figure~\ref{Fig-Sky} shows that many sources are located in the galactic plane which has been surveyed extensively for OH maser emission.
A double-peaked OH profile is a characteristic of evolved O-stars.
The target list was correlated with the OH database of \citet{EngelsOH} using a search radius of 3.5\arcsec.
OH maser sources in the MCs \citep{Goldman17,Goldman18} were also considered.
The target list is also correlated with the classification of over 11~000 sources from IRAS LRS spectra \citep{Kwok97} that contains
information on the dust species and continuum shape in the 8--23~$\mu$m region using a search radius of 12\arcsec, and
with the compilation of spectral types \citep{Skiff14} using a search radius of 3.2\arcsec.

Finally, the target list was correlated with a number of catalogues containing extra galactic sources.
There are no matches with the catalogues of
‘Quasars and Active Galactic Nuclei’  (13th ed., \citealt{Veron10}), 
the ‘Large Quasar Astrometric Catalogue 4’ \citep{Gattano18}, and 
the ‘SDSS quasar catalogue’ (DR16, \citealt{Lyke20}).
There are two matches in
‘The Million Quasars’ catalogue (version 7.2, April 2021; \citealt{Flesch15}), that lists a source at
Ra= 286.187653, Dec= +48.885826 as having a 91\% probability of being a QSO with a redshift of 0.700, and at
Ra=  80.513848, Dec= -68.322622 as having a 69\% probability of being a QSO with a redshift of 0.900.

There are more matches in catalogues listing AGN and QSO {\it candidates}, such as the 
‘Gaia DR2 quasar and galaxy classification’ (\citealt{BailerJ19}, 60 matches), 
% 60 matches within 1"
the ’QSO candidates catalogue with APOP and ALLWISE’ (\citealt{Guo18}, 3 matches), 
and the ‘The WISE AGN candidates catalogues’  (the R90 90\% reliability catalogue, \citealt{Assef18}, 2 matches).
The recent Gaia EDR3 list of AGN and QSOs \citep{GEDR3_Brown,GEDR3_Klioner} contains 3 matches within 1 arcsec.
These matches are listed in Tables~\ref{WISEREALSam}, \ref{noWISESambutREAL}, and \ref{WISESamnotREAL} from which it is clear that many are
actually stellar sources. For example, of the 60 candidate QSOs in \cite{BailerJ19} 39 show LPV-like pulsations (see Sect.~\ref{S:LPV}).
The target list was also correlated with the catalogue of \cite{Solarz17}.
They used a novel approach (one-class support vector machines, OCSVM) to identify anomalous patterns
in AllWISE colours. Their method allowed them to detect anomalies (e.g. objects with spurious photometry), and also
real sources such as a sample of heavily reddened AGN/quasar candidates.

\section{Results}

The results of the literature search and the period analysis are compiled in Tables~\ref{WISEREALSam}-\ref{WISEPernotREAL}.
There is a table listing the results of the literature search and a table listing the results of the period analysis for three classes of
objects:
1224 bona fide stellar sources with a period analysis based on WISE data (Tables~\ref{WISEREALSam}, \ref{WISEREALPer}); 
 118 bona fide stellar sources without sufficient WISE data for the LC analysis in both filters, but possibly with a period from the
literature or analysis of literature data (Tables~\ref{noWISESambutREAL}, \ref{noWISEPerbutREAL}); and
650  other sources that may contain a few extra galactic objects, but most are not bona fide sources (Tables~\ref{WISESamnotREAL}, \ref{WISEPernotREAL}).

The distinction between the bona fide stellar sources and those that (very likely) are not is based on the number of associations with a
SIMBAD object and the other external catalogues mentioned in Sect.~\ref{S-Lit}, an inspection of the LC, and the result of the LC fitting.
Signatures of a fake source are no, or only one association with an external catalogue (often close to the limit of the search radius used),
and a LC with a few points.
They are further discussed in Sect.~\ref{S-NR}.

Tables~\ref{WISEREALSam}, \ref{noWISESambutREAL}, and \ref{WISESamnotREAL} contain the results of the literature search and they include the
distance to the closest SIMBAD object and the other photometric catalogues discussed in Sect.~\ref{S-Lit}. This also includes
the blue and red velocity of any OH maser emission, the IRAS LRS classification, and the spectral type.
Tables~\ref{WISEREALPer}, \ref{noWISEPerbutREAL}, and \ref{WISEPernotREAL}  contain the periods quoted in the literature,
the periods derived from fitting literature data, and the results of fitting the WISE data (period with error, amplitude with error,
mean magnitude with error, and the reduced $\chi^2$ in the W1 and W2 filters).
Examples of the lightcurve and the fits are shown for both the WISE data (Fig.~\ref{Fig:WISE}) and the other fitted data from
the literature (Figs.~\ref{Fig:VMC}-\ref{Fig:ZTF}). The complete set of fitted LCs is available at \url{https://doi.org/10.5281/zenodo.5825878}.
Figures~\ref{Fig:DIRBE}--\ref{Fig:GDS} show the LCs for the datasets with a limited number of matches that can fit in a single figure.

\section{Discussion} 
\label{S-Dis}

\subsection{AllWISE sources that are very likely not real}
\label{S-NR}

In inspecting the coordinates of the mostly fake objects (Table~\ref{WISESamnotREAL}) it is striking to observe that they can be
very similar, and in fact many are related to very bright objects (hereafter, `clones').
The most conspicuous example are the clones of CW Leo. CW Leo is located at (Ra, Dec)= (146.989193, $+13.278768$) and it is not present
in the AllWISE catalogue.
In the target list, there are 42 sources located up to 17\arcmin\ from this position not associated with any other known source.
Other well-known IR bright sources have clones, including AFGL 3068 (Ra, Dec= 349.802533, $+17.192628$, W1= 4.7~mag, W2 is unreliable; 16 sources  up to 9\arcmin),
VY~CMa (Ra, Dec= 110.7430362, $-25.7675659$, W1 $\sim$ 1.7~mag and W2 $\sim$ 3.4~mag but both are unreliable; nine sources up to 10\arcmin), or
IRC~+10~420 (Ra, Dec= 291.700408, $+11.354634$, not in AllWISE; seven sources up to 9\arcmin).
For a slightly fainter source such as IRAS 08171$-2134$ (Ra, Dec= 124.8263077, $-21.737400$, W1=7.3~mag and W2=3.8~mag), the number is reduced to
three sources up to 1.2\arcmin\ distance.

\begin{figure}
 
\begin{minipage}{0.24\textwidth}
\resizebox{\hsize}{!}{\includegraphics[angle=-0]{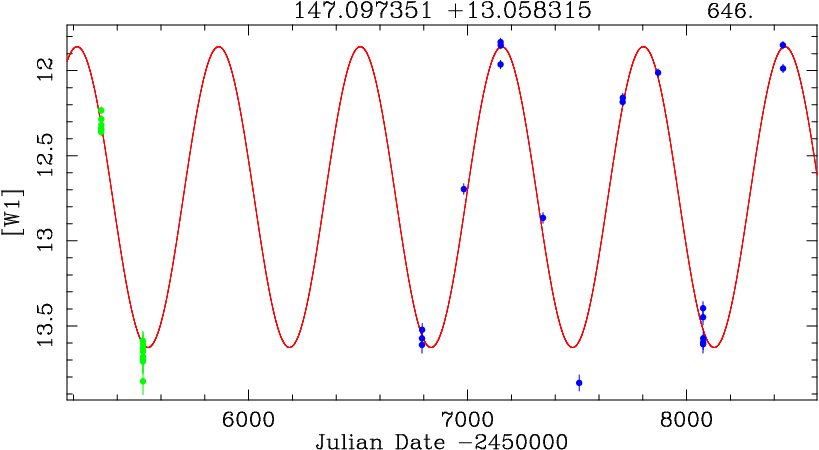}} 
\end{minipage}
\begin{minipage}{0.24\textwidth}
\resizebox{\hsize}{!}{\includegraphics[angle=-0]{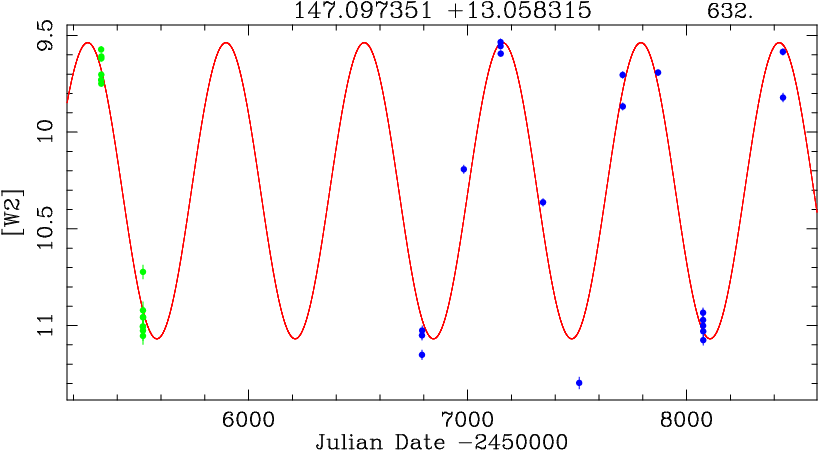}} 
\end{minipage}
\begin{minipage}{0.24\textwidth}
\resizebox{\hsize}{!}{\includegraphics[angle=-0]{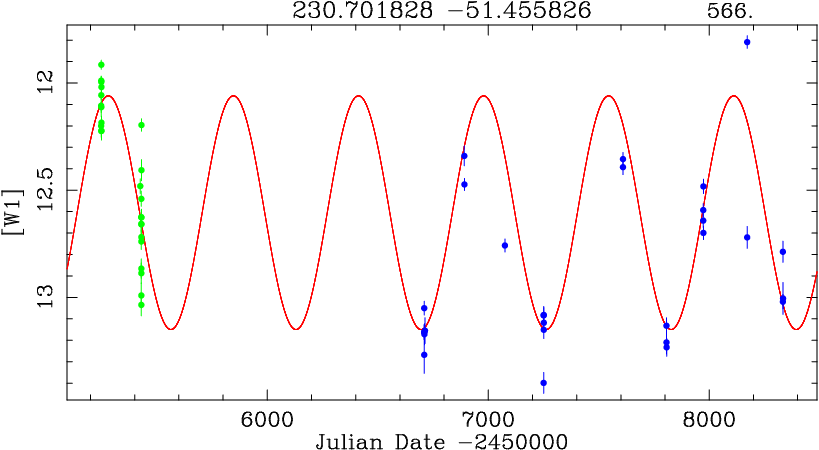}} 
\end{minipage}
\begin{minipage}{0.24\textwidth}
\resizebox{\hsize}{!}{\includegraphics[angle=-0]{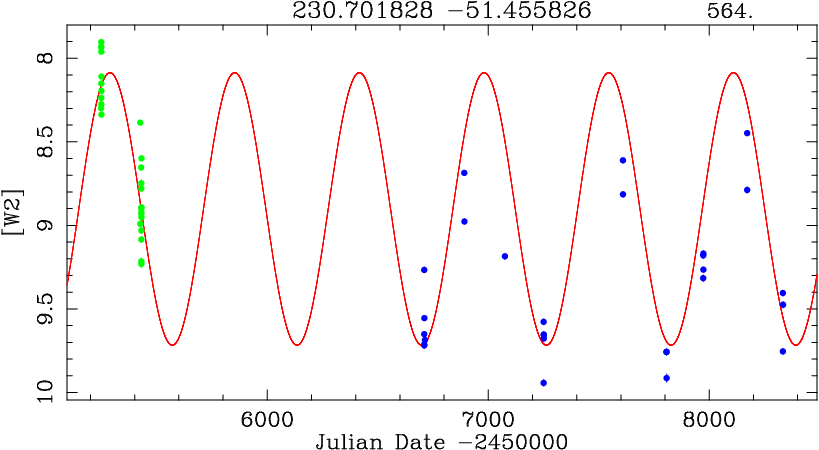}} 
\end{minipage}

\begin{minipage}{0.24\textwidth}
\resizebox{\hsize}{!}{\includegraphics[angle=-0]{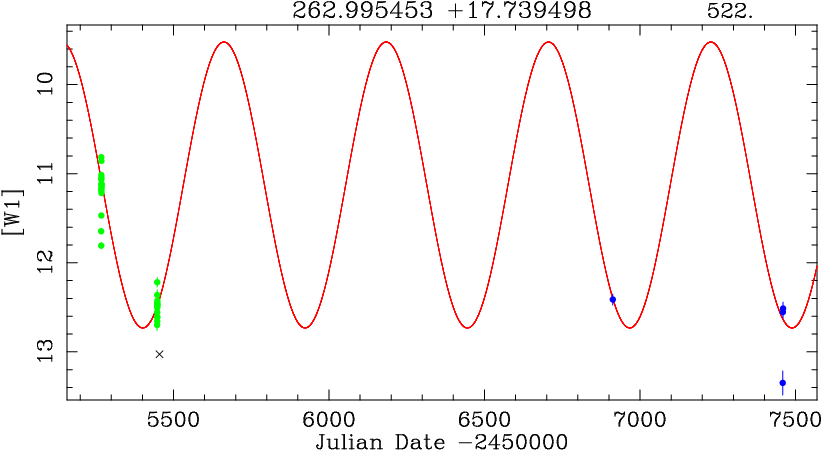}} 
\end{minipage}
\begin{minipage}{0.24\textwidth}
\resizebox{\hsize}{!}{\includegraphics[angle=-0]{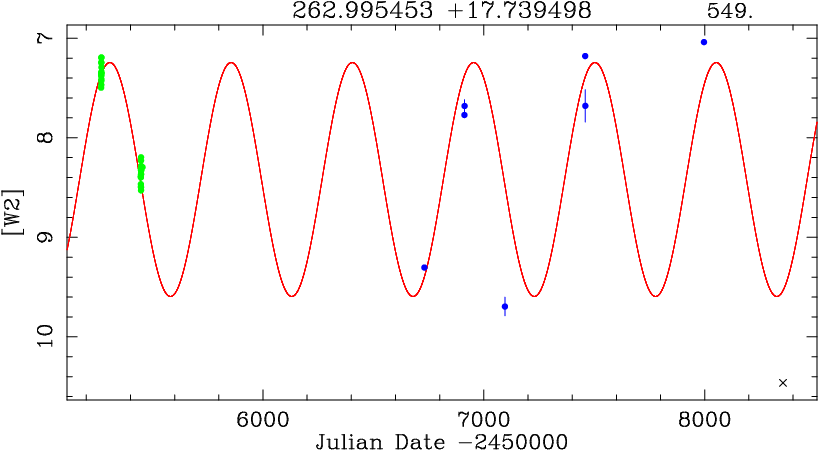}} 
\end{minipage}
\begin{minipage}{0.24\textwidth}
\resizebox{\hsize}{!}{\includegraphics[angle=-0]{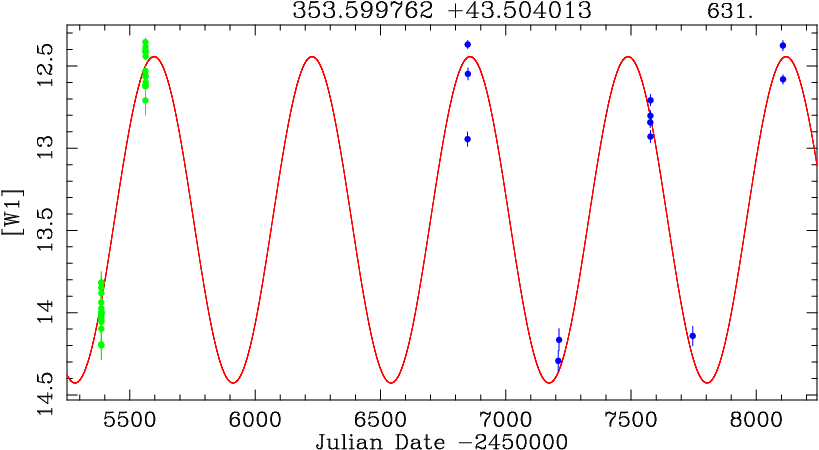}} 
\end{minipage}
\begin{minipage}{0.24\textwidth}
\resizebox{\hsize}{!}{\includegraphics[angle=-0]{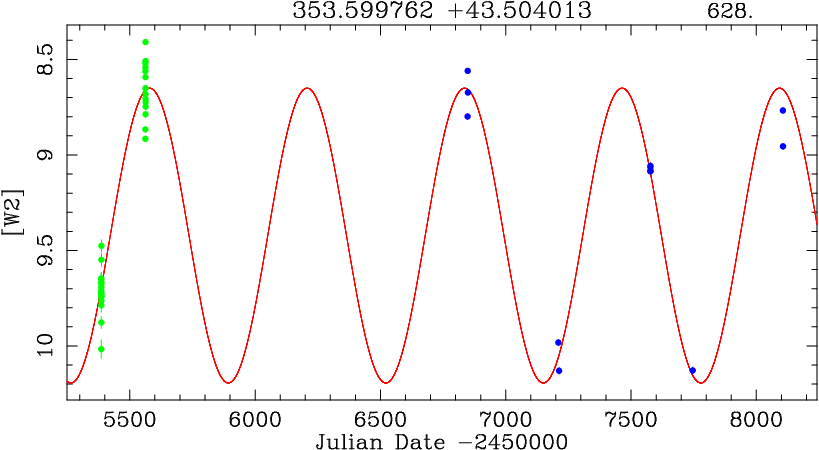}} 
\end{minipage}
 
\begin{minipage}{0.24\textwidth}
\resizebox{\hsize}{!}{\includegraphics[angle=-0]{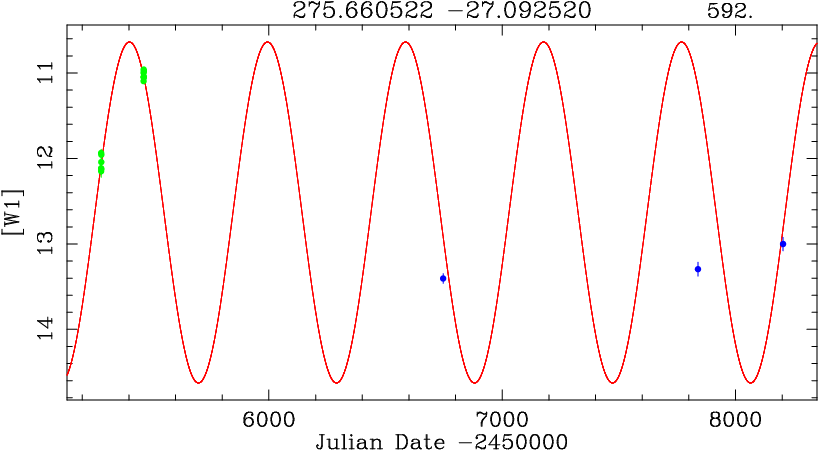}} 
\end{minipage}
\begin{minipage}{0.24\textwidth}
\resizebox{\hsize}{!}{\includegraphics[angle=-0]{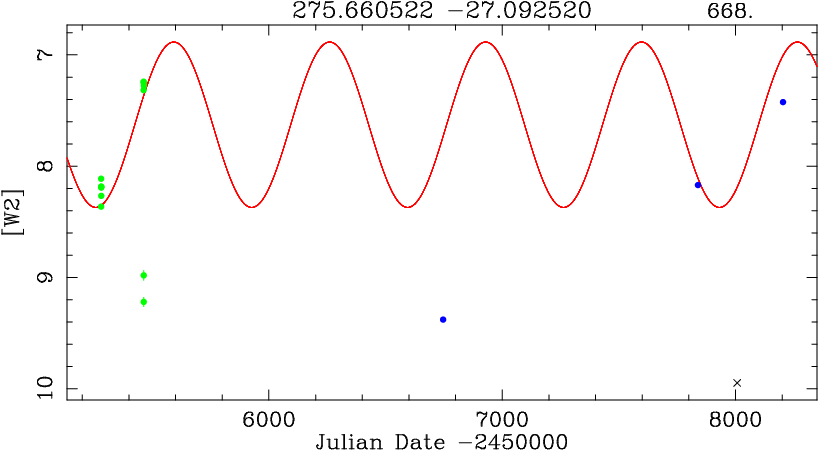}} 
\end{minipage}

% 2
\caption{Examples of LC fits (W1 in the left-hand panel and  W2 in the right-hand panel) to clones of CW Leo (Ra= 147.097351),
  IRAS 15194-5115 (Ra= 230.701828), IRC +20~326 (Ra= 262.995453), AFGL~3116 (Ra= 353.599762), and AFGL~2135 (Ra= 275.660522).
  The identifier is listed on top of each panel, with the period to the right.
}
\label{Fig:Clones}

\end{figure}

Some of these clones have LCs that are periodic (see Fig.~\ref{Fig:Clones}) with periods in agreement with the literature values.
These are 
CW~Leo $P= 643 \pm 1.4$~d (the weighted mean of the periods obtained in the W1 and W2 bands, cf. Table~\ref{WISEPernotREAL}) compared to
$P= 639 \pm 4$ (e.g. \citealt{Gr_CWL} and references therein),
IRAS~15194-5115 $P= 565 \pm 2.3$~d compared to $P= 580$ \citep{LeBertre92} and $P= 576$ \citep{Whitelock06},
IRC~+20~326 $P= 549 \pm 3.6$~d compared to $P= 540$ \citep{Uttenthaler19},
AFGL~3116 $P= 630 \pm 3.3$~d  compared to  $P= 620$ \citep{Jones90} or $P= 599$ \citep{Drake14}, and
AFGL~2135 $P= 619 \pm 13$~d  compared to $P= 655$ \citep{Whitelock06}.

The number of fake sources is almost one-third of the sample and one may wonder if these could have been filtered out using flags
available in the AllWISE catalogue.
Both \citet{Chen18} and \citet{Uchiyama19} used additional selection criteria, for example  on the photometric quality ({\it ph\_qual}),
contamination and confusion flag ({\it cc\_flags}), variability flag  ({\it var\_flag}), fraction of saturated pixels ({\it w?sat}), 
or poor PSF profile fitting ({\it w?rchi2}),  where ‘?’ stands for  ‘1’ or  ‘2’ depending on the filter.

Among the fake sources 86\% have a {\it cc\_flag} in the W1 and W2 filter which is not equal to ’00’, but so do 77\% of the bona fide sources.
The fake sources also do not necessarily have poor photometric quality flags (77\% in fact have a {\it ph\_flag} in the W1 and W2
filter of ‘AA’). 
Other flags were inspected, but in conclusion, many bona fide sources with good quality data would be eliminated by applying stricter
selection criteria, although this implies including a significant number of fake sources.

\subsection{Sources in the VVV survey}
\label{S-VVV}

Of the 122 sources analysed by \citet{FerreiraLopes20}, 51 are listed with a period of 1 day (when rounded to one digit) and
another ten with periods below 10 days.
Out of the remaining 61, only nine periods agree to within 10\% with the periods derived here; while in 32 cases, the difference in period
is more than a factor of two, and up to a factor of ten.
The reason for this large discrepancy is very likely related to the frequency range that was explored in \citet{FerreiraLopes20}, which is namely
larger than $\frac{2}{T}$ $d^{-1}$, with $T$ being the time span (see Sect.~4.2 of that paper). This time span is not explicitly given and likely
varies from source to source, but it probably leads to a lower frequency limit that is too large.
In fact, an earlier paper by the VVV team \citep{ContrerasPena17} that searched for high-amplitude infrared variable stars % ($\Delta K > 1$)   
lists different periods for a number of stars (they do not state what frequency range they searched).
Out of the ten sources in the sample with periods in \citet{ContrerasPena17}, nine have a period from WISE data of which seven agree to
within 10\% and all nine to within 20\%. The periods listed in \citet{FerreiraLopes20} for those stars are all incorrect (three have a period
of one day or periods are too small by factors of 1.7 to 3.2).

As an additional check, for a sample of 49 stars the $K$-band data from the VVV sources were reanalysed using the publically available data. 
For 42, the periods derived in this way agreed to within 10\% with the period from the WISE data.
All the revised periods and LCs are available in the Appendices, as explained at the end of Section~5.

\subsection{Non-variable OHIR stars}
\label{S-nvOH}

In a recent paper, \citet{Kamizuka20} investigated the NIR brightening of non-variable OH/IR stars.
The OH maser emission of OH/IR stars on the AGB is expected to follow the pulsation period of the
underlying star, see Sect.~\ref{S-Orich}.
However, non-variable OH/IR stars are known to exist \citep{Herman85} and this is expected to happen
in the transition from the AGB to the P-AGB phase when large-amplitude pulsations stop.

\citet{Kamizuka20} selected 16 stars from the sample in \citet{Herman85}, which had the smallest 
variability amplitudes in their OH/IR maser emission.
They established NIR multi-epoch data for six objects, based on archival data from 2MASS \citep{Cutri_2MASS},
% the UKIRT Infrared Deep Sky Survey (
UKIDSS \citep{2008MNRAS.391..136L},  and data taken with the
Okayama Astrophysical Observatory Wide Field Camera (OAOWFC; \citealt{Yanagisawa19}).
For all six stars, they derived a brightening in the $K$-band in the range $0.01-0.13$~mag/year over a 20 year period for five objects.

Of the 16 objects studied in \citet{Kamizuka20}, four are in the WISE sample: 
OH $17.7-2.0$ (ra=277.627928) and OH $31.0+0.0$ (ra=281.921407) that are not among the six for
which  \citet{Kamizuka20} determined a NIR brightening, and
OH $31.0-0.2$ (ra=282.179217) and  OH $37.1$$-0.8$ (ra=285.526137) for which
they determined a brightening of 2.04~mag over 2250~d (0.33 mag/yr) and
0.35~mag over 2170~d (0.06 mag/yr), respectively, however based on only two data points in both cases.

\begin{figure}
 
\begin{minipage}{0.24\textwidth}
\resizebox{\hsize}{!}{\includegraphics[angle=-0]{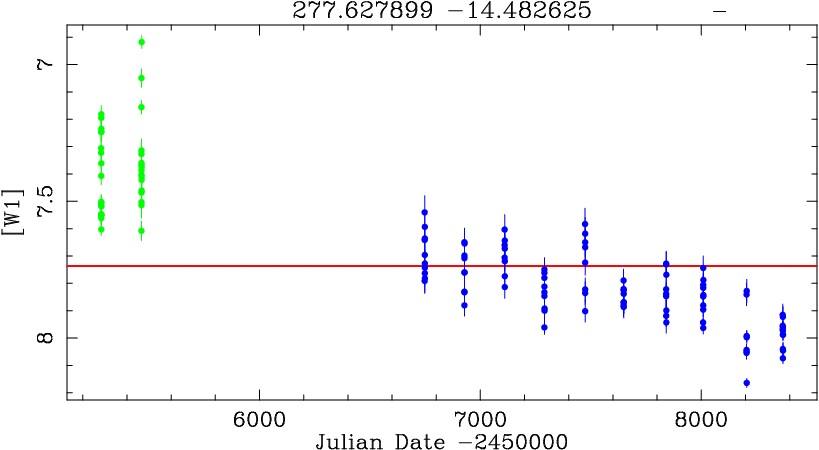}} 
\end{minipage}
\begin{minipage}{0.24\textwidth}
\resizebox{\hsize}{!}{\includegraphics[angle=-0]{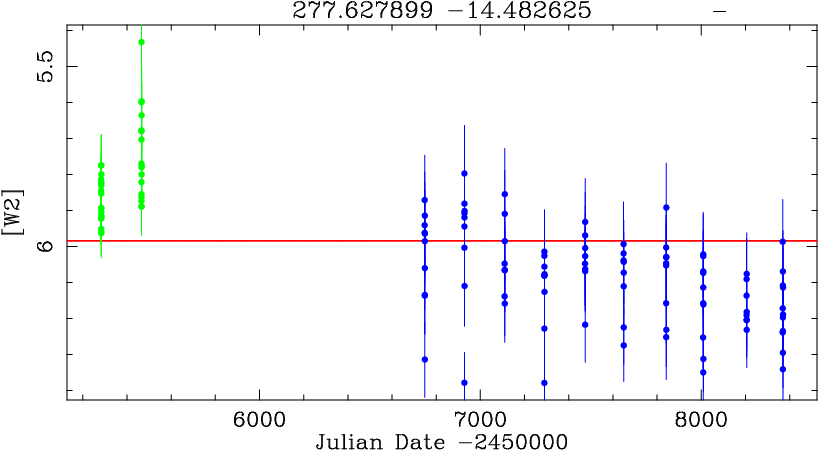}} 
\end{minipage}
\begin{minipage}{0.24\textwidth}
\resizebox{\hsize}{!}{\includegraphics[angle=-0]{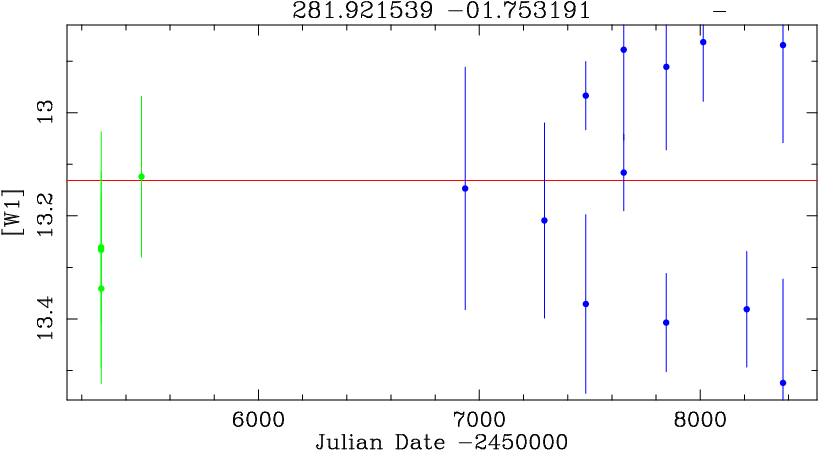}} 
\end{minipage}
\begin{minipage}{0.24\textwidth}
\resizebox{\hsize}{!}{\includegraphics[angle=-0]{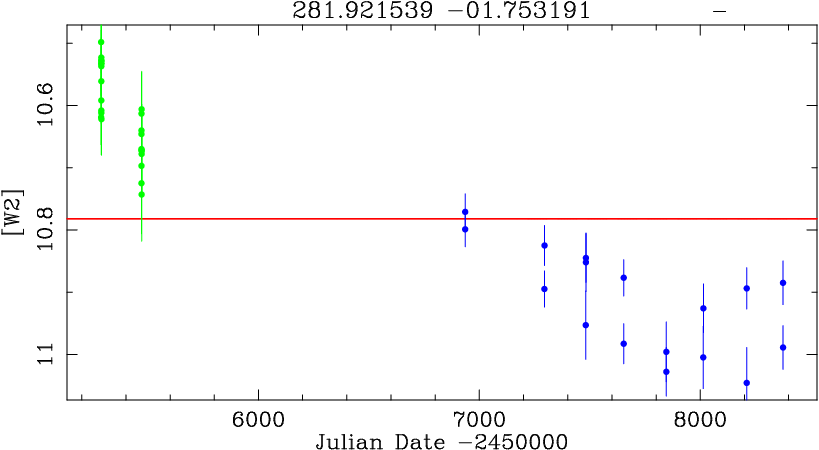}} 
\end{minipage}

\begin{minipage}{0.24\textwidth}
\resizebox{\hsize}{!}{\includegraphics[angle=-0]{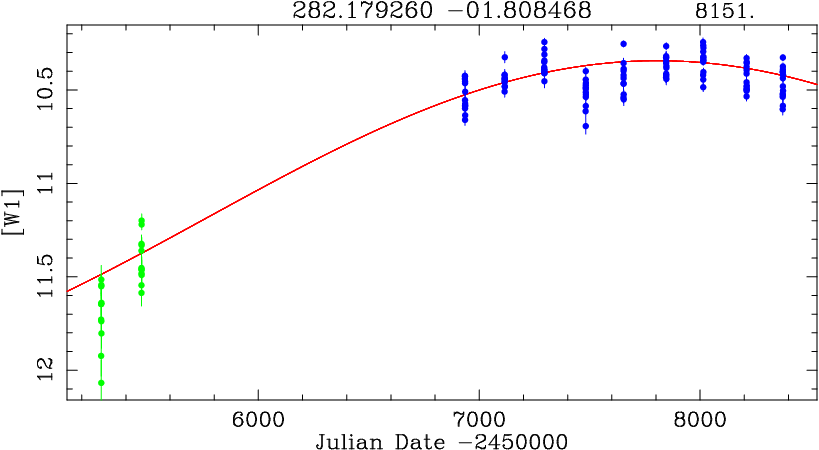}} 
\end{minipage}
\begin{minipage}{0.24\textwidth}
\resizebox{\hsize}{!}{\includegraphics[angle=-0]{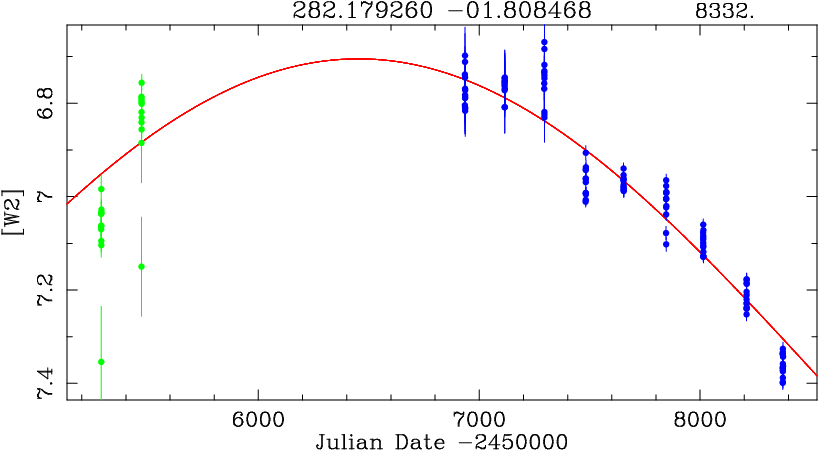}} 
\end{minipage}
\begin{minipage}{0.24\textwidth}
\resizebox{\hsize}{!}{\includegraphics[angle=-0]{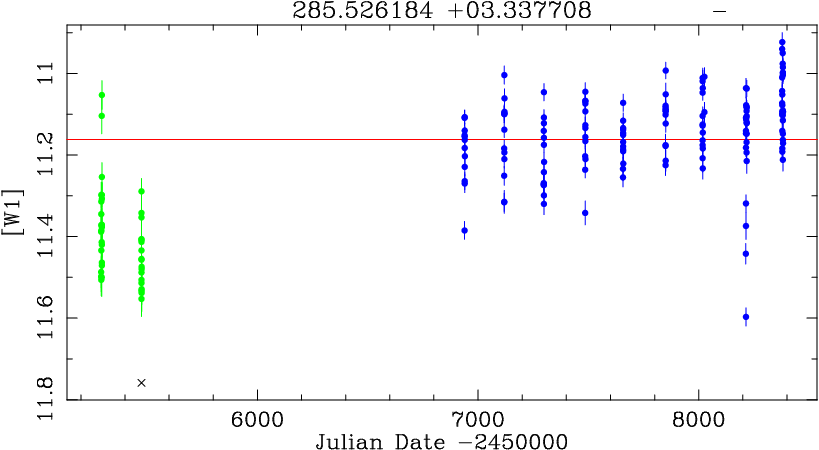}} 
\end{minipage}
\begin{minipage}{0.24\textwidth}
\resizebox{\hsize}{!}{\includegraphics[angle=-0]{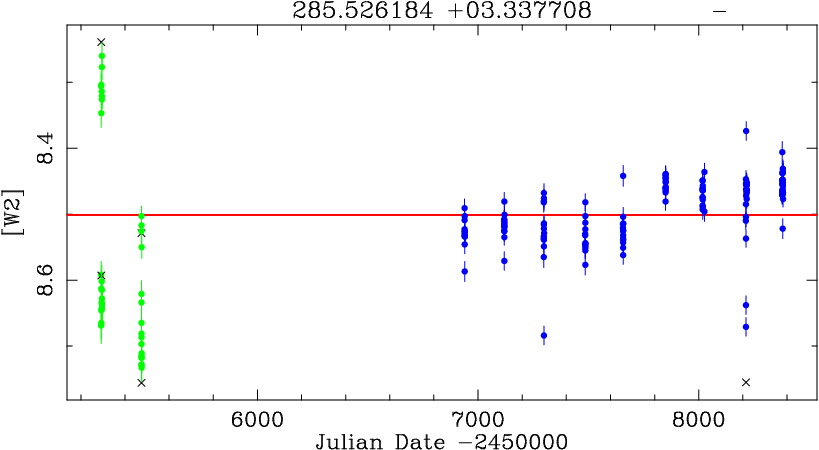}} 
\end{minipage}

\begin{minipage}{0.24\textwidth}
\resizebox{\hsize}{!}{\includegraphics[angle=-0]{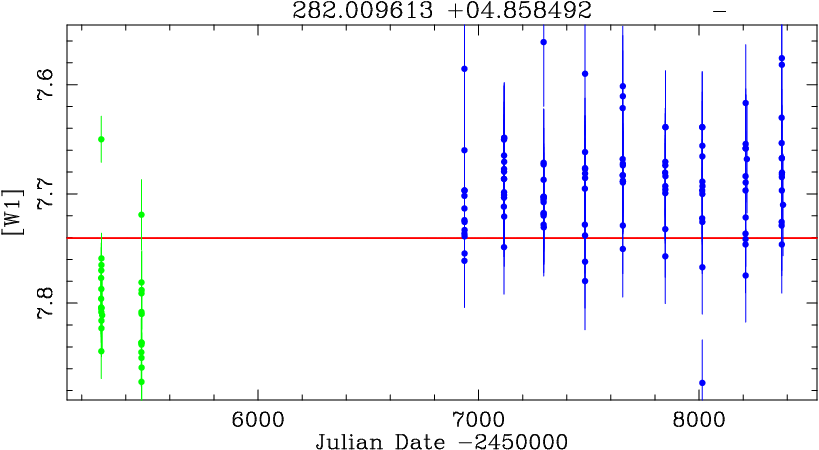}} 
\end{minipage}
\begin{minipage}{0.24\textwidth}
\resizebox{\hsize}{!}{\includegraphics[angle=-0]{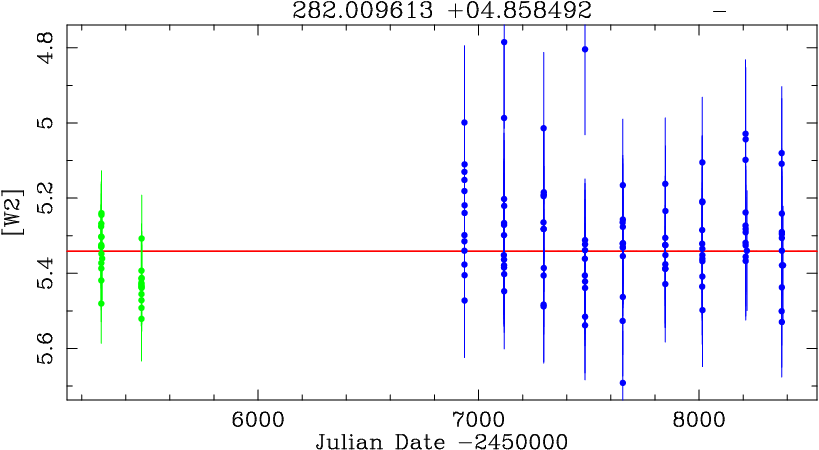}} 
\end{minipage}

\caption{WISE LCs of non-variable OH/IR stars in the sample.
}
\label{Fig:nvarOH}

\end{figure}

The W1 and W2 LCs for these four objects are shown in the top four panels of Fig.~\ref{Fig:nvarOH}.
We note that OH $17.7-2.0$ is becoming fainter by $\sim 0.7$ (W1) and $\sim 0.4$ (W2) mag over $\sim 3200$~d ($\sim 0.05 - 0.08$ mag/year).
For OH $31.0+0.0$, the situation is less clear in W1, but in the W2 filter there is a faintening by $\sim 0.4$~mag.
Also, OH $31.0-0.2$ is not clearly brightening or faintening. The formal LC fitting gives very long periods, which 
must be taken with caution as the time span of the WISE observations  covers less than half of the putative pulsation period.
The two $K$-band data points considered by \citet{Kamizuka20} were taken at epochs 2453536 and 2455787.
The WISE LCs do show a brightening between 2455200 and the last $K$-band epoch.
Furthermore, OH $37.1-0.8$ shows a marginal brightening of order 0.2~mag over 8.7 years in both filters.

Along similar lines, \citet{Lewis02} observed  the OH maser emission of 328 stars after 12 years again to find four with
undetectable emission at re-observation, and one in 'terminal decline'. One of these five stars is in the sample
(IRAS 18455+0448, ra=282.009613) and its WISE LC is shown in the bottom panels of Fig.~\ref{Fig:nvarOH}.
Its WISE emission is consistent with no variation.

To investigate this more systematically, the light curves of all stars in the sample were inspected that
either have a SIMBAD classification as an OH/IR star (Col.~5 in Tables~\ref{WISEREALSam}, \ref{noWISESambutREAL}, \ref{WISESamnotREAL})
or an entry in the database of \citet{EngelsOH} (Col.~12/13 in those tables).
All stars with a period from analysis of the WISE data, a period from the literature or from refitting literature data, as well
as stars not detected in OH and classified different from an O-rich star, were removed.
Twenty new candidate non-variable OH/IR stars were identified. They are labelled with ‘nvoh’ in
column~20 in Tables~\ref{WISEREALPer} and \ref{noWISEPerbutREAL}.
The five objects in the sample previously identified  in the literature are labelled with  ‘NVOH’ in those tables.

\subsection{Selecting LPVs}
\label{S:LPV}

The selection of (candidate) LPVs from the WISE data is based on the amplitude.
The geometric mean of the amplitude in the W1 and W2 filters (AmpW) and the errors therein ($\sigma_{\rm AmpW}$) were calculated.
LPV candidates are those with
AmpW $> 0.2$~mag, AmpW/$\sigma_{\rm AmpW} > 2.5$ and a SN $>6$ in the amplitude detection in either the W1 or W2 filter.
One well-known LPV (AFGL 3068, Ra= 349.802533) was added manually to this list.
The cut in amplitude is chosen to correspond roughly with typical cut-off values of about 0.45~mag used in the $I$-band and
of about 0.2~mag in the and $K$-band. The cuts on the SN ratio were determined empirically by visually inspecting the
LCs and fits to the LCs of stars selected in this way, and those that are not.
In this way 752, LPVs were selected, of which 356 appear to be newly classified as such.
They are marked ‘LPV’ in Tables~\ref{WISEREALSam} and \ref{noWISESambutREAL}. 
However, there always remain some borderline cases where the LC of an LPV candidate appears noisy, 
and only higher precision photometry over a sufficiently long time span may resolve the variable nature of some sources.
One hundred fourty-five sources have periods longer than 1000~days\footnote{Sources (81.850469, $-69.662488$) and (83.154874, $-67.115672$) have periods close to this limit of
  993  and 988~days, respectively.} of which
109 are new (13 had previously quoted periods below 1000, typically 160-700 days), which is a significant increase in the 16 template sources 
in Table~\ref{Tab-Known} from \citet{Menzies19}.
The referee directed us to the paper by \citet{Chen20} which analysed and classified a large number of variable stars based on ZTF data.
  A comparison with the periods derived in the present paper based on WISE data and ZTF data is presented in Appendix~\ref{App:ZTF}. 

The list of non-LPV candidates selected in this way (the complementary sample) contains interesting sources, some of which are periodic, but with
a smaller amplitude, or where a good LC fit is obtained in one filter only, or they show peculiar LCs.
These sources are marked ‘PER’ in Tables~\ref{WISEREALPer} and \ref{noWISEPerbutREAL}. 
They include known OH/IR stars, also sometimes with a period derived from the literature, but with poor WISE data in one filter, in addition to
Sakurai's object (see \citealt{Evans20} for a detailed discussion on its $K$ and WISE LC).
Examples of LCs for such sources are displayed in Fig.~\ref{Fig:WISE-PER}.

\subsection{EROs and the mass return to the ISM}
\label{S-ERO}

As the (rare) C-rich stars with the highest MLRs dominate the mass return by AGB stars to the ISM (see references in the introduction) it is of
interest to identify new objects in this class, both in the Galaxy and the MCs.
The template sample in Table~\ref{Tab-Known} of EROs is based on the shape of the spectrum in the MIR (a red flat continuum or the SiC feature
in absorption), but spectral data are generally not available, only photometric data are.
Based on the colours in Table~\ref{Tab-Known}, the spectral energy distributions (SEDs) of the 316
objects with W2$-$W3~$>$~3.0~mag were constructed using data in the literature. For a subset of 141 stars, MIR spectra were available.
The SEDs were fitted with the dust radiative transfer code  more of DUSTY (MoD, \citealt{Gr_MOD}), which is an extension of the radiative transfer code DUSTY \citep{Ivezic_D}, 
allowing the derivation of luminosities and MLRs; the details are given in Appendix~\ref{App-sed}.

Distances to the galactic sources were derived as follows.
Based on the C-rich objects in the MCs the following period-luminosity ($PL$) relation was derived (see details in Appendix~\ref{App-sed} and shown in Fig.~\ref{Fig:PLrel}),
\begin{equation}
  M_{\rm bol} = -5.07 \cdot (\log P -2.8)  -4.47~{\rm mag},
\label{Eq:PL}  
\end{equation}
based on 31 objects and with an rms of 0.31~mag.
This $PL$ relation was then applied to the Galactic objects for which a period was available, O- and C-rich alike.
The $PL$ relation was derived using stars with periods up to $\log P \approx 3.1$ (about 1260 days), while the longest period for which it has been applied
has a period of about 2600~days ($\log P \approx 3.4$).
For C-stars without a period the median luminosity of 7100~\lsol\ of the MCs objects with a period was used; for the
O-rich objects without a period an arbitrary distance of 2.0~kpc was adopted. Interstellar (IS) reddening was included (see Appendix~\ref{App-sed}).
That notion that the $PL$ relation derived for ERO C-stars in the MCs would hold for Galactic C-stars, and for O-stars, is an assumption made here.
  Data for less reddened and shorter period ($P \less 400$~d) Miras are consistent with the premise that any differences are small \citep{Whitelock08}.
  We note that, to first order, ignoring the dependence of the reddening on distance, $L \sim d^2$ and \mdot $\sim d$ if the reader prefers another distance.

Based on the MIR spectra and the fitting of the SED, the sample was divided into 197 C-rich and 119 O-rich sources.
Of the C-rich sources, 18 belong to the sample of Galactic and LMC template sources of EROs (and are labelled ERO in Tables~\ref{WISEREALSam} and \ref{noWISESambutREAL}),
65 sources (including eight in the LMC) are EROs with MIR spectra (and are labelled eroS in these Tables),
110 sources (including two in the direction of the SMC, and 26 in the direction of the LMC) are candidate EROs based on the fitting of the SEDs (and are labelled eroP in these Tables),
and the remaining four are classified as C-rich non-ERO sources (and are labelled sedC in these Tables)
The O-rich sources appear to be a mixture of O-rich AGB and P-AGB stars, H{\sc ii} regions, planetary nebulae and YSOs (and are labelled sedO in these Tables).

Table~\ref{Tab-Cres} show the results for the C-stars and Table~\ref{Tab-Ores} for the O-stars. Only the first entries are shown, and the full tables are available at the CDS.
Displayed are the adopted distance and reddening and the results of fitting the SEDs. The last column shows the total MLR, which assumes
spherical symmetry of the CSE, a dust-to-gas (DTG) ratio of 0.005, and a CSE expansion velocity of 10~\ks\ for every star.

\begin{figure}
 
\begin{minipage}{0.24\textwidth}
\resizebox{\hsize}{!}{\includegraphics[angle=-0]{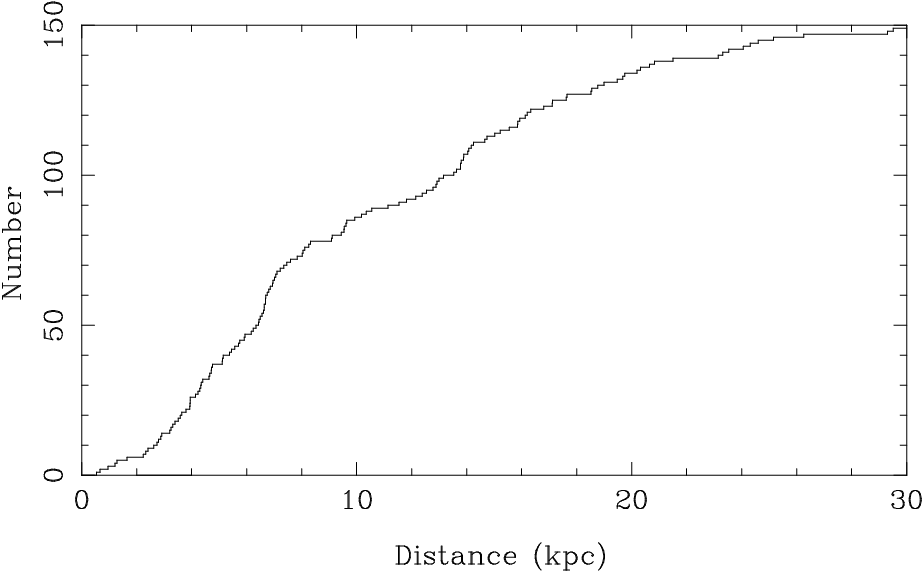}} 
\end{minipage}
\begin{minipage}{0.24\textwidth}
\resizebox{\hsize}{!}{\includegraphics[angle=-0]{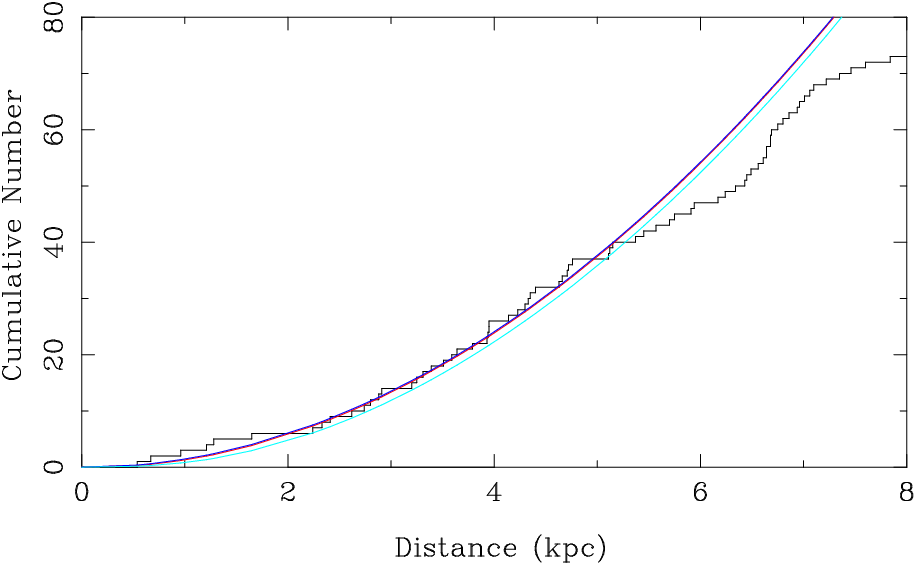}} 
\end{minipage}

\caption{Cumulative number distribution of ERO (candidates) versus distance.
  The right-hand panel is a zoom, and it shows model predictions for combinations of local space density and vertical scale height of
  1.6 kpc$^{-3}$, 150 pc (dark blue); 0.8 kpc$^{-3}$, 300 pc (red);  and 0.30 kpc$^{-3}$, 800 pc (light blue), respectively.
}
\label{Fig:NDIST}

\end{figure}

Figure~\ref{Fig:NDIST} shows the cumulative number distributions of ERO (candidates) versus distance up to 30~kpc on the left, and up to 8~kpc on the right.
When the number density of stars is assumed to depend exponentially on the height above the Galactic plane the number of stars within a certain radius can be calculated
analytically, see Eq.~19 in \citet{Groenewegen92}, while Eq.~20 in that paper can be used to determine the scale height.
The right-hand panel of Figure~\ref{Fig:NDIST} shows some models for different scale heights ($H$) and local space densities ($\rho_0$).
The number of objects scales to first order with ($H \cdot \rho_0$) and the three models that show a very similar behaviour all have $H \cdot \rho_0 = 0.24 $~kpc kpc$^{-3}$.
\citet{Groenewegen92}. based on a very small number of EROs derived $H$ = 195 $\pm$ 20~pc, and $H \cdot \rho_0 = 0.23-0.27$~kpc kpc$^{-3}$.

The predicted number of stars versus distance suggests that the sample of EROs may be complete up to 5~kpc. 
There are 36 stars in a cylinder around the Sun with this distance with a total estimated MLR of 4.1 $\cdot 10^{-4}$ \msolyr\ (or 5.2 $\cdot 10^{-6}$ \msolyr/kpc$^2$) and an
estimated scale height of 180~pc. The average MLR is this sample is 1.1 $\cdot 10^{-5}$ \msolyr.
As the DTG ratio is an assumed quantity (1/200), a more certain number is the dust-production rate (DPR) which is 
2.0 $\cdot 10^{-6}$ \msolyr\ (or 2.6 $\cdot 10^{-8}$ \msolyr/kpc$^2$).
To estimate an uncertainty on these numbers, Monte Carlo calculations were performed generating samples with other distances and MLRs,
assuming Gaussian distributions with a width of 0.3~mag in $M_{\rm bol}$ (which leads to a change in distance and in MLR), a condensation temperature
taking the estimated error from Table~\ref{Tab-Cres} (with a minimum of 50~K), and an optical depth taking the estimated error from Table~\ref{Tab-Cres}.
The 2.7\%, 97.3\%, and 50\% percentiles (corresponding to $\pm$2$\sigma$ in a Gaussian distribution and the median) indicate
a number of stars of 37 (33-41; 0.4-0.5~kpc$^{-2}$), a cumulative MLR of 4.0 (3.3-6.0  $\cdot 10^{-4}$ \msolyr), and a $H= 190$ (156-224~pc).
Changing the limiting distance to 3~kpc has some impact on the estimated DPR per unit surface area from
2.6 (2.1-3.8  $\cdot 10^{-8}$ \msolyr/kpc$^2$) to  2.3 (1.7-3.0  $\cdot 10^{-8}$ \msolyr/kpc$^2$).
The number of stars is reduced to the range 13 to 18 with a median of 16.
All (dust) MLRs quoted above are based on an average expansion velocity of 10~\ks. If this were 15~\ks\ (as assumed in e.g. \citealt{JK89}),
all MLRs would increase by a factor of 1.5.

The cumulative mass-loss return of the 45 ERO sources in the LMC is 1.0 $\cdot 10^{-3}$ \msolyr\  (or 5.0 $\cdot 10^{-6}$ \msolyr\ in dust).
Thirty-three were modelled by \citet{Nanni19}, finding a total MLR of 7.8 $\cdot 10^{-3}$ \msolyr\
(and 3.3 $\cdot 10^{-6}$ \msolyr\ in dust)\footnote{Using the J1000 set of models in \citet{Nanni19}.} and implying an average gas-to-dust ratio of 240.
Here we find  7.3 $\cdot 10^{-3}$ \msolyr\  (or 3.7 $\cdot 10^{-6}$ \msolyr\ in dust) for that sub-sample, which is in good agreement.
What is interesting and in highlighting, again, the importance of the EROs is the impact of the only 12 stars not included in the study by \citet{Nanni19}.
The total dust return by C-stars for the entire LMC is 16.0 $\cdot 10^{-6}$ \msolyr\ (J1000 models; \citealt{Nanni19}) from 8239 stars, of which
82\% (13.1 $\cdot 10^{-6}$ \msolyr) are by the 16\% (1332) classified as X-stars.
The sub-sample of 33 stars (0.4\%) already contributes 21\% to the total dust return.
Adding the other 12 stars (0.15\%) augments the total dust return by 8\% to about 17.3 $\cdot 10^{-6}$ \msolyr.

\begin{table*} 
\setlength{\tabcolsep}{1.3mm}

\caption{Fit results of the C-star sample (first entries)} 
\begin{tabular}{rrrrrrrrrrrrr} \hline \hline 
RA         & Dec       &  Period & $d$  & $A_{\rm V}$ & $T_{\rm eff}$  &  $L$       & $\tau_{0.5}$        & $T_{\rm c}$      & f & $p$          & f  &  \mdot     \\
(deg)      &  (deg)    &  (days) & (kpc) & (mag)     &    (K)        &   (\lsol)  &                    &    (K)          &   &              &    &  (\msolyr)         \\
\hline 

 19.463942 &  67.231445 & 1047 &  2.74 &  3.73 &   3300 &   13214 $\pm$     75 &  170 $\pm$   1.5 & 1000 $\pm$   0 & 0 & 2.0 $\pm$ 0.0 & 0 & 0.262E-04 \\ 
349.802521 &  17.192619 &  746 &  0.77 &  0.23 &   2700 &    6661 $\pm$    108 &   95 $\pm$   0.9 & 1000 $\pm$   0 & 0 & 2.0 $\pm$ 0.0 & 0 & 0.893E-05 \\ 
353.599762 &  43.504013 &  629 &  0.67 &  0.36 &   2500 &    4729 $\pm$     66 &   15 $\pm$   0.2 &  716 $\pm$   7 & 1 & 2.0 $\pm$ 0.0 & 0 & 0.206E-05 \\ 
124.826309 & -21.737400 &  939 &  2.80 &  0.20 &   3300 &   10621 $\pm$    219 &  136 $\pm$   3.9 & 1000 $\pm$   0 & 0 & 2.0 $\pm$ 0.0 & 0 & 0.167E-04 \\ 
287.486908 &   9.447611 & 1071 &  2.32 &  5.45 &   2800 &   13014 $\pm$    341 &   99 $\pm$   1.6 & 1201 $\pm$   0 & 0 & 2.0 $\pm$ 0.0 & 0 & 0.977E-05 \\ 
237.773834 & -56.890007 &  951 &  2.51 &  1.54 &   2500 &   10605 $\pm$    262 &   57 $\pm$   0.5 &  655 $\pm$  17 & 1 & 2.0 $\pm$ 0.0 & 0 & 0.149E-04 \\ 
323.345001 &  56.743065 &  930 &  2.03 &  2.54 &   2600 &   10334 $\pm$    187 &  109 $\pm$   2.6 & 1000 $\pm$   0 & 0 & 2.0 $\pm$ 0.0 & 0 & 0.122E-04 \\ 
 75.631233 & -68.093285 & 1884 & 50.00 &  0.22 &   2700 &    7897 $\pm$     44 &  251 $\pm$   1.8 & 1000 $\pm$   0 & 0 & 2.0 $\pm$ 0.0 & 0 & 0.304E-04 \\ 
 76.023376 & -68.394501 &    - & 50.00 &  0.22 &   3100 &    5992 $\pm$     10 &  248 $\pm$   0.9 & 1000 $\pm$   0 & 0 & 2.0 $\pm$ 0.0 & 0 & 0.262E-04 \\ 
 79.548790 & -70.507469 &    - & 50.00 &  0.22 &   3200 &    9496 $\pm$     18 &  273 $\pm$   1.0 & 1000 $\pm$   0 & 0 & 2.0 $\pm$ 0.0 & 0 & 0.330E-04 \\ 
 79.701599 & -69.559563 &    - & 50.00 &  0.22 &   3100 &    6935 $\pm$     10 &  189 $\pm$   0.6 & 1000 $\pm$   0 & 0 & 2.0 $\pm$ 0.0 & 0 & 0.199E-04 \\ 
 81.419411 & -70.140877 &    - & 50.00 &  0.22 &   3000 &    4000 $\pm$     10 &  205 $\pm$   1.1 & 1000 $\pm$   0 & 0 & 2.0 $\pm$ 0.0 & 0 & 0.167E-04 \\ 
 82.407959 & -72.831322 &  678 & 50.00 &  0.22 &   2800 &    5498 $\pm$     18 &  125 $\pm$   0.7 & 1000 $\pm$   0 & 0 & 2.0 $\pm$ 0.0 & 0 & 0.116E-04 \\ 
 87.608788 & -69.934212 & 1110 & 50.00 &  0.22 &   2600 &   10351 $\pm$     12 &  111 $\pm$   3.8 &  743 $\pm$  15 & 1 & 2.0 $\pm$ 0.0 & 0 & 0.245E-04 \\ 
 78.257469 & -69.564110 &    - & 50.00 &  0.22 &   2600 &    6381 $\pm$     21 &  274 $\pm$   1.4 & 1000 $\pm$   0 & 0 & 2.0 $\pm$ 0.0 & 0 & 0.307E-04 \\ 
 82.684006 & -71.716766 &  954 & 50.00 &  0.22 &   3600 &    8233 $\pm$     29 &   46 $\pm$   1.8 &  269 $\pm$   5 & 1 & 2.0 $\pm$ 0.0 & 0 & 0.103E-03 \\ 
 87.249886 & -70.556229 & 3434 & 50.00 &  0.22 &   3000 &   12280 $\pm$     28 &   85 $\pm$   1.4 &  283 $\pm$   3 & 1 & 2.0 $\pm$ 0.0 & 0 & 0.182E-03 \\ 
 89.161446 & -67.892776 & 1220 & 50.00 &  0.22 &   4000 &   20161 $\pm$    136 &   74 $\pm$   0.8 & 1000 $\pm$   0 & 0 & 2.0 $\pm$ 0.0 & 0 & 0.121E-04 \\ 
  5.961981 &  62.636379 & 1065 &  4.98 &  2.21 &   2800 &   13619 $\pm$    261 &   93 $\pm$   2.3 & 1000 $\pm$   0 & 0 & 2.0 $\pm$ 0.0 & 0 & 0.116E-04 \\ 
 38.251453 &  58.035065 &  827 &  2.33 &  1.69 &   2400 &    8209 $\pm$    803 &   46 $\pm$   1.5 &  642 $\pm$  33 & 1 & 2.0 $\pm$ 0.0 & 0 & 0.119E-04 \\ 
 39.529259 &  54.587803 &  905 &  4.63 &  1.04 &   2400 &    9856 $\pm$    223 &   75 $\pm$   2.7 & 1000 $\pm$   0 & 0 & 2.0 $\pm$ 0.0 & 0 & 0.770E-05 \\ 
 41.103512 &  55.187542 &  477 &  3.95 &  1.64 &   2400 &    2691 $\pm$     78 &   78 $\pm$   5.6 &  514 $\pm$  32 & 1 & 2.0 $\pm$ 0.0 & 0 & 0.187E-04 \\ 
 47.976738 &  60.956123 & 1163 &  8.11 &  3.52 &   3300 &   16391 $\pm$    266 &   68 $\pm$   1.9 & 1000 $\pm$   0 & 0 & 2.0 $\pm$ 0.0 & 0 & 0.899E-05 \\ 
 57.080860 &  44.701607 &  731 &  1.65 &  1.33 &   2400 &    6370 $\pm$    104 &   27 $\pm$   0.6 &  842 $\pm$  19 & 1 & 2.0 $\pm$ 0.0 & 0 & 0.266E-05 \\ 
 72.919174 & -68.792953 &  784 & 50.00 &  0.22 &   4000 &    5328 $\pm$     15 &   80 $\pm$   0.5 & 1000 $\pm$   0 & 0 & 2.0 $\pm$ 0.0 & 0 & 0.729E-05 \\ 
 74.691780 & -68.343849 &  777 & 50.00 &  0.22 &   5000 &    6900 $\pm$     63 &   40 $\pm$   1.5 &  736 $\pm$  16 & 1 & 2.0 $\pm$ 0.0 & 0 & 0.859E-05 \\ 
 76.270149 & -68.963379 &  938 & 50.00 &  0.22 &   3000 &    9645 $\pm$    154 &   83 $\pm$   2.0 & 1000 $\pm$   0 & 0 & 2.0 $\pm$ 0.0 & 0 & 0.850E-05 \\ 
 76.646332 & -70.280640 &  554 & 50.00 &  0.22 &   5000 &    8943 $\pm$     92 &   22 $\pm$   0.8 &  682 $\pm$  15 & 1 & 2.0 $\pm$ 0.0 & 0 & 0.592E-05 \\ 
 78.003212 & -70.540047 & 1182 & 50.00 &  0.22 &   4000 &   15047 $\pm$     77 &   55 $\pm$   0.5 & 1000 $\pm$   0 & 0 & 2.0 $\pm$ 0.0 & 0 & 0.740E-05 \\ 
 82.525955 & -70.511375 &  807 & 50.00 &  0.22 &   3200 &    9845 $\pm$    115 &   64 $\pm$   1.1 & 1200 $\pm$   0 & 0 & 2.0 $\pm$ 0.0 & 0 & 0.474E-05 \\ 
 85.336433 & -69.078796 &  895 & 50.00 &  0.22 &   2700 &    9363 $\pm$     84 &   63 $\pm$   2.7 &  764 $\pm$  23 & 1 & 2.0 $\pm$ 0.0 & 0 & 0.108E-04 \\ 
 87.485626 & -70.886604 & 1041 & 50.00 &  0.22 &   2600 &   17597 $\pm$    264 &   27 $\pm$   2.1 &  794 $\pm$  26 & 1 & 2.0 $\pm$ 0.0 & 0 & 0.543E-05 \\ 
 91.000160 &   7.431088 &  696 &  1.28 &  0.83 &   2400 &    5818 $\pm$    103 &   31 $\pm$   0.5 &  775 $\pm$  19 & 1 & 2.0 $\pm$ 0.0 & 0 & 0.353E-05 \\ 
 91.039764 &  47.795067 &  934 &  5.15 &  0.43 &   2400 &   10507 $\pm$    225 &   86 $\pm$   3.0 & 1000 $\pm$   0 & 0 & 2.0 $\pm$ 0.0 & 0 & 0.935E-05 \\ 
 95.182762 &  -4.558214 & 1795 & 23.15 &  0.77 &   2700 &   39535 $\pm$   1639 &  257 $\pm$  11.1 & 1000 $\pm$   0 & 0 & 2.0 $\pm$ 0.0 & 0 & 0.700E-04 \\ 
 99.256760 &  -1.450483 &  854 &  4.36 &  2.71 &   2400 &    8762 $\pm$    243 &   42 $\pm$   2.3 & 1000 $\pm$   0 & 0 & 2.0 $\pm$ 0.0 & 0 & 0.371E-05 \\ 

\hline 
\end{tabular}
\label{Tab-Cres}
\tablefoot{
  Columns~1 and 2 give the RA and Dec in decimal degrees,
  Column~3 gives the pulsation period in days,
  Column~4 gives the distance in kpc, 
  Column~5 gives the IS reddening $A_{\rm V}$,  
  Column~6 gives the effective temperature in Kelvin, 
  Column~7 gives the luminosity and error in \lsol, 
  Column~8 gives the optical depth at 0.5~$\mu$m, 
  Column~9 gives the condensation temperature with error in Kelvin, 
  Column~10 indicates if the condensation temperature was fitted (1), or fixed (0). In the latter case the error is set to 0~K,
  Column~11 gives the slope of the density law with error,
  Column~12 indicates if the slope was fitted (1), or fixed (0). In the latter case the error is set to 0,
  Column~13 gives the gas MLR in \msolyr. \\
The first entries are the template ERO sources from Table~\ref{Tab-Known}. The remainder are listed in order of RA.  
The full table is available at the CDS.
}
\end{table*}

\subsection{The nature of the EROs}
\label{S-nature}

Although the C-rich EROs are thought to be major contributors to the dust and mass return of AGB stars to the ISM the nature of these objects is not fully understood.
Many are clearly pulsating with large amplitudes and are LPVs. They follow a well-defined $PL$ relation up to about 1260 days (see Fig.~\ref{Fig:PLrel}).
These objects definitely show the characteristics of AGB evolution.
However, there are also objects that are classified as EROs based on the SEDs and MIR spectra that show no clear evidence for pulsation or variability,
or with different properties.

This was first recognised in \citet{Sloan16} where it is remarked that some of the embedded sources are relatively non-variable and that some have
relatively blue colours (compared to other embedded sources) at shorter wavelengths and that this can be interpreted as the central star revealing itself.
They conclude that some deeply embedded stars may be evolving off of the AGB and/or they may have non-spherical dust geometries.

One of the parameters derived from the SED fitting is the temperature at the inner radius. In most cases, a standard value (800-1200~K, consistent with the condensation
temperature of amorphous carbon dust) is sufficient to fit the data. However, for a non-negligible fraction of objects, a lower value has to be adopted,
and this can be due to non-spherical dust geometries, or a spherical shell that expands, consistent with the drop in MLR when the AGB star evolves into the P-AGB phase.
Of the 133 objects with a condensation temperature consistent with 800~K or more, 116 show a plausible pulsation period, and only 12\% show no
obvious variability or a period longer than 1300 days. For one-third of the sample (64 stars), a lower dust temperature at the inner radius is inferred
of which 73\% show no obvious variability or a period longer than 1300 days. This implies that lower temperatures at the inner radius are found for
a non-negligible number of objects and that these show, on average, less pronounced variability.

\begin{figure}
 
\begin{minipage}{0.235\textwidth}
\resizebox{\hsize}{!}{\includegraphics[angle=-0]{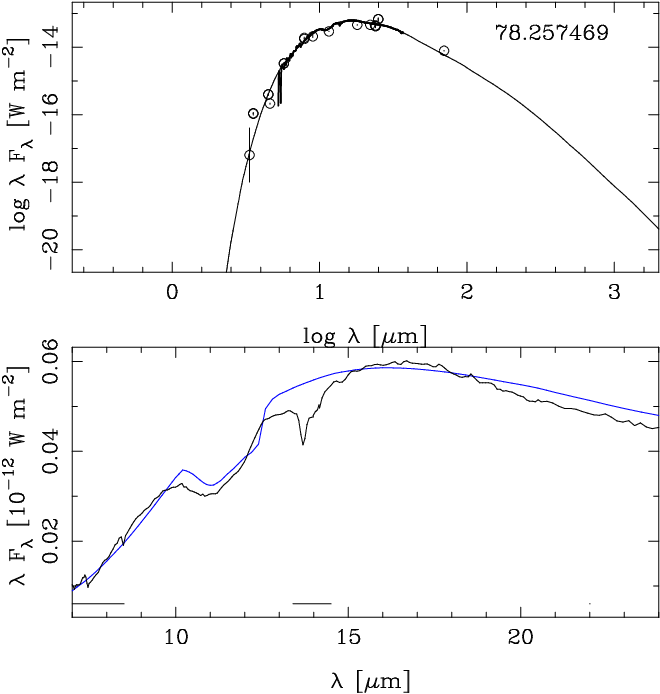}} 
\end{minipage}
\begin{minipage}{0.235\textwidth}
\resizebox{\hsize}{!}{\includegraphics[angle=-0]{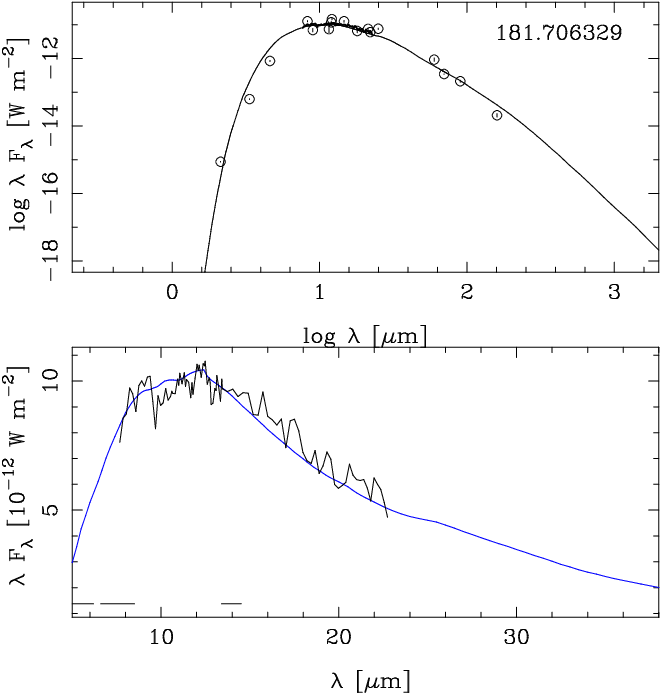}} 
\end{minipage}
\begin{minipage}{0.235\textwidth}
\resizebox{\hsize}{!}{\includegraphics[angle=-0]{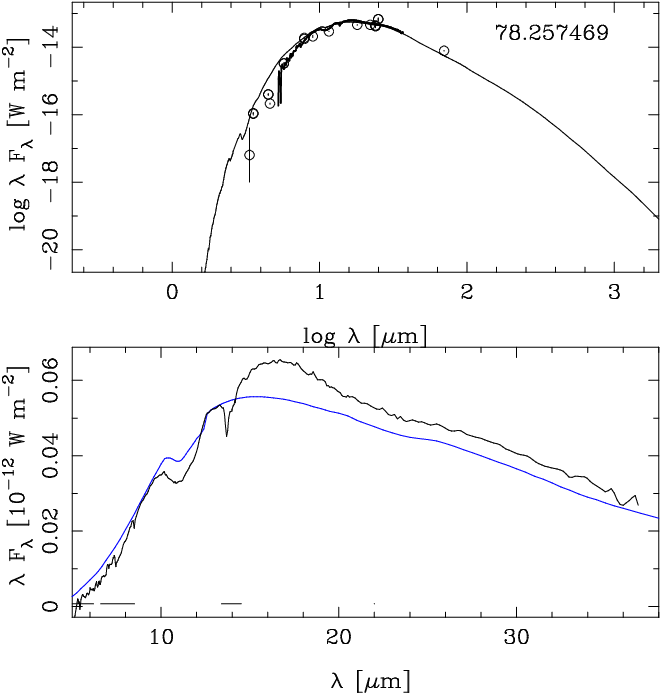}} 
\end{minipage}
\begin{minipage}{0.235\textwidth}
\resizebox{\hsize}{!}{\includegraphics[angle=-0]{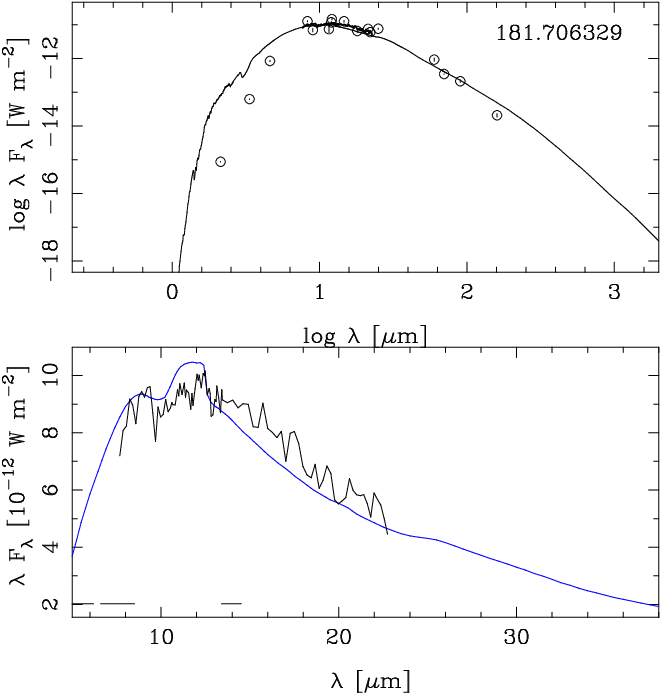}} 
\end{minipage}

\begin{minipage}{0.235\textwidth}
\resizebox{\hsize}{!}{\includegraphics[angle=-0]{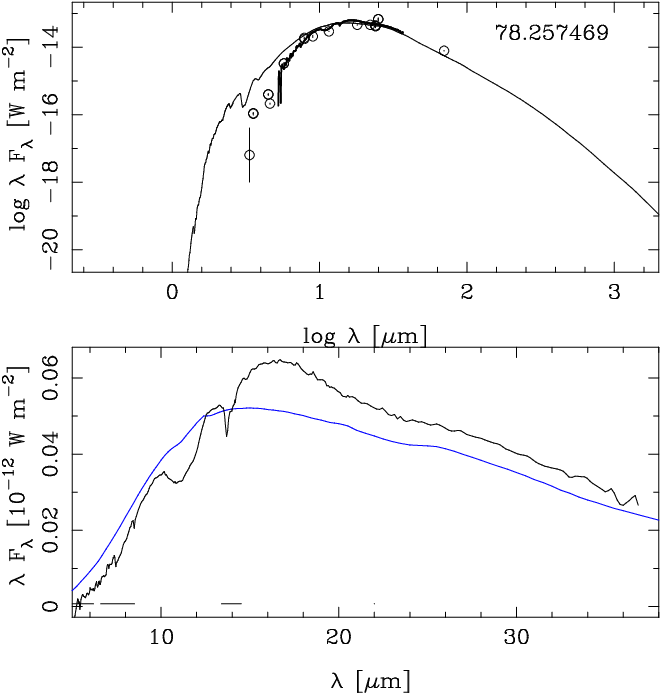}} 
\end{minipage}
\begin{minipage}{0.235\textwidth}
\resizebox{\hsize}{!}{\includegraphics[angle=-0]{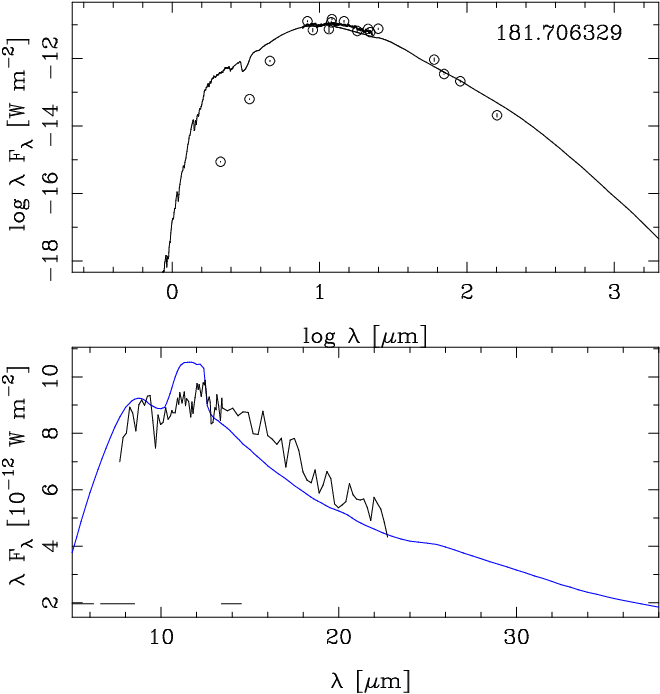}} 
\end{minipage}
\begin{minipage}{0.235\textwidth}
\resizebox{\hsize}{!}{\includegraphics[angle=-0]{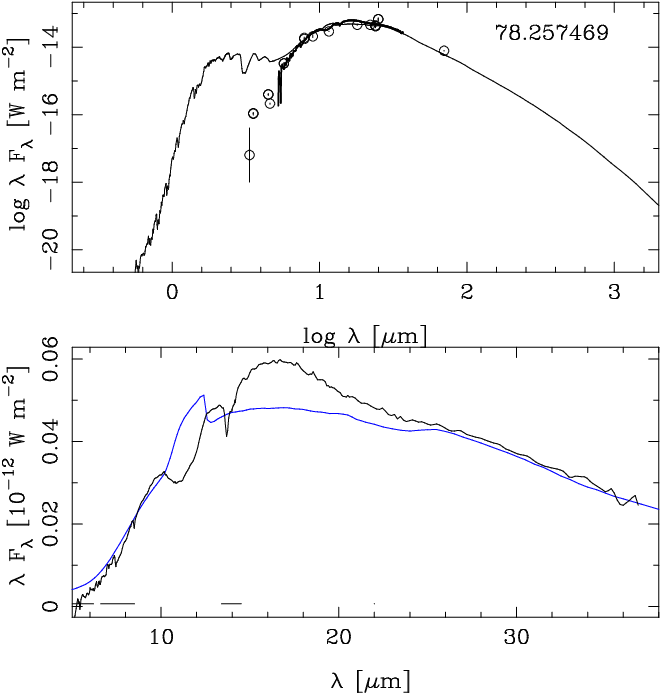}} 
\end{minipage}
\begin{minipage}{0.235\textwidth}
\resizebox{\hsize}{!}{\includegraphics[angle=-0]{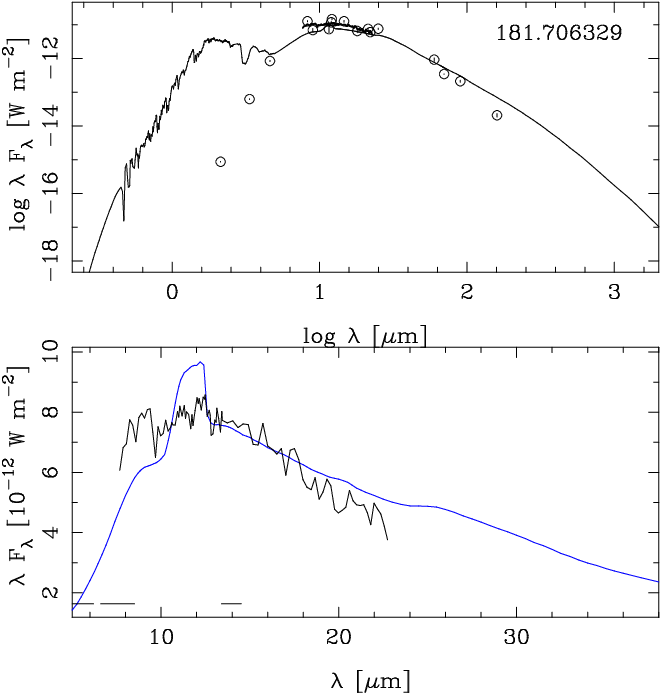}} 
\end{minipage}

\begin{minipage}{0.235\textwidth}
\resizebox{\hsize}{!}{\includegraphics[angle=-0]{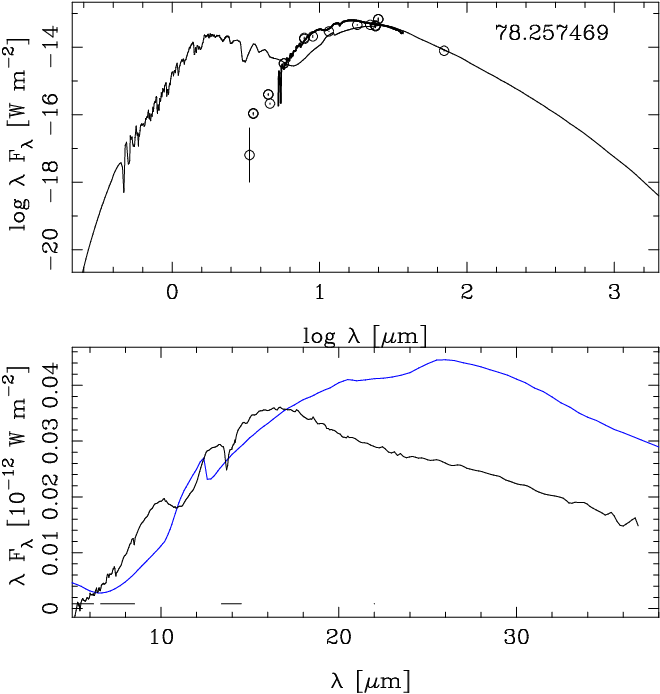}} 
\end{minipage}
\begin{minipage}{0.235\textwidth}
\resizebox{\hsize}{!}{\includegraphics[angle=-0]{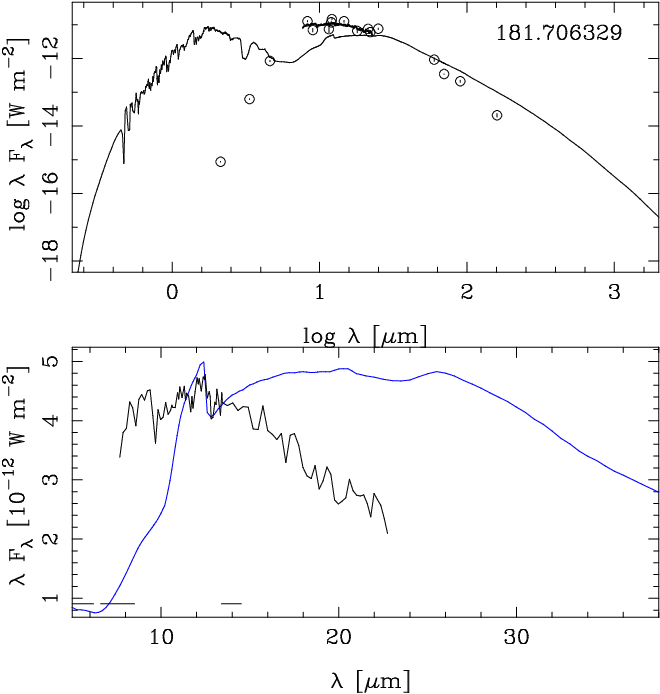}} 
\end{minipage}

\caption{Predicted SEDs and MIR spectra as a function time  for two stars (left-hand and right-hand panels) when
  the MLR stops abruptly and the CSE expands at 10~\ks. From top to bottom the SED at $t$=0, and after 30, 50, 200, and 500 years.
}
\label{Fig:PAGB}

\end{figure}

However, one issue with the interpretation of some of these stars evolving from the AGB into the P-AGB phase is the timescale.
To investigate this further the SEDs and MIR spectra were calculated for two objects under the assumption that the MLR drops abruptly to zero and that
the CSE then expands at a velocity of 10~\ks. The results are shown in Fig.~\ref{Fig:PAGB}. 
The models in Fig.~\ref{Fig:PAGB} were calculated for an effective temperatures typical for AGB stars (2600-2800~K).
The P-AGB models of \citet{M3B16M} indicate that the effective temperature of stars with an initial mass of 2 and 3~\msol\ is about 3800~K at an envelope
mass of 0.01~\msol. Similar calculation were done for $T_{\rm eff}= 3800$~K and are shown in Fig.~\ref{Fig:PAGB3800}. The differences are small.
The change in effective temperature at that phase of the evolution is also small, 0.22-0.52 K/year.
A first indication that the central star becomes slightly visible is present already after the order of 20-30 years.
When the dynamical time increases the central star becomes increasingly visible, until after about 500 years one has the classical SED of a P-AGB star
with a double-peaked SED. Important here is that the MIR bump remains bright and red, and so any selection of a sample based on MIR colours and magnitudes
would be relatively insensitive to the expanding shell. One would therefore expect more objects in the sample of (candidate) EROs which
show hints of a double-peaked SED, and this is not the case.

Recently \citet{DellAgli21} propose that binary interaction mechanisms that involve common envelope evolution (CEE) could be a possible explanation, and
that these stars could possibly hide binaries with orbital periods of the order of days. Their main argument is that single-star stellar evolution models combined
with dust formation models could not produce the location of the EROs in certain colour-colour diagrams, and that this
implies MLRs of 1-2 $\cdot 10^{-4}$ \msolyr\ or larger.
A binary scenario involving CEE might trigger the amount of dust to produce the observed colours.
For the 11 stars in Table~1 in \citet{DellAgli21}, MLRs of (0.7-3.3) $\cdot 10^{-5}$ \msolyr\ are found for nine in the present study, that is significantly lower
than $10^{-4}$ \msolyr\ (for our choice for the DTG ratio and expansion velocity).
For SSID 125 (ra=82.684006) and SSID 190 (ra=87.249886), very large MLRs of 1.0 and 1.8 $\cdot 10^{-4}$ \msolyr\ were indeed derived, respectively.
The MLR in the latter source is the largest for all (candidate) EROs in the MCs, and only two show larger MLRs in our Galaxy,
namely 2.1 $\cdot 10^{-4}$ \msolyr\ (ra=283.812347), and 3.1 $\cdot 10^{-4}$ \msolyr\ (ra=328.768372).

In the sample of EROs in the MCs that define the $PL$ relation, two objects were excluded as their periods (1884 and 3434 days but with large uncertainty)
and luminosity did not match the relation.
Similarly, among the Galactic ERO candidates, there are a few sources with (uncertain) periods in the range of 2000-5000~days where the $PL$ relation was not applied.
The longest period for which it was applied was about 2600~days. For longer periods, the implied luminosity would no longer be compatible with the AGB ($L \more 100~000~\lsol$).
For a few stars with shorter periods (1000-2000~days) located close to the Galactic plane, the $PL$ relation is also unlikely to be valid.
The implied luminosities from the $PL$ relation are compatible with the AGB, but they lead to large distances (\more 20~kpc) that imply large
reddenings ($A_{\rm V} \more 10$~mag) that are incompatible with the observed SEDs that show less extinction.

In summary, the nature of the EROs remain uncertain. Many show properties that are consistent with the properties expected for evolved AGB stars, but a
significant fraction of them do not. The P-AGB channel may apply to some, but the time evolution of an expanding (spherical) shell would predict more objects
with a classical double-peaked SED.
The CEE channel is interesting, but it will be hard to prove the predicted binary period of the order of days. The derived MLRs are in general lower than predicted,
although this depends on the assumed DTG ratio and expansion velocity, but in addition to the adopted MLR formalism in the stellar evolution models,
which \citet{DellAgli21} acknowledge might be too high.
The effect of non-spherical CSEs is also a realistic option that needs to be investigated.
Although challenging, high angular resolution observations in the MIR and the mm (with ALMA) might shed light on the morphology of the CSE.
For an typical ERO at a 3 kpc distance, the inner dust radius is predicted to be at about 10~mas, while the total CSE is of order 10\arcsec.

\subsection{MIR spectra of mass-losing carbon stars as a tracer of interstellar extinction}
\label{S-ISext}

As part of the fitting of the SEDs and MIR spectra, the observed photometry and spectra were corrected for IS extinction, including, in the MIR regime,
the local ISM model of \citet{Chiar06}, using a ratio of $A_{\rm V}/A_{\rm K}= 0.118$ to scale it to the adopted reddening law of \citet{Cardelli1989}
with the improvements by \cite{ODonnell_1994} from the UV to the NIR in MoD \citep{Gr_MOD}.
In this model, the extinction is smallest near 7.5~$\mu$m ($A_{\rm 7.5~\mu m}/A_{\rm K}= 0.38$) and has a peak at the silicate feature  ($A_{\rm 9.8~\mu m}/A_{\rm K} \approx 1.0$).
In other words, for $A_{\rm V} \approx 8.5$ the extinction becomes  $A_{\rm 9.8~\mu m} \approx 1.0$, which should be  noticeable in an MIR spectrum.

\begin{figure*}
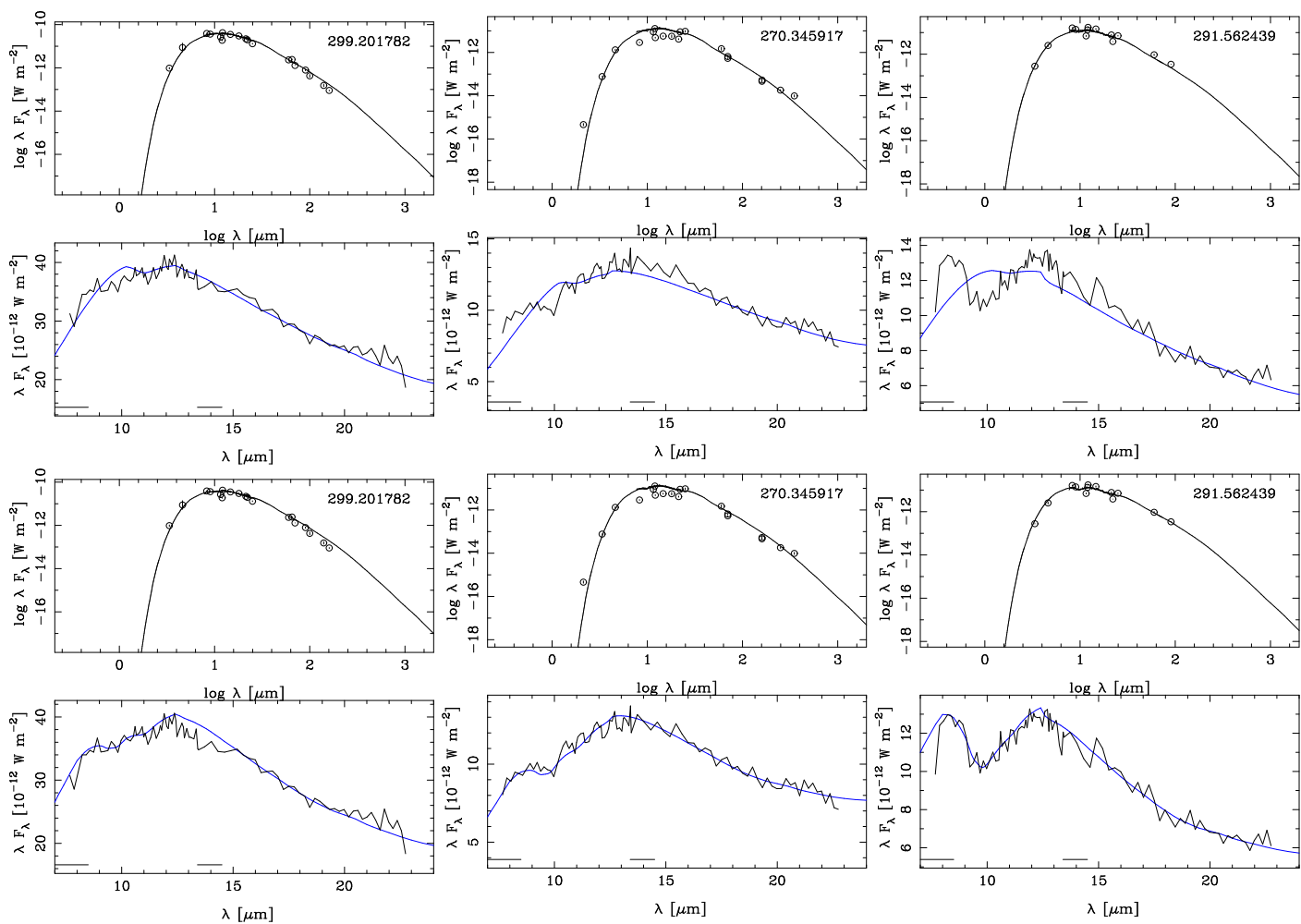


\begin{minipage}{0.33\textwidth}
\resizebox{\hsize}{!}{\includegraphics[angle=-0]{299.201782_sed.ps.Av_0.00}} 
\end{minipage}
\begin{minipage}{0.33\textwidth}
\resizebox{\hsize}{!}{\includegraphics[angle=-0]{270.345917_sed.ps.Av_0.00}} 
\end{minipage}
\begin{minipage}{0.33\textwidth}
\resizebox{\hsize}{!}{\includegraphics[angle=-0]{291.562439_sed.ps.Av_0.00}} 
\end{minipage}

\begin{minipage}{0.33\textwidth}
\resizebox{\hsize}{!}{\includegraphics[angle=-0]{299.201782_sed.ps.Av_2.42}} 
\end{minipage}
\begin{minipage}{0.33\textwidth}
\resizebox{\hsize}{!}{\includegraphics[angle=-0]{270.345917_sed.ps.Av_4.32}} 
\end{minipage}
 \begin{minipage}{0.33\textwidth}
\resizebox{\hsize}{!}{\includegraphics[angle=-0]{291.562439_sed.ps.Av_5.72}} 
\end{minipage}

\caption{Effect of 9.8~$\mu$m silicate IS extinction feature.
  The top panels show the fits to the SEDs and MIR spectra without IS extinction correction ($A_{\rm V}$ = 0), while the
  bottom panels show the fits for three stars with, from left to right, $A_{\rm V}$ = 2.4, 4.3, and 5.7~mag, respectively.
  The luminosity and optical depth were refitted in the case of $A_{\rm V}$ = 0 and that is the reason that the difference in the SEDs appears small.
}
\label{Fig:ISEXT}

\end{figure*}

Figure~\ref{Fig:ISEXT} shows this very clearly where the SEDs and MIR spectra are shown for three stars with estimated $A_{\rm V}$s
(see Appendix~\ref{App-sed} on how $A_{\rm V}$ was determined) of 2.4, 4.3, and 5.7~mag, respectively, and with no correction.
Not only is the IS 9.8~$\mu$m feature very evident, but the MIR spectrum is better fit over the entire wavelength range.
The difference in the SEDs in the upper panels appears quite small in the optical, and this is due to the fact that luminosity and optical depth were
  refitted in the models with no IS extinction.

The principle of using  MIR spectra to trace (MIR) interstellar extinction works for every type of star of course but the highly mass-losing C-stars
have an advantage as they are bright in the MIR and have no strong intrinsic features near 10~$\mu$m.
The same three stars without mass loss would be  4.0-4.6~mag fainter in the $N$-band.
The same would be the case when using normal (O-rich) red giants, while when using MIR brighter mass-losing O-rich stars one would have to distinguish between
the circumstellar and IS silicate features which would be extremely challenging.

\section{Final remarks}

The period analysis of the WISE/NEOWISE time series data of 1775 objects colour selected to include known C-rich objects with flat MIR continua or SiC in
absorption and known Galactic O-rich AGB stars with pulsation periods over 1000~days, supplemented with 217 AGB stars in the MCs previously studied
in the $K$-band, is presented.
In addition, periods from the literature and in many cases a reanalysis of time series data from other surveys is presented.
The SEDs and MIR spectra were modelled with a dust RT programme for a subset of 316 stars as well.
The results are presented in the seven subsections of Sect.~6.
This includes the detection of new C-rich EROs and new LPVs, of which 145 have periods $>$1000~days.

The nature of the C-rich EROs remains uncertain. Most of the objects in the MCs follow a $PL$ relation and this relation was used to estimate the distance
to the Galactic EROs with a period. However, a significant fraction of galactic sources show variability not consistent with an AGB nature
(implausibly long periods or no variability).
A P-AGB nature for some EROs is difficult to exclude, but given the lifetimes, one would expect a larger number of SEDs with double-peaked SEDs.
A possibility is that the shape of some of the SEDs is linked to an aspherical CSE, which is possibly linked to binarity.
High spatial resolution line and continuum observations in the MIR or (sub-)mm would be helpful to better characterise the CSE of some of these objects.

In general, 
distance determinations to these red sources are crucial as well as is investigating and understanding their nature, in particular for the
ERO sources. The $PL$ relation is the only available method at the moment, but this has clear limitations.
The targets studied here are ideal candidates for an IR astrometric mission \citep{Hobbs21}.
Of the 150 (candidate) EROs in our Galaxy about 85\% have $K <$20~mag.  %(126/150)
Going to even slightly redder wavelengths would be even more advantageous as all Galactic ERO candidates have $L< $14.3 for example.

\begin{acknowledgements}
The referee is thanked for pointing out the paper by \citet{Chen20}.
This research has made use of the SIMBAD database,  and
%
%This research has made use of
the VizieR catalogue access tool, operated at CDS, Strasbourg, France (DOI: 10.26093/cds/vizier).
The original description of the VizieR service was published in 2000, A\&AS 143, 23.
This publication makes use of data products from the Wide-field
Infrared Survey Explorer, which is a joint project of the University
of California, Los Angeles, and the Jet Propulsion Laboratory/California Institute of Technology, 
funded by the National Aeronautics and Space Administration.
This publication also makes use of data products from NEOWISE,
which is a project of the Jet Propulsion Laboratory/California Institute of Technology,
funded by the Planetary Science Division of the National Aeronautics and Space Administration.
Based on observations obtained with the Samuel Oschin 48-inch Telescope at the
Palomar Observatory as part of the Zwicky Transient Facility project.
ZTF is supported by the National Science Foundation under Grant No. AST-1440341
and a collaboration including Caltech, IPAC, the Weizmann Institute for Science,
the Oskar Klein Center at Stockholm University, the University of Maryland,
the University of Washington, Deutsches Elektronen-Synchrotron and Humboldt University,
Los Alamos National Laboratories, the TANGO Consortium of Taiwan,
the University of Wisconsin at Milwaukee, and Lawrence Berkeley National Laboratories.
Operations are conducted by COO, IPAC, and UW.

\end{acknowledgements}

\bibliographystyle{aa.bst}
\bibliography{references.bib}

\begin{appendix}

\section{Basic data and results from the period analysis}

Tables~\ref{WISEREALSam}, \ref{WISEREALPer}, \ref{noWISESambutREAL}, \ref{noWISEPerbutREAL}, \ref{WISESamnotREAL}, \ref{WISEPernotREAL}
contain the basic data and the results of the period analysis.

%%%%%%%%%%%%%%%%%%%%%%%%%%%%%%%%%%%%%%%%%%%%%%%%%%%%%%%%%%%%%%%%%%%%%%%%%%%%%%%%%%%%%%%%%%%%%%%%%%%%%%%%%%%%%%%%%%%%%%%%%%%%%%%%%%%%%%%%%%%%%%%%%%%
%/home/marting/art72/WISE/OTHERPeriods LPVSample_WISE_REAL.tex
% add tablefootnotes from,   tail  -50 LPVSample.tex

\longtab[1]{
\begin{landscape}
\footnotesize
\setlength{\tabcolsep}{1.5mm}
\begin{longtable}{rrllllllllllllllllllllllllllllllllllllllll}
\caption{\label{WISEREALSam} Sample of bona fide stellar sources with a period analysis based on WISE data, general information (first entries).}\\
\hline \hline
Ra & Dec  & Identifiers  & Dist & ObjType & 2M     &  GLP  & Aka  & MSX       &  MGL & HGL & $V_{\rm b}$ &  $V_{\rm r}$ & LRS &Spec. Type & Comment  \\
(deg) & (deg) &          & (\arcsec) &        & (\arcsec) & (\arcsec) & (\arcsec) & (\arcsec) &  (\arcsec) &  (\arcsec) & (\ks) & (\ks) & \\
\hline
\endfirsthead
\caption{continued.}\\
\hline\hline
Ra & Dec  & Identifiers  & Dist & ObjType & 2M     &  GLP  & Aka  & MSX       &  MGL & HGL & $V_{\rm b}$ &  $V_{\rm r}$ & LRS &Spec. Type & Comment \\
(deg) & (deg) &          & (\arcsec) &        & (\arcsec) & (\arcsec) & (\arcsec) & (\arcsec) &  (\arcsec) &  (\arcsec) & (\ks) & (\ks) & \\
\hline
\endhead
\hline
\endfoot
 19.463942  &  67.231448  & IRAS01144+6658 RAFGL190 &   0.22  &  C   &     &     & 1.1  & 1.7   &     &     &  nd  &  nd  & U &  &  \\
124.826307  & -21.737399  & IRAS08171-2134 RAFGL5250 &   0.30  &  C   & 0.3  &     & 0.1  &      &     &     &  -  &  -  & U &  &  \\
237.773837  & -56.890007  & IRAS15471-5644 &   0.20  &  Can.C   & 0.2  &     & 1.2  & 0.9   &     &     &  nd  &  nd  & U &  &  \\
287.486914  &   9.447611  & IRAS19075+0921 RAFGL2333 &   0.03  &  C   & 0.0  &     & 1.2  & 0.8   &     & 1.6  &  -  &  -  & U &  &  \\
323.345002  &  56.743063  & IRAS21318+5631 RAFGL5625S &   0.51  &  C   &     &     & 1.1  & 0.5   &     &     &  nd  &  nd  & U & C  &  \\
349.802533  &  17.192628  & IRAS23166+1655 RAFGL3068 &   0.33  &  C   & 0.3  &     & 0.7  &      &     &     &  -  &  -  &   & C  &  \\
 89.161446  & -67.892776  & IRAS05568-6753 &   0.23  &  C   &     &     & 1.9  & 3.4   &     &     &  -  &  -  &   &  &  \\
 82.407959  & -72.831322  & ERO0529379 IRAS05305-7251 &   0.17  &  C   &     &     & 0.7  & 1.4   &     &     &  -  &  -  &   &  & q(6) \\
 87.608788  & -69.934212  & ERO0550261 IRAS05509-6956 &   0.16  &  C   &     &     & 0.8  & 0.7   &     &     &  -  &  -  &   &  & q(6) \\
 79.701599  & -69.559563  & ERO0518484 IRAS05191-6936 &   0.30  &  C   &     &     & 1.5  & 3.1   &     &     &  -  &  -  &   &  &  \\
 87.249886  & -70.556229  & IRAS05495-7034 &   0.77  &  AGB   &     &     & 0.6  &      &     &     &  -  &  -  &   &  &  \\
 75.631233  & -68.093285  & ERO0502315 SSTISAGE1CJ050231.47-680535.9 &   0.04  &  AGB   &     &     & 1.1  & 3.3   &     &     &  -  &  -  &   &  &  \\
 76.023376  & -68.394501  & ERO0504056 IRAS05042-6827 &   0.16  &  C   &     &     & 1.1  &      &     &     &  -  &  -  &   &  &  \\
 81.419411  & -70.140877  & ERO0525406 IRAS05260-7010 &   0.15  &  C   &     &     & 0.4  & 2.3   &     &     &  -  &  -  &   &  &  \\
 78.257469  & -69.564110  & IRAS05133-6937 &   0.28  &  C   &     &     & 0.8  &      &     &     &  -  &  -  &   &  &  \\
 79.548790  & -70.507469  & ERO0518117 IRAS05187-7033 &   0.12  &  C   &     &     & 0.7  &      &     &     &  -  &  -  &   &  &  \\
 82.684006  & -71.716766  & IRAS05315-7145 &   0.32  &  AGB   &     &     & 0.6  &      &     &     &  -  &  -  &   &  & q(6) \\
281.468597  &  -1.778703  & IRAS18432-0149 OH30.7+0.4 &   1.31  &  OH/IR   &     &     & 1.3  & 0.8   &     & 1.3  & 49.5 & 84.1 & A &  &  \\
282.174774  &  -2.841357  & IRAS18460-0254 OH30.1-0.7 RAFGL5535 &   0.06  &  OH/IR   &     &     & 1.0  & 1.0   &     & 1.2  & 78.5 & 118.9 &   &  &  \\
282.859344  &  -1.064581  & IRAS18488-0107 OH32.0-0.5 &   0.36  &  OH/IR   &     &     & 0.6  & 0.8   &     & 0.7  & 55.2 & 95.8 & A &  &  \\
283.092775  &  -0.236713  & IRAS18498-0017 OH32.8-0.3 RAFGL5540 &   0.37  &  OH/IR   & 0.4  &     & 0.8  & 1.0   &     & 1.5  & 45.6 & 76.0 & H &  &  \\
287.284637  &   8.276116  & GPSR042.309-0.133 &   0.78  &  Radio   &     &     & 1.7  & 1.2   &     & 0.3  & 42.3 & 75.3 &   &  &  \\
 23.463343  &  62.448162  & IRAS01304+6211 OH127.8-0.0 RAFGL230 &   0.19  &  OH/IR   & 0.2  &     & 0.2  & 0.2   &     &     & -65.2 & -43.0 & A &  &  \\
 53.377491  &  60.335957  & IRAS03293+6010 OH141.7+3.5 RAFGL5097 &   0.61  &  Mira   & 0.1  &     & 0.8  &      &     &     & -69.1 & -44.9 & A &  &  \\
 79.197746  &  45.567860  & IRAS05131+4530 RAFGL712 &   0.22  &  OH/IR   & 0.2  &     &     &      &     &     &  nd  &  nd  & F &  &  \\
111.101738  & -20.198791  & IRAS07222-2005 &   0.45  &  Star   & 0.4  &     & 0.6  & 1.5   &     &     & 76.9 & 93.2 & E & M9:  &  \\
266.099701  & -31.927624  & IRAS17411-3154 RAFGL5379 &   4.13  &  OH/IR   &     &     &     & 0.7   &     & 0.6  & -39.2 & -2.7 &   &  &  \\
280.309784  &  -6.250149  & IRAS18385-0617 OH26.2-0.6 &   0.22  &  OH/IR   & 0.2  &     & 2.0  & 1.3   &     & 0.9  & 49.8 & 93.0 &   &  &  \\
277.128905  &  -9.970644  & IRAS18257-1000 OH21.5+0.5 &   0.00  &  OH/IR   &     &     & 1.1  & 0.7   &     & 1.8  & 97.6 & 134.0 &   &  &  \\
 51.285198  &  65.535362  & IRAS03206+6521 OH138.0+7.2 RAFGL5093 &   0.35  &  OH/IR   & 0.4  &     & 0.6  &      &     &     & -46.4 & -27.6 & A &  &  \\
  0.441592  &  62.748863  & IRAS23592+6228 &   2.27  &  Star   & 0.1  &     & 0.9  & 1.4   &     &     &  -  &  -  & F &  &  \\
  3.853554  &  54.906147  & IRAS00127+5437 &   0.21  &  Star   & 0.1  &     & 0.2  &      &     &     &  -  &  -  & E &  &  \\
  4.963700  &  65.991821  & IRAS00170+6542 &   0.20  &  Can.PAGB   & 0.1  &     & 1.2  & 1.5   &     &     & -65.0 & -37.6 & E &  &  \\
\hline
\end{longtable}
\tablefoot{
Column~16. References for extragalactic catalogues
(‘Q’ marks probable QSOs, ‘q’ marks candidate QSOs).
(1) \citet{Flesch15},  
(2) \citet{BailerJ19},
(3) \citet{Guo18},
(4) \citet{Assef18},
(5) \citet{GEDR3_Klioner}, and
(6) \citet{Solarz17}.
None of the sources is listed in the QSO catalogues of \citet{Veron10}, \citet{Gattano18}, and \citet{Lyke20}.
}
\end{landscape}
}

%%%%%%%%%%%%%%%%%%%%%%%%%%%%%%%%%%%%%%%%%%%%%%%%%%%%%%%%%%%%%%%%%%%%%%%%%%%%%%%%%%%%%%%%%%%%%%%%%%%%%%%%%%%%%%%%%%%%%%%%%%%%%%%%%%%%%%%%%%%%%%%%%%%%%%%%%%%%%%
%/home/marting/art72/WISE/OTHERPeriods LPVPeriods_WISE_REAL.tex
% tabelfootnotes from COMMENTS.dat  EN insert /home/marting/art72/WISE/OTHERPeriods/Comments.tex
%

\longtab[2]{
\begin{landscape}
%\footnotesize
\scriptsize
\setlength{\tabcolsep}{1.3mm}
\begin{longtable}{rrllrrrrccrclccrllllllllllllllllllllllll}
\caption{\label{WISEREALPer}  Sample of bona fide stellar sources with a period analysis based on WISE data, period information (first entries).}\\
\hline \hline
\centering
Ra & Dec  & Periods literature & P refit & $P1$ & $\sigma_{\rm P1}$ & $P2$ & $\sigma_{\rm P2}$ &  Amp1 & $\sigma_{\rm A1}$ &  $W1$ & $\sigma_{\rm W1}$ & $\chi_{\rm r,W1}^2$ & Amp2 & $\sigma_{\rm A2}$ &  $W2$ & $\sigma_{\rm W2}$ & $\chi_{\rm r,W2}^2$ & Com & Class \\
(deg) & (deg) &  (d) &  (d)    & (d) & (d) & (d) & (d) & (mag) & (mag) & (mag) & (mag) & & (mag) & (mag) & (mag) & (mag) & \\
\hline
\endfirsthead
\caption{continued.}\\
\hline\hline
Ra & Dec  & Periods literature & P refit & $P1$ & $\sigma_{\rm P1}$ & $P2$ & $\sigma_{\rm P2}$ & Amp1 & $\sigma_{\rm A1}$ &  $W1$ & $\sigma_{\rm W1}$ & $\chi^2_{\rm r,W1}$ & Amp2 & $\sigma_{\rm A2}$ &  $W2$ & $\sigma_{\rm W2}$ & $\chi_{\rm r,W2}^2$ & Com \\
(deg) & (deg) &  (d)  &  (d)   & (d) & (d) & (d) & (d) & (mag) & (mag) & (mag) & (mag) & & (mag) & (mag) & (mag) & (mag) & \\
\hline
\endhead
\endfoot
 19.463942  &  67.231448  & 1060 (24) &  & 1053 &  4.1 & 1042 &  4.2 & 1.85 & 0.13 &  8.874 &  0.019 &  139.9  & 1.52 & 0.05 &  4.767 &  0.026 &    2.2 &  & ERO LPV & \\
124.826307  & -21.737399  &  &  &  944 &  7.4 &  937 &  5.8 & 1.52 & 0.09 &  7.251 &  0.059 &   50.0  & 1.39 & 0.08 &  3.811 &  0.047 &    2.1 &  & ERO LPV & \\
237.773837  & -56.890007  & no (8) 453 (3) & 925 $\pm$ 3 (43) &  942 &  7.6 &  974 & 11.4 & 1.43 & 0.09 &  7.321 &  0.057 &   42.6  & 1.49 & 0.15 &  3.718 &  0.061 &    4.2 &  & ERO LPV & \\
287.486914  &   9.447611  &  &  & 1068 &  8.9 & 1075 &  9.7 & 1.19 & 0.06 &  6.144 &  0.078 &    9.7  & 1.55 & 0.12 &  2.496 &  0.062 &    1.9 &  & ERO LPV & \\
323.345002  &  56.743063  &  &  &  924 &  4.3 &  962 &  9.6 & 1.14 & 0.04 &  7.390 &  0.022 &   40.1  & 1.51 & 0.14 &  3.700 &  0.066 &    7.0 &  & ERO LPV & \\
349.802533  &  17.192628  & 696. (28) 520? (26) 700. (22) &  &  747 &  4.6 &  705 & 674.4 & 0.89 & 0.04 &  5.283 &  0.032 &    1.7  & 0.63 & 2.01 &  1.813 &  2.042 &    6.0 &  & ERO LPV & \\
 89.161446  & -67.892776  & 1197 $\pm$ 32 (1) 1209 (2) &  & 1223 &  1.2 & 1217 &  1.3 & 1.24 & 0.04 & 11.978 &  0.003 &   37.3  & 1.02 & 0.05 &  8.971 &  0.002 &   80.4 &  & ERO LPV & \\
 82.407959  & -72.831322  & 680 (2) &  &  679 &  1.0 &  677 &  1.1 & 0.62 & 0.01 & 14.119 &  0.008 &    7.9  & 0.61 & 0.02 & 10.709 &  0.006 &   38.6 &  & ERO LPV & \\
 87.608788  & -69.934212  & 1052 (2) &  & 1104 &  4.1 & 1113 &  2.6 & 0.65 & 0.01 & 15.485 &  0.012 &    2.8  & 0.72 & 0.03 & 11.466 &  0.005 &   35.4 &  & ERO LPV & \\
 79.701599  & -69.559563  & no (2) &  &      &      &      &      & 0.00 & 0.00 & 15.307 &  0.070 &    0.3  & 0.00 & 0.00 & 12.700 &  0.015 &   10.2 &  & ERO & \\
 87.249886  & -70.556229  & no (2) &  & 4453 & 1288 & 3429 & 98.6 & 0.06 & 0.05 & 15.502 &  0.011 &    2.1  & 0.12 & 0.01 & 13.502 &  0.007 &    1.6 &  & ERO & \\
 75.631233  & -68.093285  & no (2) &  & 1739 & 111.8 & 1890 & 20.7 & 0.76 & 0.26 & 15.662 &  0.125 &   19.8  & 0.31 & 0.02 & 13.074 &  0.007 &    4.4 &  & ERO LPV & \\
 76.023376  & -68.394501  & no (2) &  &      &      &      &      & 0.00 & 0.00 & 16.213 &  0.140 &    0.2  & 0.00 & 0.00 & 13.194 &  0.007 &    4.5 &  & ERO & \\
 81.419411  & -70.140877  & no (2) &  &      &      &      &      & 0.00 & 0.00 & 15.703 &  0.351 &    4.4  & 0.00 & 0.00 & 13.313 &  0.016 &    5.2 &  & ERO & \\
 78.257469  & -69.564110  & no (2) &  &      &      & 9094 & 3214 & 0.00 & 0.00 & 15.547 &  0.087 &    0.1  & 0.28 & 0.15 & 13.871 &  0.046 &    1.2 &  & ERO & \\
 79.548790  & -70.507469  & no (2) &  & 8188 & 2519 & 4666 & 81.4 & 0.63 & 0.34 & 14.815 &  0.285 &    2.5  & 0.48 & 0.01 & 11.941 &  0.004 &    4.2 &  & ERO LPV & \\
 82.684006  & -71.716766  & no (2) &  &  969 & 16.4 &  944 & 13.8 & 0.05 & 0.02 & 13.812 &  0.004 &    2.8  & 0.04 & 0.00 & 11.957 &  0.003 &    2.4 &  & ERO & \\
281.468597  &  -1.778703  & 1140 $\pm$ 30 (31) &  & 1176 & 43.1 & 1125 & 17.4 & 1.05 & 0.10 &  5.754 &  0.184 &   31.7  & 1.35 & 0.10 &  2.518 &  0.106 &    0.5 &  & LPV & \\
282.174774  &  -2.841357  & 2013 $\pm$ 243 (32) 1730 $\pm$ 200 (31) 2173 (30) &  & 1952 & 45.9 & 1524 & 16.8 & 0.67 & 0.06 &  7.430 &  0.031 &    8.0  & 0.88 & 0.08 &  2.966 &  0.053 &    1.3 &  & LPV & \\
282.859344  &  -1.064581  & 1417 $\pm$ 108 (32) 1519 (30) no (4) &  & 1632 & 14.0 & 1605 & 21.6 & 0.89 & 0.07 &  7.506 &  0.060 &    9.0  & 1.34 & 0.12 &  3.400 &  0.068 &    2.5 &  & LPV & \\
283.092775  &  -0.236713  & 1539 $\pm$ 31 (32) 1750 $\pm$ 130 (31) 1691 (30) no (4) &  & 1436 & 32.6 & 1666 & 25.8 & 0.67 & 0.07 &  7.296 &  0.048 &   10.5  & 1.32 & 0.13 &  2.694 &  0.087 &    2.8 &  & LPV & \\
287.284637  &   8.276116  &  &  & 2475 & 69.6 & 1976 & 51.4 & 0.68 & 0.09 &  7.659 &  0.056 &   73.0  & 0.88 & 0.15 &  4.566 &  0.100 &    5.0 &  & LPV & \\
 23.463343  &  62.448162  & 759 (12) 1540 $\pm$ 16 (21) 1638 $\pm$ 57 (32) 1592 (30) &  &      &      &      &      & 0.00 & 0.00 &  3.864 &  0.024 &    2.1  & 0.00 & 0.00 &  1.445 &  0.054 &    0.5 &  &  & \\
 53.377491  &  60.335957  & 1800 $\pm$ 400 (25) 2210 (30) &  & 1445 & 45.2 & 1833 & 293.3 & 0.62 & 0.09 &  4.502 &  0.097 &    2.4  & 0.33 & 0.29 &  2.418 &  0.152 &    5.7 &  & LPV & \\
 79.197746  &  45.567860  & 1100 $\pm$ 100 (25) 1051 (30) &  & 1182 & 16.0 & 1133 & 35.3 & 1.05 & 0.09 &  4.439 &  0.057 &    2.0  & 0.72 & 0.16 &  2.404 &  0.099 &    3.6 &  & LPV & \\
111.101738  & -20.198791  & 5746/1411 (6) 1200 $\pm$ 200 (25) 7 (4) &  & 1447 & 24.3 & 1488 & 21.0 & 0.62 & 0.10 &  4.016 &  0.104 &    0.8  & 1.09 & 0.11 &  2.999 &  0.082 &    3.8 &  & LPV & \\
266.099701  & -31.927624  & 1440: (25) 1440 (33) 1 (3) & 1766 $\pm$ 47 (43) & 2054 & 58.6 &      &      & 0.72 & 0.11 &  4.086 &  0.134 &    0.8  & 0.00 & 0.00 &  0.708 &  0.106 &    0.0 & 1 & PER & \\
280.309784  &  -6.250149  & 1172 $\pm$ 94 (32) 1330 $\pm$ 50 (31) &  & 1372 & 22.3 & 1206 & 66.3 & 0.62 & 0.09 &  4.485 &  0.120 &    0.4  & 0.48 & 0.16 &  2.468 &  0.132 &    2.4 &  & LPV & \\
277.128905  &  -9.970644  & 1785 $\pm$ 114 (32) &  & 1548 & 96.5 & 1702 & 67.1 & 1.59 & 0.19 &  8.564 &  0.142 &   99.5  & 1.64 & 0.15 &  3.848 &  0.075 &    3.4 &  & LPV & \\
 51.285198  &  65.535362  & 1276 $\pm$ 150 (32) 1410 (30) &  & 1431 & 26.2 & 1300 & 77.7 & 0.53 & 0.05 &  4.075 &  0.059 &    0.8  & 0.30 & 0.12 &  2.009 &  0.103 &    1.5 &  & LPV & \\
  0.441592  &  62.748863  &  &  &  731 &  1.5 &  720 &  3.8 & 1.02 & 0.03 &  5.201 &  0.015 &    0.9  & 1.06 & 0.07 &  3.209 &  0.036 &    2.5 &  & LPV & \\
  3.853554  &  54.906147  &  &  &  613 &  6.4 &  627 &  6.4 & 0.46 & 0.04 &  3.738 &  0.033 &    0.6  & 0.59 & 0.12 &  2.268 &  0.079 &    2.9 &  & LPV & \\
  4.963700  &  65.991821  &  &  &      &      &      &      & 0.00 & 0.00 &  3.746 &  0.037 &    2.8  & 0.00 & 0.00 &  2.475 &  0.047 &    3.1 &  & nvoh & \\

\hline
\end{longtable}
\tablefoot{
References for Cols~3-4:
(1)= \citet{Gr20},
(2)= \citet{GS18},
(3)= \citet{FerreiraLopes20},
(4)= \citet{Shappee14,Kochanek17},
(5)= \citet{GAIADR2},
(6)= \citet{Heinze18},
(7)= \citet{Price10},
(8)= \citet{Whitelock06},
(9)= \citet{Pojmanski02},
(10)= \citet{ContrerasPena17}
(11)= \citet{AlfonsoGarzon12}
(12)= \citet{Urago20}
(13)= \citet{Wozniak04}
(14)= \citet{Drake09}
(20)= \citet{Kiss07}
(21)= \citet{Suh02}
(22)= \citet{Whitelock94}
(23)= \citet{Kerschbaum06}
(24)= \citet{Groenewegen98}
(25)= \citet{Jimenez06}
(26)= \citet{Jones90}
(27)= \citet{LeBertre93}
(28)= \citet{LeBertre92}
(29)= \citet{Nakashima2000}
(30)= \citet{Engels15}
(31)= \citet{Engels83}
(32)= \citet{vanLangevelde90}
   (33)= \citet{Olivier2001}
(41)= Refit of VMC $K$-band data from \citet{Gr20} with initial period from the present work,
(43)= Refit of VVV $K$-band \citep{FerreiraLopes20},
(44)= Refit of ASAS-SN $V$-band data \citep{Shappee14,Kochanek17}.
(46)= Refit of ATLAS $o$-band data \citep{Heinze18}.
(47)= Refit of DIRBE 4.9~$\mu m$ data \citep{Price10}
      (For the source with Ra = 204.411850 data at 3.5~$\mu m$ data is used).
(51)= Fit of OMC data \citep{AlfonsoGarzon12}.
(54)= Fit of CSS data \citep{Drake09}.
(63)= Fit to the data from \citet{Kerschbaum06} with $K$-band photometry from the literature added.
(70)= Fit to data from the Bochum Galactic Disk Survey \citep{Hackstein15}.
(71)= Fit to $r$-type data from the ZTF \citep{Masci19}.
\\
Column.~19, Comments:
(1) W2 data is dummy.
(2) Possibly a real source. IRAS 02408+5458 is located at 3\arcsec.
(3) ASAS-SN: LC is peculiar.
(4) Palt = 572.
(5) ZTF: LC is peculiar.
(6) Clones of VY CMa (110.7430362 -25.7675659).
(7) Source is getting fainter with time.
(8) Clones of IRAS 08074-3615 (122.334335 -36.407444).
(9) Clones of HD 76220 (133.849921103 +19.6998172)?
(10) Palt = 262.
(11) Peculiar LC.
(12) Palt = 491.
(13) Source is getting brighter with time.
(14) A clone of CW Leo (146.989203 +13.278759).
(15) Source is getting brighter with time.
(16) Clones of HD 97300 (167.4575881 -76.6130701).
(17) Possibly a clone of IRAS 11145-6534, located at 75\arcsec.
(18) ASAS-SN: bizarre LC.
(19) Possibly a real source. Not an obvious clone.
(20) A clone of IRAS 15194-5115 (230.7711108 -51.4329088).
(21) Source (W2) is getting fainter with time.
(22) Source (W2) is getting fainter with time.
(23) Palt = 580, 780.
(24) Palt = 1280.
(25) Palt = 800.
(26) Source is getting fainter with time.
(27) Palt = 303.
(28) Palt = 460.
(29) $P$= 34.48 $\pm$ 0.16 days.
(30) Palt = 460.
(31) A clone of IRC +20 326 (262.9805754 +17.7558379).
(32) Palt = 480, 730.
(33) W1 data is dummy.
(34) Palt = 314.
(35) Palt = 969.
(36) Sakurai's object. Continueously brightening in $K$ (VVV) and WISE filters.
(37) Palt = 460,301.
(38) Palt = 355.
(39) A clone of AFGL 2135 (275.644226 -27.108224).
(40) ASAS-SN: Burst, then nearly constant.
(41) Source is getting fainter with time.
(42) ZTF: peculiar LC (period fixed to 600 days).
(43) ATLAS: Slowly getting brighter, with burst (?).
(44) ZTF: very peculiar LC.
(45) Palt = 1900.
(46) ATLAS, ZTF: peculiar LC. Strong drop in brightness.
(47) Source (W1) is getting brighter with time.
(48) Palt = 330.
(49) Clones of IRC +10 420 (291.700408 +11.354634).
(50) OMC, ZTF: Slow change in both datasets.
(51) ZTF: very peculiar LC.
(52) $P$= 30.58 $\pm$ 0.02 days.
(53) ASAS-SN: constant but then strong brightening.
(54) ZTF: very peculiar LC.
(55) Clones of AFGL 3068 (349.802533 +17.192628).
(56) A clone of AFGL 3116 (353.61467 +43.55036).

}
\end{landscape}
}

%%%%%%%%%%%%%%%%%%%%%%%%%%%%%%%%%%%%%%%%%%%%%%%%%%%%%%%%%%%%%%%%%%%%%%%%%%%%%%%%%%%%%%%%%%%%%%%%%%%%%%%%%%%%%%%%%%%%%%%%%%%%%%%%%%%%%%%%%%%%%%%%%%%%%%%%%%%%%%%%%%%%%%%%%%%%%%%%%%%%%%%%%%%%%%%%%%%%%
%/home/marting/art72/WISE/OTHERPeriods LPVSample_noWISE_REAL.tex
%
% EDIT spectral types to fit page width

\longtab[3]{
\begin{landscape}
\footnotesize
\setlength{\tabcolsep}{1.3mm}
\begin{longtable}{rrllllllllllllllllllllllllllllllllllllllll}
\caption{\label{noWISESambutREAL} Sample of bona fide stellar sources with a period analysis based on literature data, general information.}\\
\hline \hline
\centering
Ra & Dec  & Identifiers  & Dist & ObjType & 2M     &  GLP  & Aka  & MSX       &  MGL & HGL & $V_{\rm b}$ &  $V_{\rm r}$ & LRS &Spec. Type & Comment   \\
(deg) & (deg) &          & (\arcsec) &        & (\arcsec) & (\arcsec) & (\arcsec) & (\arcsec) &  (\arcsec) &  (\arcsec) & (\ks) & (\ks) & \\
\hline
\endfirsthead
\caption{continued.}\\
\hline\hline
Ra & Dec  & Identifiers  & Dist & ObjType & 2M     &  GLP  & Aka  & MSX       &  MGL & HGL & $V_{\rm b}$ &  $V_{\rm r}$ & LRS &Spec. Type & Comment \\
(deg) & (deg) &          & (\arcsec) &        & (\arcsec) & (\arcsec) & (\arcsec) & (\arcsec) &  (\arcsec) &  (\arcsec) & (\ks) & (\ks) & \\
\hline
\endhead
\endfoot
353.614540  &  43.550311  & IRAS23320+4316 IRC+40540 RAFGL3116 &   0.39  &  C   & 0.4  &     & 0.8  &      &     &     &  -  &  -  & C & C C8,3.5e J M8+  &  \\
284.625305  &   6.715967  & IRAS18560+0638 RAFGL2290 &   0.39  &  OH/IR   & 0.4  &     & 0.5  & 2.3   &     &     & 2.4 & 34.8 & A &  &  \\
334.864610  &  59.856052  & IRAS22177+5936 NSV25875 RAFGL2885 &   0.22  &  OH/IR   & 0.2  &     & 1.0  & 2.6   &     &     & -39.8 & -10.0 & A &  &  \\
279.385437  &  -5.399753  & IRAS18348-0526 OH26.5+0.6 RAFGL2205 &   0.10  &  OH/IR   & 0.1  &     &     & 1.0   &     & 0.5  & 12.8 & 39.9 & A &  &  \\
  6.921023  &  69.647507  & IRAS00247+6922 RAFGL67 &   0.75  &  C   & 0.6  &     & 0.6  &      &     &     &  -  &  -  & C & C:  &  \\
 75.878654  & -69.966820  & IRAS05039-7002 &   0.48  &  Star   &     &     & 0.5  &      &     &     &  -  &  -  &   &  &  \\
 77.845406  & -11.848938  & IRAS05090-1154 RAFGL702 RXLep &   0.66  &  AGB   & 0.7  &     & 0.2  &      &     &     &  nd  &  nd  &   & M6III M7 M7II:  &  \\
 79.323730  &  53.586071  & IRAS05132+5331 RAFGL715 RAur &   0.19  &  S   & 0.4  &     & 1.5  &      &     &     &  nd  &  nd  & E & M10 M6+e M7.5Se  M9e  &  \\
 82.525955  & -70.511375  &  &  &  &     &     & 1.5  & 1.3   &     &     &  -  &  -  &   &  & q(6) \\
 85.957069  &  32.701687  & IRAS05405+3240 RAFGL809 &   0.15  &  C   & 0.3  &     & 0.7  & 1.5   &     &     &  nd  &  nd  & C &  &  \\
 86.374641  &  29.118120  & IRAS05423+2905 &   0.24  &  OH/IR   & 0.2  &     & 0.4  & 0.5   &     &     &  -  &  -  & E & M10.5 M9  &  \\
 88.954704  &  20.175150  & IRAS05528+2010 RAFGL837 UOri &   0.58  &  Mira   & 0.5  &     & 0.5  & 2.4   &     &     & -41.2 & -40.7 & E & M6e M7e M8+e M9.5  &  \\
 97.072632  & -13.053068  & HD45677 IRAS06259-1301 RAFGL5195 &   0.17  &  Be*   & 0.2  &     & 0.6  &      &     &     &  nd  &  nd  & E & B0e B2e  B2IV/V[e]  &  \\
 97.397789  &   8.788045  & IRAS06268+0849 RAFGL5196 V477Mon &   0.58  &  C   & 0.5  &     & 1.3  & 1.9   &     &     &  -  &  -  & C & C  & q(2) \\
 99.136795  &  38.445480  & IRAS06331+3829 RAFGL966 UUAur &   0.13  &  C   & 0.2  &     & 0.3  &      &     &     &  -  &  -  & C & C5,3 C6,4 C7,4 N0 N3  &  \\
 99.226021  &   3.424734  & IRAS06342+0328 RAFGL971 &   0.36  &  C   & 0.4  &     & 1.5  & 3.2   &     &     &  -  &  -  & C & C5,9e  & q(2) \\
105.929901  & -11.551617  & IRAS07013-1128 RAFGL1059 ZCMa &   0.48  &  Ae   & 0.5  &     & 0.4  & 0.8   &     &     &  -  &  -  & U & Beq Bneq+F F8III/IV[e]  &  \\
108.111450  &   8.517594  & IRAS07097+0836 &   0.06  &  IR   &     &     & 0.2  &      &     &     &  -  &  -  &   &  &  \\
111.014603  & -12.874392  & IRAS07217-1246 RAFGL5230 &   0.50  &  C   & 0.5  &     & 1.0  & 0.4   &     &     &  -  &  -  & C & C  &  \\
112.697861  &  -9.776847  & IRAS07284-0940 RAFGL1135 UMon &   0.24  &  RVTau   & 0.3  &     & 0.3  & 0.9   &     &     &  -  &  -  & E & F8Ibe G0I G2Ie G5 G8 K2  &  \\
118.454765  & -53.920090  &  &  &  &     &     &     &      &     &     &  -  &  -  &   &  &  \\
139.178528  & -47.939484  & [LLN92]IRS49-1 &   1.60  &  IR   &     &     &     & 3.1   &     &     &  -  &  -  & H &  &  \\
148.891602  & -77.287849  & HD86774 IRAS09555-7702 &   0.48  &  Star   & 0.3  &     & 0.5  &      &     &     &  -  &  -  &   & M1III  &  \\
149.819809  & -56.911324  &  &  &  &     &     & 1.0  & 3.1   &     &     &  -  &  -  &   &  &  \\
159.787247  & -77.956032  & EZCha IRAS10383-7741 &   0.35  &  LPV   & 0.4  &     & 1.0  &      &     &     &  -  &  -  & E & M5/7  &  \\
161.142731  & -60.092995  &  &  &  &     &     & 1.6  & 2.6   &     &     &  -  &  -  &   &  &  \\
182.255554  & -63.264767  &  &  &  &     &     & 2.6  & 4.8   &     & 3.4  &  nd  &  nd  & H &  &  \\
184.390961  & -67.960808  & epsMus IRAS12148-6741 RAFGL4149 &   2.53  &  LPV   &     &     & 1.1  &      &     &     &  -  &  -  & S & M3 M3/5III M4III M5III  &  \\
205.221130  & -61.762733  & IRAS13374-6130 &   1.50  &  HII   &     &     & 1.5  &      &     &     &  -  &  -  &   &  &  \\
205.756912  & -62.147499  & IRAS13395-6153 RAFGL4176 &   0.39  &  IR   & 0.4  &     &     & 0.4   &     & 0.0  &  -  &  -  &   &  &  \\
207.907654  & -61.652130  & IRAS13481-6124 &   0.20  &  YSO   & 0.2  &     & 0.6  & 0.6   &     & 3.1  &  -  &  -  & A &  &  \\
211.898087  & -61.455872  & [HJF2013]G311.899+00.083 &   1.02  &  denseCore   & 0.6  & 0.4  &     &      &     &     &  -  &  -  & H &  &  \\
216.240234  & -60.378918  & PMNJ1424-6022 &   2.17  &  HII   &     & 3.0  & 1.7  & 1.9   &     &     &  -  &  -  & H &  &  \\
220.506821  & -60.506161  & [MHL2007]G316.1386-00.50091 &   2.06  &  Can.YSO   &     & 2.6  &     & 2.1   &     &     &  -  &  -  & H &  &  \\
227.501114  & -58.294167  &  &  &  &     &     & 0.9  & 4.6   &     &     &  -  &  -  & H &  &  \\
237.410202  & -54.639774  & IRAS15457-5429 &   2.82  &  HII   &     &     &     & 2.4   &     & 4.3  &  -  &  -  & H &  &  \\
237.682434  & -53.345341  & IRAS15469-5311 &   0.18  &  PAGB   & 0.3  &     & 0.4  & 0.3   &     & 1.6  & -45.8 & -36.3 & F &  &  \\
240.578812  & -52.924671  & IRAS15584-5247  &   2.32  &  HII   &     & 1.4  & 2.8  & 2.7   &     &     &  -  &  -  & P &  &  \\
242.593582  & -52.100525  & MSX5CG330.8849-00.3703 &   2.39  &  IR   &     &     & 2.0  &      &     &     &  -  &  -  & H &  &  \\
242.660965  & -49.097557  & SFO76 &   0.85  &  Cloud   &     &     &     & 1.3   &     &     &  -  &  -  & P &  &  \\
242.891373  & -48.330719  & IRAS16079-4812 &   0.35  &  C   & 0.3  &     & 0.6  & 0.4   &     &     &  -  &  -  & C & C  &  \\
244.379456  & -50.541546  & IRAS16137-5025 &   2.20  &  HII   &     &     &     & 3.0   &     & 3.6  &  -  &  -  & P &  &  \\
\hline
\end{longtable}
\tablefoot{
  See the notes in Table~\ref{WISEREALSam}.
}
\end{landscape}
}

%%%%%%%%%%%%%%%%%%%%%%%%%%%%%%%%%%%%%%%%%%%%%%%%%%%%%%%%%%%%%%%%%%%%%%%%%%%%%%%%%%%%%%%%%%%%%%%%%%%%%%%%%%%%%%%%%%%%%%%%%%%%%%%%%%%%%%%%%%%%%%%%%%%%%%%%%%%%%%
%/home/marting/art72/WISE/OTHERPeriods LPVPeriods_noWISE_REAL.tex

\longtab[4]{
\begin{landscape}
%\footnotesize
\scriptsize
\setlength{\tabcolsep}{1.3mm}
\begin{longtable}{rrllrrrrccrclccrllllllllllllllllllllllllll}
\caption{\label{noWISEPerbutREAL}  Sample of bona fide stellar sources with a period analysis based on literature data, period information (first entries).}\\
\hline \hline
\centering
Ra & Dec  & Periods literature & P refit & $P1$ & $\sigma_{\rm P1}$ & $P2$ & $\sigma_{\rm P2}$ &  Amp1 & $\sigma_{\rm A1}$ &  $W1$ & $\sigma_{\rm W1}$ & $\chi_{\rm r,W1}^2$ & Amp2 & $\sigma_{\rm A2}$ &  $W2$ & $\sigma_{\rm W2}$ & $\chi_{\rm r,W2}^2$ & Com & Class \\
(deg) & (deg) &  (d) &  (d)    & (d) & (d) & (d) & (d) & (mag) & (mag) & (mag) & (mag) & & (mag) & (mag) & (mag) & (mag) & \\
\hline
\endfirsthead
\caption{continued.}\\
\hline\hline
Ra & Dec  & Periods literature & P refit & $P1$ & $\sigma_{\rm P1}$ & $P2$ & $\sigma_{\rm P2}$ & Amp1 & $\sigma_{\rm A1}$ &  $W1$ & $\sigma_{\rm W1}$ & $\chi^2_{\rm r,W1}$ & Amp2 & $\sigma_{\rm A2}$ &  $W2$ & $\sigma_{\rm W2}$ & $\chi_{\rm r,W2}^2$ & Com & Class \\
(deg) & (deg) &  (d)  &  (d)   & (d) & (d) & (d) & (d) & (mag) & (mag) & (mag) & (mag) & & (mag) & (mag) & (mag) & (mag) & \\
\hline
\endhead
\endfoot

353.614540  &  43.550311  & 620. (26) 59 (4) & 686 $\pm$ 21 (71) &      &      &      &      &      &      &        &        &         &      &      &        &        &        &  &  & \\
284.625305  &   6.715967  & 1430 $\pm$ 27 (32) 1260 (30) &  &      &      &      &      &      &      &        &        &         &      &      &        &        &        &  &  & \\
334.864610  &  59.856052  & 1460 $\pm$ 24 (32) 1749 (30) &  &      &      &      &      &      &      &        &        &         &      &      &        &        &        &  &  & \\
279.385437  &  -5.399753  & 1556 (27) 1589 $\pm$ 42 (32) 1559 $\pm$ 7 (21) (1500) (33) 1592 (30) &  &      &      &      &      &      &      &        &        &         &      &      &        &        &        &  &  & \\
  6.921023  &  69.647507  & 650. (26) & 637 $\pm$ 55 (71) &      &      &      &      &      &      &        &        &         &      &      &        &        &        &  &  & \\
 75.878654  & -69.966820  &  &  &      &      &      &      &      &      &        &        &         &      &      &        &        &        &  & ero & \\
 77.845406  & -11.848938  &/827 (7) &  &      &      &      &      &      &      &        &        &         &      &      &        &        &        &  &  & \\
 79.323730  &  53.586071  & 514/819 (7) 461 (4) 456 $\pm$ 36 (5) &  &      &      &      &      &      &      &        &        &         &      &      &        &        &        &  &  & \\
 82.525955  & -70.511375  & 808 $\pm$ 4 (1) 806 (2) &  &      &      &      &      &      &      &        &        &         &      &      &        &        &        &  & ero & \\
 85.957069  &  32.701687  & 683: $\pm$ 160 (29) 780. (26) & 606 $\pm$ 85 (71) &      &      &      &      &      &      &        &        &         &      &      &        &        &        &  &  & \\
 86.374641  &  29.118120  & 508/508 (7) 708/498 (6) 449 $\pm$ 69 (29) 455 $\pm$ 50 (5) &  &      &      &      &      &      &      &        &        &         &      &      &        &        &        &  &  & \\
 88.954704  &  20.175150  & 370/122 (7) 376 (11) 378 (9) 369 (4) &  &      &      &      &      &      &      &        &        &         &      &      &        &        &        &  &  & \\
 97.072632  & -13.053068  &  &  &      &      &      &      &      &      &        &        &         &      &      &        &        &        &  &  & \\
 97.397789  &   8.788045  & 422: (23) no (4) & 769 $\pm$ 75 (71) &      &      &      &      &      &      &        &        &         &      &      &        &        &        &  &  & \\
 99.136795  &  38.445480  & 429/426 (7) &  &      &      &      &      &      &      &        &        &         &      &      &        &        &        &  &  & \\
 99.226021  &   3.424734  & 653. (28) &  &      &      &      &      &      &      &        &        &         &      &      &        &        &        &  &  & \\
105.929901  & -11.551617  & no (11) 785 (9) no (4) &  &      &      &      &      &      &      &        &        &         &      &      &        &        &        &  &  & \\
108.111450  &   8.517594  &  &  &      &      &      &      &      &      &        &        &         &      &      &        &        &        &  &  & \\
111.014603  & -12.874392  & 10 (4) & - (44) &      &      &      &      &      &      &        &        &         &      &      &        &        &        &  &  & \\
112.697861  &  -9.776847  &  &  &      &      &      &      &      &      &        &        &         &      &      &        &        &        &  &  & \\
118.454765  & -53.920090  &  & 451 $\pm$ 1 (54) &      &      &      &      &      &      &        &        &         &      &      &        &        &        &  &  & \\
139.178528  & -47.939484  &  &  &      &      &      &      &      &      &        &        &         &      &      &        &        &        &  &  & \\
148.891602  & -77.287849  &  &  &      &      &      &      &      &      &        &        &         &      &      &        &        &        &  &  & \\
149.819809  & -56.911324  &  &  &      &      &      &      &      &      &        &        &         &      &      &        &        &        &  &  & \\
159.787247  & -77.956032  & no (7) 62 (9) &  &      &      &      &      &      &      &        &        &         &      &      &        &        &        &  &  & \\
161.142731  & -60.092995  &  &  &      &      &      &      &      &      &        &        &         &      &      &        &        &        &  &  & \\
182.255554  & -63.264767  &  &  &      &      &      &      &      &      &        &        &         &      &      &        &        &        &  &  & \\
184.390961  & -67.960808  & no (7) &  &      &      &      &      &      &      &        &        &         &      &      &        &        &        &  &  & \\
205.221130  & -61.762733  &  &  &      &      &      &      &      &      &        &        &         &      &      &        &        &        &  &  & \\
205.756912  & -62.147499  &  &  &      &      &      &      &      &      &        &        &         &      &      &        &        &        &  &  & \\
207.907654  & -61.652130  &  &  &      &      &      &      &      &      &        &        &         &      &      &        &        &        &  &  & \\
211.898087  & -61.455872  &  &  &      &      &      &      &      &      &        &        &         &      &      &        &        &        &  &  & \\
216.240234  & -60.378918  &  &  &      &      &      &      &      &      &        &        &         &      &      &        &        &        &  &  & \\
220.506821  & -60.506161  &  &  &      &      &      &      &      &      &        &        &         &      &      &        &        &        &  &  & \\
227.501114  & -58.294167  &  &  &      &      &      &      &      &      &        &        &         &      &      &        &        &        &  &  & \\
237.410202  & -54.639774  &  &  &      &      &      &      &      &      &        &        &         &      &      &        &        &        &  &  & \\
237.682434  & -53.345341  &  &  &      &      &      &      &      &      &        &        &         &      &      &        &        &        &  &  & \\
240.578812  & -52.924671  & 1 (3) &  &      &      &      &      &      &      &        &        &         &      &      &        &        &        &  &  & \\
242.593582  & -52.100525  & no (7) &  &      &      &      &      &      &      &        &        &         &      &      &        &        &        &  &  & \\
242.660965  & -49.097557  &  &  &      &      &      &      &      &      &        &        &         &      &      &        &        &        &  &  & \\
242.891373  & -48.330719  & 710 (8) no (4) &  &      &      &      &      &      &      &        &        &         &      &      &        &        &        &  &  & \\
244.379456  & -50.541546  &  &  &      &      &      &      &      &      &        &        &         &      &      &        &        &        &  &  & \\
\hline
\end{longtable}
\tablefoot{
 See the notes in Table~\ref{WISEREALPer}.
}
\end{landscape}
}

%%%%%%%%%%%%%%%%%%%%%%%%%%%%%%%%%%%%%%%%%%%%%%%%%%%%%%%%%%%%%%%%%%%%%%%%%%%%%%%%%%%%%%%%%%%%%%%%%%%%%%%%%%%%%%%%%%%%%%%%%%%%%%%%%%%%%%%%%%%%%%%%%%%%%%%%%%%%%%
%/home/marting/art72/WISE/OTHERPeriods LPVSample_notREAL.tex

\longtab[5]{
\begin{landscape}
\footnotesize
\setlength{\tabcolsep}{1.3mm}
\begin{longtable}{rrllllllllllllllllllllllllllllllllllllllll}
\caption{\label{WISESamnotREAL} Sample of likely non bona fide sources, general information (first entries).}\\
\hline \hline
\centering
Ra & Dec  & Identifiers  & Dist & ObjType & 2M     &  GLP  & Aka  & MSX       &  MGL & HGL & $V_{\rm b}$ &  $V_{\rm r}$ & LRS &Spec. Type & Comment  \\
(deg) & (deg) &          & (\arcsec) &        & (\arcsec) & (\arcsec) & (\arcsec) & (\arcsec) &  (\arcsec) &  (\arcsec) & (\ks) & (\ks) & \\
\hline
\endfirsthead
\caption{continued.}\\
\hline\hline
Ra & Dec  & Identifiers  & Dist & ObjType & 2M     &  GLP  & Aka  & MSX       &  MGL & HGL & $V_{\rm b}$ &  $V_{\rm r}$ & LRS &Spec. Type & Comment \\
(deg) & (deg) &          & (\arcsec) &        & (\arcsec) & (\arcsec) & (\arcsec) & (\arcsec) &  (\arcsec) &  (\arcsec) & (\ks) & (\ks) & \\
\hline
\endhead
\endfoot
  1.697865  &  43.074245  &  &  &  &     &     &     &      &     &     &  -  &  -  &   &  &  \\
  2.079027  &  -0.576736  &  &  &  &     &     &     &      &     &     &  -  &  -  &   &  &  \\
  2.224788  &  -0.267857  &  &  &  &     &     &     &      &     &     &  -  &  -  &   &  &  \\
  2.242542  &  -0.750332  &  &  &  &     &     &     &      &     &     &  -  &  -  &   &  &  \\
  2.248379  &  -0.725155  &  &  &  &     &     &     &      &     &     &  -  &  -  &   &  &  \\
  2.259252  &  -0.380092  &  &  &  &     &     &     &      &     &     &  -  &  -  &   &  &  \\
  2.262807  &  -0.495763  &  &  &  &     &     &     &      &     &     &  -  &  -  &   &  &  \\
  2.266214  &  -0.497638  &  &  &  &     &     &     &      &     &     &  -  &  -  &   &  &  \\
  2.293089  &  -0.716659  &  &  &  &     &     &     &      &     &     &  -  &  -  &   &  &  \\
  2.800782  &  60.514320  &  &  &  &     &     &     &      &     &     &  -  &  -  & E &  &  \\
  6.898109  &  69.647568  &  &  &  &     &     &     &      &     &     &  -  &  -  &   &  &  \\
  6.945048  &  69.648048  &  &  &  &     &     &     &      &     &     &  -  &  -  &   &  &  \\
 16.626133  &  12.590879  &  &  &  &     &     &     &      &     &     &  -  &  -  &   &  &  \\
 19.419439  &  67.217979  &  &  &  &     &     &     &      &     &     &  -  &  -  &   &  &  \\
 19.440771  &  67.231651  &  &  &  &     &     &     &      &     &     &  -  &  -  &   &  &  \\
 19.454063  &  67.239548  &  &  &  &     &     &     &      &     &     &  -  &  -  &   &  & q(6) \\
 19.457726  &  67.212669  &  &  &  &     &     &     &      &     &     &  -  &  -  &   &  &  \\
 19.466803  &  67.238457  &  &  &  &     &     &     &      &     &     &  -  &  -  &   &  &  \\
 19.478022  &  67.238297  &  &  &  &     &     &     &      &     &     &  -  &  -  &   &  &  \\
 34.503338  &  28.621672  &  &  &  &     &     &     &      &     &     &  -  &  -  &   &  &  \\
 34.506149  &  28.606207  &  &  &  &     &     &     &      &     &     &  -  &  -  &   &  &  \\
 34.545307  &  28.618717  &  &  &  &     &     &     &      &     &     &  -  &  -  &   &  &  \\
 36.434948  &  62.061813  &  &  &  &     &     &     &      &     &     &  -  &  -  &   &  &  \\
 36.437519  &  62.056862  &  &  &  &     &     &     &      &     &     &  -  &  -  &   &  &  \\
 36.439510  &  62.058491  &  &  &  &     &     &     &      &     &     &  -  &  -  &   &  &  \\
 36.440208  &  62.060390  &  &  &  &     &     &     &      &     &     &  -  &  -  &   &  &  \\
 36.685730  &  62.272881  &  &  &  &     &     &     &      &     &     &  -  &  -  &   &  &  \\
 36.703152  &  62.258183  &  &  &  &     &     &     &      &     &     &  -  &  -  &   &  &  \\
 36.706955  &  62.260796  &  &  &  &     &     &     &      &     &     &  -  &  -  &   &  &  \\
 36.715996  &  62.270233  &  &  &  &     &     &     &      &     &     &  -  &  -  &   &  &  \\
 36.720409  &  62.263744  &  &  &  &     &     &     &      &     &     &  -  &  -  &   &  &  \\
 36.722710  &  62.260414  &  &  &  &     &     &     &      &     &     &  -  &  -  &   &  &  \\
 36.732460  &  62.265739  &  &  &  &     &     &     &      &     &     &  -  &  -  &   &  &  \\
 36.739769  &  62.265717  &  &  &  &     &     &     &      &     &     &  -  &  -  &   &  &  \\
 38.204800  &  58.044067  &  &  &  &     &     &     &      &     &     &  -  &  -  &   &  &  \\
 38.217022  &  58.041882  &  &  &  &     &     &     &      &     &     &  -  &  -  &   &  &  \\
 38.236389  &  58.013664  &  &  &  &     &     &     &      &     &     &  -  &  -  &   &  &  \\
 38.238350  &  58.016823  &  &  &  &     &     &     &      &     &     &  -  &  -  &   &  &  \\
 38.242611  &  58.022480  &  &  &  &     &     &     &      &     &     &  -  &  -  &   &  &  \\
 38.263523  &  58.052654  &  &  &  &     &     &     &      &     &     &  -  &  -  &   &  &  \\
 38.264610  &  58.054668  &  &  &  &     &     &     &      &     &     &  -  &  -  &   &  &  \\
 38.264866  &  58.018578  &  &  &  &     &     &     &      &     &     &  -  &  -  &   &  &  \\
 38.268902  &  58.031715  &  &  &  &     &     &     &      &     &     &  -  &  -  &   &  &  \\
 38.269836  &  58.038555  &  &  &  &     &     &     &      &     &     &  -  &  -  &   &  &  \\
\hline
\end{longtable}
\tablefoot{
   See the notes in Table~\ref{WISEREALSam}.
}
\end{landscape}
}

%%%%%%%%%%%%%%%%%%%%%%%%%%%%%%%%%%%%%%%%%%%%%%%%%%%%%%%%%%%%%%%%%%%%%%%%%%%%%%%%%%%%%%%%%%%%%%%%%%%%%%%%%%%%%%%%%%%%%%%%%%%%%%%%%%%%%%%%%%%%%%%%%%%%%%%%%%%%%%
%/home/marting/art72/WISE/OTHERPeriods LPVPeriods_notREAL.tex

\longtab[6]{
\begin{landscape}
\footnotesize
\setlength{\tabcolsep}{1.3mm}
\begin{longtable}{rrllrrrrccrclccrllllllllllllllllllllllllll}
\caption{\label{WISEPernotREAL}  Sample of likely non bona fide sources, period information (first entries).}\\
\hline \hline
\centering
Ra & Dec  & Periods literature & P refit & $P1$ & $\sigma_{\rm P1}$ & $P2$ & $\sigma_{\rm P2}$ &  Amp1 & $\sigma_{\rm A1}$ &  $W1$ & $\sigma_{\rm W1}$ & $\chi_{\rm r,W1}^2$ & Amp2 & $\sigma_{\rm A2}$ &  $W2$ & $\sigma_{\rm W2}$ & $\chi_{\rm r,W2}^2$ & Com & Class  \\
(deg) & (deg) &  (d)           &  (d)    & (d)  & (d)             &  (d) &  (d) & (mag) & (mag) & (mag) & (mag) & & (mag) & (mag) & (mag) & (mag) & \\
\hline
\endfirsthead
\caption{continued.}\\
\hline\hline
Ra & Dec  & Periods literature & P refit & $P1$ & $\sigma_{\rm P1}$ & $P2$ & $\sigma_{\rm P2}$ & Amp1 & $\sigma_{\rm A1}$ &  $W1$ & $\sigma_{\rm W1}$ & $\chi^2_{\rm r,W1}$ & Amp2 & $\sigma_{\rm A2}$ &  $W2$ & $\sigma_{\rm W2}$ & $\chi_{\rm r,W2}^2$ & Com  & Class \\
(deg) & (deg) &  (d)            &  (d)    & (d) & (d) &  (d) &  (d) & (mag) & (mag) & (mag) & (mag) & & (mag) & (mag) & (mag) & (mag) & \\
\hline
\endhead
\endfoot
  1.697865  &  43.074245  &  &  &      &      &      &      &      &      &        &        &         &      &      &        &        &        &  &  & \\
  2.079027  &  -0.576736  &  &  &      &      &      &      &      &      &        &        &         &      &      &        &        &        &  &  & \\
  2.224788  &  -0.267857  &  &  &      &      &      &      &      &      &        &        &         &      &      &        &        &        &  &  & \\
  2.242542  &  -0.750332  &  &  &      &      &      &      &      &      &        &        &         &      &      &        &        &        &  &  & \\
  2.248379  &  -0.725155  &  &  &      &      &      &      &      &      &        &        &         &      &      &        &        &        &  &  & \\
  2.259252  &  -0.380092  &  &  &      &      &      &      &      &      &        &        &         &      &      &        &        &        &  &  & \\
  2.262807  &  -0.495763  &  &  &      &      &      &      &      &      &        &        &         &      &      &        &        &        &  &  & \\
  2.266214  &  -0.497638  &  &  &      &      &      &      &      &      &        &        &         &      &      &        &        &        &  &  & \\
  2.293089  &  -0.716659  &  &  &      &      &      &      &      &      &        &        &         &      &      &        &        &        &  &  & \\
  2.800782  &  60.514320  &  &  &      &      &      &      &      &      &        &        &         &      &      &        &        &        &  &  & \\
  6.898109  &  69.647568  &  &  &      &      &      &      & 0.00 & 0.00 &  7.883 &  0.069 &  351.9  & 0.00 & 0.00 &  5.455 &  0.093 &   24.7 &  &  & \\
  6.945048  &  69.648048  &  &  &      &      &      &      &      &      &        &        &         &      &      &        &        &        &  &  & \\
 16.626133  &  12.590879  &  &  &      &      &      &      & 0.00 & 0.00 &  8.506 &  0.077 &  500.3  & 0.00 & 0.00 &  4.976 &  0.290 &  123.8 &  &  & \\
 19.419439  &  67.217979  &  &  &      &      &      &      &      &      &        &        &         &      &      &        &        &        &  &  & \\
 19.440771  &  67.231651  &  &  &      &      &      &      & 0.00 & 0.00 & 11.707 &  0.013 &   10.6  & 0.00 & 0.00 &  9.898 &  0.024 &   38.0 &  &  & \\
 19.454063  &  67.239548  &  &  &      &      &      &      & 0.00 & 0.00 & 12.996 &  0.008 &    3.3  & 0.00 & 0.00 & 10.938 &  0.088 &  387.0 &  &  & \\
 19.457726  &  67.212669  &  &  &      &      &      &      &      &      &        &        &         &      &      &        &        &        &  &  & \\
 19.466803  &  67.238457  &  &  &      &      &      &      &      &      &        &        &         &      &      &        &        &        &  &  & \\
 19.478022  &  67.238297  &  &  &      &      &      &      &      &      &        &        &         &      &      &        &        &        &  &  & \\
 34.503338  &  28.621672  &  &  &      &      &      &      &      &      &        &        &         &      &      &        &        &        &  &  & \\
 34.506149  &  28.606207  &  &  &      &      &      &      &      &      &        &        &         &      &      &        &        &        &  &  & \\
 34.545307  &  28.618717  &  &  &      &      &      &      &      &      &        &        &         &      &      &        &        &        &  &  & \\
 36.434948  &  62.061813  &  &  &      &      &      &      &      &      &        &        &         &      &      &        &        &        &  &  & \\
 36.437519  &  62.056862  &  &  &      &      &      &      &      &      &        &        &         &      &      &        &        &        &  &  & \\
 36.439510  &  62.058491  &  &  &      &      &      &      &      &      &        &        &         &      &      &        &        &        &  &  & \\
 36.440208  &  62.060390  &  &  &      &      &      &      &      &      &        &        &         &      &      &        &        &        &  &  & \\
 36.685730  &  62.272881  &  &  &      &      &      &      & 0.00 & 0.00 & 11.126 &  0.013 &   27.4  & 0.00 & 0.00 & 10.470 &  0.018 &   44.3 &  &  & \\
 36.703152  &  62.258183  &  &  &      &      &      &      &      &      &        &        &         &      &      &        &        &        &  &  & \\
 36.706955  &  62.260796  &  &  &      &      &      &      &      &      &        &        &         &      &      &        &        &        &  &  & \\
 36.715996  &  62.270233  &  &  &      &      &      &      &      &      &        &        &         &      &      &        &        &        &  &  & \\
 36.720409  &  62.263744  &  &  &      &      &      &      &      &      &        &        &         &      &      &        &        &        &  &  & \\
 36.722710  &  62.260414  &  &  &      &      &      &      &      &      &        &        &         &      &      &        &        &        &  &  & \\
 36.732460  &  62.265739  &  &  &      &      &      &      & 0.00 & 0.00 &  9.977 &  0.006 &    1.3  & 0.00 & 0.00 &  9.114 &  0.015 &    7.8 &  &  & \\
 36.739769  &  62.265717  &  &  &      &      &      &      & 0.00 & 0.00 &  9.833 &  0.005 &    1.6  & 0.00 & 0.00 &  9.234 &  0.007 &    2.2 &  &  & \\
 38.204800  &  58.044067  &  &  &      &      &      &      &      &      &        &        &         &      &      &        &        &        &  &  & \\
 38.217022  &  58.041882  &  &  &      &      &      &      & 0.00 & 0.00 & 15.121 &  0.071 &    6.9  & 0.00 & 0.00 & 10.567 &  0.047 &  114.5 &  &  & \\
 38.236389  &  58.013664  &  &  &      &      &      &      & 0.00 & 0.00 & 14.121 &  0.024 &    1.7  & 0.00 & 0.00 & 11.015 &  0.052 &  106.1 &  &  & \\
 38.238350  &  58.016823  &  &  &      &      &      &      & 0.00 & 0.00 & 14.989 &  0.062 &    8.1  & 0.00 & 0.00 & 10.483 &  0.062 &  226.9 &  &  & \\
 38.242611  &  58.022480  &  &  &      &      &      &      & 0.00 & 0.00 & 13.310 &  0.057 &   50.5  & 0.00 & 0.00 &  9.325 &  0.061 &  443.6 &  &  & \\
 38.263523  &  58.052654  &  &  &      &      &      &      &      &      &        &        &         &      &      &        &        &        &  &  & \\
 38.264610  &  58.054668  &  &  &      &      &      &      & 0.00 & 0.00 & 14.313 &  0.029 &    6.2  & 0.00 & 0.00 & 10.924 &  0.051 &   88.0 &  &  & \\
 38.264866  &  58.018578  &  &  &      &      &      &      &      &      &        &        &         &      &      &        &        &        &  &  & \\
 38.268902  &  58.031715  &  &  &      &      &      &      &      &      &        &        &         &      &      &        &        &        &  &  & \\
 38.269836  &  58.038555  &  &  &      &      &      &      &      &      &        &        &         &      &      &        &        &        &  &  & \\
\hline
\end{longtable}
\tablefoot{
   See the notes in Table~\ref{WISEREALPer}.
}
\end{landscape}
}

%%%%%%%%%%%%%%%%%%%%%%%%%%%%%%%%%%%%%%%%%%%%%%%%%%%%%%%%%%%%%%%%%%%%%%%%%%%%%%%%%%%%%%%%%
\FloatBarrier
\section{Figures of LCs}

\begin{figure}[b]

\begin{minipage}{0.24\textwidth}
\resizebox{\hsize}{!}{\includegraphics[angle=-0]{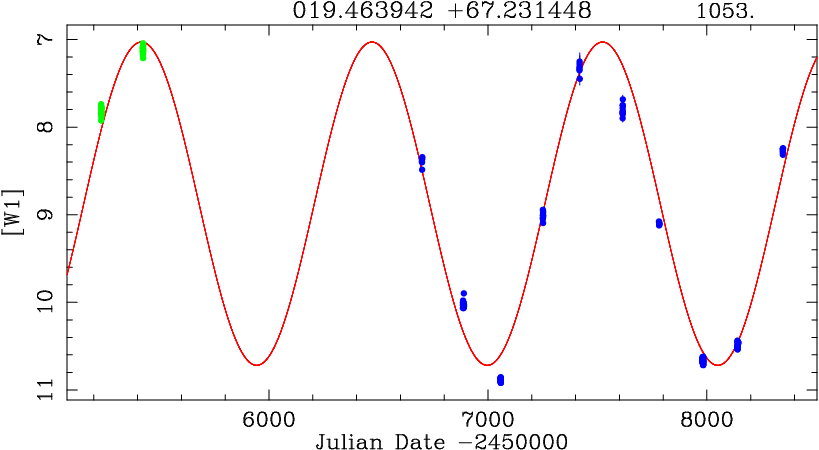}} 
\end{minipage}
\begin{minipage}{0.24\textwidth}
\resizebox{\hsize}{!}{\includegraphics[angle=-0]{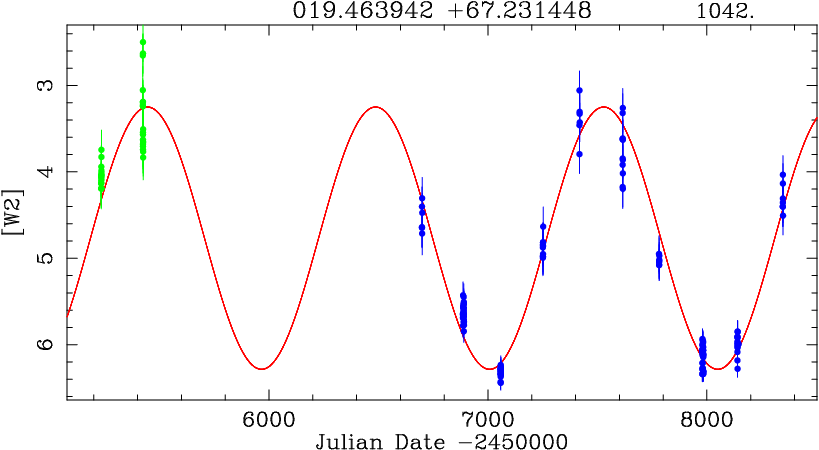}} 
\end{minipage}
 \begin{minipage}{0.24\textwidth}
\resizebox{\hsize}{!}{\includegraphics[angle=-0]{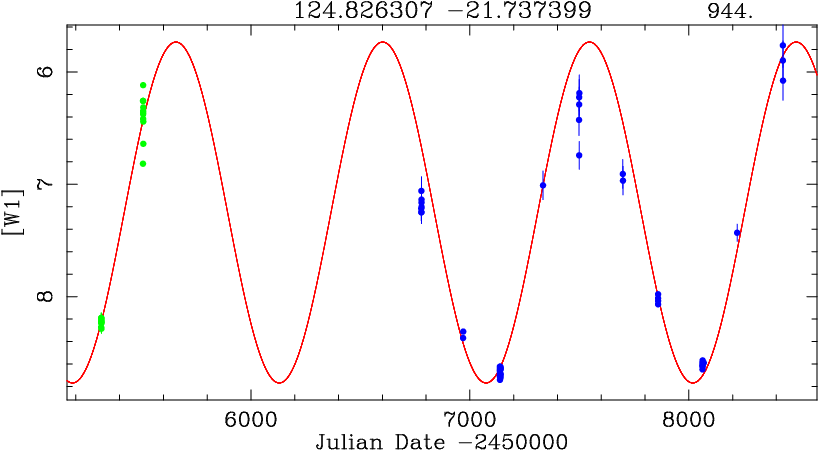}} 
\end{minipage}
\begin{minipage}{0.24\textwidth}
\resizebox{\hsize}{!}{\includegraphics[angle=-0]{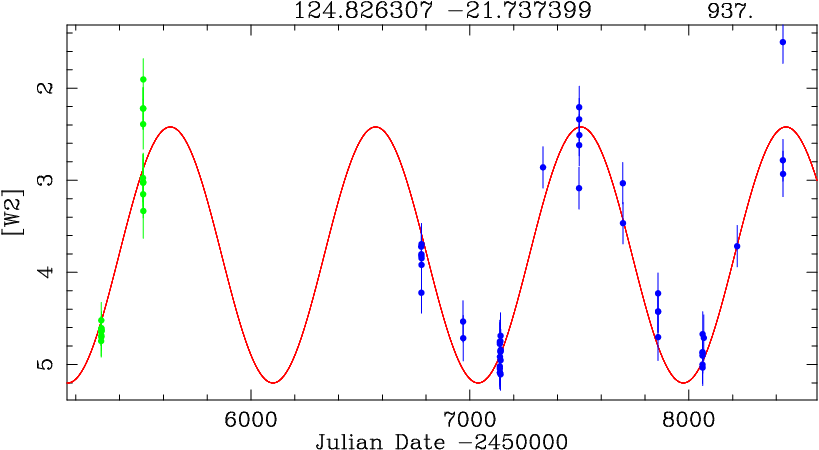}} 
\end{minipage}
 
\begin{minipage}{0.24\textwidth}
\resizebox{\hsize}{!}{\includegraphics[angle=-0]{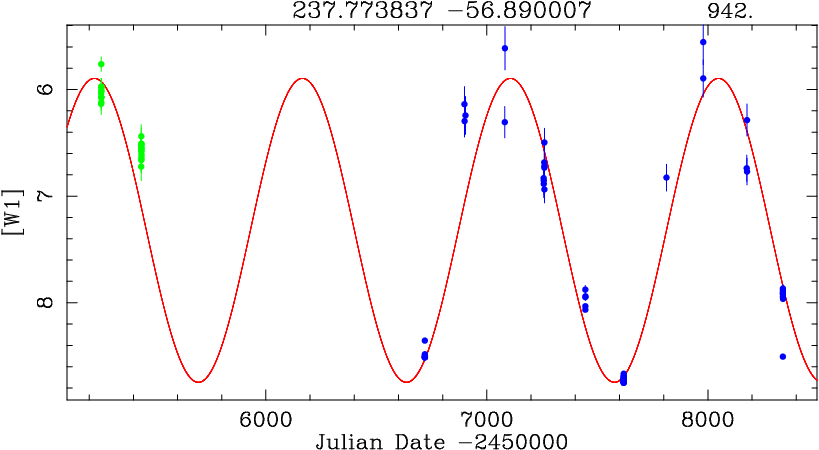}} 
\end{minipage}
\begin{minipage}{0.24\textwidth}
\resizebox{\hsize}{!}{\includegraphics[angle=-0]{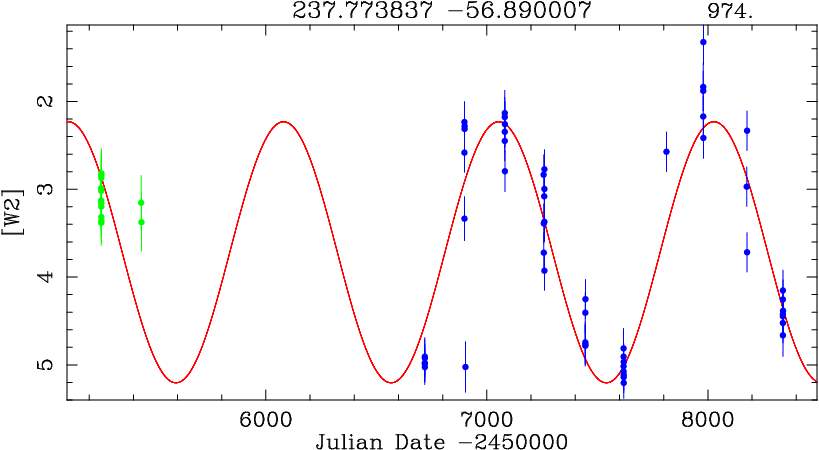}} 
\end{minipage}
\begin{minipage}{0.24\textwidth}
\resizebox{\hsize}{!}{\includegraphics[angle=-0]{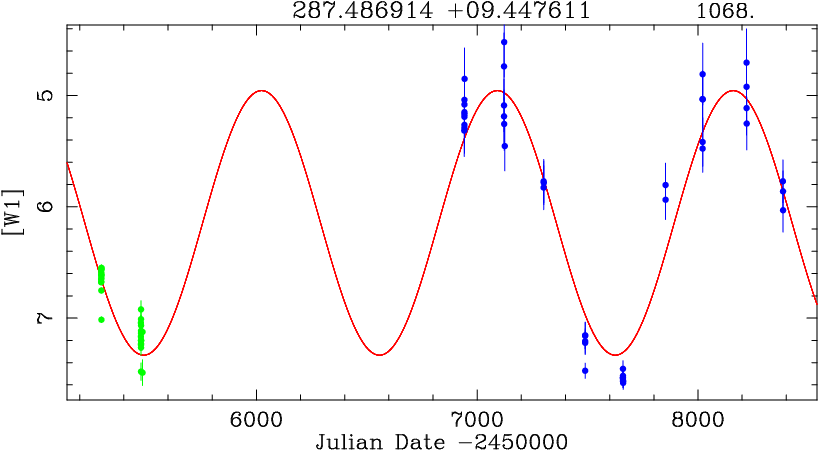}} 
\end{minipage}
\begin{minipage}{0.24\textwidth}
\resizebox{\hsize}{!}{\includegraphics[angle=-0]{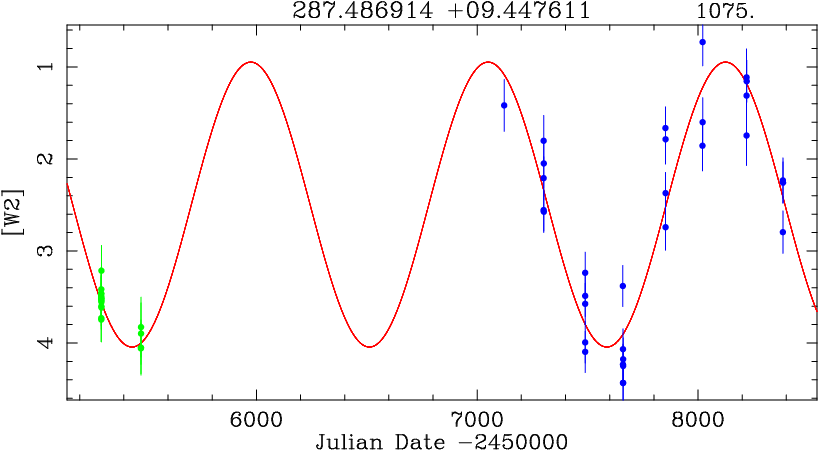}} 
\end{minipage}
 
\begin{minipage}{0.24\textwidth}
\resizebox{\hsize}{!}{\includegraphics[angle=-0]{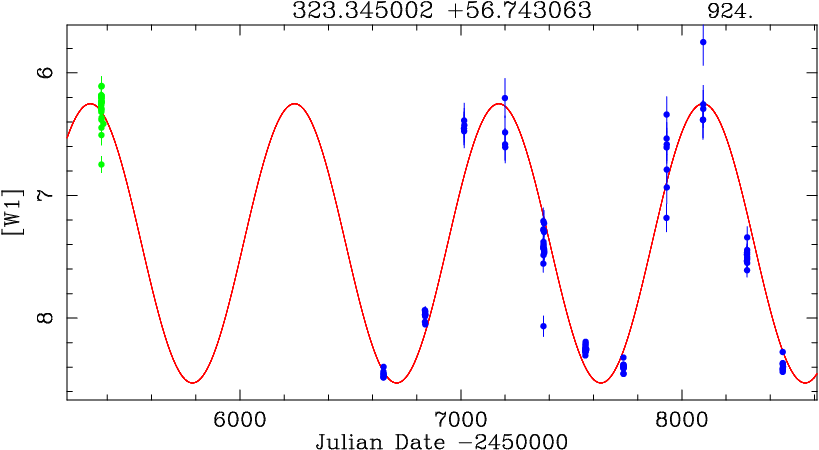}} 
\end{minipage}
\begin{minipage}{0.24\textwidth}
\resizebox{\hsize}{!}{\includegraphics[angle=-0]{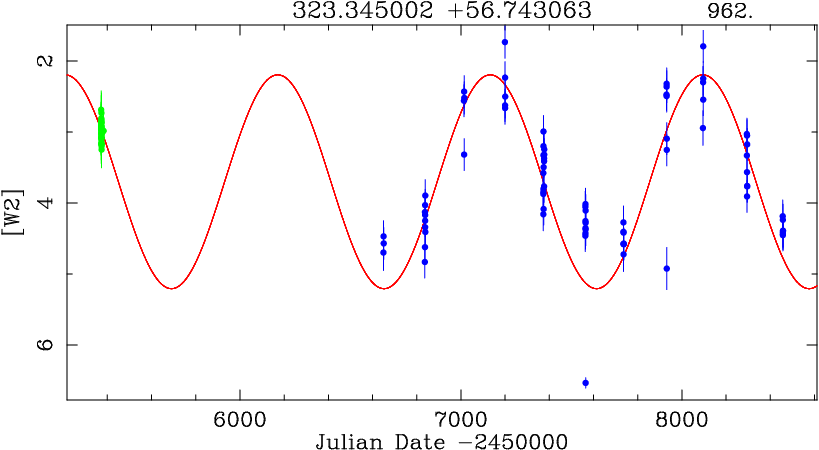}} 
\end{minipage}
\begin{minipage}{0.24\textwidth}
\resizebox{\hsize}{!}{\includegraphics[angle=-0]{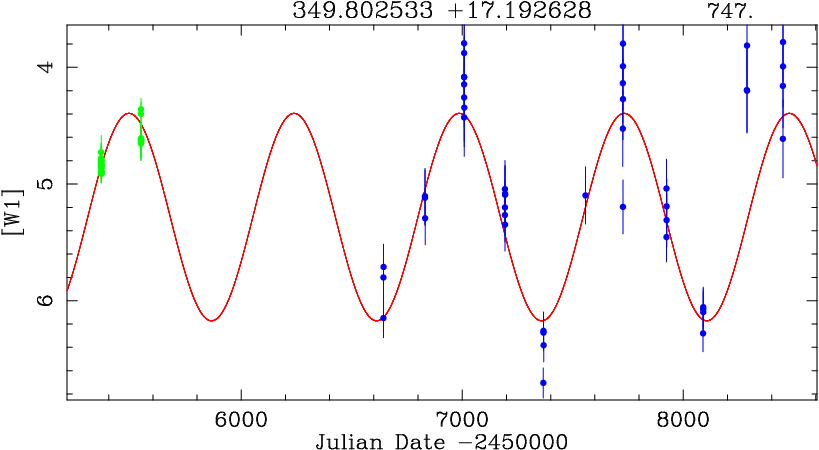}} 
\end{minipage}
\begin{minipage}{0.24\textwidth}
\resizebox{\hsize}{!}{\includegraphics[angle=-0]{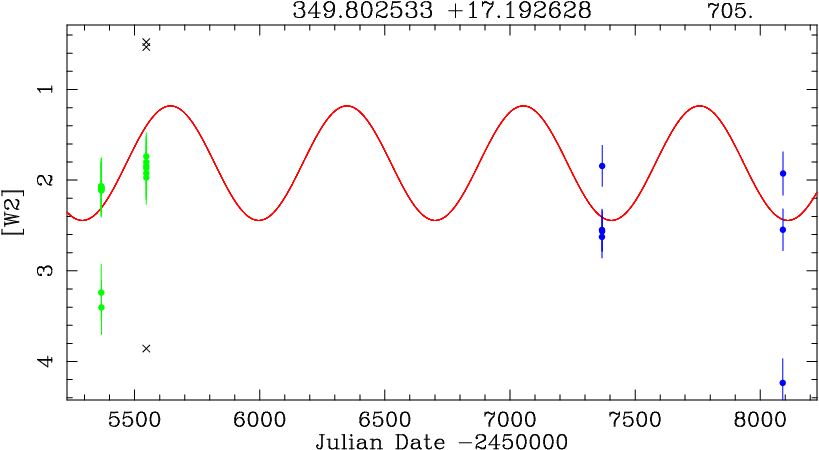}} 
\end{minipage}

% 2
\caption{Examples of fits to WISE data (W1 on the left, W2 on the right).
  The identifier is listed on the top of each panel, with the period to the right.
  Green points refer to WISE, and blue points to NEOWISE data.
  Crosses indicate points excluded from the fitting.
  The complete set of LCs is available at \protect\url{https://doi.org/10.5281/zenodo.5825878}.
}
\label{Fig:WISE}

\end{figure}

\begin{figure}

\begin{minipage}{0.24\textwidth}
\resizebox{\hsize}{!}{\includegraphics[angle=-0]{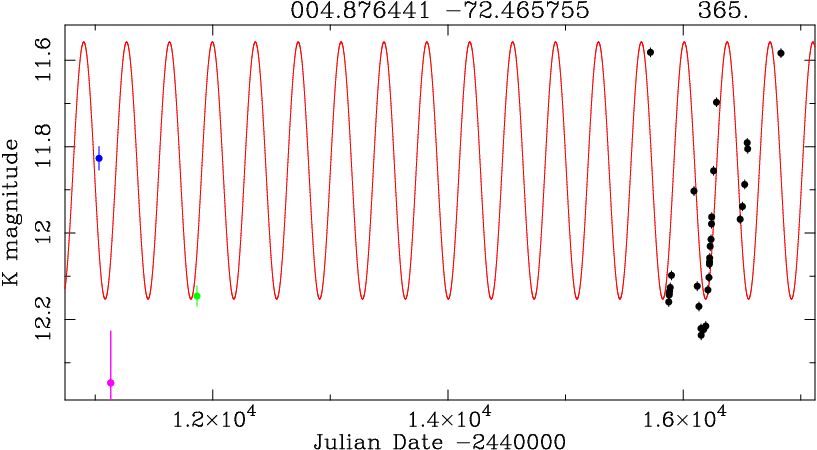}} 
\end{minipage}
\begin{minipage}{0.24\textwidth}
\resizebox{\hsize}{!}{\includegraphics[angle=-0]{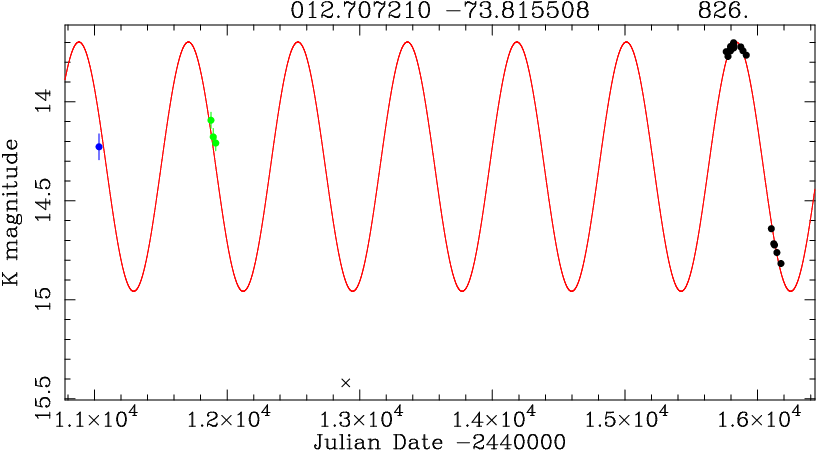}} 
\end{minipage}
\begin{minipage}{0.24\textwidth}
\resizebox{\hsize}{!}{\includegraphics[angle=-0]{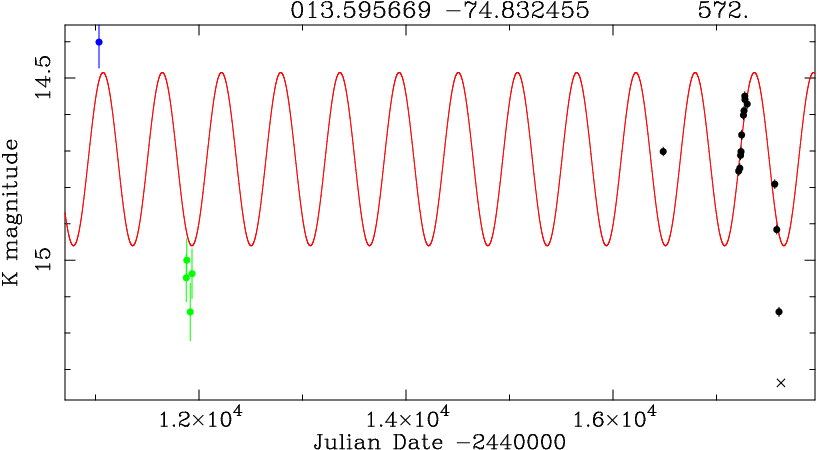}} 
\end{minipage}
\begin{minipage}{0.24\textwidth}
\resizebox{\hsize}{!}{\includegraphics[angle=-0]{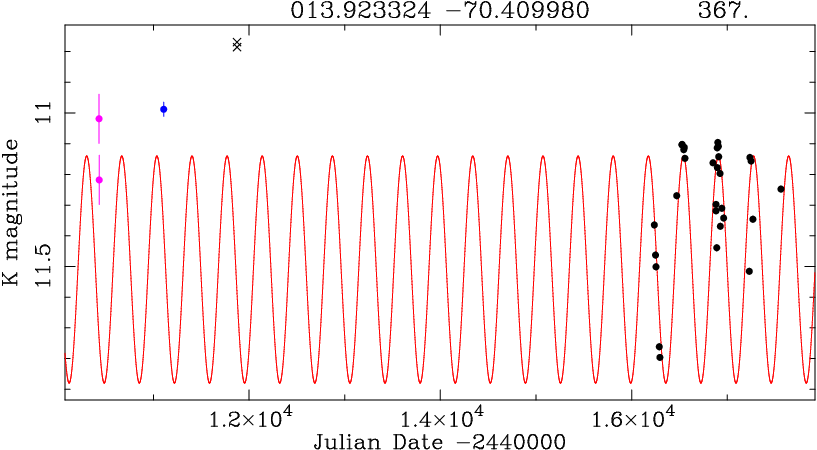}} 
\end{minipage}
 
\begin{minipage}{0.24\textwidth}
\resizebox{\hsize}{!}{\includegraphics[angle=-0]{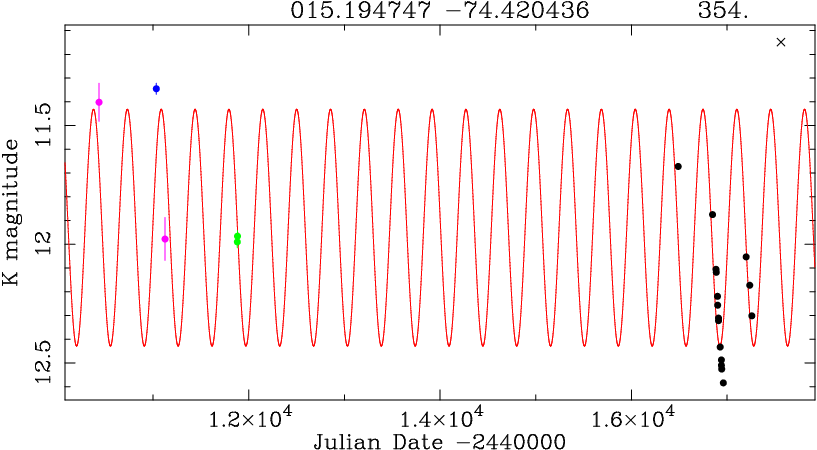}} 
\end{minipage}
\begin{minipage}{0.24\textwidth}
\resizebox{\hsize}{!}{\includegraphics[angle=-0]{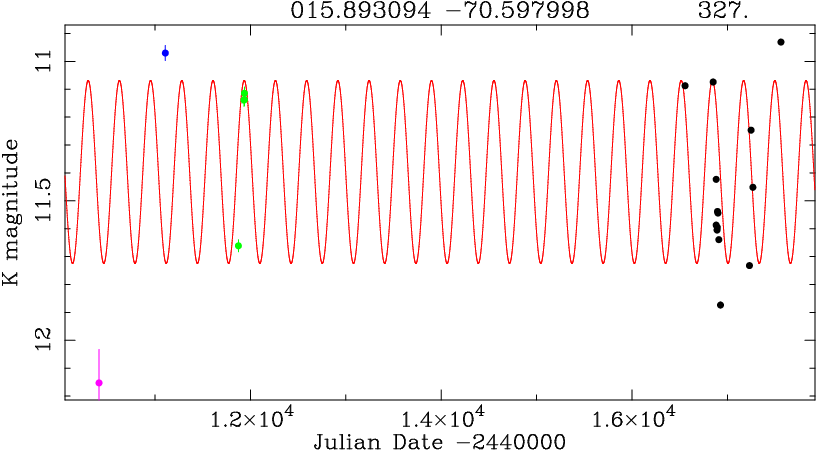}} 
\end{minipage}
\begin{minipage}{0.24\textwidth}
\resizebox{\hsize}{!}{\includegraphics[angle=-0]{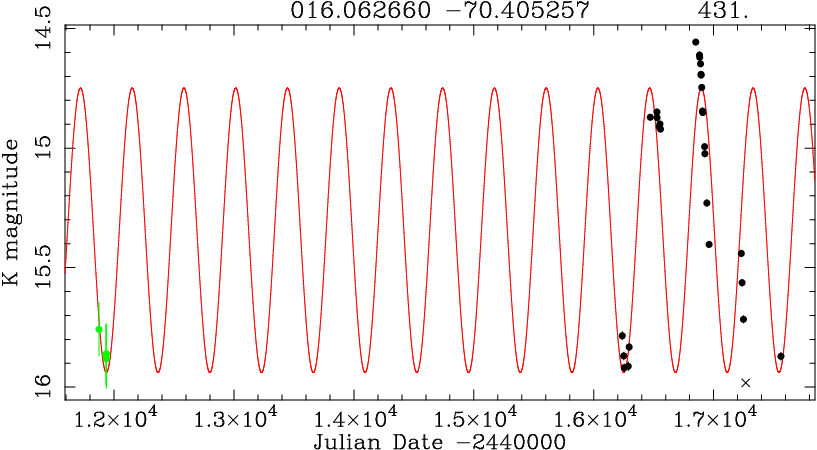}} 
\end{minipage}
\begin{minipage}{0.24\textwidth}
\resizebox{\hsize}{!}{\includegraphics[angle=-0]{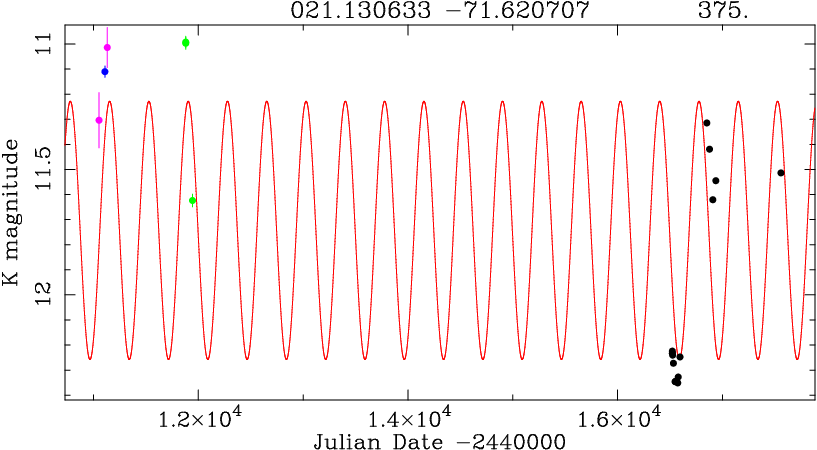}} 
\end{minipage}
 
\begin{minipage}{0.24\textwidth}
\resizebox{\hsize}{!}{\includegraphics[angle=-0]{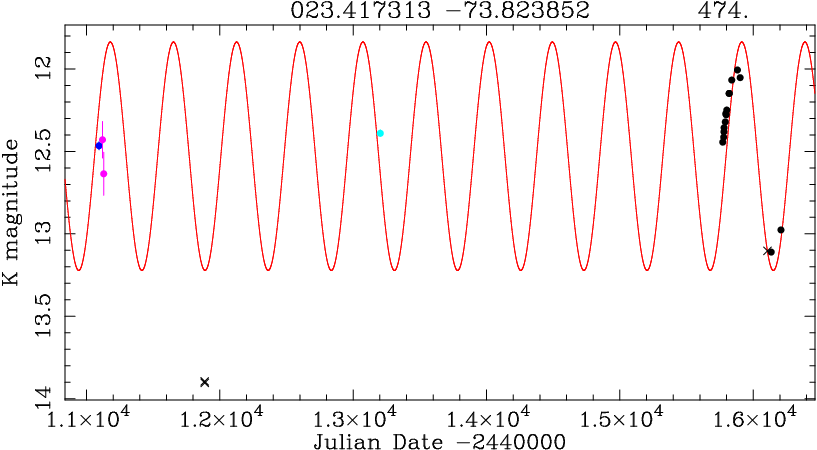}} 
\end{minipage}
\begin{minipage}{0.24\textwidth}
\resizebox{\hsize}{!}{\includegraphics[angle=-0]{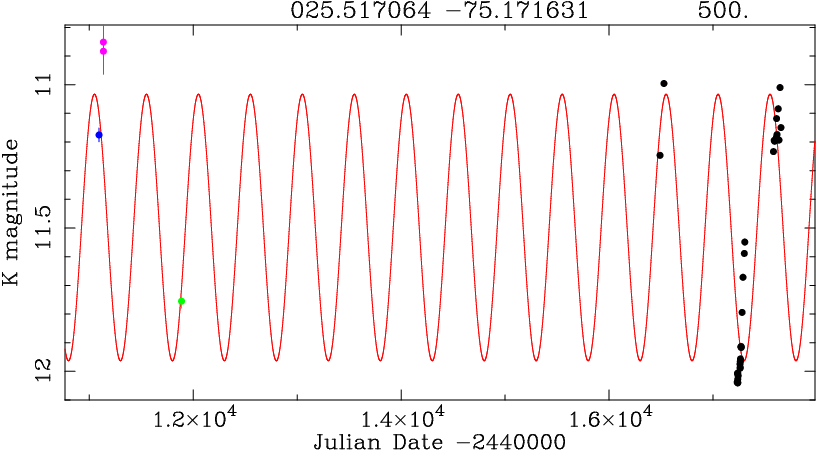}} 
\end{minipage}
 \begin{minipage}{0.24\textwidth}
\resizebox{\hsize}{!}{\includegraphics[angle=-0]{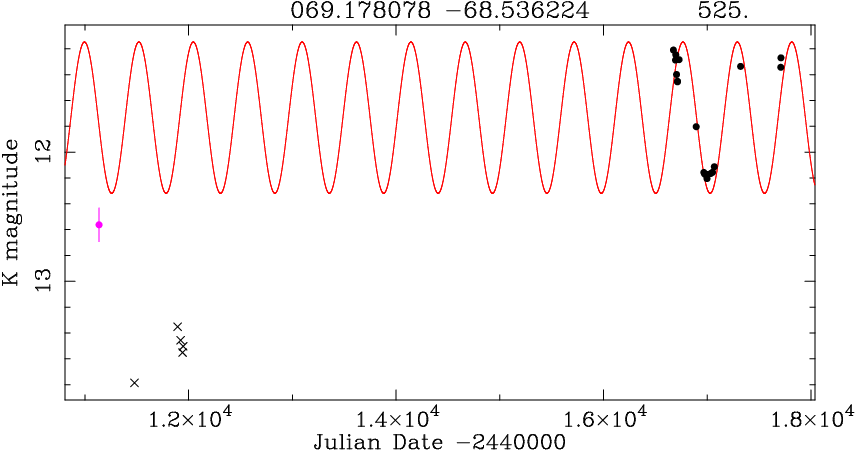}} 
\end{minipage}
\begin{minipage}{0.24\textwidth}
\resizebox{\hsize}{!}{\includegraphics[angle=-0]{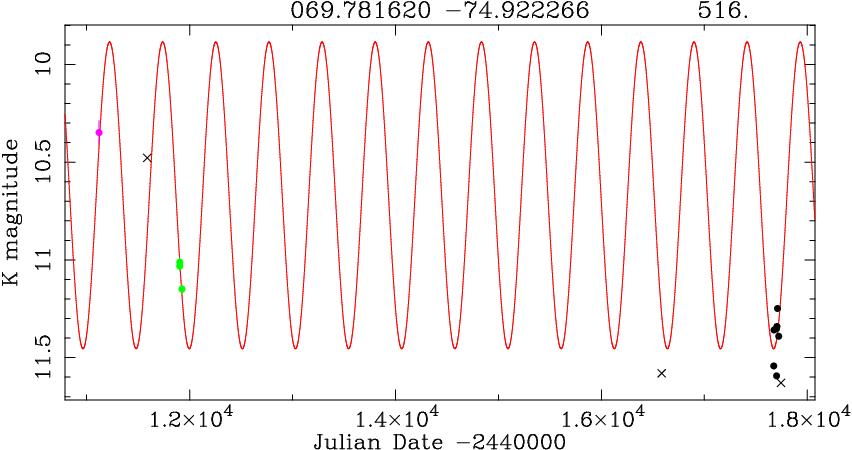}} 
\end{minipage}
 
% 2
\caption{Examples of fits to VMC data $K$-band data.
  The identifier is listed on the top of each panel, with the period to the right.
Data points (with error bars) in black show the VMC, green points represent 2MASS,  dark blue points are for 2MASS-6X,  
light blue points are for the IRSF, and magenta corresponds to DENIS (see \citealt{Gr20}).
}
\label{Fig:VMC}

\end{figure}

\begin{figure}
\begin{minipage}{0.24\textwidth}
\resizebox{\hsize}{!}{\includegraphics[angle=-0]{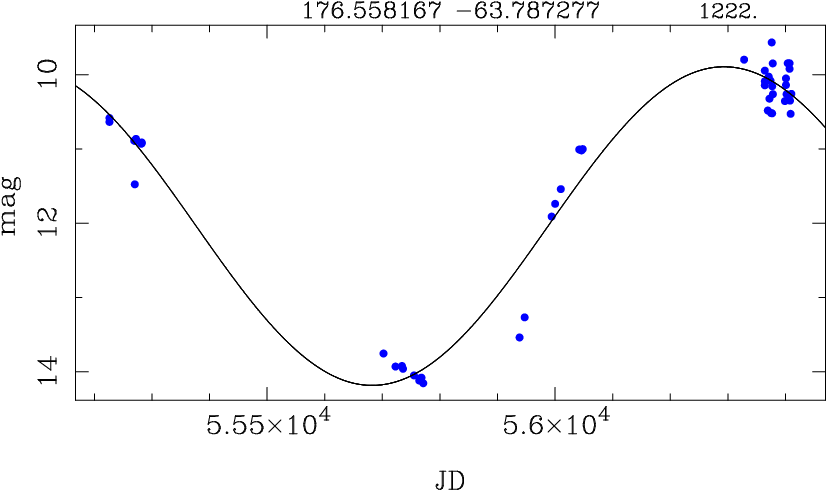}} 
\end{minipage}
\begin{minipage}{0.24\textwidth}
\resizebox{\hsize}{!}{\includegraphics[angle=-0]{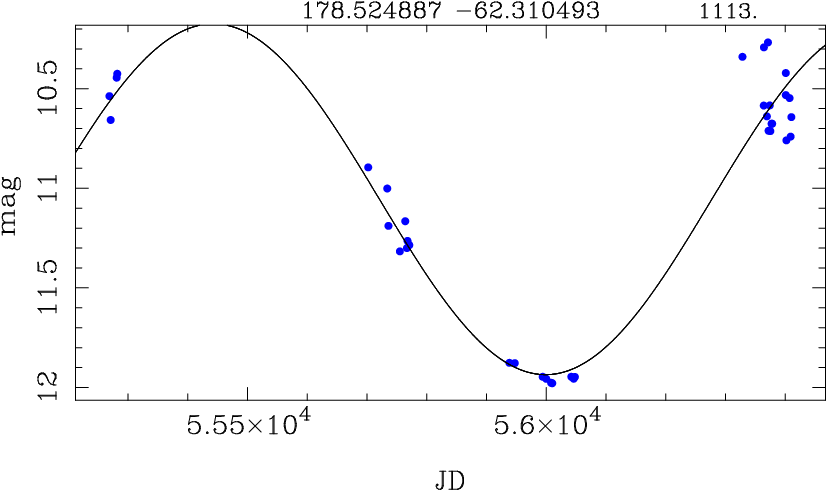}} 
\end{minipage}
\begin{minipage}{0.24\textwidth}
\resizebox{\hsize}{!}{\includegraphics[angle=-0]{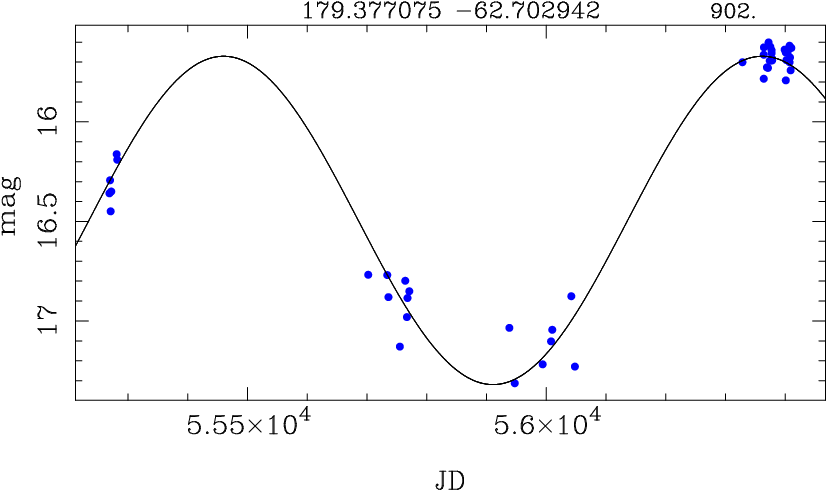}} 
\end{minipage}
\begin{minipage}{0.24\textwidth}
\resizebox{\hsize}{!}{\includegraphics[angle=-0]{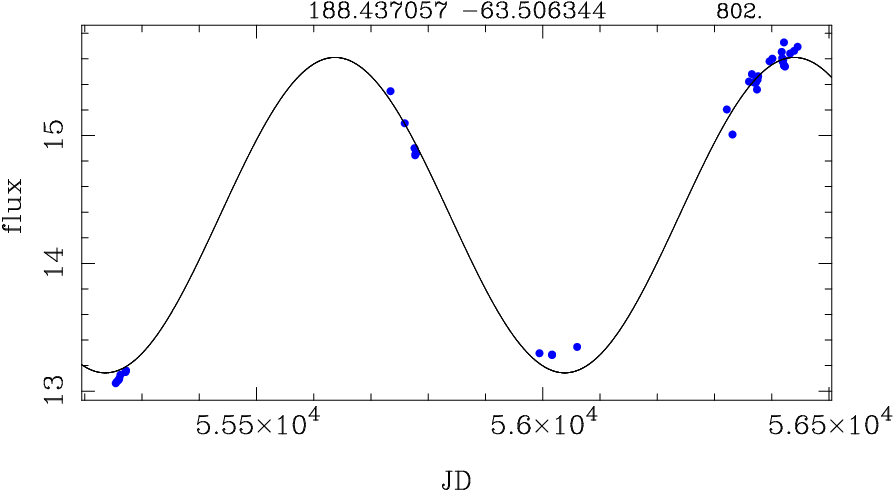}} 
\end{minipage}
 
\begin{minipage}{0.24\textwidth}
\resizebox{\hsize}{!}{\includegraphics[angle=-0]{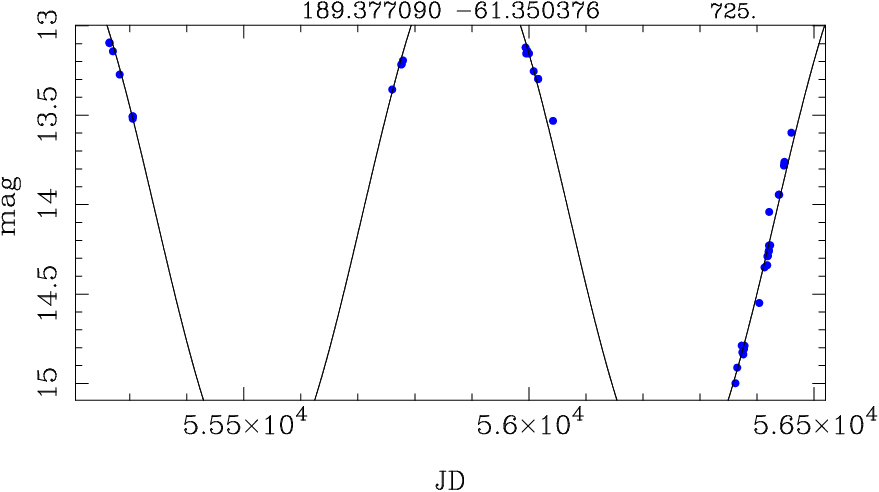}} 
\end{minipage}
\begin{minipage}{0.24\textwidth}
\resizebox{\hsize}{!}{\includegraphics[angle=-0]{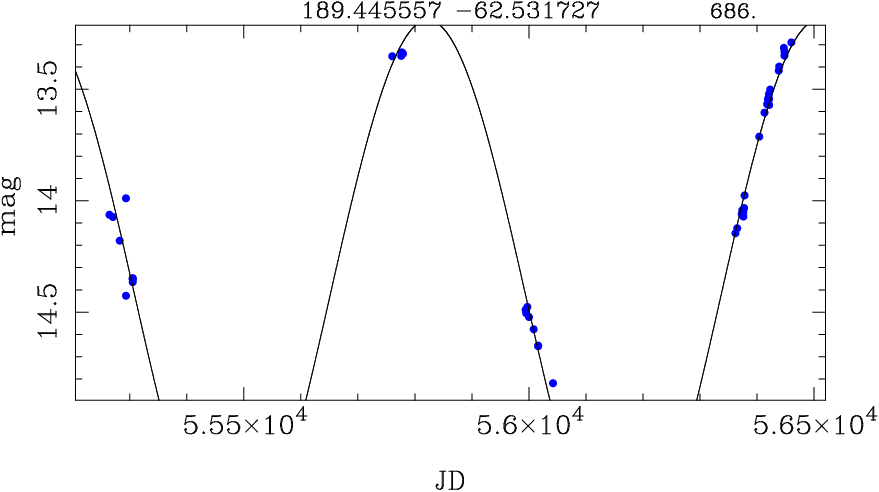}} 
\end{minipage}
\begin{minipage}{0.24\textwidth}
\resizebox{\hsize}{!}{\includegraphics[angle=-0]{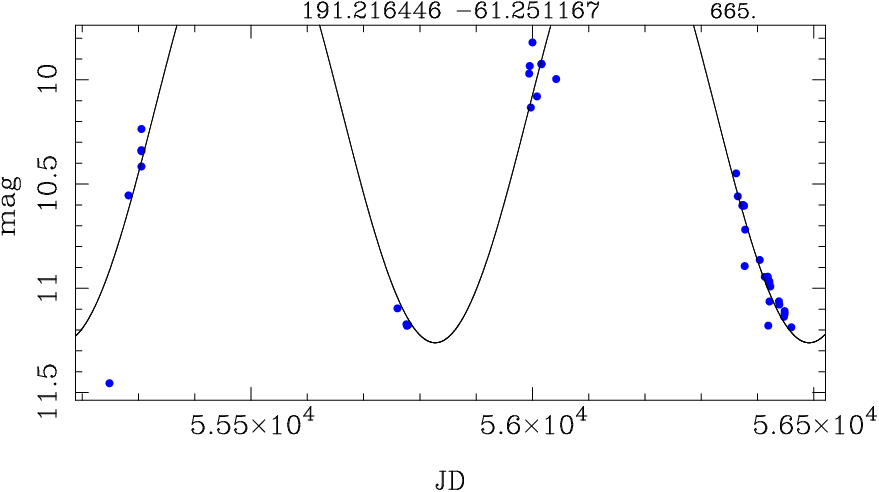}} 
\end{minipage}
\begin{minipage}{0.24\textwidth}
\resizebox{\hsize}{!}{\includegraphics[angle=-0]{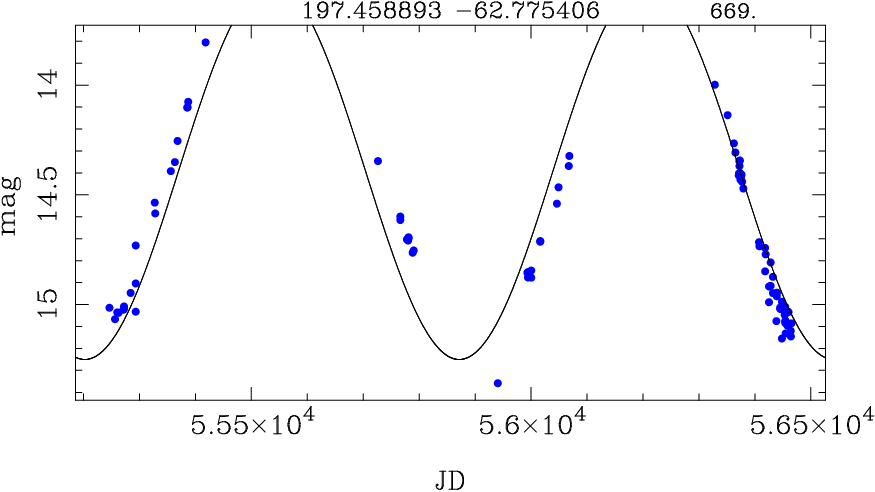}} 
\end{minipage}
 
\begin{minipage}{0.24\textwidth}
\resizebox{\hsize}{!}{\includegraphics[angle=-0]{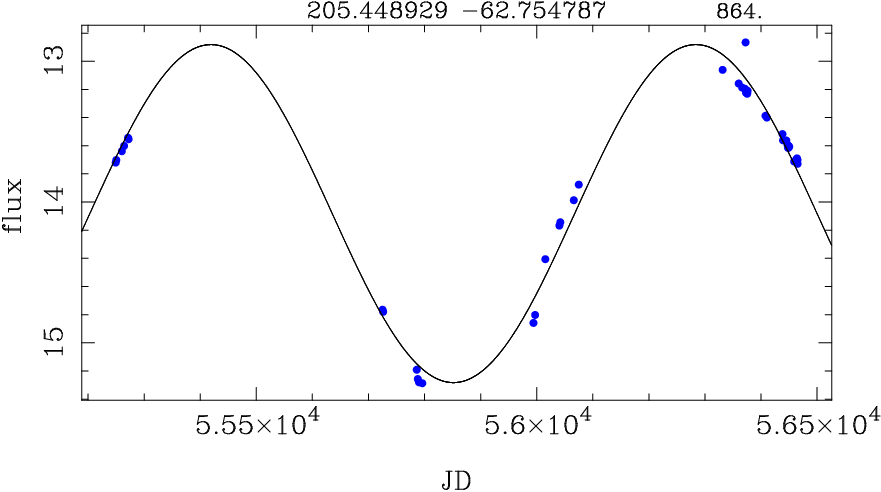}} 
\end{minipage}
\begin{minipage}{0.24\textwidth}
\resizebox{\hsize}{!}{\includegraphics[angle=-0]{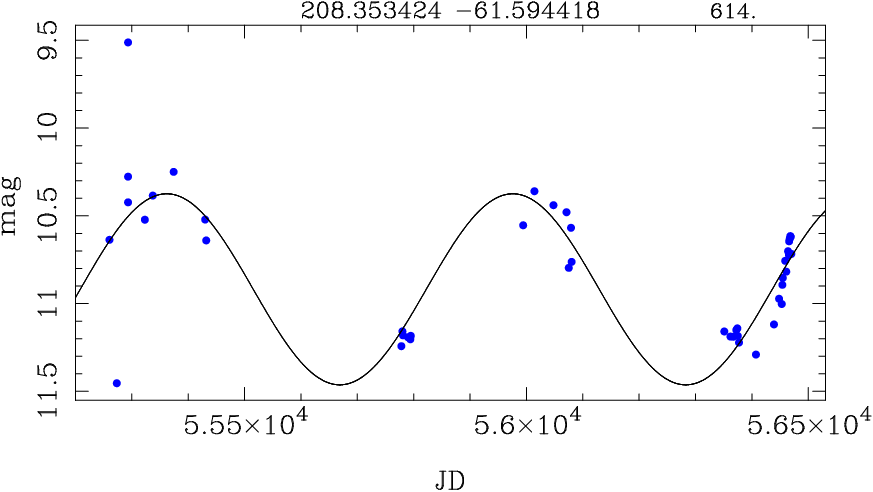}} 
\end{minipage}
\begin{minipage}{0.24\textwidth}
\resizebox{\hsize}{!}{\includegraphics[angle=-0]{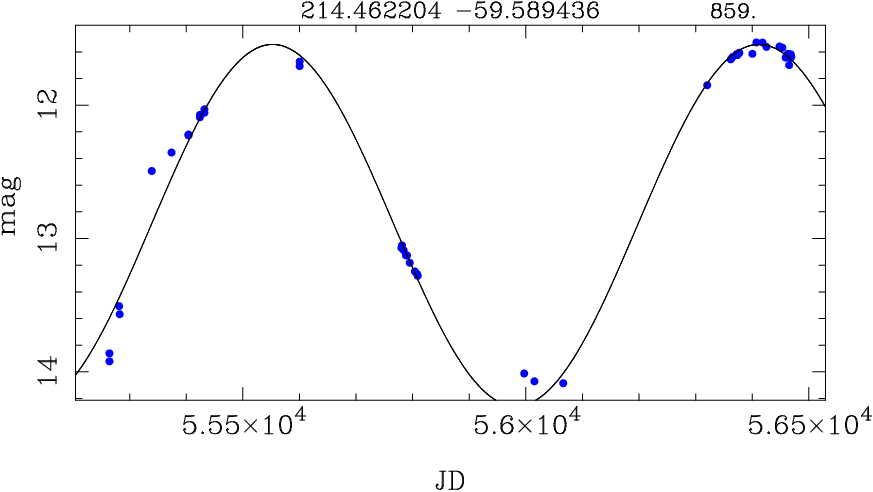}} 
\end{minipage}
\begin{minipage}{0.24\textwidth}
\resizebox{\hsize}{!}{\includegraphics[angle=-0]{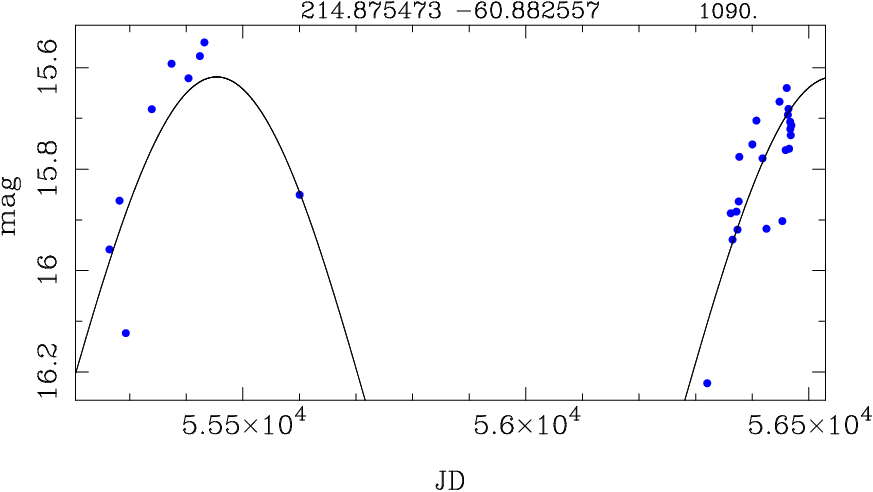}} 
\end{minipage}
 
% 2
\caption{Examples of fits to VVV $K$-band data.
  The identifier is listed on the top of each panel, with the period to the right.
}
\label{Fig:VVV}

\end{figure}

\begin{figure}
\begin{minipage}{0.24\textwidth}
\resizebox{\hsize}{!}{\includegraphics[angle=-0]{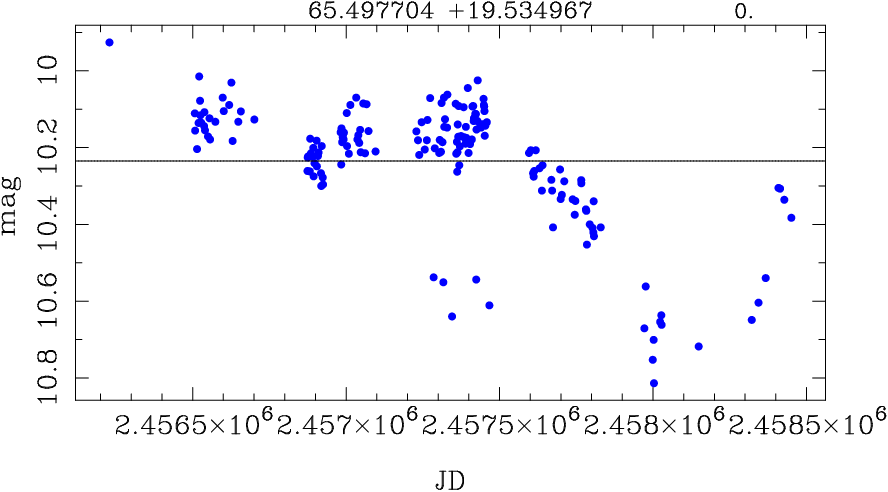}} 
\end{minipage}
\begin{minipage}{0.24\textwidth}
\resizebox{\hsize}{!}{\includegraphics[angle=-0]{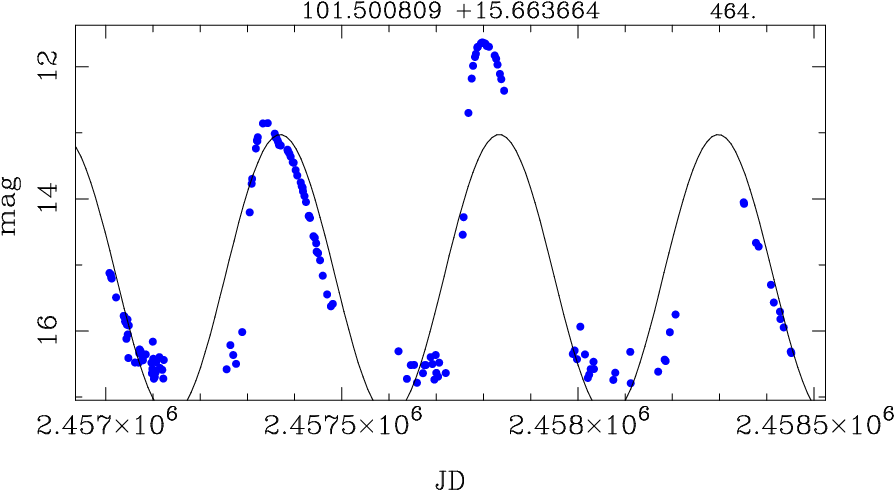}} 
\end{minipage}
 \begin{minipage}{0.24\textwidth}
\resizebox{\hsize}{!}{\includegraphics[angle=-0]{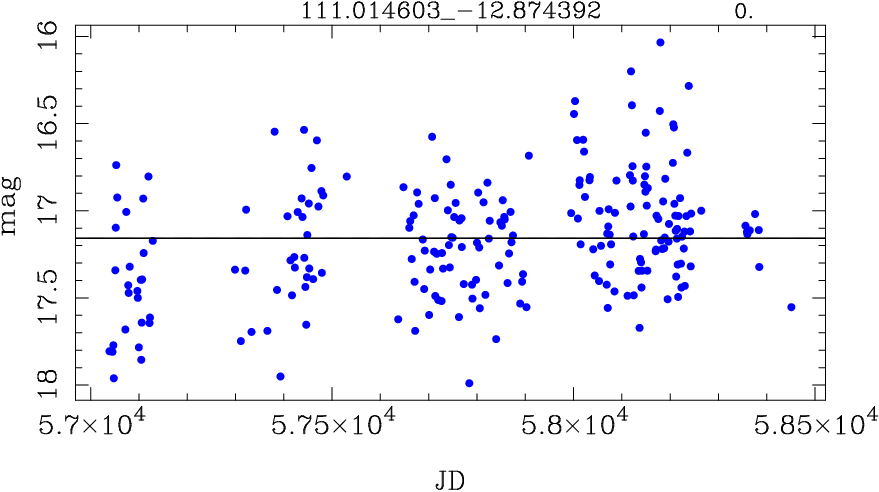}} 
\end{minipage}
\begin{minipage}{0.24\textwidth}
\resizebox{\hsize}{!}{\includegraphics[angle=-0]{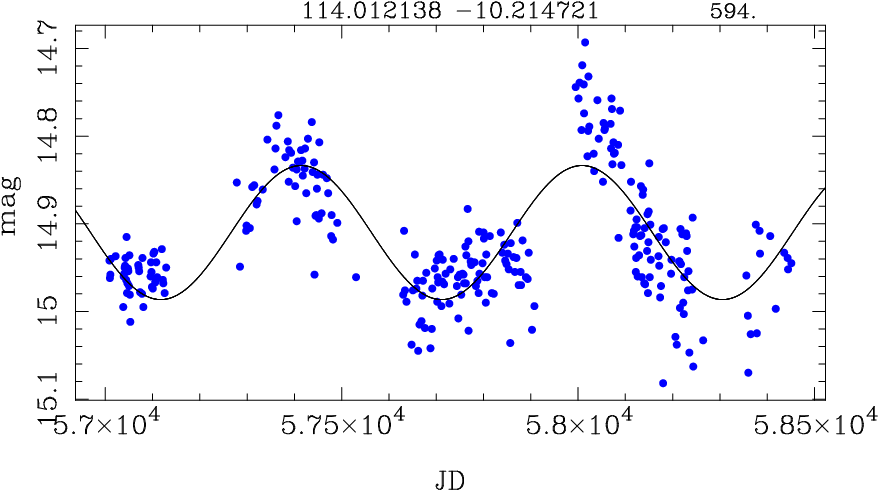}} 
\end{minipage}
 
\begin{minipage}{0.24\textwidth}
\resizebox{\hsize}{!}{\includegraphics[angle=-0]{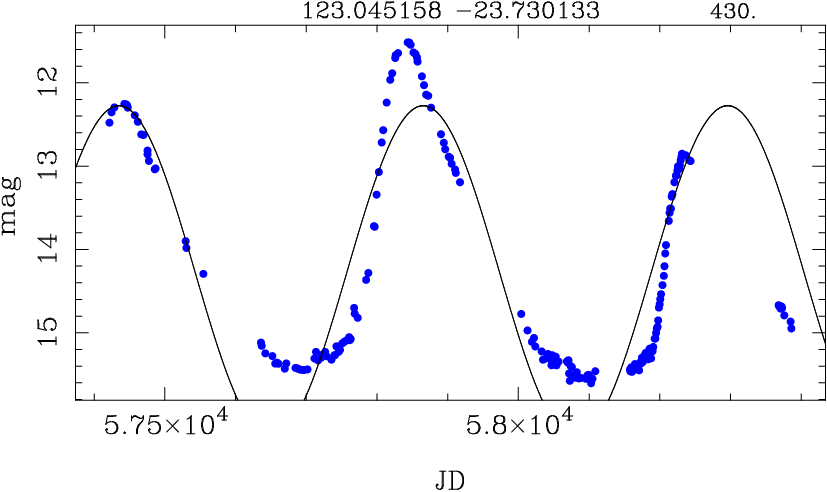}} 
\end{minipage}
\begin{minipage}{0.24\textwidth}
\resizebox{\hsize}{!}{\includegraphics[angle=-0]{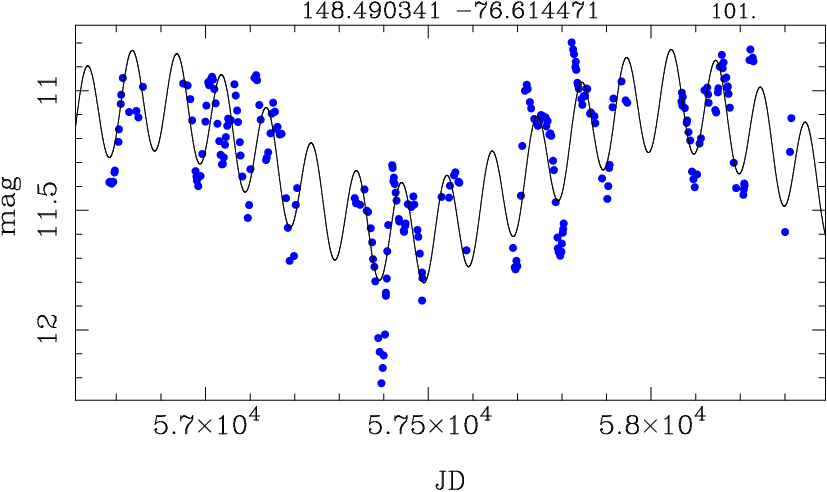}} 
\end{minipage}
\begin{minipage}{0.24\textwidth}
\resizebox{\hsize}{!}{\includegraphics[angle=-0]{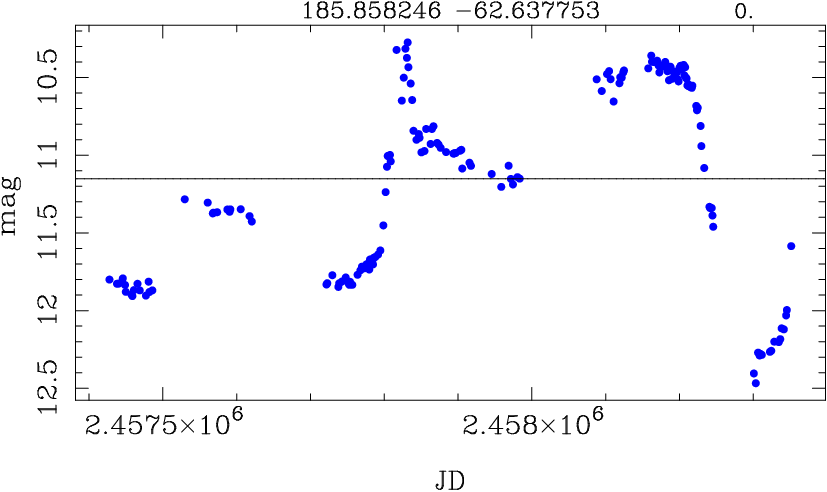}} 
\end{minipage}
\begin{minipage}{0.24\textwidth}
\resizebox{\hsize}{!}{\includegraphics[angle=-0]{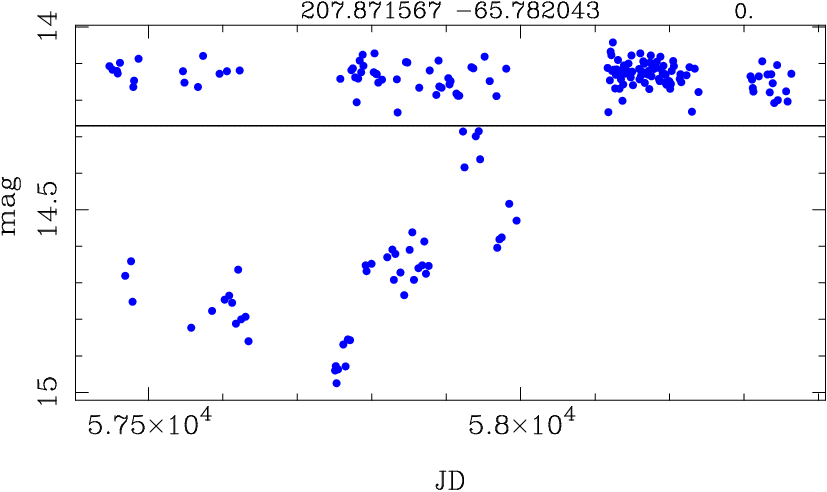}} 
\end{minipage}
 
\begin{minipage}{0.24\textwidth}
\resizebox{\hsize}{!}{\includegraphics[angle=-0]{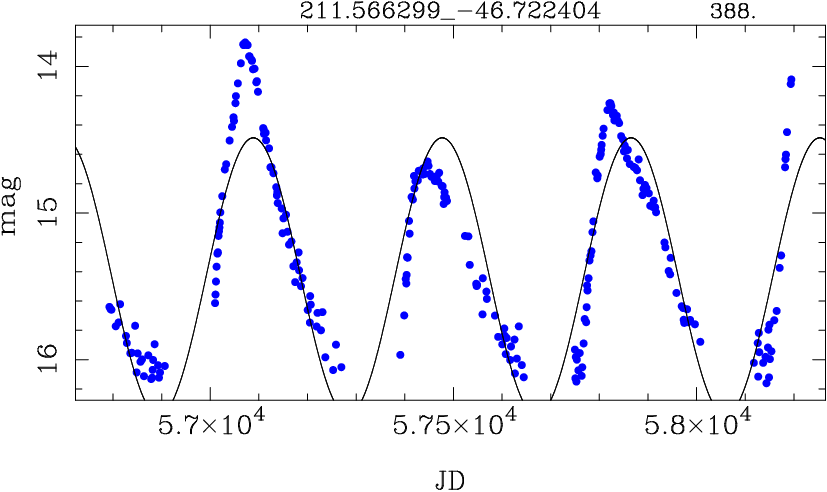}} 
\end{minipage}
\begin{minipage}{0.24\textwidth}
\resizebox{\hsize}{!}{\includegraphics[angle=-0]{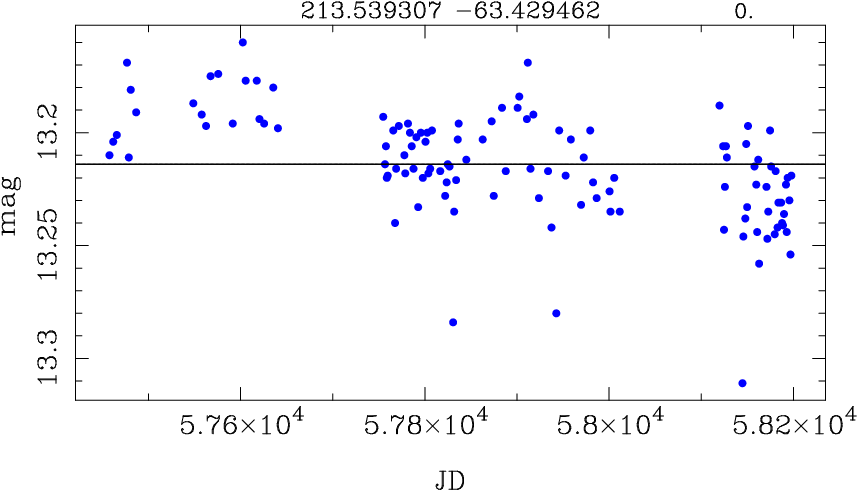}} 
\end{minipage}
\begin{minipage}{0.24\textwidth}
\resizebox{\hsize}{!}{\includegraphics[angle=-0]{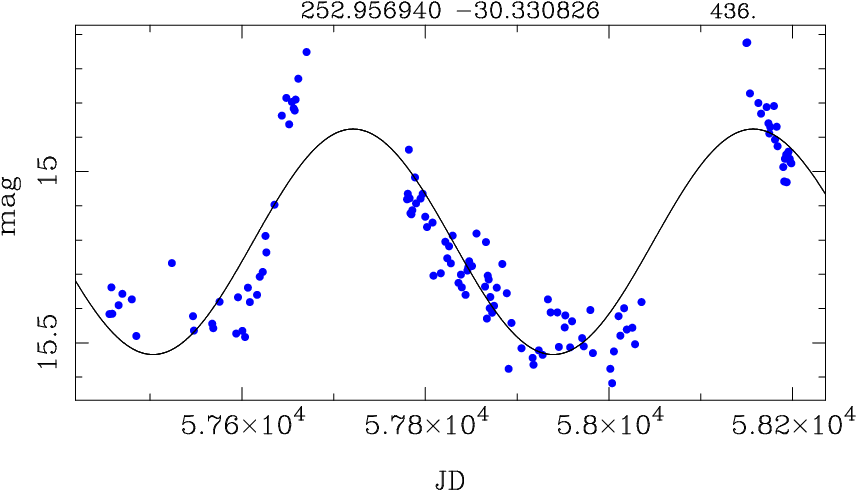}} 
\end{minipage}
\begin{minipage}{0.24\textwidth}
\resizebox{\hsize}{!}{\includegraphics[angle=-0]{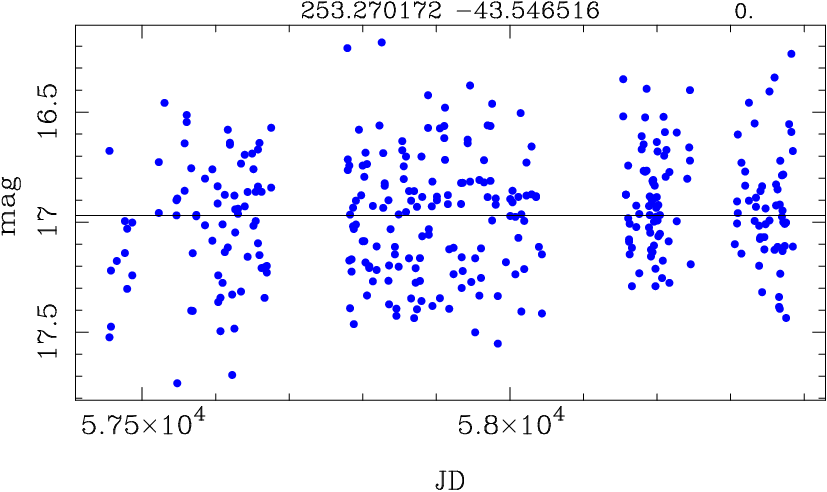}} 
\end{minipage}
 
% 2
\caption{Examples of fits to ASAS-SN $V$-band data.
  The identifier is listed on the top of each panel, with the period to the right.
}
\label{Fig:ASASSN}

\end{figure}

\begin{figure}
\begin{minipage}{0.24\textwidth}
\resizebox{\hsize}{!}{\includegraphics[angle=-0]{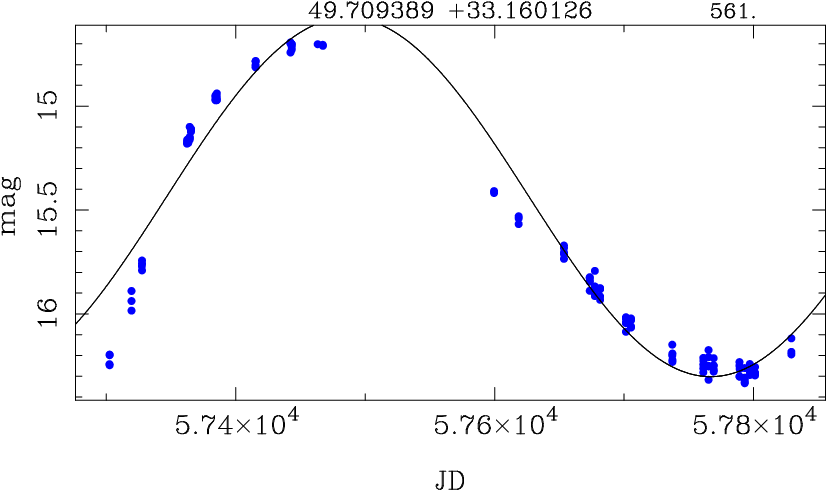}} 
\end{minipage}
\begin{minipage}{0.24\textwidth}
\resizebox{\hsize}{!}{\includegraphics[angle=-0]{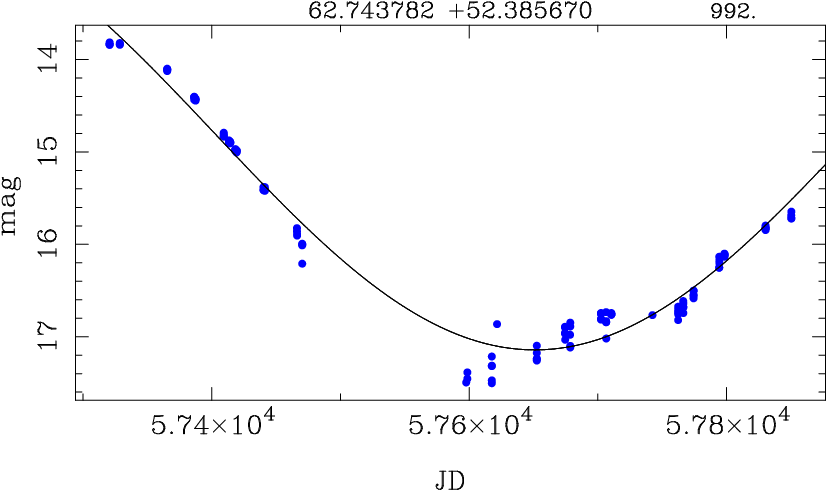}} 
\end{minipage}
 \begin{minipage}{0.24\textwidth}
\resizebox{\hsize}{!}{\includegraphics[angle=-0]{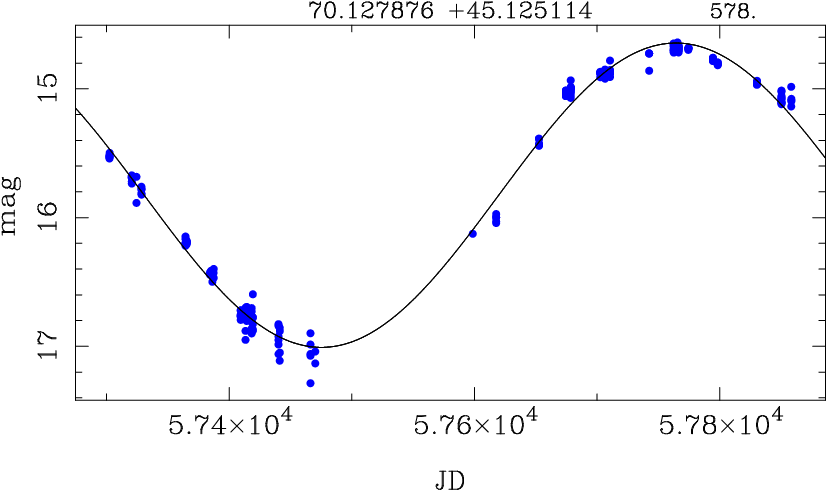}} 
\end{minipage}
\begin{minipage}{0.24\textwidth}
\resizebox{\hsize}{!}{\includegraphics[angle=-0]{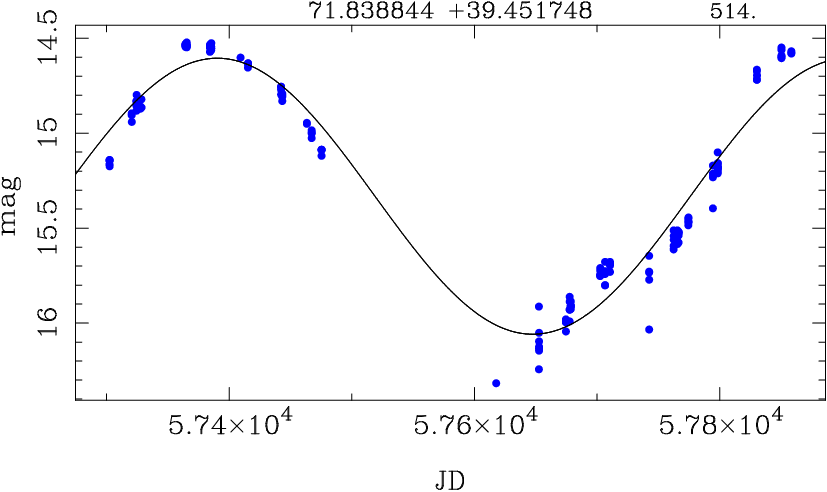}} 
\end{minipage}
 
\begin{minipage}{0.24\textwidth}
\resizebox{\hsize}{!}{\includegraphics[angle=-0]{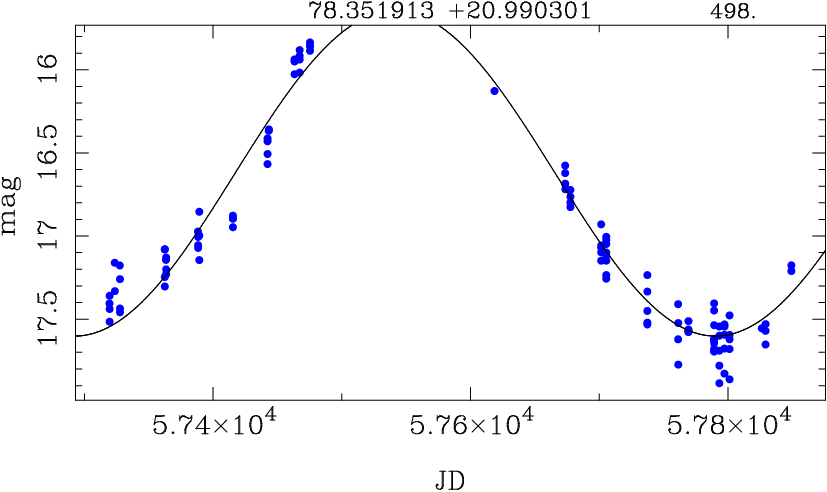}} 
\end{minipage}
\begin{minipage}{0.24\textwidth}
\resizebox{\hsize}{!}{\includegraphics[angle=-0]{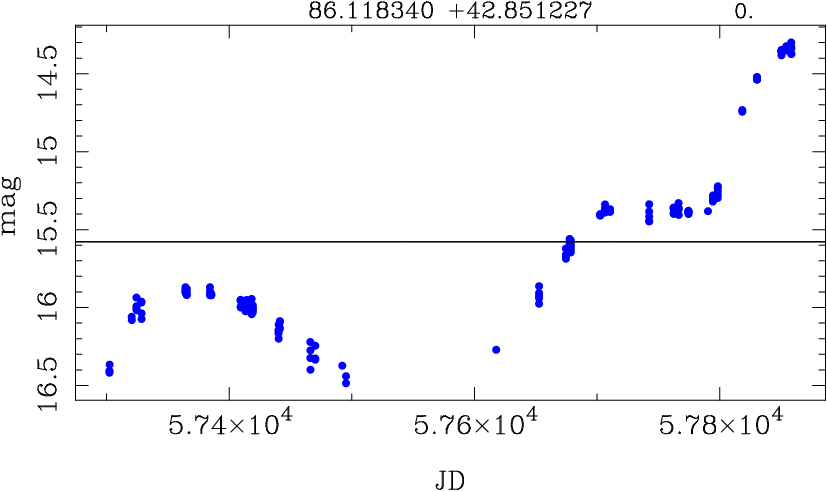}} 
\end{minipage}
\begin{minipage}{0.24\textwidth}
\resizebox{\hsize}{!}{\includegraphics[angle=-0]{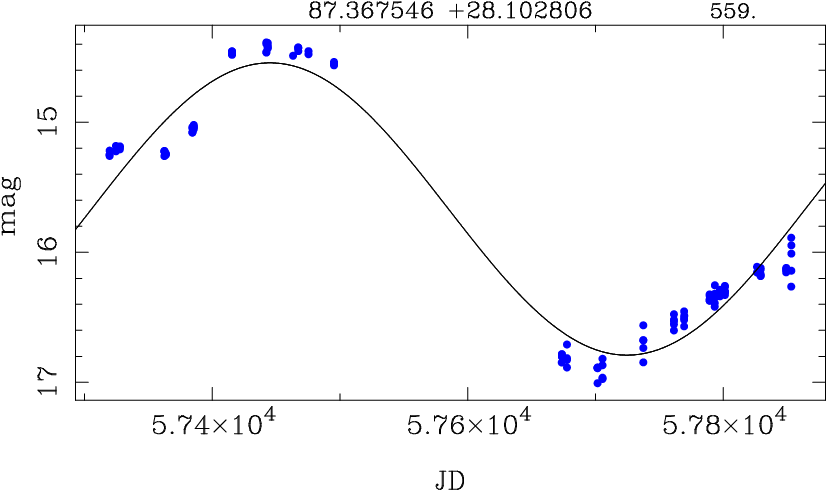}} 
\end{minipage}
\begin{minipage}{0.24\textwidth}
\resizebox{\hsize}{!}{\includegraphics[angle=-0]{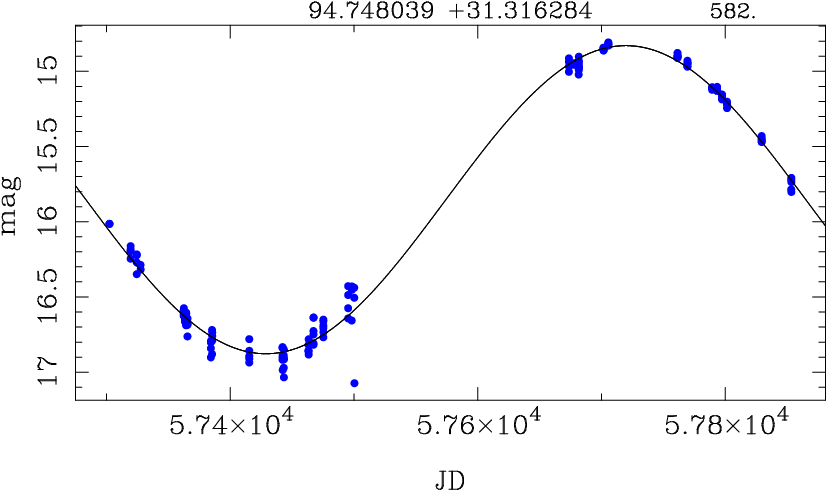}} 
\end{minipage}
 
\begin{minipage}{0.24\textwidth}
\resizebox{\hsize}{!}{\includegraphics[angle=-0]{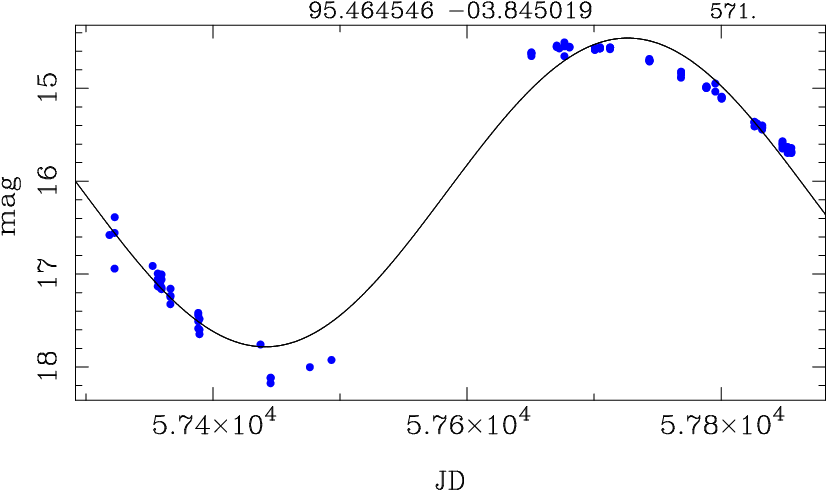}} 
\end{minipage}
\begin{minipage}{0.24\textwidth}
\resizebox{\hsize}{!}{\includegraphics[angle=-0]{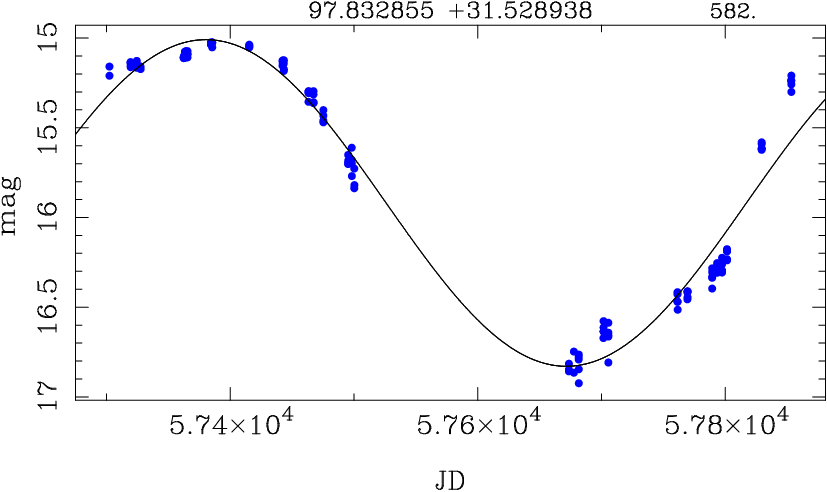}} 
\end{minipage}
\begin{minipage}{0.24\textwidth}
\resizebox{\hsize}{!}{\includegraphics[angle=-0]{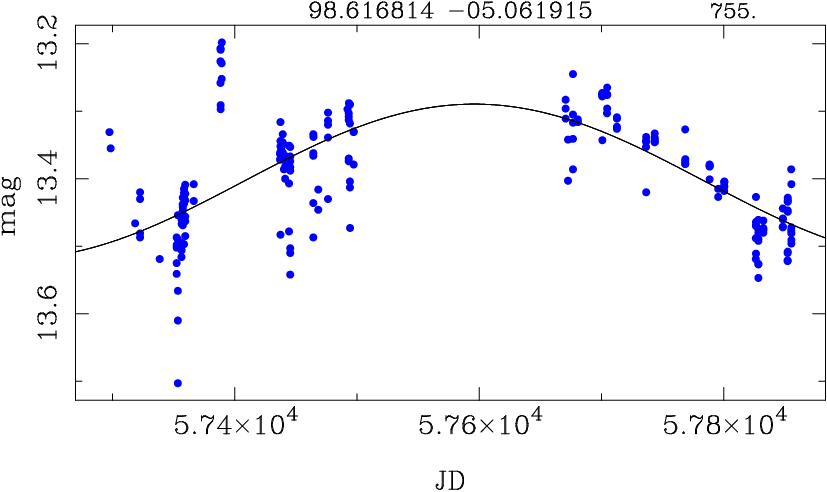}} 
\end{minipage}
\begin{minipage}{0.24\textwidth}
\resizebox{\hsize}{!}{\includegraphics[angle=-0]{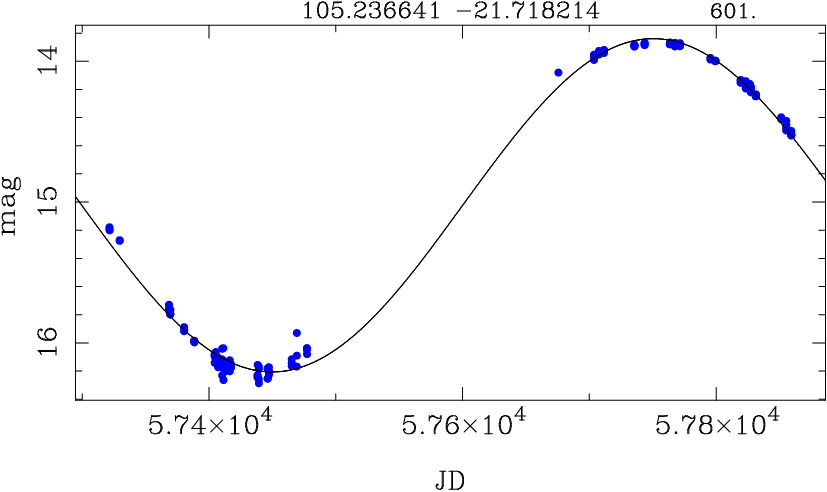}} 
\end{minipage}
 
% 2
\caption{Examples of fits to ATLAS $o$-band data.
  The identifier is listed on the top of each panel, with the period to the right.
}
\label{Fig:ATLAS}

\end{figure}

\begin{figure}

\begin{minipage}{0.24\textwidth}
\resizebox{\hsize}{!}{\includegraphics[angle=-0]{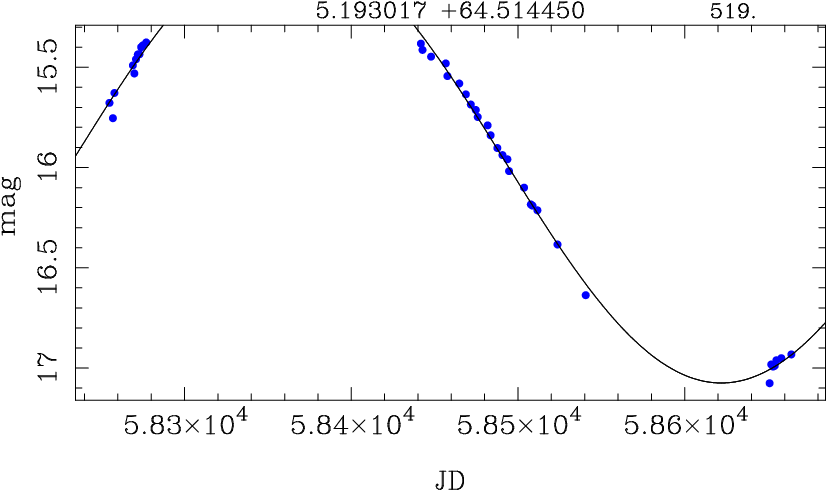}} 
\end{minipage}
\begin{minipage}{0.24\textwidth}
\resizebox{\hsize}{!}{\includegraphics[angle=-0]{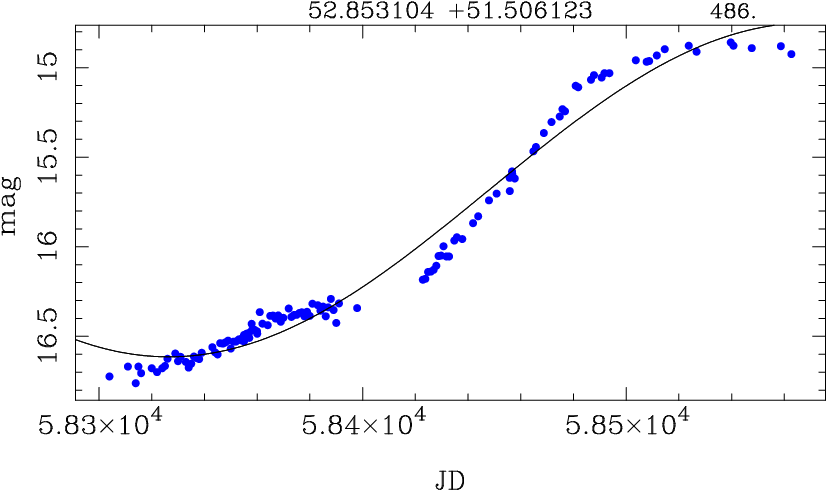}} 
\end{minipage}
 \begin{minipage}{0.24\textwidth}
\resizebox{\hsize}{!}{\includegraphics[angle=-0]{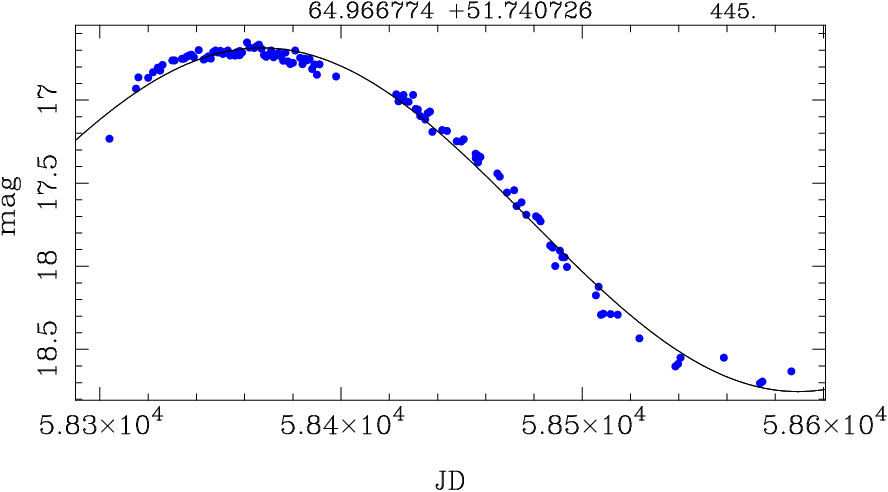}} 
\end{minipage}
\begin{minipage}{0.24\textwidth}
\resizebox{\hsize}{!}{\includegraphics[angle=-0]{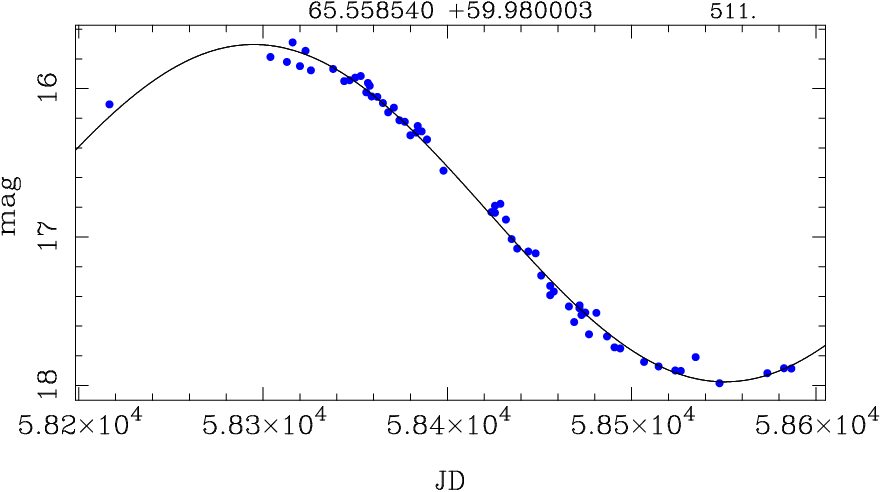}} 
\end{minipage}
 
\begin{minipage}{0.24\textwidth}
\resizebox{\hsize}{!}{\includegraphics[angle=-0]{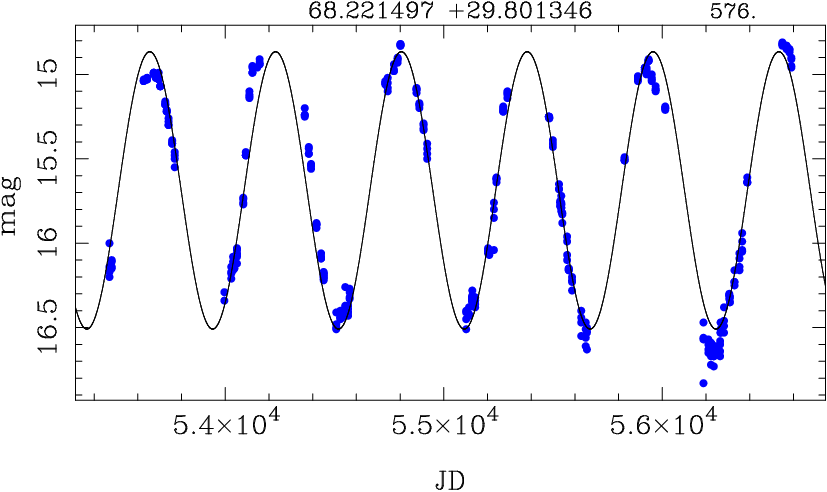}} 
\end{minipage}
\begin{minipage}{0.24\textwidth}
\resizebox{\hsize}{!}{\includegraphics[angle=-0]{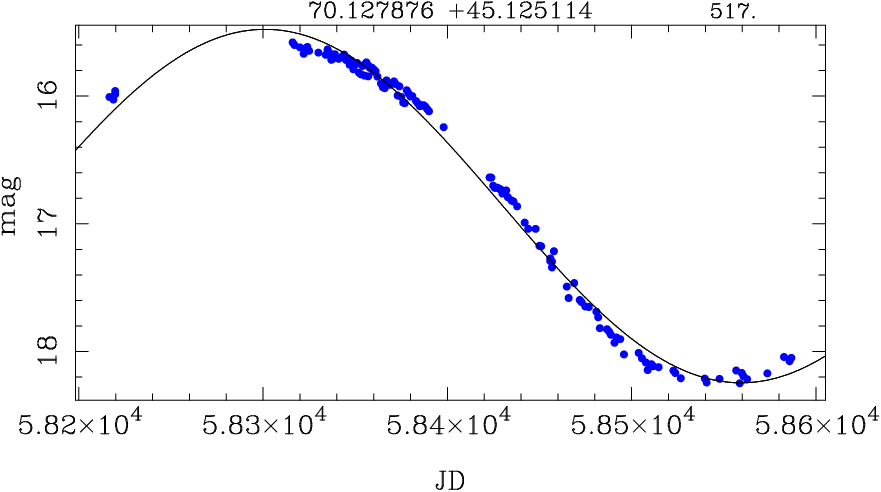}} 
\end{minipage}
\begin{minipage}{0.24\textwidth}
\resizebox{\hsize}{!}{\includegraphics[angle=-0]{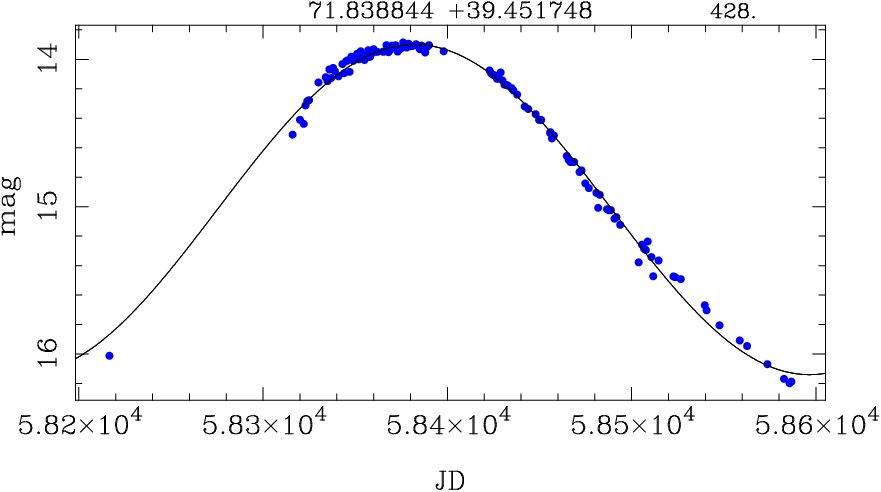}} 
\end{minipage}
\begin{minipage}{0.24\textwidth}
\resizebox{\hsize}{!}{\includegraphics[angle=-0]{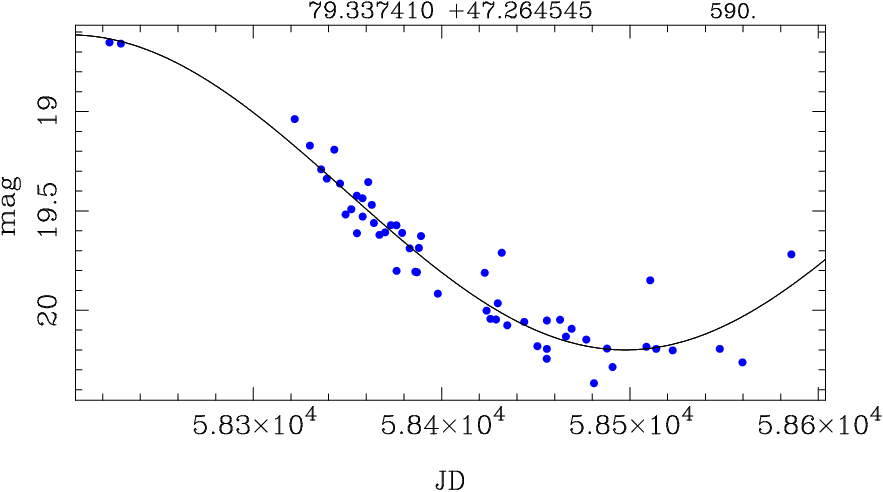}} 
\end{minipage}
 
\begin{minipage}{0.24\textwidth}
\resizebox{\hsize}{!}{\includegraphics[angle=-0]{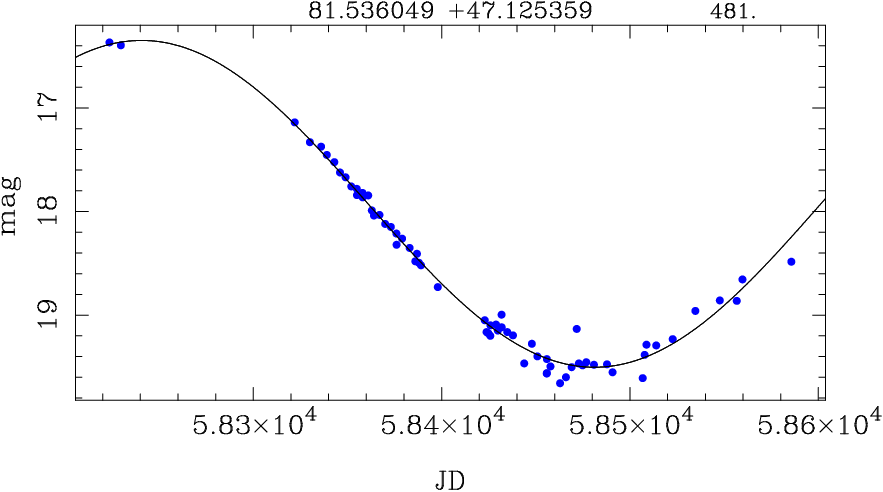}} 
\end{minipage}
\begin{minipage}{0.24\textwidth}
\resizebox{\hsize}{!}{\includegraphics[angle=-0]{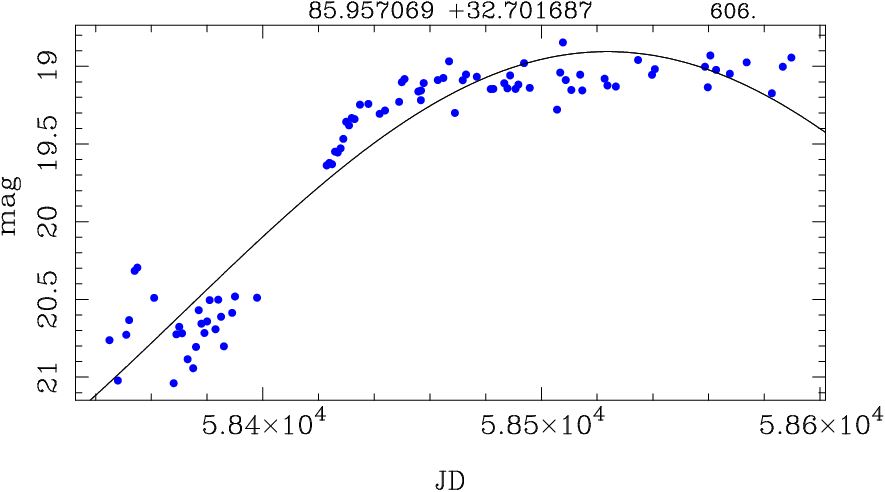}} 
\end{minipage}
\begin{minipage}{0.24\textwidth}
\resizebox{\hsize}{!}{\includegraphics[angle=-0]{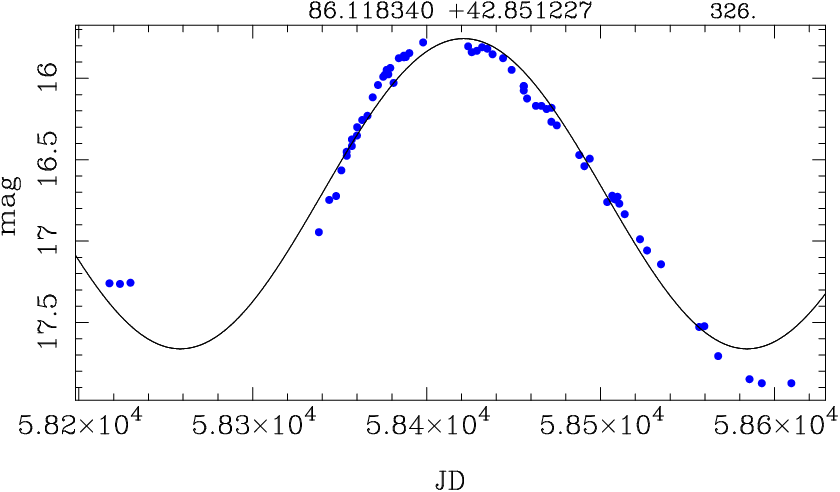}} 
\end{minipage}
\begin{minipage}{0.24\textwidth}
\resizebox{\hsize}{!}{\includegraphics[angle=-0]{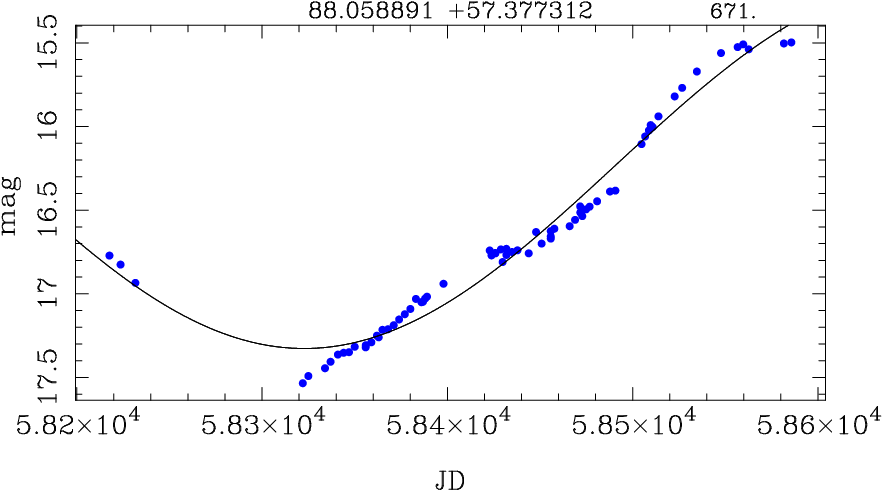}} 
\end{minipage}
 
% 2
\caption{Examples of fits to ZTF $r$-band LCs.
  The identifier is listed on the top of each panel, with the period to the right.
}
\label{Fig:ZTF}

\end{figure}

\begin{figure}

\begin{minipage}{0.24\textwidth}
\resizebox{\hsize}{!}{\includegraphics[angle=-0]{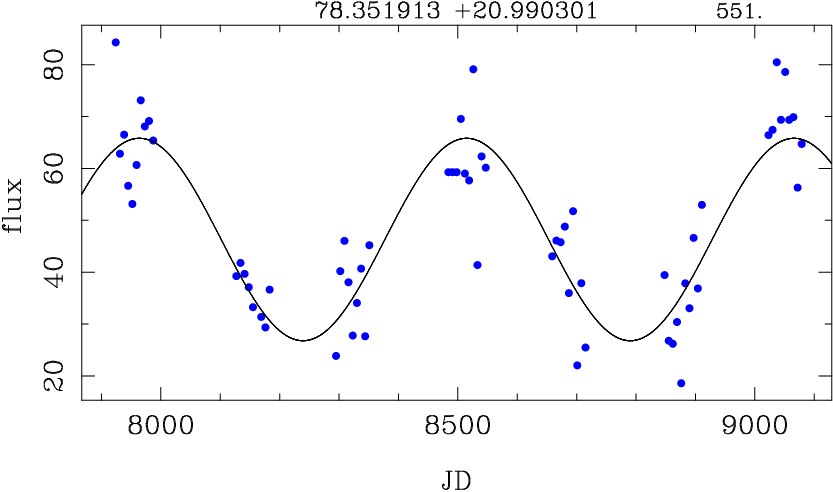}} 
\end{minipage}
\begin{minipage}{0.24\textwidth}
\resizebox{\hsize}{!}{\includegraphics[angle=-0]{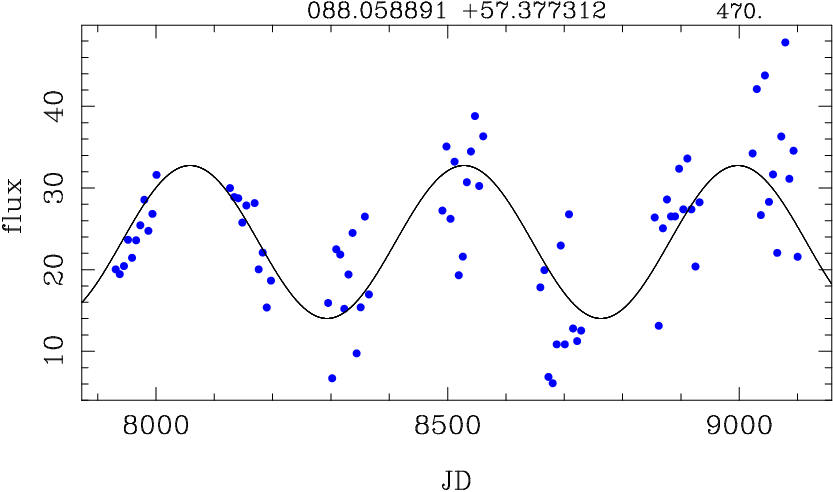}} 
\end{minipage}
\begin{minipage}{0.24\textwidth}
\resizebox{\hsize}{!}{\includegraphics[angle=-0]{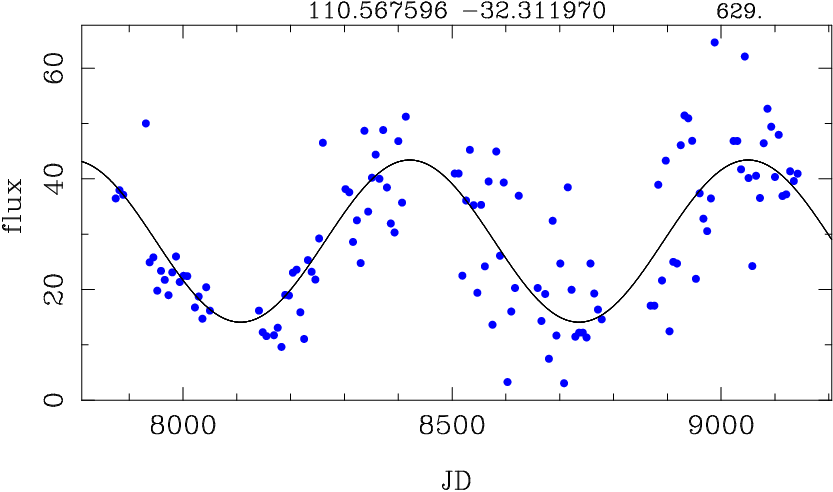}} 
\end{minipage}
\begin{minipage}{0.24\textwidth}
\resizebox{\hsize}{!}{\includegraphics[angle=-0]{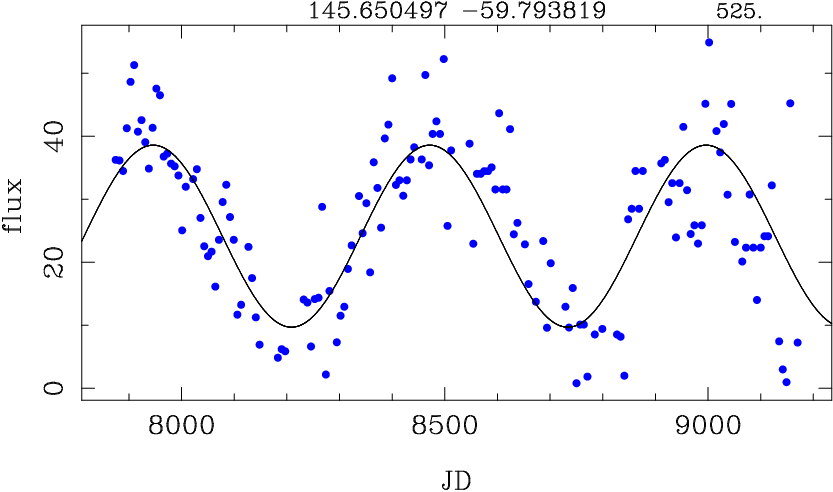}} 
\end{minipage}
 
\begin{minipage}{0.24\textwidth}
\resizebox{\hsize}{!}{\includegraphics[angle=-0]{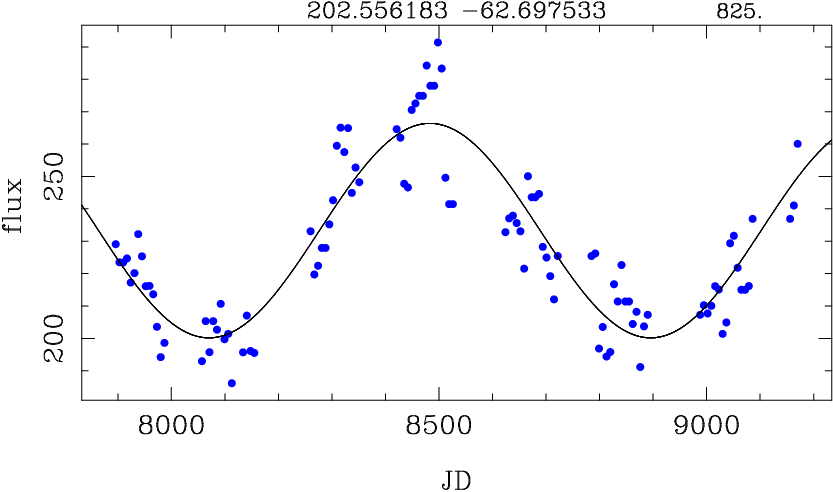}} 
\end{minipage}
\begin{minipage}{0.24\textwidth}
\resizebox{\hsize}{!}{\includegraphics[angle=-0]{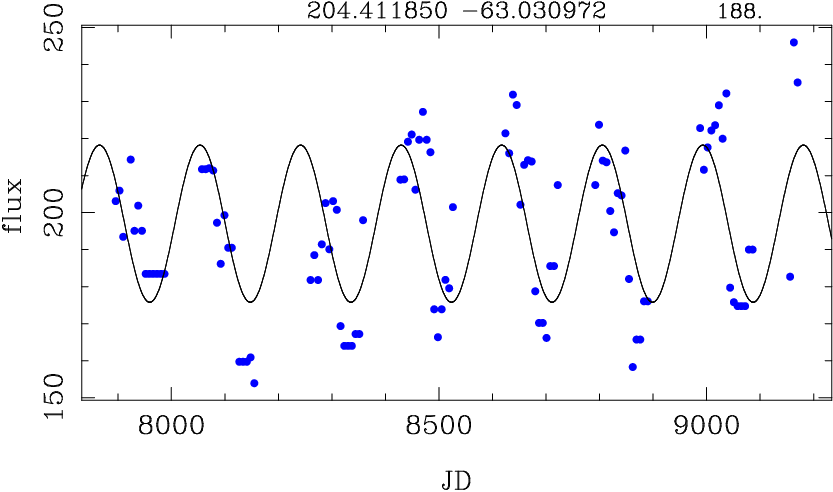}} 
\end{minipage}
\begin{minipage}{0.24\textwidth}
\resizebox{\hsize}{!}{\includegraphics[angle=-0]{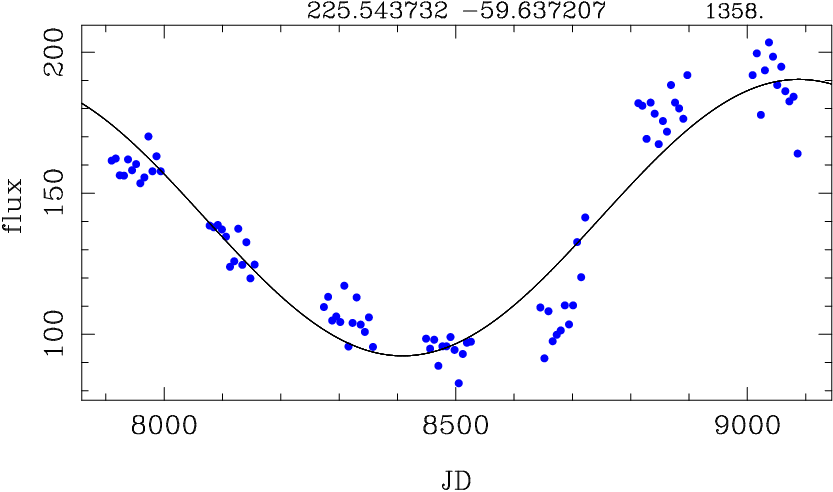}} 
\end{minipage}
\begin{minipage}{0.24\textwidth}
\resizebox{\hsize}{!}{\includegraphics[angle=-0]{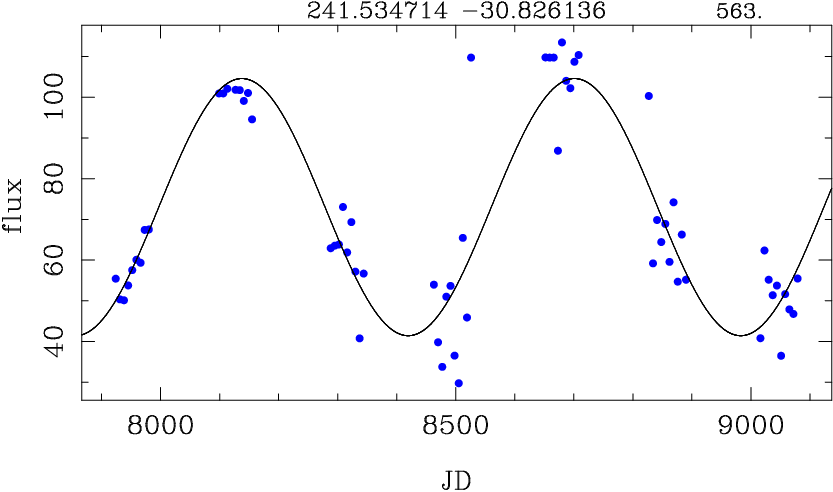}} 
\end{minipage}
 
\begin{minipage}{0.24\textwidth}
\resizebox{\hsize}{!}{\includegraphics[angle=-0]{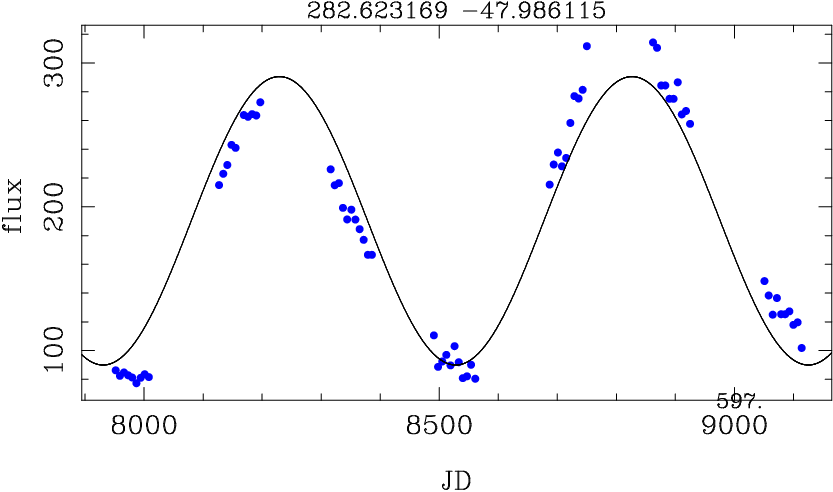}} 
\end{minipage}
\begin{minipage}{0.24\textwidth}
\resizebox{\hsize}{!}{\includegraphics[angle=-0]{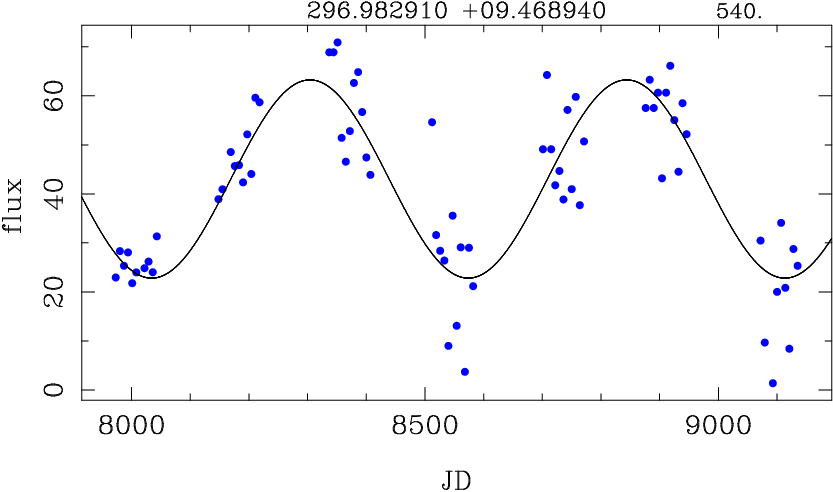}} 
\end{minipage}
 \begin{minipage}{0.24\textwidth}
\resizebox{\hsize}{!}{\includegraphics[angle=-0]{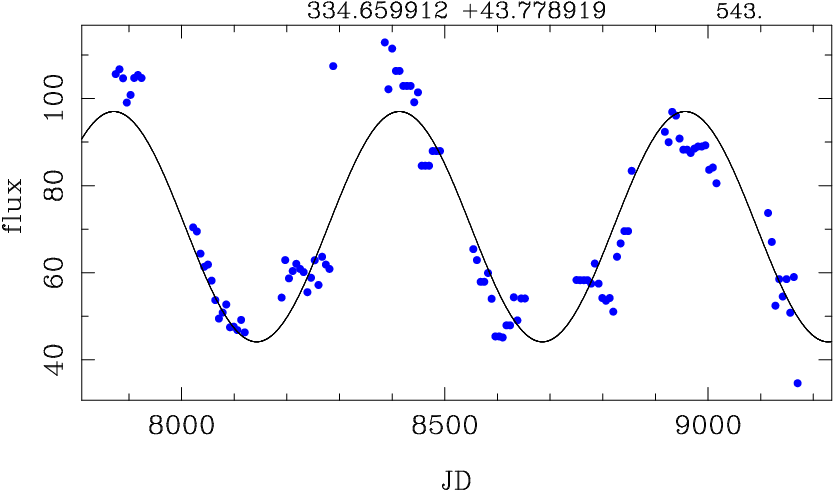}} 
\end{minipage}

  % 2
\caption{Fits to DIRBE  4.9~$\mu$m flux (in Jy).
  The identifier is listed on top of each panel, with the period to the right.
}
\label{Fig:DIRBE}

\end{figure}

\begin{figure}
\begin{minipage}{0.48\textwidth}
\resizebox{\hsize}{!}{\includegraphics[angle=-0]{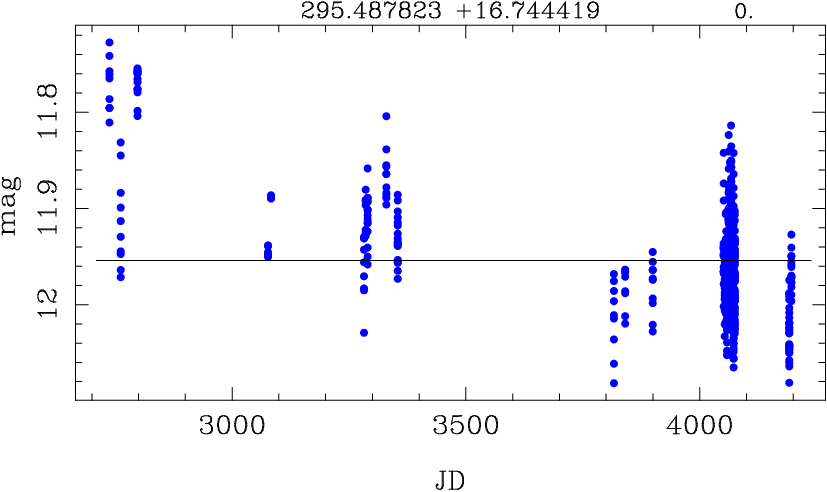}} 
\end{minipage}

  % 2
\caption{Fit to one OMC source in the $V$-band.
  The identifier is listed on top of each panel, with the period to the right.
}
\label{Fig:OMC}

\end{figure}

\begin{figure}
\begin{minipage}{0.48\textwidth}
\resizebox{\hsize}{!}{\includegraphics[angle=-0]{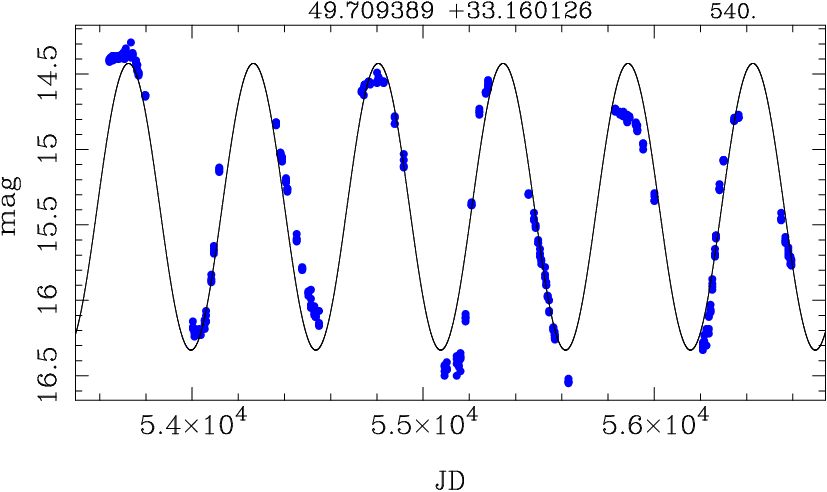}} 
\end{minipage}
\begin{minipage}{0.48\textwidth}
\resizebox{\hsize}{!}{\includegraphics[angle=-0]{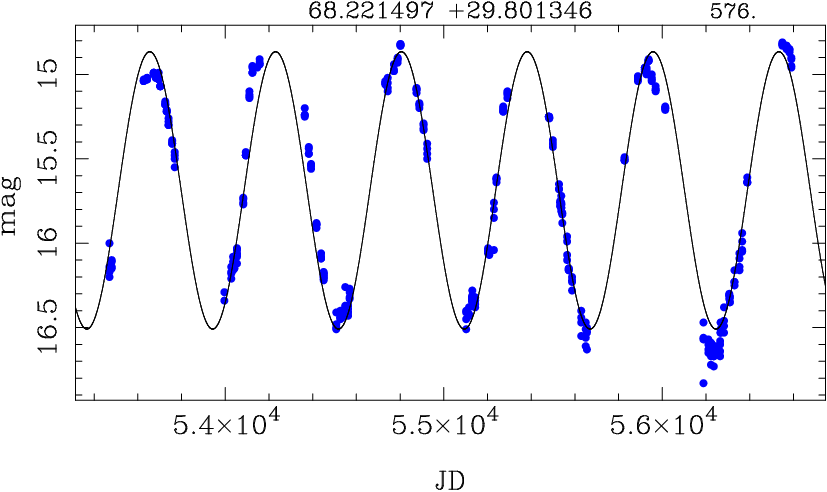}} 
\end{minipage}
\begin{minipage}{0.48\textwidth}
\resizebox{\hsize}{!}{\includegraphics[angle=-0]{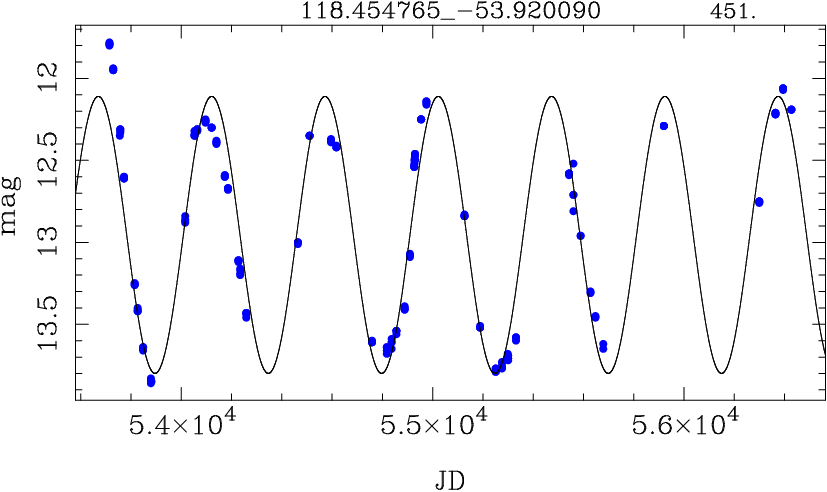}} 
\end{minipage}

  % 2
\caption{Fits to three CSS sources in the $V$-band.
  The identifier is listed on top of each panel, with the period to the right.
}
\label{Fig:CSS}

\end{figure}

\begin{figure}

\begin{minipage}{0.47\textwidth}
\resizebox{\hsize}{!}{\includegraphics[angle=-0]{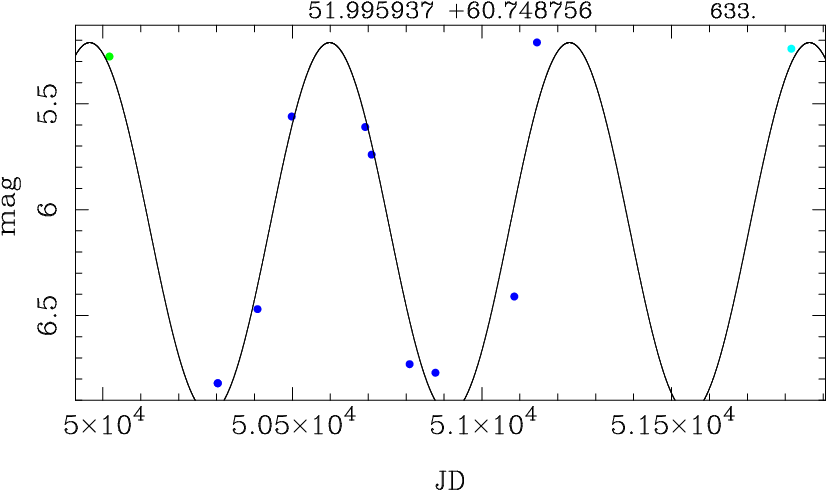}} 
\end{minipage}
\begin{minipage}{0.47\textwidth}
\resizebox{\hsize}{!}{\includegraphics[angle=-0]{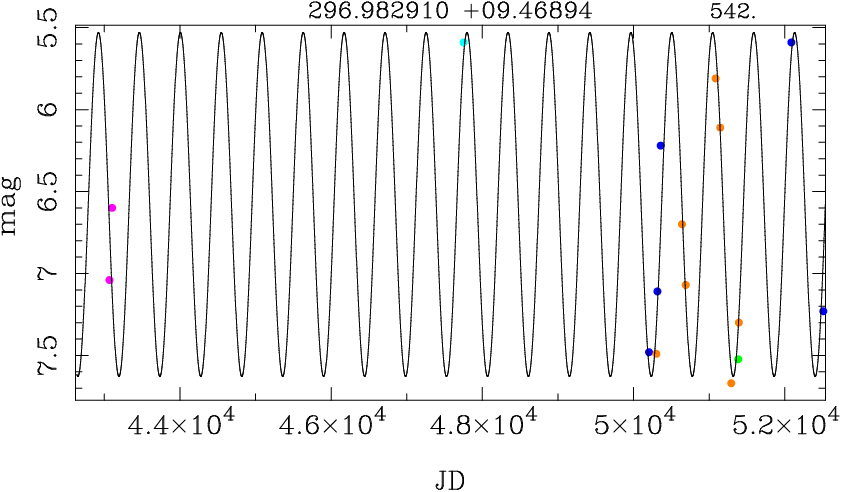}} 
\end{minipage}

\caption{Fits to $K$-band data from \citet{Kerschbaum06}.
  The identifier is listed on top of each panel, with the period to the right.
}
\label{Fig:KGL}

\end{figure}

\begin{figure}

\begin{minipage}{0.47\textwidth}
\resizebox{\hsize}{!}{\includegraphics[angle=-0]{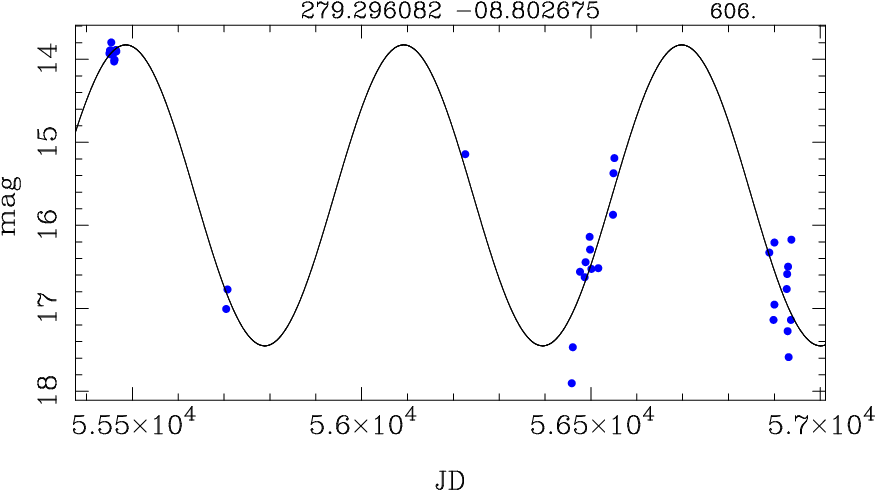}} 
\end{minipage}
\begin{minipage}{0.47\textwidth}
\resizebox{\hsize}{!}{\includegraphics[angle=-0]{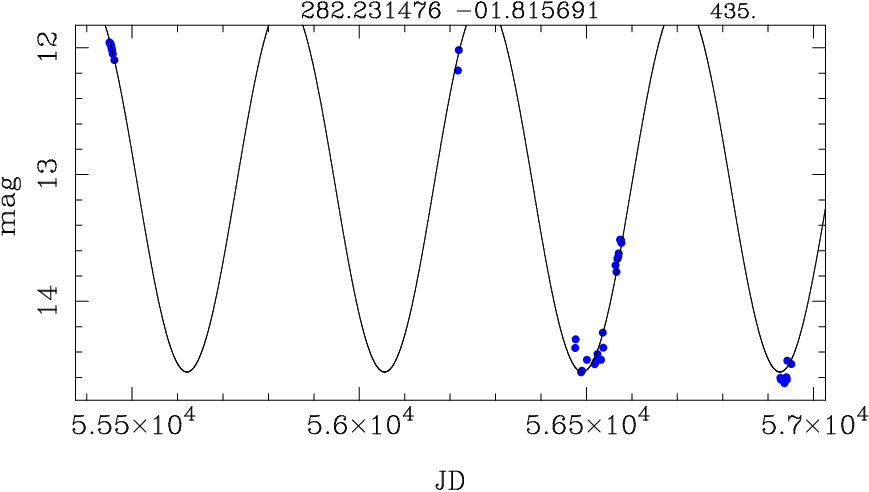}} 
\end{minipage}

\caption{Fits to two GDS  sources in the $I$-band.
  The identifier is listed on top of each panel, with the period to the right.
}
\label{Fig:GDS}

\end{figure}

\begin{figure}

\begin{minipage}{0.24\textwidth}
\resizebox{\hsize}{!}{\includegraphics[angle=-0]{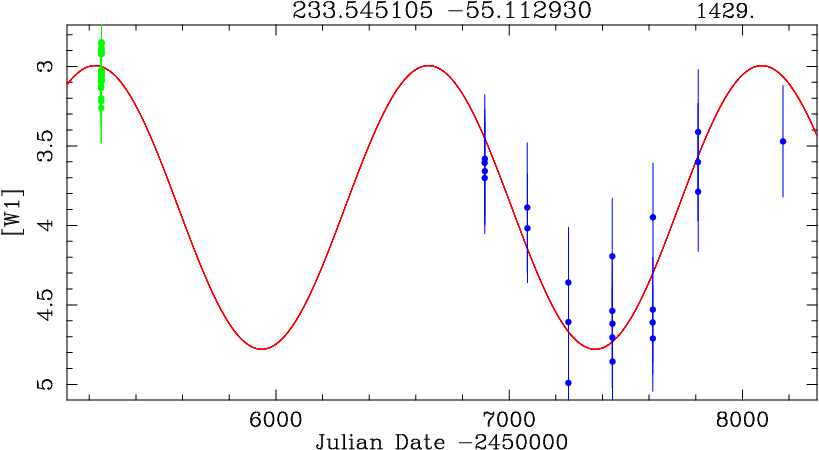}} 
\end{minipage}
\begin{minipage}{0.24\textwidth}
\resizebox{\hsize}{!}{\includegraphics[angle=-0]{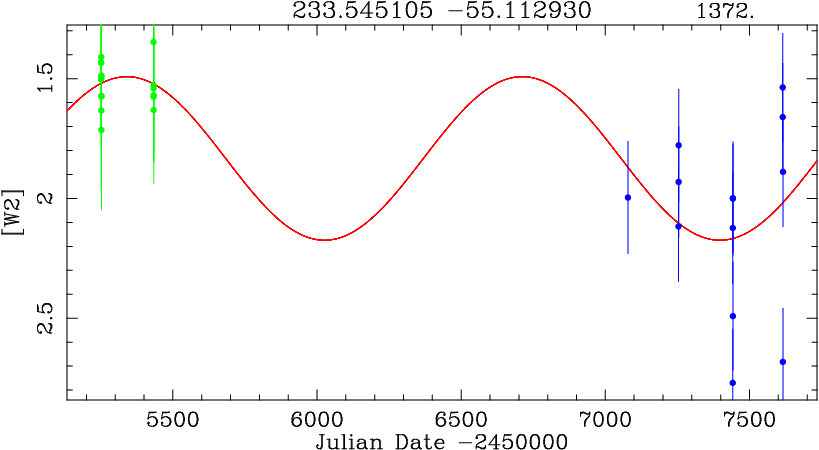}} 
\end{minipage}

\begin{minipage}{0.24\textwidth}
\resizebox{\hsize}{!}{\includegraphics[angle=-0]{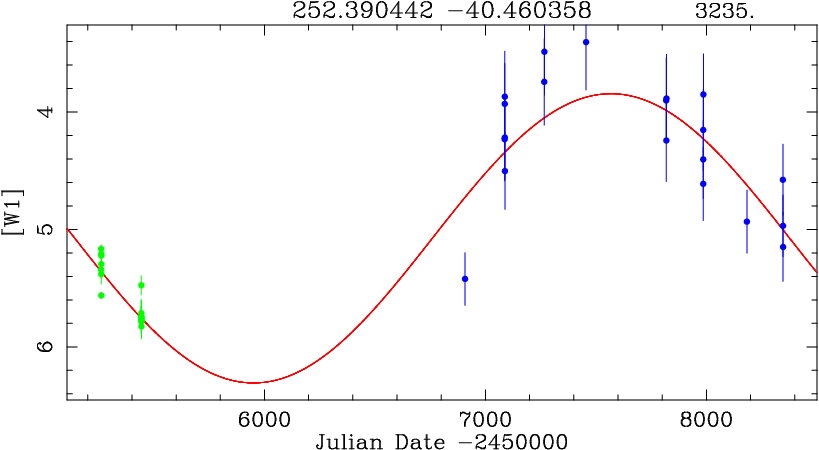}} 
\end{minipage}
\begin{minipage}{0.24\textwidth}
\resizebox{\hsize}{!}{\includegraphics[angle=-0]{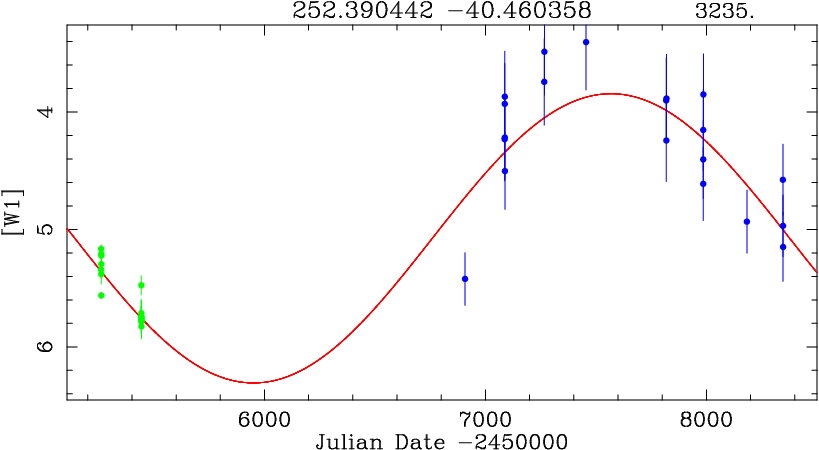}} 
\end{minipage}

\begin{minipage}{0.24\textwidth}
\resizebox{\hsize}{!}{\includegraphics[angle=-0]{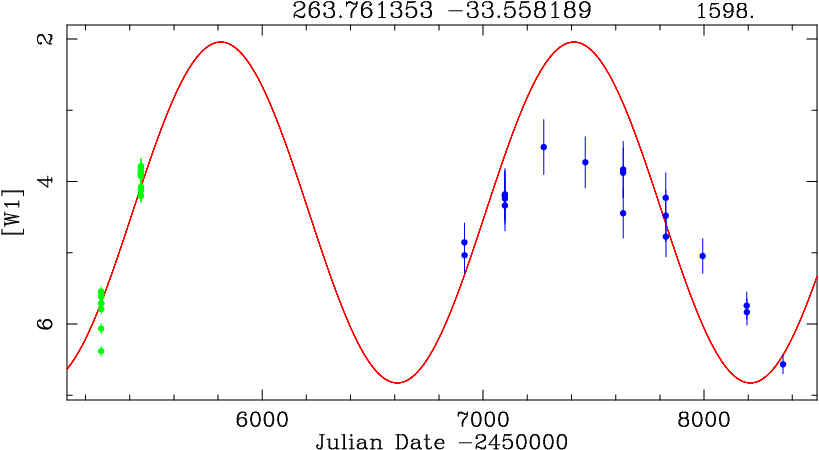}} 
\end{minipage}
\begin{minipage}{0.24\textwidth}
\resizebox{\hsize}{!}{\includegraphics[angle=-0]{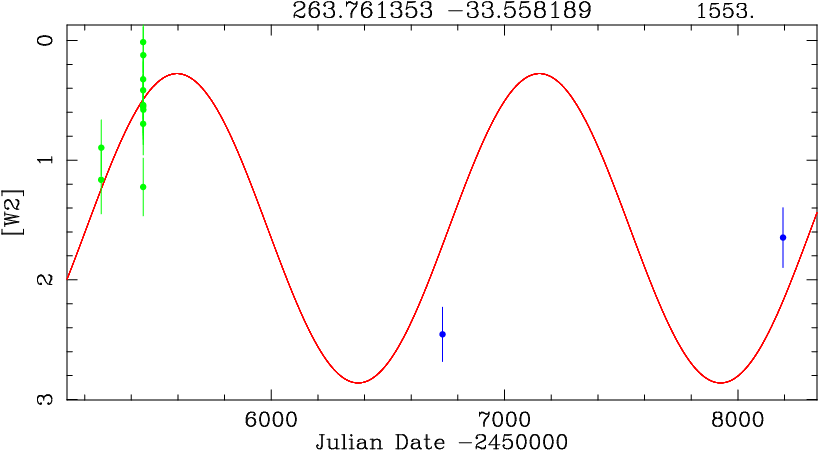}} 
\end{minipage}

\begin{minipage}{0.24\textwidth}
\resizebox{\hsize}{!}{\includegraphics[angle=-0]{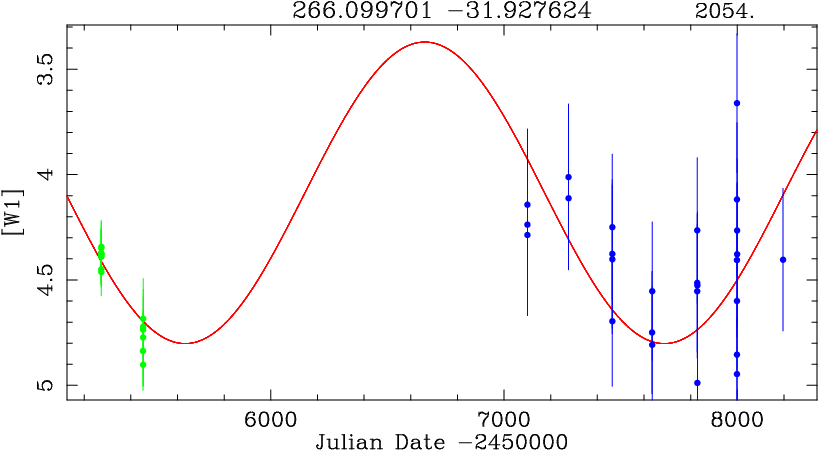}} 
\end{minipage}
\begin{minipage}{0.24\textwidth}
\resizebox{\hsize}{!}{\includegraphics[angle=-0]{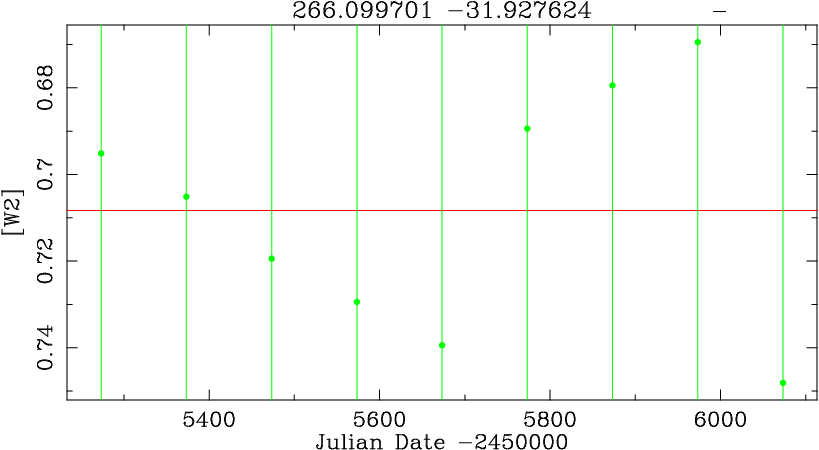}} 
\end{minipage}

\begin{minipage}{0.24\textwidth}
\resizebox{\hsize}{!}{\includegraphics[angle=-0]{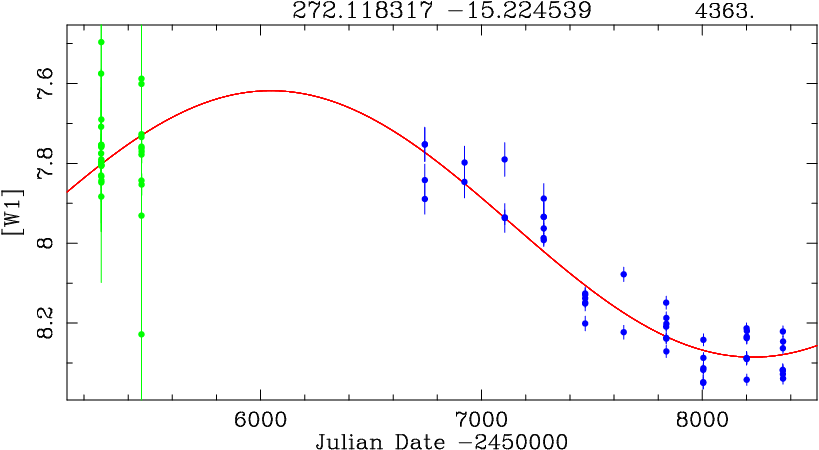}} 
\end{minipage}
\begin{minipage}{0.24\textwidth}
\resizebox{\hsize}{!}{\includegraphics[angle=-0]{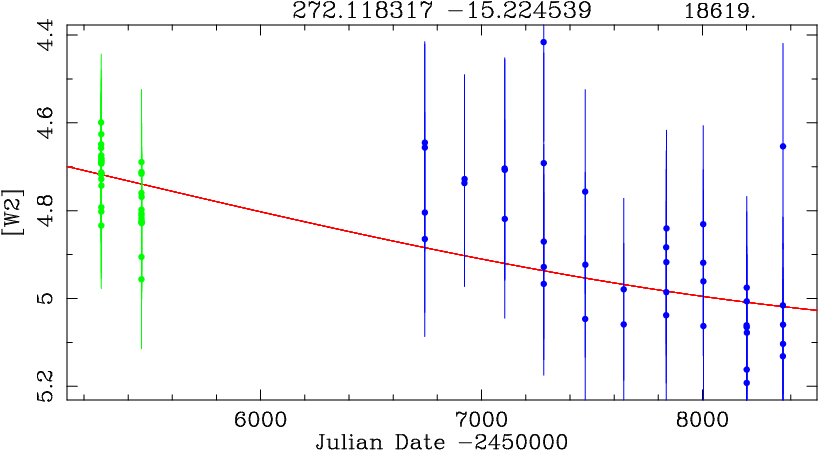}} 
\end{minipage}

\begin{minipage}{0.24\textwidth}
\resizebox{\hsize}{!}{\includegraphics[angle=-0]{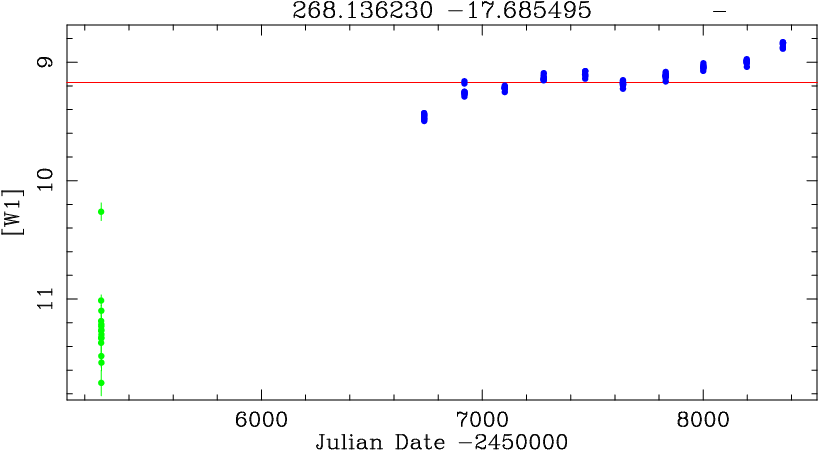}} 
\end{minipage}
\begin{minipage}{0.24\textwidth}
\resizebox{\hsize}{!}{\includegraphics[angle=-0]{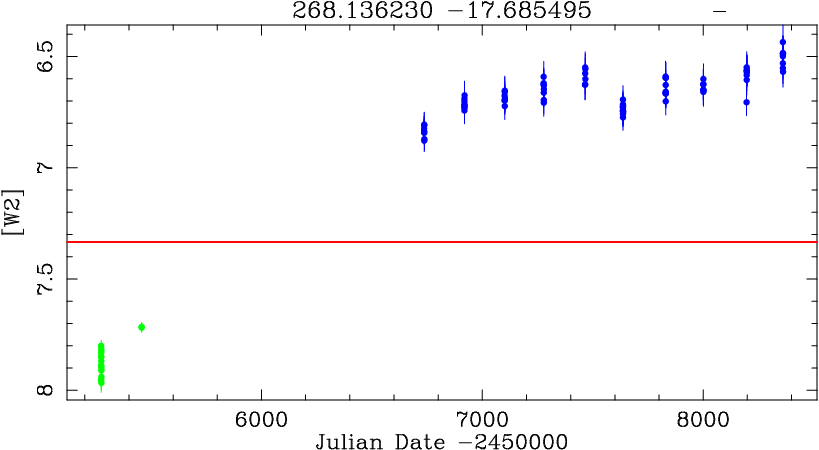}} 
\end{minipage}

\caption{Examples of fits to WISE data of some interesting LCs (see Sect.~\ref{S:LPV}.
  The identifier is listed on the top of each panel, with the period to the right.
}
\label{Fig:WISE-PER}

\end{figure}

Figures~\ref{Fig:WISE}--\ref{Fig:GDS} contain the observed data and fitted LCs, with one figure per separate dataset.
Figure~\ref{Fig:WISE-PER} shows the LCs of some interesting sources (see Sect.~\ref{S:LPV}).

%%%%%%%%%%%%%%%%%%%%%%%%%%%%%%%%%%%%%%%%%%%%%%%%%%%%%%%%%%%%%%%%%%%%%%%%%%%%%%%%%%%%%%%%%
% /home/marting/art72/WISE/OTHERPeriods/Refits /VMC.tex

%%%%%%%%%%%%%%%%%%%%%%%%%%%%%%%%%%%%%%%%%%%%%%%%%%%%%%%%%%%%%%%%%%%%%%%%%%%%%%%%%%%%%%%%%
% /home/marting/art72/WISE/OTHERPeriods/Refits /VVV.tex

%%%%%%%%%%%%%%%%%%%%%%%%%%%%%%%%%%%%%%%%%%%%%%%%%%%%%%%%%%%%%%%%%%%%%%%%%%%%%%%%%%%%%%%%%
% /home/marting/art72/WISE/OTHERPeriods/Refits /ASASSN.tex

%%%%%%%%%%%%%%%%%%%%%%%%%%%%%%%%%%%%%%%%%%%%%%%%%%%%%%%%%%%%%%%%%%%%%%%%%%%%%%%%%%%%%%%%%
% /home/marting/art72/WISE/OTHERPeriods/Refits /ATLAS.tex

%%%%%%%%%%%%%%%%%%%%%%%%%%%%%%%%%%%%%%%%%%%%%%%%%%%%%%%%%%%%%%%%%%%%%%%%%%%%%%%%%%%%%%%%%%%%%%%%%%%%
% /home/marting/art72/WISE/OTHERPeriods/Refits /ZTF.tex

%%%%%%%%%%%%%%%%%%%%%%%%%%%%%%%%%%%%%%%%%%%%%%%%%%%%%%%%%%%%%%%%%%%%%%%%%%%%%%%%%%%%%%%%%%%%%%%%%%%%%%%%%%%%%%%%%%%%%%%%%%%%%%%%%%%%%%%%%%%%
% /home/marting/art72/WISE/OTHERPeriods/Refits /DIRBE.tex

%%%%%%%%%%%%%%%%%%%%%%%%%%%%%%%%%%%%%%%%%%%%%%%%%%%%%%%%%%%%%%%%%%%%%%%%%%%%%%%%%%%%%%%%%%%%%%%%%%%%%%%%%%%%%%%%%%%%%%%%%%%%%%%%%%%%%%%%%%%%
% /home/marting/art72/WISE/OTHERPeriods/Refits /OMC.tex

%%%%%%%%%%%%%%%%%%%%%%%%%%%%%%%%%%%%%%%%%%%%%%%%%%%%%%%%%%%%%%%%%%%%%%%%%%%%%%%%%%%%%%%%%%%%%%%%%%%%%%%%%%%%%%%%%%%%%%%%%%%%%%%%%%%%%%%%%%%%
% /home/marting/art72/WISE/OTHERPeriods/Refits /CSS.tex

%%%%%%%%%%%%%%%%%%%%%%%%%%%%%%%%%%%%%%%%%%%%%%%%%%%%%%%%%%%%%%%%%%%%%%%%%%%%%%%%%%%%%%%%%%%%%%%%%%%%%%%%%%%%%%%%%%%%%%%%%%%%%%%%%%%%%%%%%%%%
% /home/marting/art72/WISE/OTHERPeriods/Refits /KGL.tex

%%%%%%%%%%%%%%%%%%%%%%%%%%%%%%%%%%%%%%%%%%%%%%%%%%%%%%%%%%%%%%%%%%%%%%%%%%%%%%%%%%%%%%%%%%%%%%%%%%%%%%%%%%%%%%%%%%%%%%%%%%%%%%%%%%%%%%%%%%%%
% /home/marting/art72/WISE/OTHERPeriods/Refits /GDS.tex

%%%%%%%%%%%%%%%%%%%%%%%%%%%%%%%%%%%%%%%%%%%%%%%%%%%%%%%%%%%%%%%%%%%%%%%%%%%%%%%%%%%%%%%%%
%examples of 'Interesting' sources  (marked PER)

%%%%%%%%%%%%%%%%%%%%%%%%%%%%%%%%%%%%%%%%%%%%%%%%%%%%%%%%%%%%%%%%%%%%%%%%%%%%%%%%%%%%%%%%%%%%%%%%%%%%%%%%%%%%%%%%%%%%%%%%%%%%%%%%%%%%%%%%%%%
\FloatBarrier
\section{Constructing and fitting spectral energy distributions}
\label{App-sed}

This appendix describes the construction and fitting of the spectral energy distributions (SEDs) of a subset of stars in order to find
new C-rich EROs and derive their MLRs.
As described in the main text, the intial subsample consisted of
stars with W2$-$W3 $>$3.0 based on the colours of the known C-rich EROs in Table~\ref{Tab-Known}.
Sixty-seven stars with a blue and red detection on the OH maser database of \citet{EngelsOH} were removed as these are confirmed O-rich sources.

For the remaining sample of 316 objects, the following databases were queried to collect photometric data: 
AllWISE \citep{Cutri_Allwise},   {\it Akari} FIS \citep{AkariFIS},  {\it Akari} IRC \citep{AkariIRC},
{\it Akari} NIR PSC \citep{Kato_AkariLMC},
IRAC and MIPS observations from the SAGE program\footnote{Vizier catalogue II/305/archive for IRAC data and
  \url{http://irsa.ipac.caltech.edu/applications/Gator/} for MIPS data.} \citep{Meixner06}
and \citet{Gruendl09}, 
GLIMPSE \citep{GLIMPSE}, MIPSGAL \citep{MIPSGAL}, MSX \citep{MSXgp}, IRAS Point and Faint Source Catalogue \citep{IRASPSC, IRASFSC}, 
{\it Herschel} PACS \citep{PACSpsc}, % at 70, 110, and 160~$\mu$m
{\it Herschel} PACS and SPIRE observations from the Heritage \citep{Seale14} %J/AJ/148/124/LMC_cat J/AJ/148/124/SMC_cat
and the Hi-GAL programme \citep{Molinari16,HIGAL17}, % (Herschel InfraRed Galactic Plane Survey) J/MNRAS/471/100/hcatalog
VVV (VISTA Variable in the Via Lactea Survey, datarelease 2; \citealt{VVVDR2}), 
VMC (VISTA Magellanic Survey, datarelease 4; \citealt{Cioni11}),
the Large Magellanic Cloud Near-Infrared Synoptic Survey \citep{Macri15},
the IRSF Magellanic Clouds Point Source Catalogue \citep{Kato_IRSF},
the JCMT Plane Survey \citep{Eden17}, and
ATLASGAL  \citep{Csengeri14}.
After inspecting the SEDs and an initial round of model fitting (see below) the literature was searched for additional photometric
data for selected sources. Details on the SEDs of individual sources are available upon reasonable request to the author.

In addition, MIR spectra were collected from
the IRAS LRS \citep{VK89}\footnote{\url{http://isc83.astro.unc.edu/iraslrs/getlrs_test.html}}, 
the {\it Infrared Space Observatory} short-wavelength spectrograph (SWS)
from \citet{Sloan03}\footnote{\url{https://users.physics.unc.edu/~gcsloan/library/2003/swsatlas.html}}, and the
SST IRS.
In the latter case the spectra were retrieved through the CASSIS\footnote{\url{http://cassis.astro.cornell.edu}}
service \citep{Lebouteiller11}.

The SEDs and MIR spectra were modelled with MoD \citep{Gr_MOD}.
The central stars of C-stars are represented by model atmospheres from \citet{Aringer09}, and those for 
O-stars with MARCS model atmospheres \citep{Gustafsson_MARCS}.
For stars hotter than AGB stars PHOENIX model atmospheres\footnote{\url{http://phoenix.ens-lyon.fr/Grids/BT-NextGen/SPECTRA/}}
\citep{Hauschildt1999} were used.

For C-stars, the dust composition is a mixture of amorphous carbon (AmC), silicon carbide (SiC), and magnesium sulfide (MgS).
For O-stars, the dust composition is a mixture of amorphous silicates %, corundum,
and metallic iron.
When MIR spectra are available, the distinction between C-rich and O-rich chemistry is in most cases clear and the ratio of
SiC/AmC and MgS/AmC is estimated by eye. %, and similarly for the ratios of corundum and metallic iron relative to silicates.
If no MIR spectra are available, typical ratios (SiC/AMC= 0.05, MgS/AMC= 0.10) are adopted in the fitting.
For the O-rich sources, a ratio of metallic iron to amorphous silicates of 0.15 is adopted, which seems to be a reasonable fit in most cases.
The actual dust composition may be more complicated, but it is not our aim to study this in detail here.
The distinction between C- and O-rich chemistry is then made by fitting the photometry alone, and adopting the best fit.

For a given set of photometry and spectra as input data, 
MoD determines the best-fitting luminosity (for a given distance), %($L$)
dust optical depth, %($\tau$, at 0.55 $\mu$m),
dust temperature at the inner radius, %($T_{\rm c}$),
and slope of the density profile. % ($\rho \sim r^{-p}$).
Any of these parameters can also be fixed.

\begin{figure}

\begin{minipage}{0.49\textwidth}
\resizebox{\hsize}{!}{\includegraphics[angle=-0]{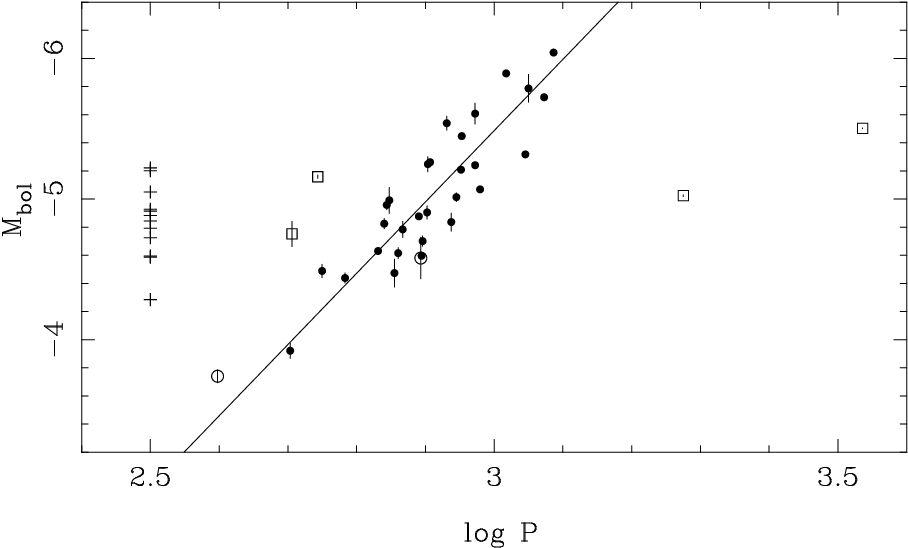}} 
\end{minipage}

\caption{Bolometric $PL$ relation for red C-rich objects  in the Magellanic Clouds.
      Stars without a period are plotted as plus signs at $\log P = 2.5$.
      Stars that are excluded from the fit are plotted as open symbols.
      Circles (two objects) represent stars in the SMC, and the squares are objects in the LMC.
}
\label{Fig:PLrel}
\end{figure}

Canonical distances to the LMC of 50~kpc and 61~kpc to the SMC were adopted, which are well within the error bars of the 
current best estimates \citep{deGrijs2014,deGrijs2015}.
A priori, the distances to the objects in our Galaxy are unknown.
However, we make use of the fact that LPVs are expected to follow a $PL$ relation, and that many ERO candidates
in the sample are located in the LMC (and a few in the SMC).
Figure~\ref{Fig:PLrel} shows the derived $PL$ relation based on 31 objects and with an rms of 0.31~mag.
The functional form is given in the main text, Eq.~\ref{Eq:PL}.

%  cp /home/marting/DUSTY/EROs/PL.ps .

This $PL$ relation was then applied to the ERO candidates in the Galaxy which have a period, to obtain a luminosity, for both C- and O-rich sources.
For C-rich ERO candidates in the Galaxy without a period a luminosity of 7100~\lsol was adopted, which is the median luminosity of C-rich sources in the MCs.
O-rich ERO candidates in the Galaxy without a period were placed at an arbitrary distance of 2 kpc.
For a few sources with exceptionally long (and or uncertain periods) the $PL$ relation is not applied\footnote{These sources are identified with
  ra= 85.435890, 294.898102, 274.757111, 256.972198, 130.866959, and 270.724915 in this paper.} as it would lead to unrealistically large luminosities
incompatible with an AGB status, or, for a few sources at low galactic latitudes, it would lead to large distances with correspondingly large reddenings that are incompatible with the SED.
These sources were treated as if no period were available, that is $L=$ 7100~\lsol\ was assumed if it is C-rich, and a distance of 2.0~kpc if it was O-rich.
In these cases, the objects may indeed not be LPVs, or the periodicity is not related to pulsation, or the period analysis has led to a spurious result.

\smallskip
Interstellar reddening also needs to be taken into account in the fitting.
For the sources in the MCs the recent average values from \citet{Skowron21} were adopted, that is $A_{\rm V}= 0.22$ and $0.10$~mag
for LMC and SMC, respectively, adopting $A_{\rm V}= 3.1 \; E(B-V)$ and $E(B-V)= E(V-I)/1.4$.
For the Galactic sources two recent 3D reddening models were used to estimate the reddening in the direction of the stars in the sample.
The first is described in \cite{Lallement18}\footnote{\url{https://stilism.obspm.fr/} (version 4.1).
  % Creation date : 2018-03-19  queried September 2019
} (hereafter STILISM) and is based on {\it Gaia}, 2MASS and APOGEE-DR14 data.
For a given galactic longitude, latitude and distance, the tool returns the value of $E(B-V)$ and an error, as well as the distance to which these values refer.
If this distance is smaller than the input distance the returned value for the reddening is a lower limit.
In these cases, a simple estimate of the reddening at the distance of the source was made.
A second reddening value was queried at a distance 0.75 times the maximum distance available in the grid in that direction.
Based on this, the first derivative (with an error bar) was determined and the reddening at the distance of the target estimated.
Since the total reddening flattens with distance, this estimate is an (severe) overestimate if the distance to the object is (much) larger than
the last available grid point in the grid. Therefore, the extrapolated reddening is limited to twice that at the last available grid point (see below).

The second reddening model is that described in \citet{Green2019}\footnote{\url{https://argonaut.skymaps.info} The `Bayestar19' dataset.}
and is based on {\it Gaia} DR2 data, 2MASS, and Pan-STARRS~1 data.
Reddening values are provided out to `several' kpc for stars north of declination $-30\degr$.
For 36 sources in overlap and where the distance is available in the STILISM grid, the ratio of the reddening value between
STILISM and the Bayestar19 models is 0.94 with quite some spread (a median absolute deviation of 0.18).
Given this scatter, no attempt was made to scale the two reddening maps.
For over a hundred sources where the reddening is available in the `Bayestar19' map and is extrapolated based on STILISM, it was decided to limit the
extrapolated reddening to twice that at the last available grid point.
The reddening model of \citet{Green2019} is the preferred one as it extends to larger distances, and it is available for 147 of the Galactic sources.
The values from STILISM were adopted for the remaining 120 Galactic sources (with the linear extrapolation of the reddening and the limit to it as just described).

\begin{figure}

\begin{minipage}{0.24\textwidth}
\resizebox{\hsize}{!}{\includegraphics[angle=-0]{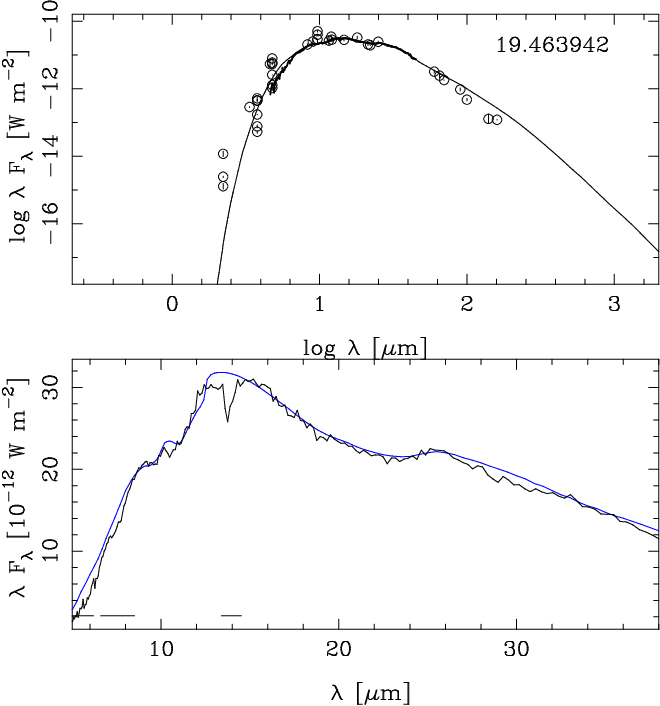}} 
\end{minipage}
\begin{minipage}{0.24\textwidth}
\resizebox{\hsize}{!}{\includegraphics[angle=-0]{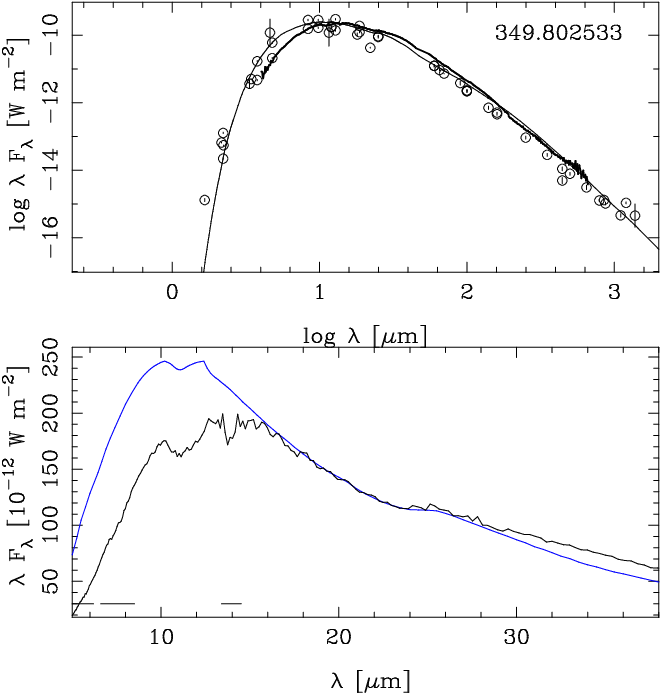}} 
\end{minipage}

\begin{minipage}{0.24\textwidth}
\resizebox{\hsize}{!}{\includegraphics[angle=-0]{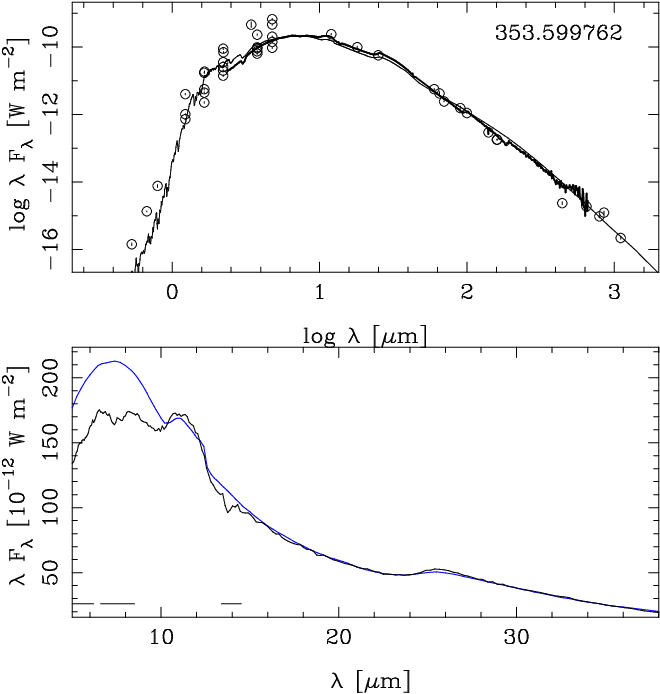}} 
\end{minipage}
\begin{minipage}{0.24\textwidth}
\resizebox{\hsize}{!}{\includegraphics[angle=-0]{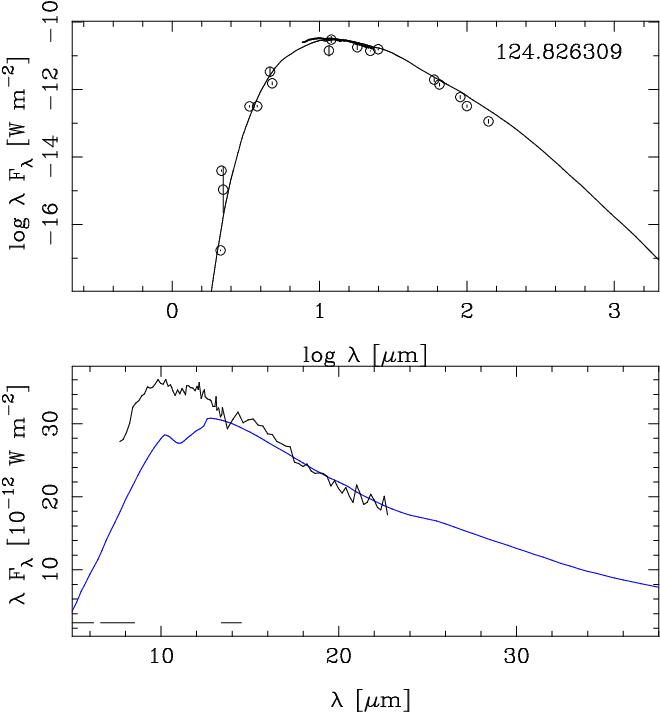}} 
\end{minipage}

\begin{minipage}{0.24\textwidth}
\resizebox{\hsize}{!}{\includegraphics[angle=-0]{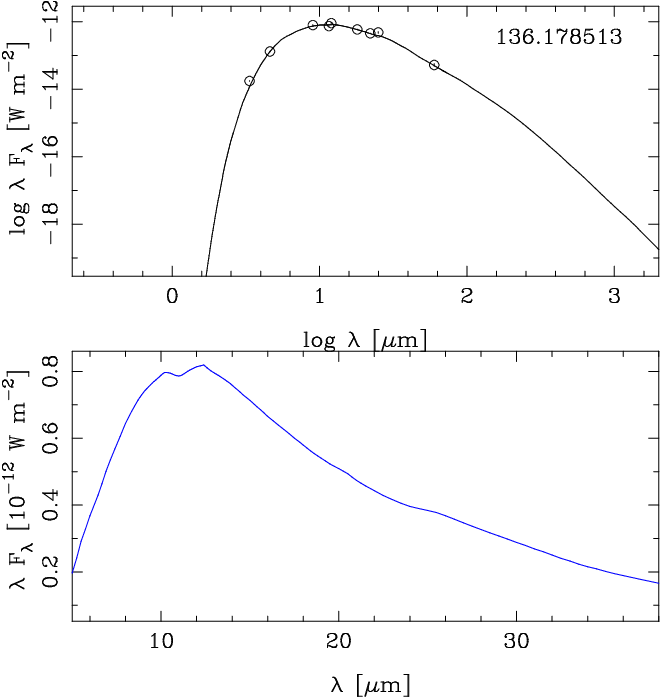}} 
\end{minipage}
\begin{minipage}{0.24\textwidth}
\resizebox{\hsize}{!}{\includegraphics[angle=-0]{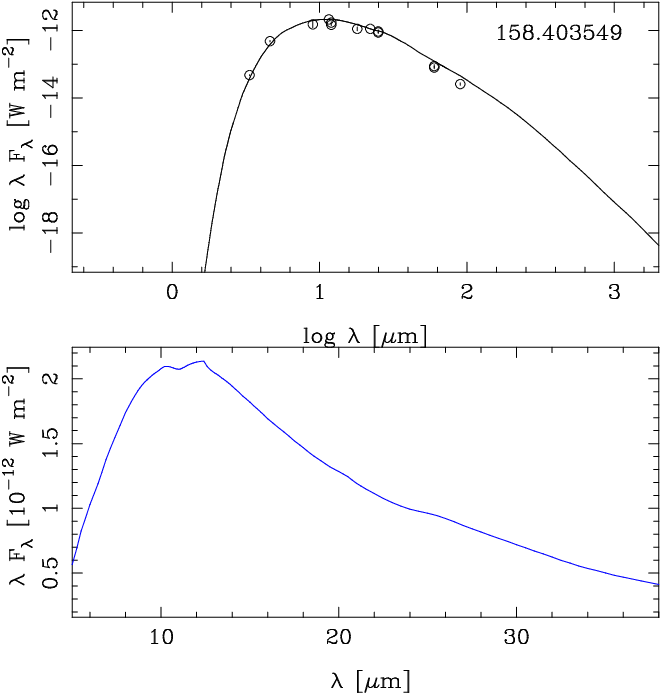}} 
\end{minipage}

\begin{minipage}{0.24\textwidth}
\resizebox{\hsize}{!}{\includegraphics[angle=-0]{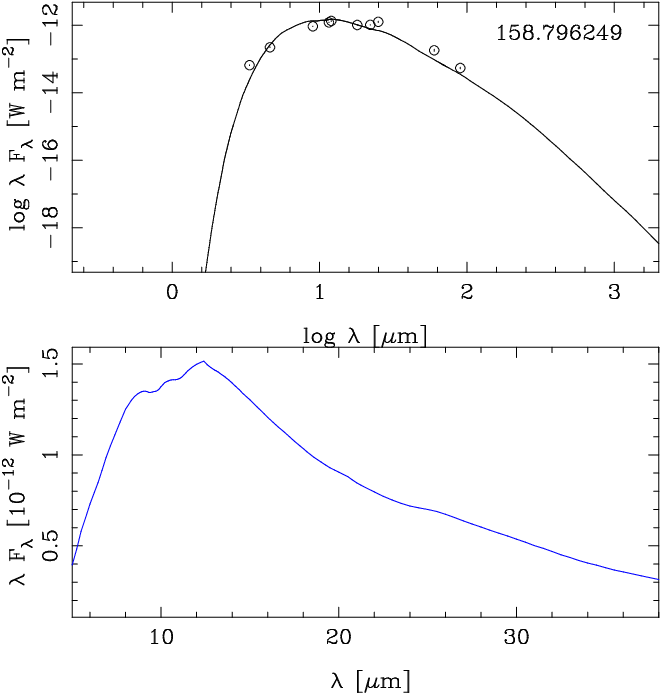}} 
\end{minipage}
\begin{minipage}{0.24\textwidth}
\resizebox{\hsize}{!}{\includegraphics[angle=-0]{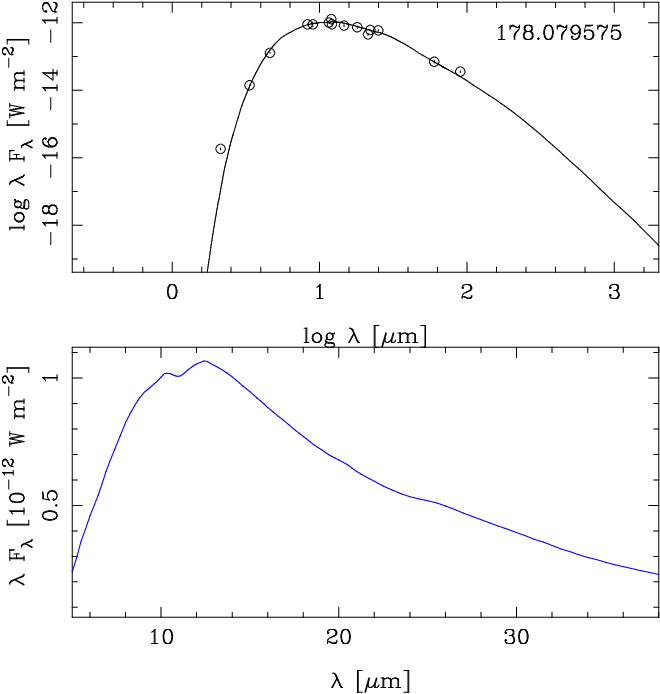}} 
\end{minipage}

\caption{Examples of fits to the SEDs and MIR spectra of some C-stars.
The complete set of SED fits for the O- and C-stars is available at \protect\url{https://doi.org/10.5281/zenodo.5825878}.
To highlight the difference in the dust features, the observations were scaled to the models based on the
average flux in the 16-20~$\mu$m region for the C-stars and the 13-15~$\mu$m region for the O-stars in the MIR panels.  
}
\label{Fig-SED-Cstars}

\end{figure}

The results of fitting the SEDs and MIR spectra are shown is several Tables and Figures.
Figure~\ref{Fig-SED-Cstars} shows some examples of fits for the C-stars in cases when there is, or not, an MIR spectrum available.
The complete set of SED fits for the O- and C-stars is available at \url{https://doi.org/10.5281/zenodo.5825878}.
Tables~\ref{Tab-Cres} and \ref{Tab-Ores} show the result of the fitting. The results for the C-stars are discussed in detail in Section~\ref{S-ERO} in connection
with the mass return of C-rich EROs in the solar neighbourhood. % The results for the O-star are given for completeness in this Appendix.
For another distance, $L \sim d^2$ and \mdot $\sim d$ to first order (ignoring the dependence of the reddening on distance).
A cautionary note is made that the MLRs quoted in the last column assume spherical symmetry.
The SEDs of many of the O-rich sources cannot be fitted very well and are incompatible with the spherical model rendering the MLRs estimates highly uncertain.

%%%%%%%%%%%%%%%%%%%%%%%%%%%%%%%%%%%%%%%%%%%%%%%%%%%%%%%%%%%%%%%%%%%%%%%%%%%%%%%%%%%

\begin{table*} 
\setlength{\tabcolsep}{1.2mm}

\caption{Fit results of the O-star sample (first entries)} 
\begin{tabular}{rrrrrrrrrrrrr} \hline \hline 
RA         & Dec       &  Period & $d$  & $A_{\rm V}$ & $T_{\rm eff}$  &  $L$       & $\tau_{0.5}$        & $T_{\rm c}$      & f & $p$          & f  &  \mdot     \\
(deg)      &  (deg)    &  (days) & (kpc) & (mag)     &    (K)        &   (\lsol)  &                    &    (K)          &   &              &    &  (\msolyr)         \\
\hline 

 32.021637 &  60.767200 &    - &  2.00 &  2.37 &   7200&    1726 $\pm$    341&   29 $\pm$   2.1 &  312 $\pm$  16& 1 & 1.8 $\pm$ 0.4 & 1 & 0.351E-04 \\ 
 83.559708 & -69.789062 &  664 & 50.00 &  0.22 &  30000&  253682 $\pm$  12232&   37 $\pm$   1.6 &  373 $\pm$   5& 1 & 1.8 $\pm$ 0.0 & 0 & 0.571E-03 \\ 
 84.930687 & -69.647270 &    - & 50.00 &  0.22 &  10000& 1424E3 $\pm$ 132E3 &   34 $\pm$   4.0 &  239 $\pm$   5& 1 & 2.0 $\pm$ 0.0 & 0 & 0.319E-02 \\ 
 85.435890 &  -2.268356 &    - &  2.00 &  7.47 &  12500&    1158 $\pm$     12&   23 $\pm$   0.4 &  522 $\pm$   8& 1 & 1.8 $\pm$ 0.0 & 1 & 0.105E-04 \\ 
119.982277 & -41.122845 & 1977 & 26.94 &  0.52 &   5000&   48073 $\pm$   9013&  868 $\pm$ 104 & 1200 $\pm$   0& 0 & 2.2 $\pm$ 0.0 & 0 & 0.740E-03 \\ 
130.866959 & -46.111153 &    - &  2.00 &  1.22 &  10000&     379 $\pm$     18&    8 $\pm$   0.2 &  250 $\pm$   8& 1 & 1.8 $\pm$ 0.0 & 0 & 0.816E-05 \\ 
137.809174 & -45.586235 &    - &  2.00 &  1.01 &   7000&    1837 $\pm$    225&   34 $\pm$   0.8 &  470 $\pm$  19& 1 & 1.1 $\pm$ 0.1 & 1 & 0.392E-05 \\ 
140.372116 & -55.011105 &    - &  2.00 &  0.94 &   7600&    1604 $\pm$    130&   79 $\pm$   7.1 &  552 $\pm$  36& 1 & 1.4 $\pm$ 0.1 & 1 & 0.183E-04 \\ 
155.062988 & -58.053551 &  936 &  4.25 &  0.66 &   3000&   10552 $\pm$   2697&  551 $\pm$  92.3 & 1000 $\pm$   0& 0 & 2.0 $\pm$ 0.0 & 0 & 0.182E-03 \\ 
159.833176 & -59.665565 &    - &  2.00 &  0.87 &   7600&    7549 $\pm$   2324&   91 $\pm$   8.9 &  166 $\pm$  11& 1 & 1.5 $\pm$ 0.0 & 0 & 0.626E-03 \\ 
165.436462 & -60.957218 &    - &  2.00 &  0.90 &   7600 &    1902 $\pm$    326 &   80 $\pm$  12.3 &  270 $\pm$  18 & 1 & 1.5 $\pm$ 0.1 & 1 & 0.976E-04 \\ 
168.056854 & -61.279396 &    - &  2.00 &  0.92 &  25000 &    2068 $\pm$    641 &  205 $\pm$  45.8 &  800 $\pm$   0 & 0 & 1.6 $\pm$ 0.0 & 0 & 0.495E-04 \\ 
169.139252 & -61.498360 &    - &  2.00 &  0.91 &   3400 &    2764 $\pm$    435 &  570 $\pm$  34.6 & 1000 $\pm$   0 & 0 & 2.0 $\pm$ 0.0 & 0 & 0.100E-03 \\ 
175.132355 & -64.307816 &    - &  2.00 &  1.31 &  15000 &     627 $\pm$     47 &   26 $\pm$   0.8 &  277 $\pm$   7 & 1 & 2.0 $\pm$ 0.0 & 0 & 0.429E-04 \\ 
183.846237 & -62.923225 & 1665 &  3.54 &  2.10 &   4000 &   33933 $\pm$  12302 & 1138 $\pm$ 184 & 1200 $\pm$   0 & 0 & 2.0 $\pm$ 0.0 & 0 & 0.691E-03 \\ 
188.649780 & -64.304626 &    - &  2.00 &  1.52 &  40000 &     740 $\pm$     96 &    9 $\pm$   1.1 &  318 $\pm$  28 & 1 & 1.0 $\pm$ 0.0 & 0 & 0.169E-05 \\ 
189.445557 & -62.531727 &  723 &  9.19 &  1.62 &  17500 &    6251 $\pm$    375 &   26 $\pm$   1.1 &  486 $\pm$  19 & 1 & 2.0 $\pm$ 0.0 & 0 & 0.453E-04 \\ 
196.041290 & -63.172348 & 1835 &  8.49 &  2.35 &  30000 &   41329 $\pm$   3026 &  256 $\pm$  11.7 & 1000 $\pm$   0 & 0 & 1.5 $\pm$ 0.0 & 0 & 0.141E-03 \\ 
197.123367 & -62.254200 &    - &  2.00 &  2.50 &   7600 &    1814 $\pm$    401 &   37 $\pm$   1.2 &  190 $\pm$  49 & 1 & 1.5 $\pm$ 0.0 & 0 & 0.900E-04 \\ 
198.607941 & -62.741917 &    - &  2.00 &  1.81 &  40000 &   18613 $\pm$   3344 &  161 $\pm$  18.6 & 1000 $\pm$   0 & 0 & 1.4 $\pm$ 0.0 & 0 & 0.481E-04 \\ 
207.647903 & -61.672104 &    - &  2.00 &  1.12 &  20000 &   12519 $\pm$    750 &   46 $\pm$   0.9 &  397 $\pm$  16 & 1 & 1.5 $\pm$ 0.0 & 0 & 0.872E-04 \\ 
208.393326 & -61.347855 &    - &  2.00 &  1.08 &   2600 &     269 $\pm$     19 &  658 $\pm$  28.5 & 1000 $\pm$   0 & 0 & 2.0 $\pm$ 0.0 & 0 & 0.372E-04 \\ 
217.075897 & -58.612061 &  797 & 11.87 &  1.48 &   2700 &    7617 $\pm$    614 &   91 $\pm$   2.7 &  646 $\pm$  53 & 1 & 2.0 $\pm$ 0.0 & 0 & 0.414E-04 \\ 
222.278412 & -60.765858 &  925 &  7.86 &  2.04 &   3000 &   10316 $\pm$    407 &  198 $\pm$  10.2 & 1000 $\pm$   0 & 0 & 2.0 $\pm$ 0.0 & 0 & 0.465E-04 \\ 
222.906799 & -60.005619 &    - &  2.00 &  1.57 &   7600 &    8175 $\pm$    444 &  119 $\pm$   3.5 & 1422 $\pm$  90 & 1 & 1.0 $\pm$ 0.0 & 1 & 0.207E-05 \\ 
226.084015 & -58.357189 &  842 &  3.80 &  2.96 &   2600 &    8516 $\pm$    332 &  215 $\pm$   4.0 & 1000 $\pm$   0 & 0 & 2.0 $\pm$ 0.0 & 0 & 0.466E-04 \\ 
226.338486 & -57.535931 &  788 &  2.37 &  1.76 &   2700 &    7234 $\pm$   2192 &  174 $\pm$  14.3 &  190 $\pm$  12 & 1 & 2.0 $\pm$ 0.0 & 0 & 0.998E-03 \\ 
233.154190 & -60.618160 &    - &  2.00 &  0.98 &   4200 &     424 $\pm$     65 &  165 $\pm$  26.4 &  341 $\pm$  37 & 1 & 2.0 $\pm$ 0.0 & 0 & 0.847E-04 \\ 
238.526169 & -53.194221 &    - &  2.00 &  1.26 &  20000 &   30574 $\pm$   1881 &   53 $\pm$   1.1 &  377 $\pm$  17 & 1 & 1.8 $\pm$ 0.0 & 1 & 0.279E-03 \\ 
240.232468 & -52.607029 &    - &  2.00 &  1.29 &   7600 &    5593 $\pm$    272 &   46 $\pm$   3.4 &  295 $\pm$   9 & 1 & 1.5 $\pm$ 0.0 & 0 & 0.763E-04 \\ 
240.560760 & -52.643375 &    - &  2.00 &  1.31 &   7600 &    2021 $\pm$    216 &   62 $\pm$   6.4 &  315 $\pm$  14 & 1 & 1.4 $\pm$ 0.1 & 1 & 0.410E-04 \\ 
241.908493 & -52.518345 &    - &  2.00 &  1.47 &   9000 &   20801 $\pm$   1611 &   59 $\pm$   3.1 &  168 $\pm$   6 & 1 & 2.0 $\pm$ 0.0 & 0 & 0.144E-02 \\ 
242.561188 & -51.198288 & 1091 &  8.57 &  1.99 &   4200 &   14398 $\pm$    849 &  274 $\pm$  30.8 & 1000 $\pm$   0 & 0 & 2.0 $\pm$ 0.0 & 0 & 0.999E-04 \\ 
244.260864 & -50.784843 &    - &  2.00 &  1.56 &   4000 &   24084 $\pm$   2285 &  109 $\pm$   4.8 &  301 $\pm$  13 & 1 & 2.0 $\pm$ 0.0 & 0 & 0.489E-03 \\ 
246.458908 & -48.687710 & 1122 &  8.71 &  1.75 &   4200 &   15240 $\pm$    634 &  127 $\pm$  13.3 & 1000 $\pm$   0 & 0 & 2.0 $\pm$ 0.0 & 0 & 0.418E-04 \\ 
246.971359 & -39.095798 &    - &  2.00 &  2.42 &   4600 &     877 $\pm$    190 &  155 $\pm$  12.2 &  433 $\pm$  40 & 1 & 2.0 $\pm$ 0.0 & 0 & 0.758E-04 \\ 
248.134094 & -44.925129 &    - &  2.00 &  1.47 &   4200 &     803 $\pm$     10 &  100 $\pm$   4.6 & 1211 $\pm$  68 & 1 & 1.5 $\pm$ 0.0 & 0 & 0.226E-05 \\ 
248.302139 & -45.228764 &    - &  2.00 &  1.43 &  12500 &     333 $\pm$     12 &   21 $\pm$   0.4 &  272 $\pm$   4 & 1 & 2.0 $\pm$ 0.0 & 0 & 0.244E-04 \\ 
248.374420 & -48.059303 &    - &  2.00 &  1.60 &   6000 &    2573 $\pm$    377 &   58 $\pm$   1.8 &  279 $\pm$   9 & 1 & 1.2 $\pm$ 0.1 & 1 & 0.265E-04 \\ 
248.375641 & -48.056137 &    - &  2.00 &  1.60 &  12500 &   11772 $\pm$   1260 &   43 $\pm$   3.5 &  237 $\pm$   6 & 1 & 0.8 $\pm$ 0.0 & 0 & 0.188E-04 \\ 
249.180099 & -47.524422 &    - &  2.00 &  1.38 &   7600 &   68829 $\pm$   7556 &  124 $\pm$   6.6 &  209 $\pm$   9 & 1 & 2.4 $\pm$ 0.1 & 1 & 0.451E-02 \\

\hline 
\end{tabular}
\label{Tab-Ores}
\tablefoot{
  The meaning of the columns is as in Tab.~\ref{Tab-Cres}.
The entries are listed in order of RA.  
The full table is available at the CDS.
}
\end{table*}

%%%%%%%%%%%%%%%%%%%%%%%%%%%%%%%%%%%%%%%%%%%%%%%%%%%%%%%%%%%%%%%%%%%%%%%%%%%%%%%%%%%%%%%%%%%%%%%%%%%%%%%%%%%%%%%%%%%%%%%%%%%%%%%%%%%%%%%%%%%%%%%%%%%%%%%%%%%%%%%%%%%%%%%%%%%%%%%%%%%%%%
\FloatBarrier
\section{SEDs of P-AGB configurations}
\label{SS-PAGB}

\begin{figure}
 
\begin{minipage}{0.235\textwidth}
\resizebox{\hsize}{!}{\includegraphics[angle=-0]{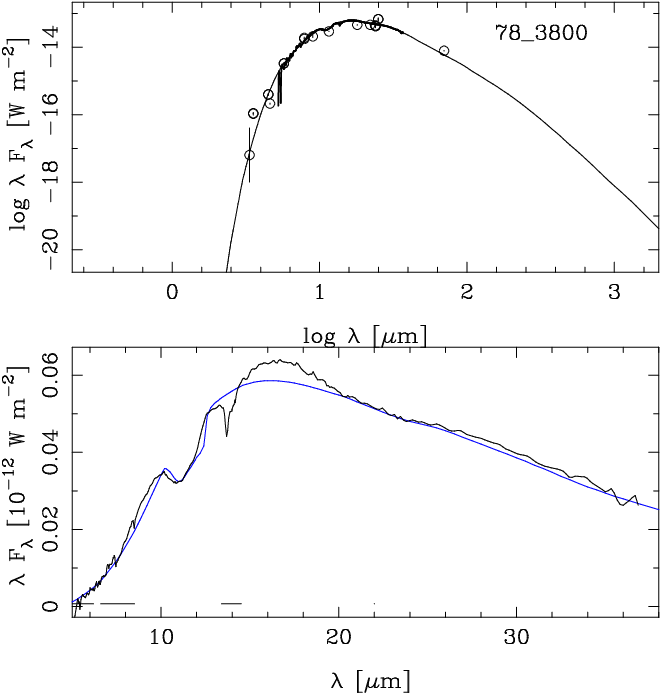}} 
\end{minipage}
\begin{minipage}{0.235\textwidth}
\resizebox{\hsize}{!}{\includegraphics[angle=-0]{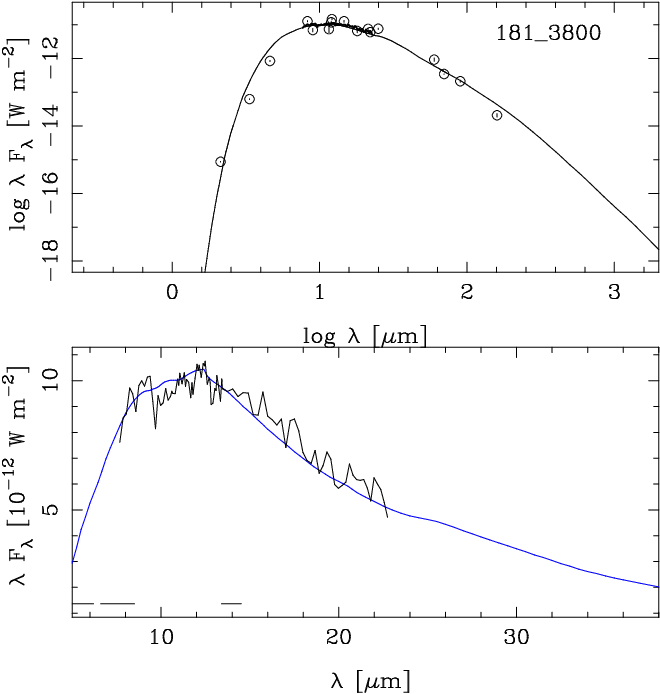}} 
\end{minipage}
\begin{minipage}{0.235\textwidth}
\resizebox{\hsize}{!}{\includegraphics[angle=-0]{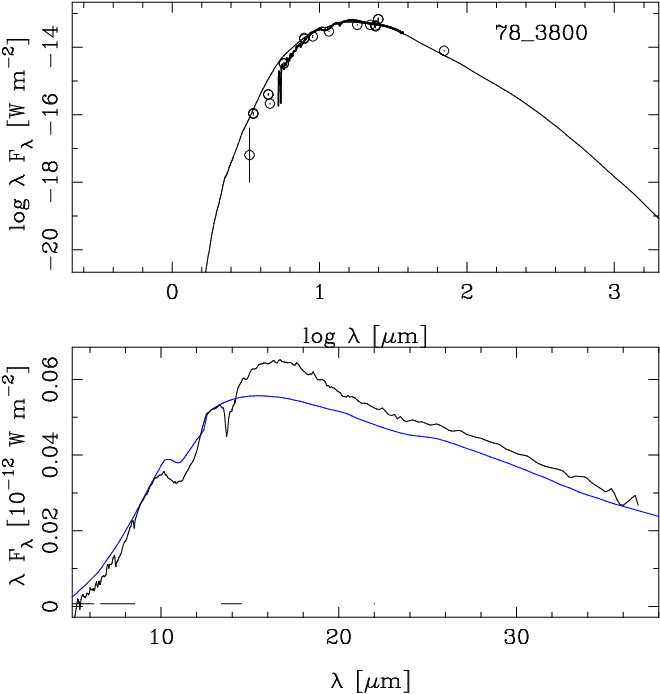}} 
\end{minipage}
\begin{minipage}{0.235\textwidth}
\resizebox{\hsize}{!}{\includegraphics[angle=-0]{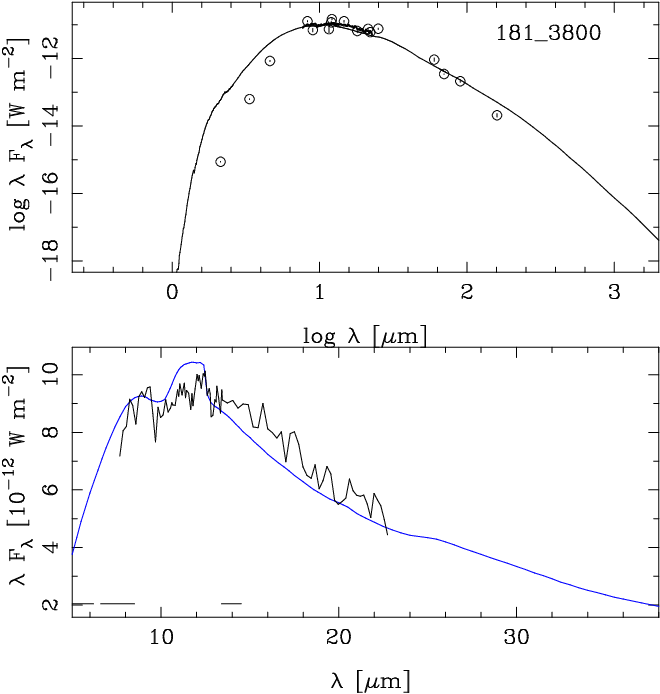}} 
\end{minipage}

\begin{minipage}{0.235\textwidth}
\resizebox{\hsize}{!}{\includegraphics[angle=-0]{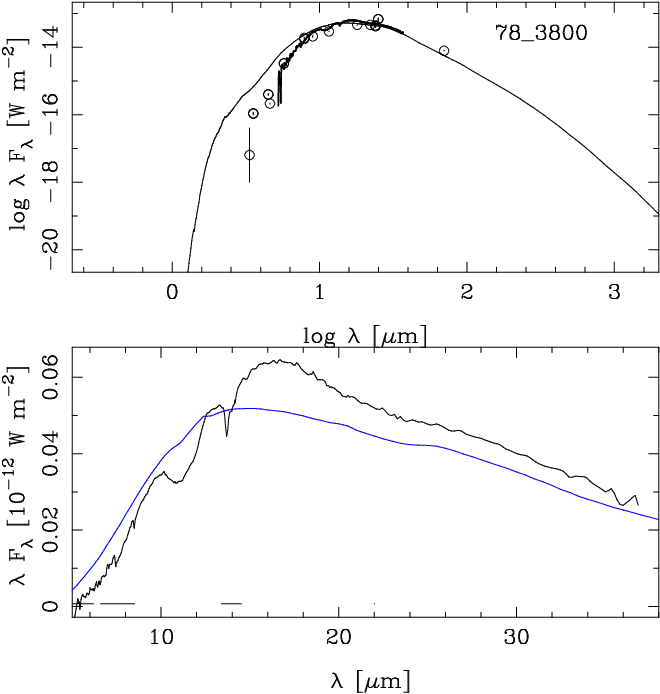}} 
\end{minipage}
\begin{minipage}{0.235\textwidth}
\resizebox{\hsize}{!}{\includegraphics[angle=-0]{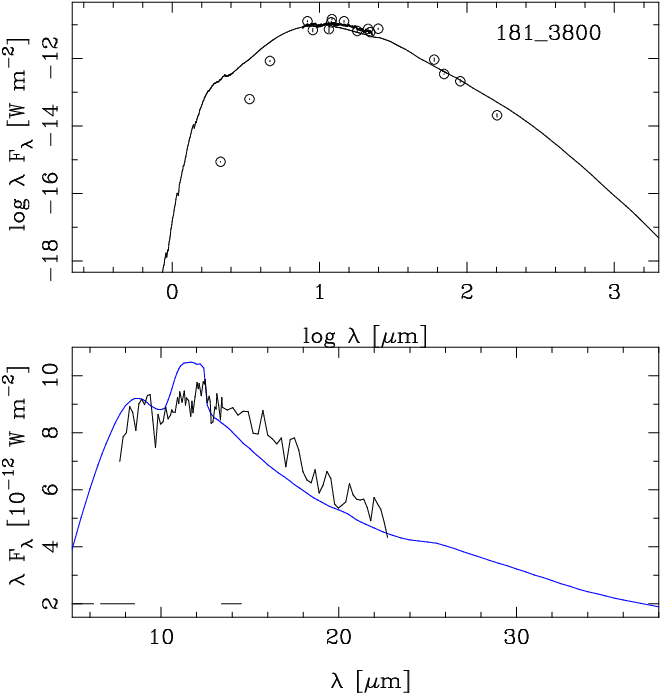}} 
\end{minipage}

\begin{minipage}{0.235\textwidth}
\resizebox{\hsize}{!}{\includegraphics[angle=-0]{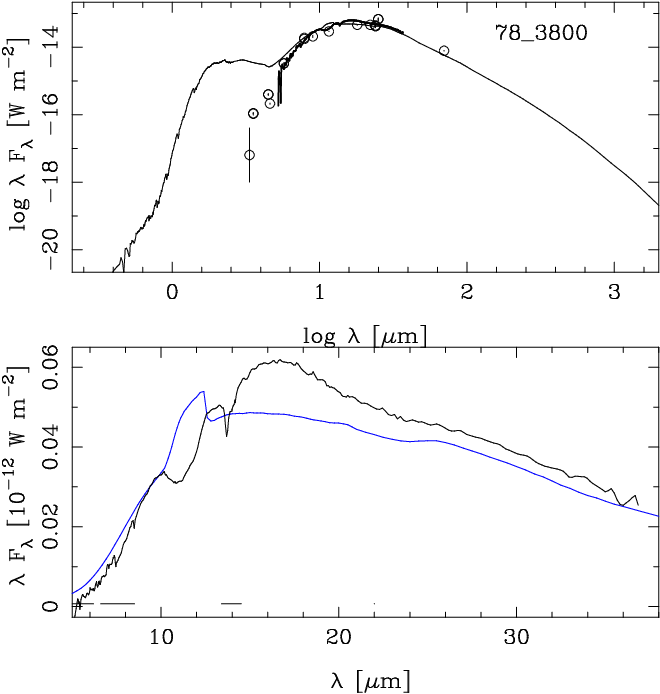}} 
\end{minipage}
\begin{minipage}{0.235\textwidth}
\resizebox{\hsize}{!}{\includegraphics[angle=-0]{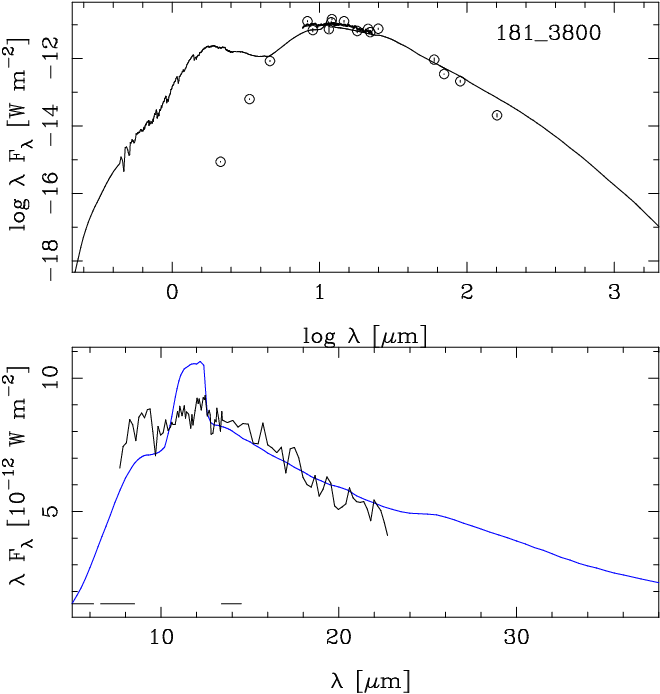}} 
\end{minipage}

\begin{minipage}{0.235\textwidth}
\resizebox{\hsize}{!}{\includegraphics[angle=-0]{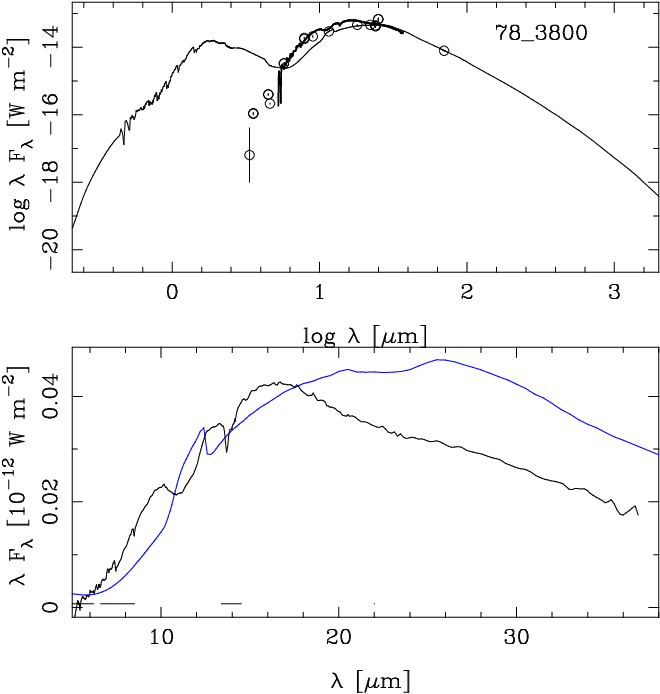}} 
\end{minipage}
\begin{minipage}{0.235\textwidth}
\resizebox{\hsize}{!}{\includegraphics[angle=-0]{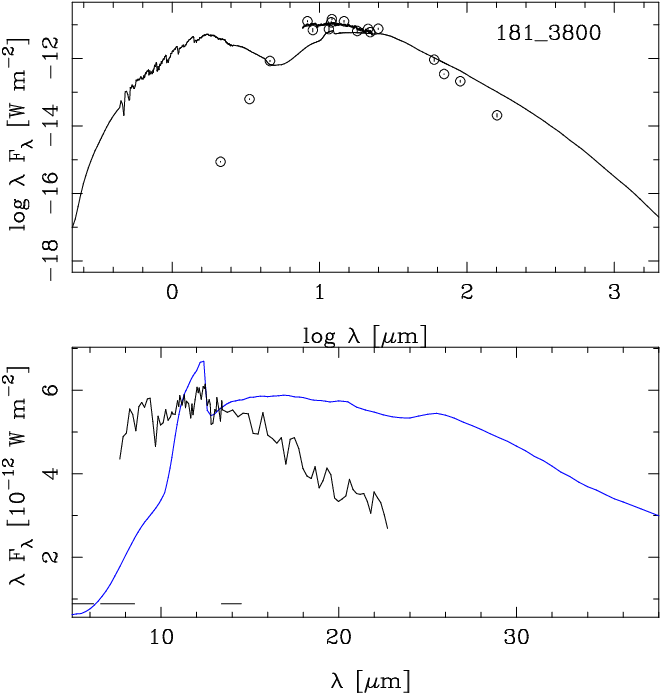}} 
\end{minipage}

\caption{Same as for figure~\ref{Fig:PAGB}, but when the effective temperature of the central star is 3800~K.
}
\label{Fig:PAGB3800}

\end{figure}

Figure~\ref{Fig:PAGB3800} is the same as figure~\ref{Fig:PAGB}, but  in the case the central star has an effective temperature of 3800~K.

\FloatBarrier
\section{Comparison to ZTF}
\label{App:ZTF}

The referee pointed out the paper of \citet{Chen20}, who classified about 780~000 periodic variables into 11 classes using ZTF datarelease 2 data.
The sample of 1992 objects was correlated with this database using a search radius of 1\arcsec.
Sixty-three matches were found, with 50 stars being classified as Miras and 12 as semi-regulars by \citet{Chen20}.
Table~\ref{AppTabZTF} compiles the periods found from WISE and our analysis of the ZTF data, as well as the periods found by  \citet{Chen20}.
We note that periods found in the literature or from refitting other datasets in the present paper are not repeated in the Table.

The periods found by \citet{Chen20} compare rather poorly to the periods derived in the present paper from WISE data or ZTF data.
Only for 20 do the periods agree within  10 percent. For many, the periods derived in the present paper are longer, sometimes by a factor of 2.
The reason is that the ZTF data only cover a time span of 470~days, and so, citing \citet{Chen20}, 
‘the periods in our catalogue are most accurate for P $<$ 100 days. This limitation causes problems for the periods of SRs and Miras’.
The advantage of the manual fitting and visual inspection, as was done in the present paper for the external data, is that periods
longer than the time span of the dataset can be assigned with confidence.

\longtab[7]{
\begin{landscape}
\vfill
\begin{longtable}{rrlllcrrccrclccrllllllllllllllllllllllllll}
\caption{\label{AppTabZTF}Comparison with \citet{Chen20}.}\\
\hline
\hline
\centering
Ra & Dec  &  $P_1$    & $P_2$ & $P_{\rm ZTF}$ &  ZTF Id & Period  &  ZTF class \\
(deg) & (deg) &  (d) &  (d) & (d)        &             &  (d)  &  \\
\hline
\endfirsthead
\caption{continued.}\\
\hline\hline
%\footnotesize
Ra & Dec  &  $P_1$    & $P_2$ & $P_{\rm ZTF}$ &  ZTF Id & Period  &  ZTF class \\
(deg) & (deg) &  (d) &  (d) & (d)        &             &  (d)  &  \\
\hline
\endhead
  5.193017  &  64.514450  &  549  &  270  &  519 $\pm$ 30   & ZTFJ002046.32+643051.9 &  284  & SR & \\
  6.921023  &  69.647507  &   &   &  637 $\pm$ 55   & ZTFJ002741.12+693851.5 &  425  & Mira & \\
 23.869953  &  49.378334  &   &   &  491 $\pm$ 11   & ZTFJ013528.77+492242.0 &  440  & Mira & \\
 62.743782  &  52.385670  &  920  &  928  &   & ZTFJ041058.47+522308.2 &  282  & SR & \\
 64.966774  &  51.740726  &  463  &  477  &  445 $\pm$ 34   & ZTFJ041952.04+514426.4 &  400  & Mira & \\
 65.558540  &  59.980003  &  573  &   &  511 $\pm$ 19   & ZTFJ042214.10+595848.0 &  440  & Mira & \\
 66.168434  &  48.123432  &   &   &   & ZTFJ042440.46+480724.0 &  794  & Mira & \\
 70.127876  &  45.125114  &  538  &  565  &  517 $\pm$ 5    & ZTFJ044030.68+450730.3 &  603  & Mira & \\
 71.838844  &  39.451748  &  512  &  482  &  428 $\pm$ 71    & ZTFJ044721.35+392706.2 &  540  & Mira & \\
 78.351913  &  20.990301  &   &   &    & ZTFJ051324.44+205924.9 &  454  & Mira & \\
 81.536049  &  47.125359  &  585  &  466  &  480 $\pm$ 7   & ZTFJ052608.61+470731.2 &  295  & Mira & \\
 85.957069  &  32.701687  &   &   &  606 $\pm$ 85   & ZTFJ054349.66+324206.5 &  487  & Mira & \\
 86.118340  &  42.851227  &   &   &  326 $\pm$ 4    & ZTFJ054428.38+425104.2 &  443  & Mira & \\
 90.742348  &  46.461830  &  453  &  430  &  406 $\pm$ 21   & ZTFJ060258.13+462742.4 &  374  & Mira & \\
 97.397789  &   8.788045  &   &   &  769 $\pm$ 75   & ZTFJ062935.45+084716.5 &  483  & Mira & \\
 97.832855  &  31.528938  &  537  &  559  &  666 $\pm$ 18   & ZTFJ063119.87+313144.0 &  521  & Mira & \\
 98.616814  &  -5.061915  &   &   &  $-$ & ZTFJ063428.06-050343.0 &  171  & SR & \\
101.275536  &  -8.471849  &  638  &  671  &  558 $\pm$ 13   & ZTFJ064506.11-082818.8 &  363  & Mira & \\
101.500809  &  15.663664  &   &   &  552 $\pm$ 10    & ZTFJ064600.17+153948.9 &  385  & Mira & \\
103.519295  &  -4.326680  &   &   &  377 $\pm$ 8   & ZTFJ065404.65-041936.3 &  187  & SR & \\
104.773621  &   3.632370  &  564  &  550  &  424 $\pm$ 9   & ZTFJ065905.66+033756.2 &  181  & SR & \\
107.645561  &  -1.190360  &  627  &  675  &  539 $\pm$ 11   & ZTFJ071034.93-011125.5 &  406  & Mira & \\
107.780273  &  -0.386771  &  623  &  577  &   & ZTFJ071107.28-002312.6 &  318  & SR & \\
110.524391  &  -3.546792  &  399  &  362  &  380 $\pm$ 6   & ZTFJ072205.85-033248.4 &  367  & SR & \\
114.012138  & -10.214721  &   &   &    & ZTFJ073602.90-101253.1 &  480  & Mira & \\
274.574738  & -10.595833  &  602  &  553  &  402 $\pm$ 9   & ZTFJ181817.92-103545.0 &  457  & Mira & \\
280.056274  &  -5.703139  &   &   &  557 $\pm$ 10   & ZTFJ184013.49-054211.4 &  540  & Mira & \\
282.231476  &  -1.815691  &  429  &  432  &  397 $\pm$ 2    & ZTFJ184855.50-014856.9 &  394  & Mira & \\
282.813995  &   1.652694  &  524  &  480  &   & ZTFJ185115.41+013909.6 &  451  & Mira & \\
283.761780  &  15.786965  &  450  &  447  &  458 $\pm$ 6    & ZTFJ185502.79+154713.2 &  445  & Mira & \\
285.819824  &   9.203329  &  496  &  468  &   & ZTFJ190316.78+091211.9 &  216  & SR & \\
287.820007  &  17.864273  &  445  &  422  &  422 $\pm$ 3   & ZTFJ191116.77+175150.7 &  440  & Mira & \\
288.179382  &   5.886957  &   &   & $-$   & ZTFJ191243.03+055312.9 &  169  & SR & \\
288.304840  &  12.003804  &  446  &  430  &  417 $\pm$ 12   & ZTFJ191313.18+120013.7 &  375  & Mira & \\
289.141144  &  18.381018  &  437  &  417  &  407 $\pm$ 5   & ZTFJ191633.90+182251.8 &  409  & Mira & \\
291.094452  &  32.319031  &  560  &  540  &  520 $\pm$ 211   & ZTFJ192422.61+321908.4 &  578  & Mira & \\
291.536896  &  23.480146  &  586  &  570  &  681 $\pm$ 25   & ZTFJ192608.79+232848.3 &  215  & SR & \\
291.658112  &  18.266577  &  520  &  590  &  490 $\pm$ 15   & ZTFJ192637.94+181559.0 &  500  & Mira & \\
291.916656  &  23.792528  &  612  &  589  &  455 $\pm$ 9   & ZTFJ192739.97+234732.3 &  517  & Mira & \\
292.580017  &  17.180479  &  694  &  651  &   & ZTFJ193019.22+171049.8 &  373  & SR & \\
293.548218  &  19.974556  &  441  &  427  &  421 $\pm$ 16   & ZTFJ193411.56+195828.1 &  427  & Mira & \\
294.350159  &  20.616007  &  521  &  494  &  358 $\pm$ 12   & ZTFJ193723.99+203657.7 &  436  & Mira & \\
295.230194  &  15.337878  &  550  &  533  &  493 $\pm$ 19   & ZTFJ194055.25+152015.9 &  408  & Mira & \\
295.535187  &  47.382542  &  460  &  453  &  394 $\pm$ 6    & ZTFJ194208.42+472257.0 &  304  & Mira & \\
296.269775  &   6.959763  &  455  &  408  &  399 $\pm$ 6   & ZTFJ194504.72+065735.4 &  385  & Mira & \\
297.452881  &  35.820549  &  453  &  445  &  491 $\pm$ 7   & ZTFJ194948.67+354913.8 &  447  & Mira & \\
297.872650  &  27.149763  &  906  &  873  &  793 $\pm$ 122   & ZTFJ195129.43+270858.8 &  350  & Mira & \\
298.620667  &  24.372654  &   &   &  597 $\pm$ 37   & ZTFJ195428.94+242221.5 &  527  & Mira & \\
299.963959  &  32.536037  &   &   &  408 $\pm$ 4   & ZTFJ195951.33+323209.8 &  408  & Mira & \\
302.033997  &  31.716911  &   &   &  468 $\pm$ 7   & ZTFJ200808.16+314300.9 &  422  & Mira & \\
305.475128  &  36.560566  &   &   &  $-$ & ZTFJ202154.04+363338.0 &  556  & Mira & \\
309.914825  &  50.204674  &  462  &  502  &  442 $\pm$ 2   & ZTFJ203939.51+501216.4 &  457  & Mira & \\
315.385437  &  38.756172  &  517  &  607  &  462 $\pm$ 4   & ZTFJ210132.52+384522.3 &  523  & Mira & \\
315.992218  &  50.244583  &  2411  &  1523  &  321 $\pm$ 3   & ZTFJ210358.11+501440.2 &  318  & SR & \\
317.444366  &  18.413706  &  368  &  371  &  445 $\pm$ 6   & ZTFJ210946.62+182449.2 &  425  & Mira & \\
317.659851  &  45.979378  &  542  &  548  &  442 $\pm$ 2    & ZTFJ211038.36+455845.9 &  424  & Mira & \\
319.133759  &  36.358849  &  476  &  541  &  624 $\pm$ 20    & ZTFJ211632.10+362132.0 &  460  & Mira & \\
325.987213  &  58.596138  &  506  &  503  &  596 $\pm$ 48   & ZTFJ214356.93+583546.1 &  480  & Mira & \\
334.659912  &  43.778919  &       &       &  $-$              & ZTFJ221838.38+434644.5 &  580  & Mira & \\
339.322784  &  59.454674  &  590  &  587  &  519 $\pm$ 28   & ZTFJ223717.44+592717.0 &  613  & Mira & \\
352.573853  &  53.883614  &  506  &  553  &  573 $\pm$ 46   & ZTFJ233017.75+535301.1 &  520  & Mira & \\
353.614540  &  43.550311  &        &      &  686 $\pm$ 21   & ZTFJ233427.48+433300.8 &  510  & Mira & \\
357.904877  &  63.010296  &  611  &  626  &  558 $\pm$ 20   & ZTFJ235137.16+630036.9 &  292  & SR & \\
\hline
\end{longtable}
\tablefoot{
Column~1 and 2: Right ascension and declination,
Column~3 and 4: Period derived in the W1 and W2 filters in the present paper,
Column~5: Period derived from ZTF data in the present paper (a $-$ means no period could be determined),
Column~6: ZTF identifier,
Column~7: Period from \citet{Chen20} rounded to the nearest integer,
Column~8: classification in  \citet{Chen20}.
}
\vfill
\end{landscape}
}

\end{appendix}

\end{document}